\documentstyle[epsf,12pt]{book}
\begin{document}
\setlength{\baselineskip}{18pt}
\jot=8pt
%
\gdef\journal#1,#2,#3,#4.{{\it #1~}{\bf #2} (#4) #3}
\def\prd{\journal Phys. Rev. D,}
\def\prl{\journal Phys. Rev. Lett.,}
\def\npb{\journal Nucl. Phys. B,}
\def\plb{\journal Phys. Lett. B,}
\def\apj{\journal Ap. J.,}
\def\apjl{\journal Ap. J. Lett.,}
\def\MNRAS{\journal MNRAS,}
\def\be{\begin{equation}}\def\bea{\begin{eqnarray}}\def\beaa{\begin{eqnarray*}}
  \def\ee{\end{equation}}  \def\eea{\end{eqnarray}}  \def\eeaa{\end{eqnarray*}}
\def\double{\baselineskip 24pt \lineskip 10pt}
\def\re#1{{[\ref{#1}]}}
\def\fun#1#2{\lower3.6pt\vbox{\baselineskip0pt\lineskip.9pt
        \ialign{$\mathsurround=0pt#1\hfill##\hfil$\crcr#2\crcr\sim\crcr}}}
\def\half{{\textstyle{ 1\over 2}}}
\def\frac#1#2{{\textstyle{#1\over #2}}}
\def\gsim{\mathrel{\raise.3ex\hbox{$>$\kern-.75em\lower1ex\hbox{$\sim$}}}}
\def\lsim{\mathrel{\raise.3ex\hbox{$<$\kern-.75em\lower1ex\hbox{$\sim$}}}}
\def\la{\bigl\langle} \def\ra{\bigr\rangle}
\def\cd{\!\cdot\!}
\def\a{\hat a}      \def\b{\hat b}      \def\c{\hat c}
\def\ab{\a\cd\b}    \def\ac{\a\cd\c}    \def\bc{\b\cd\c}
\def\cg{\cos\gamma} \def\ca{\cos\alpha} \def\cb{\cos\beta}
\def\gg{\hat\gamma}    \def\go{ \hat\gamma_1}
\def\gt{\hat\gamma_2}  \def\gth{\hat\gamma_3}
\def\gf{\hat\gamma_4}
\def\got{ \hat\gamma_1\cd\hat\gamma_2} \def\ggo{ \hat\gamma\cd\hat\gamma_1}
\def\goth{\hat\gamma_1\cd\hat\gamma_3} \def\ggt{ \hat\gamma\cd\hat\gamma_2}
\def\gtth{\hat\gamma_2\cd\hat\gamma_3} \def\ggth{\hat\gamma\cd\hat\gamma_3}

\def\n{\hat n}       \def\no{\hat n_1}   \def\nt{\hat n_2}  \def\nth{\hat n_3}
\def\nont{\no\cd\nt} \def\nonth{\no\cd\nth} \def\ntnth{\nt\cd\nth}

\def\nogo{\no\cd\hat\gamma_1} \def\nogt{\no\cd\hat\gamma_2}
\def\nogth{\no\cd\hat\gamma_3}
\def\ntgo{\nt\cd\hat\gamma_1} \def\ntgt{\nt\cd\hat\gamma_2}
\def\ntgth{\nt\cd\hat\gamma_3}
\def\nthgo{\nth\cd\hat\gamma_1} \def\nthgt{\nth\cd\hat\gamma_2}
\def\nthgth{\nth\cd\hat\gamma_3}

\def\D{ {\Delta T \over T} }   \def\dO{d\Omega}
\def\d{ {\delta T \over T} }

\def\etal{{\sl et al.}}
\def\eg{{\sl e.g.}}
\def\ie{{\sl i.e.}}

\pagestyle{empty}

\begin{center}
{\bf \huge      Cosmic Microwave Background    \\
                Anisotropies and               \\
                Theories of the Early Universe \\}

\vspace{4.0cm}

{A dissertation submitted in satisfaction of the final requirement
for the degree of}

\vspace{0.5cm}

{\bf \large Doctor Philosophiae}

\vspace{1.0cm}

{{\sl SISSA} -- International School for Advanced Studies}

\vspace{0.5cm}

{Astrophysics Sector}

\vspace{4.5cm}

\vspace*{0.5truecm}
\begin{tabular}{lll}
            Candidate:        &~~~~~~~~~~~~~~~~~~~&
                                             Supervisor:       \\
  & & \\
{\bf \large Alejandro Gangui} &~~~~~~~~~~~~~~~~~~~&
                                 {\bf \large Dennis W. Sciama} \\
\end{tabular}

\vspace{1.0cm}

{October 1995}

\end{center}

\clearpage
\pagestyle{empty}

{}~~~~~~~

{}~~~~~~~

\vspace{6.0cm}

\begin{center}
\large \bf TO DENISE
\end{center}

\clearpage
\pagestyle{plain}
\pagenumbering{roman}

\begin{center}
\large \bf Abstract
\end{center}
\addcontentsline{toc}{chapter}{Abstract}

In this thesis I present recent work aimed at showing how
currently competing theories of the early universe
leave their imprint on the temperature anisotropies of the
cosmic microwave background (CMB) radiation.

After some preliminaries, where we review
the current status of the field, we consider the three--point
correlation function of the temperature anisotropies,
as well as the inherent theoretical uncertainties associated with it,
for which we derive explicit analytic formulae.

These tools are of general validity and we apply them in the study
of possible non--Gaussian features that may arise on large angular
scales in the framework of both inflationary and topological defects
models.

In the case where we consider possible deviations
of the CMB from Gaussian statistics within inflation, we develop a
perturbative analysis for the study of spatial correlations in the
inflaton field in the context of the stochastic approach to inflation.

We also include an analysis of a particular geometry of the
CMB three--point function (the so--called `collapsed' three--point function)
in the case of post--recombination integrated effects,
which arises generically whenever the mildly non--linear growth
of perturbations is taken into account.

We also devote a part of the thesis to the study
of recently proposed analytic models for topological defects,
and implement them in the analysis of both the CMB excess kurtosis
(in the case of cosmic strings) and the CMB collapsed three--point
function and skewness (in the case of textures).

Lastly, we present a study of the CMB anisotropies on the
degree angular scale in the framework of the global texture scenario,
and show the relevant features that arise; among these, the Doppler peaks.

\tableofcontents

\newpage

\begin{center}
\large \bf Acknowledgements
\end{center}
\addcontentsline{toc}{chapter}{Acknowledgements}

First and foremost I would like to thank my supervisor
Dennis Sciama for his constant support and encouragement.
He gave me complete freedom to chose the topics of my research
as well as my collaborators, and he always followed my work
with enthusiasm.

I would also like to thank the whole group of astrophysics
for creating a nice environment
suited for pursuing my studies;
among them I warmly thank Antonio Lanza
for being always so kind and ready--to--help us in any respect.

For the many interactions we had and specially for the
fact that working with them was lots of fun, I want to express
my acknowledgement to
Ruth Durrer, Francesco Lucchin, Sabino Matarrese,
Silvia Mollerach, Leandros Perivolaropoulos, and Mairi Sakellariadou.
Our fruitful collaboration form the basis of much of the
work presented here.

My officemates, Mar Bastero, Marco Cavagli\`a, Luciano Rezzolla,
and Luca Zampieri $+$ `Pigi' Monaco also deserve some credits
for helping me to improve on my italian and better appreciate
the local {\sl caff\`e},
$\ldots$
and specially for putting up with me during our many hours together.

I am also grateful to my colleagues in Buenos Aires
for their encouragement to pursue graduate studies abroad,
and for their being always in contact with me.

Richard Stark patiently devoted many 5--minutes{\sl es} to
listen to my often directionless questions, and also helped me sometimes
as an online English grammar.

I also thank the computer staff@sissa: Marina Picek, Luisa Urgias
and Roberto Innocente for all their help during these years, and
for making computers friendly to me.

Finally, I want to express my gratitude to my wife,
for all her love and patience; I am happy to dedicate this thesis to her.
And last but not least, I would like to say thanks to our families
and friends; they always kept close to us and helped us better enjoy our
stay in Trieste.

\newpage

\begin{center}
\large \bf Units and Conventions
\end{center}
\addcontentsline{toc}{chapter}{Units and Conventions}

In this thesis we are concerned with the study of specific
predictions for the cosmic microwave background (CMB)
anisotropies as given by different early universe models.
Thus, discussions will span a wide variety of length scales, ranging
from the Planck length, passing through the grand unification scale
(\eg, when inflation is supposed to have occurred and cosmological
phase transitions might have led to the existence of topological
defects),
up to the size of the universe as a whole (as is the case when
probing horizon--size perturbation scales, e.g., from the
{\sl COBE}--DMR maps).\footnote{{\sl COBE} stands for
	COsmic Background Explorer satellite, and DMR
	is the Differential Microwave Radiometer on board.}

As in any branch of physics, it so happens that the chosen units and
conventions are almost always dictated by the problem
at hand
(whether we are considering the microphysics of a cosmic string
or the characteristic thickness of the wake of accreted matter
that the string leaves behind
due to its motion).
It is not hard to imagine then that also here disparate units
come into the game. Astronomers will not always feel comfortable
with those units employed
by particle physicists, and vice versa, and so it is worthwhile
to spend some words in order to fix notation.

The so--called natural units, namely $\hbar = c = k_B = 1$, will
be employed, unless otherwise indicated.
$\hbar$ is Planck's constant divided by $2\pi$, $c$ is the speed of
light, and $k_B$ is Boltzmann's constant.
Thus, all dimensions can be
expressed in terms of one energy--unit which is usually chosen
as GeV $=10^9$eV, and so
$$[{\rm Energy}]=[{\rm Mass}]=[{\rm Temperature}]
   =[{\rm Length}]^{-1}=[{\rm Time}]^{-1} ~.$$
Some conversion factors that will be useful in what follows are

\vspace*{0.5truecm}
\begin{tabular}{llll}
{}~~~~~~~~~~~~
&1 {\rm GeV} & = 1.60$\times 10^{-3}$ {\rm erg}       & ({\rm Energy}) \\
{}~~~~~~~~~~~~
&            &                                        &                \\
{}~~~~~~~~~~~~
&1 {\rm GeV} & = 1.16$\times 10^{13}$ {\rm K}    & ({\rm Temperature}) \\
{}~~~~~~~~~~~~
&            &                                        &                \\
{}~~~~~~~~~~~~
&1 {\rm GeV} & = 1.78$\times 10^{-24}$ {\rm g}        & ({\rm Mass})   \\
{}~~~~~~~~~~~~
&            &                                        &                \\
{}~~~~~~~~~~~~
&1 {\rm GeV}$^{-1}$ & = 1.97$\times 10^{-14}$ {\rm cm}  & ({\rm Length}) \\
{}~~~~~~~~~~~~
&                 &                                   &                \\
{}~~~~~~~~~~~~
&1 {\rm GeV}$^{-1}$ & = 6.58$\times 10^{-25}$ {\rm sec} & ({\rm Time})   \\
\end{tabular}
\vspace*{0.5truecm}


\noindent
$m_P$ stands for the Planck mass, and associated quantities are
given (both in natural and cgs units) by
$m_P = 1.22 \times  10^{19}$GeV        $= 2.18 \times 10^{-5 }$g,
$l_P = 8.2  \times 10^{-20}$GeV$^{-1}$ $= 1.62 \times 10^{-33}$cm,
and
$t_P = 5.39 \times 10^{-44}$sec.
In these units Newton's constant is given by
$G = 6.67\times 10^{-8} {\rm cm}^3 {\rm g}^{-1} {\rm sec}^{-2} = m_P^{-2}$.

These fundamental units will be replaced by other, more suitable,
ones when studying issues on large--scale structure formation.
In that case it is better to employ astronomical units, which
for historical reasons are based on solar system quantities.
Thus, the {\sl parsec} (the distance at which the Earth--Sun mean distance
(1 a.u.) subtends one second of arc) is given by
1 pc = 3.26 light years.
Of course, even this is much too small for describing the
typical distances an astronomer has to deal with
and thus many times distances are quoted in megaparsecs,
1 Mpc = 3.1 $\times 10^{24}$cm.
Regarding masses, the standard unit is given by the solar mass,
$M_{\odot} = 1.99\times 10^{33}$g.

It is worthwhile also recalling some cosmological parameters
that will be used extensively in this thesis, like for example
the present Hubble time
$H_0^{-1} = 3.09\times 10^{17} h^{-1}$ sec $= 9.78 h^{-1}$ Gyr,
where $h$ ({\sl little} $H$) parameterises the uncertainty
in the present value of the Hubble parameter
$H_0 = 100 ~h ~{\rm km}~ {\rm sec}^{-1} {\rm Mpc}^{-1}$.
Agreement with the predictions of the hot big bang model
plus direct observations yield the (conservative) constraint
$0.4 \lsim h \lsim 1$.
The Hubble distance is
$c H_0^{-1} = 2997.9 h^{-1}$ Mpc, and the critical density
$\rho_c = 3 H_0^2 / (8\pi G) = 1.88 h^2\times 10^{-29}$ g cm$^{-3}$.
Lastly, the mean value (the monopole) of the CMB radiation
temperature today is  2.726 K.

Throughout we adopt the following conventions:
Greek letters denote spacetime indices and run from 0 to 3;
spatial indices run from 1 to 3 and are given by Latin letters
from the middle of the alphabet, while those from the beginning
of the alphabet stand for group indices (unless stated otherwise).
The metric signature is taken to be ($+, -, -, -$).

Many more conversion factors, as well as the values of fundamental
constants and physical parameters (both astrophysical and cosmological)
can be found in the appendices of the monograph by Kolb \& Turner [1990].

Some abbreviations used in this thesis are the following.

\vspace*{0.5truecm}
\begin{tabular}{lllll}
CMB &= cosmic microwave background  &CDM &= cold dark matter  &  \\
FRW &= Friedmann--Robertson--Walker &HDM &= hot dark matter   &  \\
GUT &= grand unified theory         &    &                    &  \\
\end{tabular}
\vspace*{0.5truecm}

\noindent
and, lastly, {\sl COBE} and DMR that we mentioned before.

\clearpage
\pagenumbering{arabic}
\setcounter{page}{1}
\setcounter{chapter}{0}
\pagestyle{myheadings}

\chapter*{Introduction} 
\markboth{INTRODUCTION}{INTRODUCTION}
\addcontentsline{toc}{chapter}{Introduction}

In 1964 Penzias and Wilson discovered the Cosmic Microwave
Background (CMB) radiation.
Since then researchers have been studying the radiation background
by all possible means, using ground, balloon, rocket, and satellite based
experiments [Readhead \& Lawrence, 1992].
This relic radiation is a remnant of an early hot phase of the universe,
emitted when ionised hydrogen and electrons combined at a
temperature of about 3000 K. Within the simplest models this
occurs at a redshift $z \approx 1100 - 1300$, although it is also
possible that the hydrogen was reionised as recently as redshift
$z \approx 100$ [Jones \& Wyse, 1985].
The CMB radiation is a picture of our
universe when it was much smaller and hotter than it is today.

It is well known that the CMB has a thermal (blackbody) spectrum,
and is remarkably isotropic and uniform;
in fact, after spurious effects are removed from the maps,
the level of anisotropies is smaller than a part in $10^4$.
Only recently have experiments reliably detected such perturbations
away from perfect isotropy.
These perturbations were expected:
a couple of years after the discovery of the CMB,
Sachs \& Wolfe [1967] showed how variations in the density of the
cosmological fluid and gravitational wave perturbations result in
CMB temperature fluctuations, even if the surface of last scattering
was perfectly uniform in temperature.
In the following years several other mechanisms for the generation of
anisotropies, ranging all angular scales, were unveiled.

Our aim in this thesis is to report a study of
the structure of the CMB anisotropies and of the
possible early universe models behind it.
In this {\sl Introduction} we briefly review the research done
and we cite the references where the original parts of the thesis are
published.
More detailed accounts of the different topics treated here
are given in the various Chapters below.

In the first Chapter we include a brief
account of the basics of what is considered to be
the `standard cosmology', the why's and how's of its success
as well as of its pressing problems.

The second (rather lengthy) Chapter is devoted to an overview of the
salient features of the two presently favourite theories of the
early universe
(or should we call them (just) `ideas' or (pompously) `scenarios'?),
which were devised to complement the standard cosmology and
render it more satisfactory, both from the theoretical and
observational viewpoint.
Inflationary models and `seed' models with topological defects have
been widely studied both in the past and recently, specially after the
detection of the large--scale CMB anisotropies of cosmological origin that
we call $\D$. Their predictions have been confronted against a
large bulk of observations and their performance is satisfactory,
albeit not perfect. Both classes of models have their strong and weak
points and we hope to have given a general flavour of them in Chapter 2
(you can always consult [Linde, 1990] and [Vilenkin \& Shellard, 1994]
if not satisfied).

The different sources of CMB anisotropies are then explained in
detail in Chapter 3, both on large as well as on small angular scales.
We also give here a brief overview of the predictions from
standard cold dark matter (CDM) plus inflationary scenarios and confront
them with the present status of the observations on various scales.

A {\sl quantitative} account of the spectrum and the higher--order
correlation functions of the CMB temperature anisotropies, together
with the necessary introductory formulae is given in Chapter 4.
Part of it, specially the analysis in \S\ref{sec-3pointfun},
has been published in [Gangui, Lucchin, Matarrese \& Mollerach, 1994].

Having given the general expressions for the spatial correlations
of the anisotropies, we apply them in Chapter 5 to the
study of the predictions from
generalised models of inflation (\ie, a period of quasi--exponential
expansion in the early universe).
We use the {\sl stochastic approach} to inflation [\eg, Starobinskii, 1986;
Goncharov, Linde \& Mukhanov, 1987], as the natural framework
to self--consistently account for all second--order effects
in the generation of scalar field fluctuations during inflation and their
giving rise to curvature perturbations.
We then account for the non--linear relation between the inflaton
fluctuation and the peculiar gravitational potential $\Phi$,
ending \S\ref{sec-StochasticStuff} with the computation of the three--point
correlation function for $\Phi$.
We then concentrate on large angular scales, where the Sachs--Wolfe effect
dominates the anisotropies, and compute (\S\ref{sec-angularbispectrum})
the CMB temperature anisotropy bispectrum.
{}From this, we calculate the collapsed three--point function
(a particular geometry of the three--point function defined in Chapter 4),
and show its behaviour with the varying angle lag
(\S\ref{sec-angularbispectrum}).
We specialise these results in \S\ref{sec-skewness} to a long list of
interesting inflaton potentials.
We confront these primordial predictions
(coming from the non--linearities in the dynamics of the inflaton field)
with the theoretical uncertainties in \S\ref{sec-skewcosmicvariance}.
These are given by the {\sl cosmic variance} for the skewness,
for which we show the dependence for a wide range of the primordial
spectral index of density fluctuations.
We end Chapter 5 with a discussion of our results; what we report in
this Chapter has been published in
[Gangui, Lucchin, Matarrese \& Mollerach, 1994] and [Gangui, 1994].

Even for the case of primordial Gaussian curvature fluctuations, the
non--linear gravitational evolution gives rise to a non--vanishing
three--point correlation function of the CMB.
In Chapter 6 we calculate this contribution
from the non--linear evolution of Gaussian initial perturbations, as
described by the Rees--Sciama (or integrated Sachs--Wolfe) effect.
In \S\ref{sec-integrated1}
we express the collapsed three--point function in
terms of multipole amplitudes and we are then able to calculate its
expectation value for any power spectrum and for any experimental
setting on large angular scales.
We also give an analytic expression for the {\sl rms}
collapsed three--point function (\S\ref{sec-integrated2})
arising from the cosmic variance of a Gaussian fluctuation field.
We apply our analysis to the case of {\sl COBE}--DMR in
\S\ref{sec-integrated3}, where we also briefly discuss our results.
The contents of this Chapter
correspond to the study performed in
[Mollerach, Gangui, Lucchin, \& Matarrese, 1995].

The rest of the thesis is devoted to the study of possible
signatures that topological defects may leave in the CMB radiation
anisotropies.
In Chapter 7 we introduce an analytic model
for {\sl cosmic strings} and calculate the excess CMB kurtosis that
the network of strings produce on the relic photons after the era of
recombination though the Kaiser--Stebbins
effect (\S\ref{sec-FourPointStrings}).
As we did in previous Chapters, here also we quantify a measure of
the uncertainties inherent in large angular--scale analyses by
computing the {\sl rms} excess kurtosis as predicted by a Gaussian
fluctuation field; we confront it with the mean value predicted
from cosmic strings in \S\ref{sec-rmsKurto}.
This Chapter contains essentially what was published in
[Gangui \& Perivolaropoulos, 1995].

In Chapter 8 we introduce another recently proposed analytic model
for defects, but this time devised for treating {\sl textures}.
After a brief overview of the basic features of the model
(\S\ref{sec-TextuIntrud})
we calculate the multipole coefficients for the expansion
of the CMB temperature anisotropies in spherical harmonics on
the microwave sky (\S\ref{sec-TexSpotStat}).
At this point one may make use of the whole machinery developed
in previous Chapters to compute texture--spot correlations
(\S\ref{sec-SpotCorr})
as well as the angular bispectrum and, from this, also
the collapsed three--point correlation function.
We also estimate the theoretical uncertainties through
the computation of the cosmic variance for the collapsed three--point
function, for which we present an explicit analytic expression
(\S\ref{sec-TexAndThree}).
The main aim of this study is to find out whether the predicted
CMB non--Gaussian features may provide a useful
tool to constrain the analytic texture model.
The work presented in this Chapter is in progress.
We provide, however, a preliminary
discussion of the expected results in \S\ref{sec-TexDiscuFi}.

The last Chapter is devoted to the study of CMB anisotropies
from textures, this time on small angular scales ($\sim 1^\circ$).
As is well known, both inflation and defect models provide us with
adequate mechanisms for the generation of large--scale structure
and for producing the level of anisotropies that have been detected
by {\sl COBE} and other experimental settings on smaller scales.
The small--scale angular power spectrum is most probably
the first handle on the problem of discriminating between models, and
this will be the focus of our study in Chapter 9.
We first discuss (\S\ref{sec-AcouOs}) which
contributions to the anisotropy effectively probe those perturbations
with characteristic scale of order of the horizon distance at the
era of recombination.
In \S\ref{sec-linearspectra} we set up the basic system of linearised
perturbation equations describing the physics of our problem, as well as
the global scalar (texture) field source terms.
In \S\ref{sec-inicon} we discuss the issue of initial conditions
based on the causal nature of defects; we then compute the
angular power spectrum and show the existence of
the Doppler peaks (\S\ref{sec-textucls}).
Finally, we comment on the implications of our
analysis in \S\ref{sec-disctextu}.
This Chapter is a somewhat slightly inflated version of what was
reported in [Durrer, Gangui \& Sakellariadou, 1995], and
recently submitted for publication.

During my graduate studies here at {\sl SISSA} I also did some
work not directly related to the CMB radiation anisotropies
[Castagnino, Gangui, Mazzitelli \& Tkachev, 1993],
a natural continuation of
[Gangui, Mazzitelli \& Castagnino, 1991]
on the semiclassical limit of quantum cosmology;
this research, however, will not be reported in this thesis.

\chapter{Standard cosmology: Overview}
\label{chap-StandCosmo}
\markboth{Chapter 1. ~STANDARD COSMOLOGY: OVERVIEW}
         {Chapter 1. ~STANDARD COSMOLOGY: OVERVIEW}

\section{The expanding universe}
\label{sec-Theexpandinguniverse}

One of the cornerstones of the standard cosmology is
given by the large--scale homogeneity and isotropy of our
universe. This is based in the cosmological principle,
a simple assumption requiring that the universe should look
the same in every direction from every point in space, namely
that our position is not preferred.

The metric for a space with homogeneous and isotropic spatial
sections is of the Friedmann--Robertson--Walker (FRW) form
\be
\label{frw}
ds^2 = dt^2 - a^2(t)
\left[
{dr^2\over 1 - kr^2} + r^2 d\theta^2 + r^2 \sin^2\theta ~ d\phi^2
\right] ~,
\ee
where $r$, $\theta$ and $\phi$ are spatial comoving coordinates,
$t$ is physical time measured by an observer at rest in the comoving
frame $(r, \theta, \phi)=$ const, and $a(t)$ is the scale factor.
After appropriate rescaling of the coordinates,
$k$ takes the values 1,  $-1$ or 0 for respectively
constant positive, negative or zero three--space curvature.

In 1929 Hubble announced
a linear relation between the
recessional velocities of nebulae and their distances from us.
This is just one of the many kinematic effects that we may
`derive' from the metric (\ref{frw}).
If we take two comoving observers (or particles, if you want)
separated by a distance $d$, the FRW
metric
tells us that
this distance will grow in proportion to the scale factor, the
recessional velocity given by $v = H d$, where
$H \equiv \dot a / a$ is the Hubble parameter (dots stand for
$t$--derivatives).
Given the present uncertainties on its value\footnote{The
	uncertainties are mainly due to the difficulty in
	measuring distances in astronomy
	[see, \eg, Rowan--Robinson, 1985].},
$H$ is usually written
as $H = 100 h$ km sec$^{-1}$Mpc$^{-1}$,  with
$0.4 \lsim h \lsim 1$.
In addition to this, the expansion will make the measured
wavelengths, $\lambda_{obs}$, of the light emitted by stars,
$\lambda_{em}$, to redshift to lower frequencies by the usual
Doppler effect:
$1+z = a(t_{obs})/a(t_{em}) = \lambda_{obs} / \lambda_{em}$.
This expression defines the redshift $z$, which according
to observations is $z \geq 0$ in the expanding universe.
For objects with receding velocities much smaller than the
velocity of light we have $z \sim v / c$ and thus we
can estimate their distances from us as $d \sim c z / H$
(assuming negligible peculiar velocities).

Hence, we may calculate how far away nearby objects are
(say, within our local group of galaxies).
For instance , the Virgo cluster\footnote{The case of the
	Andromeda nebula (M31) is different in two aspects:
	first, its distance  $d \sim 0.65$ Mpc can be determined by direct
	means and so the $h^{-1}$ factor does not appear.
	Secondly, the negative velocity $v \sim -270$ km sec$^{-1}$
	($z \sim -0.0009$) means that it is {\sl not} receding from us but
	in fact approaching us
	(and thus its peculiar velocity wins,
	like in most of the other members of the local group of galaxies
	[Weinberg, 1972].}
has a redshift $z \sim 0.004$ which means a distance
$d \sim 12 h^{-1}$ Mpc from us and a receding velocity
$v \sim 1150$ km sec$^{-1}$.
Clearly distances as calculated above are valid for $z << 1$.
When $z \to 1$ Hubble's law is given by a somewhat more complicated
formula [see, \eg, Kolb, 1994].

\section{The Einstein field equations}
\label{sec-Einsteinfieldequations}

So far we have dealt just with the the properties of the FRW metric.
For it to be an adequate representation of the line element,
the Einstein field equations should be satisfied. These are
\be
\label{EE}
R_{\mu\nu} - {1\over 2} R g_{\mu\nu} = 8\pi G T_{\mu\nu} ~,
\ee
where $R_{\mu\nu}$, $R$, $g_{\mu\nu}$ and $T_{\mu\nu}$ are
the Riemann tensor, the Ricci scalar, the metric and the
stress--energy--momentum tensor (for all the fields present),
respectively\footnote{We are here {\sl not} including the
	$+\Lambda g_{\mu\nu}$ on the right--hand side, corresponding
	to a cosmological constant ($\Lambda$) term.}.

In order to derive the dynamical evolution of the scale factor $a(t)$,
the form of $T_{\mu\nu}$ must be specified.
Consistent with the symmetries of the metric, the energy--momentum tensor
must be diagonal with equal spatial components (isotropy).
Thus $T_{\mu\nu}$ takes the perfect fluid form, characterised by
a time--dependent energy density $\rho (t)$ and pressure $p (t)$
\be
\label{PF}
T_{\mu\nu} = (\rho + p) u_\mu u_\nu - p g_{\mu\nu} ~.
\ee
Here, $u^\mu$ is the four--velocity of the comoving matter, with
$u^0 = 1$ and $u^i = 0$.
The local conservation equation for the energy--momentum tensor,
$T^{\mu\nu}_{~~;\nu} = 0$ yields
\be
\label{FistLaw}
\dot\rho + 3 H (\rho + p) = 0 ~,
\ee
where the second term corresponds to the dilution of $\rho$ due to
the expansion ($H = \dot a / a$) and the third stands for the work
done by the pressure of the fluid (first law).

With regard to the equation of state for the fluid, we need to
specify $p = p (\rho)$. It is standard to assume the form
$p = w \rho$ and consider different types of components by
choosing different values for $w$.
In the case $w =$ const we get $\rho \propto a^{-3 (1+w)}$
(use Eq. (\ref{FistLaw})).
If the universe is filled with pressureless non--relativistic
matter (`dust') we are in the case where $p << \rho$ and thus
$\rho \propto a^{-3}$.
Instead, for radiation, the ideal relativistic gas equation of state
$p = \rho / 3$ is the most adapt, and therefore\footnote{Consider
	for example photons: not only
	their density diminishes due to the growth of the volume
	($\propto a^{-3}$), but the expansion also {\sl stretches}
	their wavelength {\sl out}, which corresponds to lowering
	their frequency, \ie, they redshift
	(hence the additional factor $\propto a^{-1}$).}
$\rho \propto a^{-4}$.
Another interesting equation of state is $p = - \rho$
corresponding to $\rho =$ const.
This is the case of `vacuum energy' and will be the relevant form of
energy during the so--called inflationary epoch; we will give
a brief review of inflation in \S\ref{sec-inflation} below.
Note from the different ways in which dust and radiation
scale with $a(t)$ that there will be a moment in the history
of the expanding universe when pressureless matter will
dominate (if very early on it was radiation the dominant component
-- as is thought to be the case within the hot big bang).
In fact, $\rho_{mat} / \rho_{rad} \propto a(t)$.
This moment is known as the time of `equality' between matter
and radiation, and is usually denoted as $t_{eq}$.

The main field equations for the FRW model
with arbitrary equation of state (arbitrary matter content) are
given by
\be
\label{elli33}
3\dot H + 3 H^2 = - {1\over 2} 8\pi G (\rho + 3 p) ~,
\ee
namely, the Raychaudhuri equation for a perfect fluid
(with vanishing shear and vorticity, no cosmological constant
and where the fluid elements have geodesic motion),
and the Friedmann equation
\be
\label{elli34}
3 H^2 + 3 K = 8\pi G \rho ~.
\ee
This last can be seen as the equation satisfied by the
Ricci scalar $^3R = 6 K$ of the three--dimensional space slices
(again for vanishing shear and cosmological
constant).\footnote{The Friedmann equation (\ref{elli34}) follows
	from the $0$--$0$ Einstein equation. Combining Eq. (\ref{elli34})
	and the $i$--$i$ Einstein equation we get (\ref{elli33}).}
The spatial curvature
is written as $K = k / a^2(t)$, with $k$ constant\footnote{If no
	point and no direction is
	preferred, the possible geometry of the universe becomes
	very simple, the curvature has to be the same everywhere
	(at a given time).}
(cf. Eq. (\ref{frw})), and has time evolution given by
$\dot K = - 2 H K$.

It is convenient at this point to define the density parameter
$\Omega \equiv \rho / \rho_c$, where $\rho_c = 3 H^2 / (8\pi G)$
is the (critical) density necessary for `closing'
the universe.\footnote{The present value of the critical density
	is $\rho_c = 1.88 h^2\times 10^{-29}$ g cm$^{-3}$, and taking
	into account the range of permitted values for $h$, this is
	$\sim (3 - 12) \times 10^{-27}$ kg m$^{-3}$ which in either case
	corresponds to a few hydrogen atoms per cubic meter.
	Just to compare, a `really good' vacuum (in the laboratory)
	of $10^{-9}$ N$/$m$^{2}$ at 300 K contains
	$\sim 2\times 10^{11}$ molecules per cubic meter.
	The universe seems to be empty indeed!}
In fact, from Eq. (\ref{elli34}) we get the relation to be
satisfied by $\Omega$,
\be
\label{elly1}
\Omega = 1 + {k\over a^2 H^2}~,
\ee
and so, for $k = -1$ (open three--hypersurfaces) we have $\Omega < 1$,
while $k = 1$ (closed three--hypersurfaces) corresponds to $\Omega > 1$
with the flat $k=0$ case given by $\Omega = 1$ (transition between
open and closed $t=$ const slices).

{}From (\ref{elly1}) we see that, when values other than the critical one
$\Omega = 1$ are considered, we may write
$H^2 = K / (\Omega - 1)$, and from this
$\dot a^2 = k / (\Omega - 1)$, for $k \not= 0$.
These relations apply for all times and all equations of state
(including an eventual inflationary era) provided
$\Omega$ includes the energy density of all sources.
We take $\rho > 0$, which is the case for ordinary matter and scalar
fields.
{}From the Raychaudhuri equation we may express the deceleration
parameter $q \equiv - \ddot a / (a H^2)$ as follows
\be
q = {3\over 2} ({1\over 3} + w) \Omega ~,
\ee
where we recall $w = p / \rho$.
This shows that $w = - {1\over 3}$ is a critical
value, separating qualitatively different models.
A period of evolution such that $w < - {1\over 3}$ ($q<0$),
and hence when the usual energy inequalities are violated,
is called `inflationary'.

{}From  (\ref{elli33}) and the conservation equation (\ref{FistLaw})
we get the evolution
of the density parameter [Madsen \& Ellis, 1988]
\be
\label{evoOmega}
{d\Omega\over da} = {3\over a} (w + {1\over 3}) \Omega (\Omega - 1)~.
\ee
Note that $\Omega = 1$ and $\Omega = 0$ are solutions regardless of
the value of $w$.
This and the previous equations imply that if $\Omega > 1$ at any
time, then it will remain so
for all times.
The same applies for the open case: if $0 < \Omega < 1$ at any given time,
this will be true for all times.
We will make use of Eq. (\ref{evoOmega}) below, when we will come to
consider the drawbacks of the standard cosmology
in \S\ref{sec-drawbacks}.

\section{Thermal evolution and nucleosynthesis}
\label{sec-TherEvol}

According to the standard hot big bang model, the early universe
was well described by a state of thermal equilibrium.
In fact, the interactions amongst the constituents of the primordial
plasma should have been so effective that nuclear statistical
equilibrium was established.
This makes the study simple, mainly because the system may be
fully described (neglecting chemical potentials for the time being)
in terms of its temperature $T$.
In a radiation dominated era the energy density and pressure
are given by
$\rho = 3 p = \pi^2 g_* T^4 / 30$, where the $T$--dependent
$g_*$ gives the effective number of distinct helicity states of
bosons plus fermions with masses $<< T$.
For particles with masses much larger than $T$ (in natural units)
the density in equilibrium is suppressed exponentially
[see, \eg, Weinberg, 1972; the notation however is borrowed from
Kolb \& Turner, 1990].

With the expansion the density in each component diminishes and
so it gets more difficult for the above mentioned effective interactions
to keep `working' as before.
It is clear that there will be a point at which the interaction rate
$\Gamma$ of a certain species with other particles will
fall below the characteristic expansion
rate, given by $H$.
Approximately at that moment that species is said to `decouple' from
the thermal fluid, and the temperature $T_{dec}$ at which this
happens is called the decoupling temperature\footnote{Massless
	neutrinos decouple when their interaction rate
	$\Gamma_\nu \simeq G_F^2 T^5$
	gets below $H \simeq T^2 / m_P$
	($G_F$ is the Fermi constant,
	characteristic strength of the weak interactions).
	This happens around $T_{dec}^\nu \sim 1$ MeV.
	Photons, on the other hand, depart from thermal equilibrium
	when
	free electron abundance is too low for maintaining
	Compton scattering equilibrium. This happens
	around $T_{dec}^\gamma \sim 0.3$ eV.
	(If you want to know why this is smaller than the
	binding energy of hydrogen, 13.6 eV,
	consult Ref. \cite{Weinberg72}).
	In the Chapters to come, by $T_{dec}$ we will always mean
	$T_{dec}^\gamma$, namely the decoupling temperature for
	the CMB radiation, the `messenger' that gives us direct
	information of how the universe was 400,000 years after the bang.
	`Direct' information of earlier times we cannot get
	($\ldots$ if only we could detect neutrinos as easily as we can
	detect CMB photons $\ldots$)}.

Let us follow the history of the universe backwards in time.
Very early on all the matter in the universe was ionised and radiation
was the dominant component.
With the expansion, the ambient temperature cools down below
$\sim 13.6$ eV and the {\sl recombination} of electrons and protons
to form hydrogen takes place, which diminishes the abundance of
free electrons, making Compton scattering not so effective.
This produces the {\sl decoupling} of CMB radiation from matter
and, assuming the universe is matter--dominated at this time,
this occurs at about
$t_{dec}\sim 5.6\times 10^{12} (\Omega h^2)^{-1/2}$ sec.
The corresponding redshift is about $z_{dec}\simeq 1100$
and the temperature $T_{dec}\simeq 3000$ K $= 0.3$ eV.

Going still backwards in time we reach the redshift
$z_{eq} \simeq 2\times 10^4 \Omega h^2$ when matter and radiation
(\ie, very light particles) densities are comparable.
By this time the universe was
$t_{eq} \simeq 1.4\times 10^3 (\Omega h^2)^{-2}$ years old,
with a temperature $T_{eq} \simeq 5.5 \Omega h^2$ eV and a
density $\rho_{eq} \simeq 3.2\times 10^{-16} (\Omega h^2)^{4}$
g$/$cm$^3$.

At even earlier times we reach densities and temperatures high
enough for the synthesis of the lightest elements:
when the age of universe was between 0.01 sec and 3 minutes
and its temperature was around 10 -- 0.1 MeV
the synthesis of D, $^3$He, $^4$He, $^7$Li took place.
The calculation of the abundance of these elements from
cosmological origin is one of the most useful probes of the
standard hot big bang model (and certainly the earliest probe we can
attain)
[refer to Malaney \& Mathews, 1993 for a recent review].

The outcome of primordial nucleosynthesis is very sensitive to
the baryon to photon ratio $\eta = n_{bar} / n_{photon}$,
and agreement with all four observed abundances limit it in the
range $\eta \simeq (3 - 5) \times 10^{-10}$.
Given the present value
$\eta \simeq 2.68 \times 10^{-8} \Omega_B h^2$,
this implies the important constraint
$0.011 \leq \Omega_B h^2 \leq 0.019$
[Walker \etal, 1991; Smith \etal, 1993].
The consequences of this are well--known. Given the uncertainties
on the Hubble parameter (buried in $h$) we derive
$0.011 \leq \Omega_B \leq 0.119$.
Recall that luminous matter contribute less than about 0.01
of critical density, hence there should be `dark' matter
(in particular, dark baryons).
On the other hand, dynamical determinations of $\Omega$ point
towards $\Omega\simeq 0.15 - 0.2$. This implies there should
also exist `non--baryonic' dark matter.

The compatibility between nucleosynthesis predictions and the
observed abundances is one of the successes of the hot big bang
model and gives confidence that the standard cosmology
provides an accurate accounting of the universe at least as
early as 0.01 sec after the bang.

\section{The CMB radiation and gravitational instability}
\label{sec-Gravinstab}

The CMB radiation provides another fundamental piece of
evidence in favour of the hot beginning of our universe.
After its discovery in 1965, the next feature that surprised
people was its near--perfect black--body distribution
(over more than three decades in wavelength
$\lambda \sim 0.03 - 100$ cm) with temperature $\sim$ 2.7 K.
Recently the {\sl COBE}--FIRAS detector measured it to be
$2.726 \pm 0.01$ K [Mather \etal, 1994].

Once decoupled, the background radiation propagates freely,
keeping its Planck spectrum and redshifting as
$T \propto (1 + z)$.
If the universe became reionised at a lower
redshift (\eg, due to early star or quasar formation)
then the `last scattering surface' may be closer to
us\footnote{Throughout this thesis we will be considering scenarios
	where early reionisation does not take place
	[see, \eg, Sugiyama, Silk \& Vittorio, 1993 for the
	effects that reionisation has on the CMB anisotropies].}.
Once that we know the temperature of the relic radiation we may
easily compute its number density and energy density
[\eg, Kolb, 1994]
\be
\label{kulb4}
n_\gamma \simeq 411~ {\rm cm}^{-3} ~~ ; ~~
\rho_\gamma \simeq 4.71\times 10^{-34} {\rm g}~ {\rm cm}^{-3} ~.
\ee

In 1992 the {\sl COBE} collaboration announced the discovery of
the long sought--after anisotropies on angular scales ranging
from about $7^\circ$ through to $90^\circ$, at a magnitude
of about 1 part in $10^5$ [Smoot \etal, 1992].
One way of characterising the level of anisotropies is by
the {\sl rms} temperature variation, which the {\sl COBE} team
found to be $30\pm 5 \mu$K on a sky averaged with a $10^\circ$
FWHM beam.
The anisotropy of the CMB radiation will be the topic of
Chapter \ref{chap-CMBanis} below.

Let us move on now to the issue of large--scale structure
formation.
The favourite picture today is that of structures developing
by gravitational instability from an initial spectrum
of primordial density perturbations.
One usually expands all quantities (like the density perturbation)
in Fourier modes (we are working in flat space).
Once a particular mode becomes smaller than the horizon
two competing effects will determine the future of the
fluctuation.
The dynamical time scale for gravitational collapse
is $\tau_{grav}\sim (G\rho)^{-1/2}$ and, unless we consider
effectively pressureless fluids, there will be an analogue
characteristic time of {\sl pressure response} given by
$\tau_{press}\sim \lambda / v_s$, where $\lambda$ is the
physical wavelength of the perturbation and $v_s$ is the sound
speed of the fluid.
If $\tau_{grav} \lsim \tau_{press}$ then pressure forces cannot overcome
the gravitational attraction and the collapse is inevitable.
This occurs for
$\lambda \gsim \lambda_J \sim v_s / (G\rho)^{1/2}$.
Hence, the Jeans length, $\lambda_J$, and associated mass,
$M_J \sim \pi \rho \lambda_J^3 / 6$,
define the scales above which structures become unstable to
collapse.\footnote{We will define the mass associated to a given
	perturbation scale in the next section.}

The subsequent evolution of fluctuations depend very much on
the kind of matter that dominates after $t_{eq}$.
The first type of matter we may think of is, of course,
baryonic dark matter.
Photon difussion [Silk, 1968] plays a relevant r\^ole
in this scenario, since it will dissipate small--scale fluctuations.
Hence, large structures (in the form of pancakes) will form
first; galaxies and smaller structure will be formed from
fragmentation of these pancakes afterwards.
This model, however, runs into problems since it cannot
get the structure we now observe formed without
generating too much anisotropies in the CMB radiation.
This is mainly due to the fact that initial perturbations
in the baryon component can only begin to grow after the era of
recombination. Before that, baryons are not free to move through
the radiation plasma to collapse. Hence, having `less' time to grow
to a certain `fixed' level, the amplitude of fluctuations at
horizon crossing have to be much larger.
As we said, this generates too much anisotropies [Primack, 1987].

If baryons will not help,
cold dark matter (CDM) particles may be possible candidates; roughly,
these are relics with very small internal velocity dispersion,
namely, heavy particles that decouple early in the history of
the universe and are non--relativistic by now\footnote{The possible
	exception being the axion, the Goldstone boson arising from
	a $U(1)$ global Peccei--Quinn symmetry breaking phase transition,
	that acquires a small mass below $\Lambda_{QCD}\sim$ few MeV, due
	to instanton effects.
	In many simple models the axion has never been in thermal
	equilibrium [see, \eg, Turner, 1990 and Raffelt, 1990 for
	reviews].}.
Other candidates are hot dark matter (HDM) particles:
these are light particles
($m \lsim$ 100 eV) that decouple late and are still relativistic
when galactic scales cross the horizon.
Example of this are massive neutrinos with $m_\nu$ a few tens of eV.

In a CDM universe and due to the small growth of the perturbations
that takes place between horizon crossing and the time of equality
(for scales less than $\lambda_{eq}\simeq 13 (\Omega h^2)^{-1}$ Mpc)
the density contrast $\delta\rho / \rho$ increases going to smaller
scales. Then the first objects to form are of sub--galactic size
leading to the `botton--up' hierarchical scenario. These small--mass
systems are subsequently clustered into larger systems that become
non--linear afterwards. The hierarchical clustering begins with
masses of order the baryon Jeans mass at recombination
($M_J(t_{rec}) \sim 10^5 M_\odot$) and continues until the present.

The situation is very different in the case of HDM.
As we mentioned before, imagine we have neutrinos
of mass $m_\nu \sim$ a few tens of eV.
They become non--relativistic approximately at $t_{eq}$.
However, small--scale fluctuations are prevented
from growing due to `free--streaming', namely, the high thermal
velocities endowed by neutrinos make them simply stream away
from the overdense regions; in so doing they erase the fluctuations.
We can define an `effective' Jeans length to this effect,
which we will call the free--streaming length, $\lambda_{fs}$;
the corresponding mass is given by $M_{fs} \simeq 10^{15} M_\odot$.
Hence, in order for perturbations (in a HDM scenario)
to survive free--streaming and collapse to form a bound structure,
the scale has to be that corresponding to superclusters.
Galaxies and smaller structure form by fragmentation in this
`top--down' picture.

\section{The perturbation spectrum at horizon crossing}
\label{sec-ThePertSpecAtHorCross}

It is standard in any treatment of the evolution of cosmic structures
to assume a statistically homogeneous and isotropic density field
$\rho(\vec x)$, and its fluctuations to be the seed needed by gravity
for the subsequent clumping of matter.

These fluctuations are given by the density contrast
$\delta(\vec x) = (\rho(\vec x) - \la\rho\ra ) / \la\rho\ra$,
and are calculated as the departures of the density from the mean
value $\la\rho\ra$. By definition the mean value of $\delta$ over the
statistical ensemble is zero. However, its mean squared value is not,
and it is called the variance $\sigma^2$,
representing a key quantity in the analysis.

It is straightforward to give an expression for $\sigma^2$ in terms
of the power spectral density function of the field $\delta$
(or, more friendly, {\sl power spectrum}) $P(k)$,
\be
\label{varia}
\sigma^2 = {1\over 2\pi^2}\int_0^\infty P(k) k^2 dk =
\int_{- \infty}^{+\infty} {\cal P}(k) d(\ln k) .
\ee
The variance gives no information of the spatial structure of the
density field. It however depends on time due to the evolution of
the Fourier modes $\delta_k \propto P^{1/2}(k)$,
and therefore encodes useful information on the amplitude of the
perturbations.

The second equality of Eq. (\ref{varia}) defines
${\cal P}(k) = P(k) k^3 / (2\pi^2)$, representing the contribution to
the variance per unit logarithmic interval in $k$: this quantity
lends itself well for comparison of large--scale galaxy clustering
[Coles \& Lucchin, 1995].

The variance as given above suffers one main drawback: it contains no
information on the relative contribution of the different modes.
It may even yield infrared or ultraviolet divergencies, depending
on the form of $P(k)$.
In practice what people do is to introduce some kind of resolution
scale, say $r$,
in the form of a window function, which acts as a filter smearing
information on the modes smaller than $r$.
The mean mass contained inside a spherical volume $\sim r^3$ is
$\la M \ra \sim \la\rho\ra r^3$.
One can then define the mass variance in this volume as
\be
\label{MassVar}
\sigma_M^2 = \la \left({\delta M \over M }\right)^{\! 2} \ra =
{1\over 2\pi^2}\int_0^\infty P(k) W^2(k r) k^2 dk  .
\ee
The second equality follows after some straightforward steps
[Kolb \& Turner, 1990].
$\sigma_M$ is the {\sl rms} mass fluctuation on the given scale $r$ and thus
depends on this scale, and through it on the mass $M$.
As already mentioned, the window function $W^2(k r)$ makes the dominant
contribution to $\sigma_M$ to come from wavelengths greater than
$r$.\footnote{This is a generic property of any $W^2(k r)$, regardless
of the specific form, \ie, top--hat, Gaussian, etc.}

Spurious effects (leading to incorrect results for $\sigma_M$)
due to boundaries in the window function may be prevented by taking a
Gaussian ansatz $W^2(k r) \sim \exp (- k^2 r^2)$.
If we also consider a featureless power
spectrum $P(k) \propto k^n$, with $n$ the so--called spectral index
of density perturbations, we easily get
$\sigma_M \propto r^{-(n+3)/2}$, with the amplitude of
$\sigma_M$  posing no problems provided we take $n > -3$.

Let us write this result for the {\sl rms} mass fluctuation as
$\sigma_M(t) \propto M^{-(n+3)/6}$,
where the $t$--dependence is to emphasize the fact that we are
computing the fluctuation at a particular time $t$, in contrast
to the one at horizon crossing time, as we will see below.
Of particular importance is the choice
$n=1$, and we can see heuristically why this is so:
The perturbation in the metric, as given by the gravitational potential
$\Phi$, may be expressed as
$\Phi(r) \simeq G ~\delta M / r \simeq G ~\delta\rho(r)~ r^2
\simeq G\rho \sigma_M r^2 \propto M^{-(n+3)/6} M^{2/3}$,
and for $n=1$ this result is independent of the mass scale $M$,
hence the name {\sl scale invariant} spectrum
[Peebles \& Yu, 1970; Harrison, 1970; Zel'dovich, 1972].

It is useful to define the mass associated to a given perturbation
$\lambda$ as the total mass contained within a sphere of radius
$\lambda /2$, \ie, $M = \frac{\pi}{6}\rho \lambda^3$.
According to this, we may write the horizon mass in cold particles
as $M_H = \frac{\pi}{6}\rho_{cdm} H^{-3}$, and given that the density in the
cold component scales as $\rho_{cdm} \propto (1 + z)^3$ and that
 $H^{-1} \propto a^2(t)     \propto (1 + z)^{-2}  $
($H^{-1} \propto a^{3/2}(t) \propto (1 + z)^{-3/2}$ )
during radiation (matter) domination, we get
 $M_H \propto (1 + z)^{-3}$
($M_H \propto (1 + z)^{-3/2}$ )
for $z > z_{eq}$ ($z < z_{eq}$ and $z >> \Omega^{-1}$).

The horizon crossing time (or redshift $z_H(M)$) of a mass scale $M$
is commonly defined as that time or redshift at which $M$ coincides
with the mass inside the horizon, $M_H$. Thus we have $M_H(z_H(M)) = M$.
{}From our discussion of the last paragraph we easily find
the dependence on the scale $M$ of the horizon crossing redshift, namely
 $z_H(M) \propto M^{-1/3}$
($z_H(M) \propto M^{-2/3}$)
for $z_H(M) > z_{eq}$ ($z_H(M) < z_{eq}$ and $z_H(M) >> \Omega^{-1}$).

Let us turn back now to the {\sl rms} mass fluctuation $\sigma_M$.
Any perturbation on super--horizon scales grows purely kinematically
from the time $t$ of its generation until the time it enters the
particle horizon, and its mass fluctuation is given
(at least in the liner regime) by
 $\sigma_M \propto t       \propto (1 + z)^{-2}$
($\sigma_M \propto t^{2/3} \propto (1 + z)^{-1}$)
before (after) the time of equivalence.
Then, we have (for $z_H(M) > z_{eq}$)
\be
\label{rmsMass1}
\sigma_M ( z_H(M) ) \simeq
\sigma_M(t) \left( { 1+z_H(M)  \over 1+z(t) } \right)^{-2} \simeq
\sigma_M(t) \left( { M_H(z(t)) \over M      } \right)^{-2/3} ,
\ee
while for $z_H(M) < z_{eq}$ we have
\bea
\label{rmsMass2}
\lefteqn{
\sigma_M ( z_H(M) ) \simeq
\sigma_M(t) \left( { 1+z_{eq}  \over 1+z(t)   } \right)^{-2}
            \left( { 1+z_H(M)  \over 1+z_{eq} } \right)^{-1}
}
\nonumber \\
&  &
\simeq
\sigma_M(t) \left( { 1+z_{eq}  \over 1+z(t)   } \right)^{-2}
            \left( { M_H(z(t)) \over M        } \right)^{-2/3} .
\eea
Recalling now that the fluctuation at its origin $t$ was given by
$\sigma_M(t) \propto M^{-(n+3)/6}$
we find in both cases that the {\sl rms} mass fluctuation at the time of
horizon crossing is given by
$\sigma_M(z_H(M)) \propto M^{(1-n)/6}$.
This shows that mass fluctuations with the Harrison--Zel'dovich
spectrum ($n=1$) cross the horizon with amplitudes independent of the
particular scale.

\section{Drawbacks of the standard model}
\label{sec-drawbacks}

In previous sections we have briefly reviewed the standard hot big bang
cosmology and emphasised its
remarkable successes in explaining a host of observational facts:
among these,
the dynamical nature (the expansion) of our universe,
the origin of the cosmic microwave background from the decoupling
between matter and radiation as a relic of the initial hot thermal phase.
It also provides a natural framework for understanding how the
large--scale structure developed, and for
the accurate prediction of light elements abundance produced during
cosmological nucleosynthesis.

In this section we will focus, instead, on its shortcomings
(or at least, some of them).
These are seen not as inconsistencies of the model but just
issues that the model cannot explain, when we extrapolate
its highly accurate predictions back in time,
beyond, say, the era of nucleosynthesis and before.
We leave for \S\ref{sec-cannot} the discussion
on how the inflationary scenario yields a well--defined,
albeit somewhat speculative, solution to these drawbacks,
based on early universe microphysics.
We are not discussing here the `cosmological constant' problem,
namely, why the present value of $\Lambda$ (or equivalently, the
present energy of the vacuum) is so small compared to any other
physical scale\footnote{In fact, observational limits impose
$\rho_\Lambda \equiv \Lambda / (8\pi G)$ to be smaller than the
critical density, and so $|\rho_\Lambda| < 10^{-48}$ GeV$^4$,
in natural units, corresponding to $|\Lambda| < 10^{-55}$ cm$^{-2}$.
Alternatively, if we construct
$m_\Lambda \equiv [|\rho_\Lambda| (\hbar / c)^3]^{1/4}$ which has
dimensions of mass, the above limits constrain
$m_\Lambda < 10^{-32}$ eV.
These unexplained very small values are undesired in cosmology,
hence the $\Lambda$--problem.};
this mainly because it keeps on being a problem even after the
advent of inflation: in fact inflation makes use of the virtues
of a vacuum dominated period in the early universe.
While offering solutions to the problems listed below, inflation
sheds no light on the problem of the cosmological constant
[see Weinberg, 1989 for an authoritative review; also Carrol \etal, 1992
and Ng, 1992 for more recent accounts].

\begin{flushleft}
{\bf The horizon (or large--scale smoothness) problem}
\end{flushleft}

The relic CMB radiation we detect today was emitted at the time of
recombination.
It is uniform to better than a part in $10^4$, which implies
that the universe on the largest scales
(greater than, say, 100 $h^{-1}$ Mpc) must have been very smooth,
since otherwise larger density inhomogeneities would have produced
a higher level of anisotropies, which is not observed.
The existence of particle horizons in standard cosmology
precludes microphysic events from explaining this observed smoothness.
The causal horizon at the time of the last scattering subtends an apparent
angle of order 2 degrees today, and yet the radiation we receive
from all directions of the sky have the same features.
If there was no correlation between distant regions,
how then very distant (causally disconnected) spots of the sky got in
agreement to produce the same radiation features, \eg, the same level of
anisotropies and temperature?
This is just one way of formulating the horizon problem.

$~$

\begin{flushleft}
{\bf The flatness/oldness problem}
\end{flushleft}

We saw in \S\ref{sec-Einsteinfieldequations} how
Eq. (\ref{evoOmega}) implies that the density parameter
$\Omega$ does not remain constant while the universe expands,
but instead evolves away from 1.
Assuming the standard evolution according to the big bang model
extrapolated to very early times and given that observations indicate that
$\Omega$ is very close to 1 today, we conclude that
it should have been much closer to 1 in the past.
Going to the Planck time we get
$|\Omega (10^{-43} {\rm sec}) - 1| \sim {\cal O} (10^{-60})$,
and even at the time of nucleosynthesis we get
$|\Omega (1 {\rm sec}) - 1| \sim {\cal O} (10^{-16})$.
These very small numbers are but one aspect of the flatness
problem.

This implies that the universe was very close to critical in the past
and that the radius of curvature of the universe
$a_{curv}=H^{-1} / |\Omega - 1|^{1/2} = a / \sqrt{k}$
was much much greater than the Hubble radius $H^{-1}$.
Were this not the case, \ie,  suppose
$(H^{-1} / a_{curv})^2 = (|k|/a^2)/(8\pi G \rho / 3) \sim {\cal O}(1)$
at the Planck time, then, if closed ($k>0$)
the universe would have collapsed after a few Planck times
(clearly not verified -- our universe is {\sl older} than this),
while if open ($k<0$) it would have become curvature dominated,
with the scale factor going like $a\propto t$ (coasting), and
cooling down so quickly that it would have
acquired the present CMB radiation temperature of $\sim$ 3 K at the
age of $10^{-11}$ sec, which clearly is at
variance with the age of the universe inferred from observations
[Kolb \& Turner, 1990].
This is just another face
of the same problem, namely the difficulty in answering the question
of `why' the universe is so old.

\begin{flushleft}
{\bf The unwanted relics problem}
\end{flushleft}

Baryogenesis is an example of the
virtues of getting together cosmology and grand unified theories
(GUTs).
However, the overproduction of
unwanted relics, arising from phase transitions in the early universe,
destroys this `friendship'.
The overproduction traces to the smallness of the horizon at very
early times.
Defects (\eg, magnetic monopoles) are produced at an abundance of
about 1 per horizon volume
(cf. the Kibble mechanism, see \S\ref{sec-Kibbbb}).
This yields a monopole to photon ration of order
$(T_{GUT} / m_P)^3$ and a present $\Omega_{monop}$ far in
excess of 1, clearly intolerable cosmologically speaking;
see \S\ref{sec-monoANDdo} for details.

We will see in \S\ref{sec-topo} below that the simplest
GUTs generically predict the formation of topological defects,
and why many of these defects are a disaster to cosmology;
in particular the low--energy standard model of particle physics
cannot be reached (as the last step of a chain of phase
transitions from a larger symmetry group)
without producing local monopoles [Preskil, 1979].
The standard cosmology has no means of ridding the universe of
these overproduced relics; hence the problem.


\chapter{Theories of the early universe}
\label{chap-theories}
\markboth{Chapter 2. ~THEORIES OF THE EARLY UNIVERSE}
         {Chapter 2. ~THEORIES OF THE EARLY UNIVERSE}

A particularly interesting cosmological issue is the origin of
structure in the universe. This structure is believed to have emerged
from the growth of primordial matter--density fluctuations
amplified by gravity.
The link of cosmology to particle physics theories has led to
the generation of two classes of theories which attempt to provide physically
motivated solutions to this problem of the origin of structure in the universe.

The first class of theories are those based on a mechanism called
{\sl inflation}; according to it, the very early universe underwent a brief
epoch of extraordinarily rapid expansion.
Primordial ripples in all forms of matter--energy perturbations
at that time were
enormously amplified and stretched to cosmological sizes and, after the
action of gravity, became the large--scale structure that we see today.
The initial idea that an early epoch of accelerated expansion
would have interesting implications for cosmology is due to Guth [1981].

According to the second class of theories, those based on
{\sl topological defects},
primordial fluctuations were produced by a superposition of seeds made of
localised distributions of energy density trapped during a symmetry
breaking phase transition in the early universe.
This idea was first proposed by Kibble [1976], and has been fully worked
out since then by many authors.

This Chapter is devoted to a general review of these two competing
models of large--scale structure formation and CMB anisotropy generation.
The Chapter is divided into two `big' sections, each of which deals
with one of these scenarios.
Let us begin first (in alphabetical order) with inflationary models.

\section{Inflation}
\label{sec-inflation}
\markboth{Chapter 2. ~THEORIES OF THE EARLY UNIVERSE}
			        {\S ~2.1. ~INFLATION}

As we have discussed in Chapter \ref{chap-StandCosmo}
the standard big bang cosmology is remarkably successful.
It explains the expansion of the universe,
the origin of the CMB radiation
and it allows us to follow with good accuracy the
development of our universe from the time of nucleosynthesis
(around 1 sec after the bang) up to the present time
($\sim$ 15 Gyr).
However, it also presents some shortcomings, namely,
the flatness/oldness,
the horizon/large--scale smoothness,
the cosmological constant, and the unwanted relics
problems.

Until the 1980's the was seemingly no foreseeable solution to
these problems. With the advent of the inflationary scenario
the way of thinking the early universe changed drastically.
The scenario, which is based on causal microphysical events
that might have occurred at times $\sim 10^{-34}$ sec in the history
of the universe (and well after the Planck era $\sim 10^{-43}$ sec),
offers a framework within which it is possible
to explain some of the above mentioned problems.

In the following sections we give a brief account of inflation.
First we review a bit of the history leading to the inflationary idea,
namely that it was worthwhile to study a period of exponential expansion
in the early evolution of our universe.
Then we concentrate in the basic facts related to the dynamics of
the inflaton field, responsible for driving the quasi--exponential
expansion.
We then show how inflation explains the handful
of cosmological facts with which the standard model alone cannot cope.
{}From this we go on to one of the nicest features that inflation predicts,
namely, the origin of density and metric perturbations, whose
consequences for structure formation and gravitational wave generation
the reader may already appreciate.
After a succinct explanation on how these perturbations evolve,
we move on to a brief survey of inflationary models, trying to
put them in perspective and explaining what features make them
attractive (or why they proved unsuitable) as a realisation of
the scenario.
We finish the section with a short account of the stochastic approach
to inflation.
The results of this will be helpful in understanding the generation
of space correlations in the primordial inflaton field that,
on horizon entry, will lead to non--Gaussian features in the
CMB radiation on large angular scales
(cf. Chapter \ref{chap-primor} below).

\subsection{The paradigm: some history}

Soon after entering the subject one realises that currently
there are many inflationary universe models, and some of them
involve very different underlying high energy physics.
Of course, something in common among them all is the existence of
an early stage of exponential (or quasi--exponential) expansion
while the universe was dominated by vacuum
energy\footnote{In most models the way of implementing the idea
	of a vacuum energy dominated universe is by assuming a scalar
	field whose initial state was displaced from its
	true vacuum (lowest energy) state.}
and filled just by almost homogeneous fields and nearly no other
form of energy.
This feature, shared by all  simple models, is what
sometimes goes under the name of the {\sl inflationary paradigm}.
The end of inflation is signaled by the decay of the vacuum
energy into lighter particles, the interaction of which with
one another leads to a state of hot thermal equilibrium.
{}From that point onwards the universe is well described by
the hot big bang model.

The inflationary models (while not yet carrying this name)
have a long history that traces back to the papers by
Hoyle, Gold, Bondi, Sato and others
[Lindley, 1985, cited in Ref. \cite{KT90},
see also Gliner, 1965, 1970], where the possible existence of an
accelerated expansion stage was first envisaged.
A few years later Linde [1974, 1979] came up with the idea
that homogeneous classical scalar fields (which appear in virtually
all GUTs) could play the r\^ole of an unstable vacuum state, whose
decay may give rise to enormous entropy production and heat
the universe.
Soon afterwards it was realised that quantum corrections in the
theory of gravity also led to an exponential expansion:
the Starobinskii model [Starobinskii, 1979].

But the major breakthrough came with Guth [1981] paper,
where the real power of inflation for resolving the
shortcomings of the standard model was spelled out.
In spite of the fact that his original model did not
succeed (essentially due to the extremely inhomogeneous universe
that was produced after the transition -- see \S\ref{sec-Closett} below)
it laid down the main idea and it took just months for people to
propose the `new' (or `slow--rollover')
scenario where the drawbacks of Guth's
model were solved [Linde, 1982a; Albrech \& Steinhardt, 1982].
However, in the search for simple and powerful models, there is
little doubt that the first prize goes to Linde's chaotic models
[Linde, 1983a]. In these models the inflaton is there
just to implement inflation, not being part of any unified theory.
Moreover, no special potential is required: an ordinary
$\lambda \phi^4$ or even a mass autointeraction term will do the job.
There is neither phase transition nor spontaneous symmetry breaking
process involved, and the only requirement is that the inflaton field
be displaced from the minimum of its potential initially.
Different regions of the universe have arbitrarily different
(`chaotically' distributed) initial values for $\phi$.

\subsection{Scalar field dynamics}
\label{sec-Scalarfielddynamics}

The basic idea of inflation is that there was an epoch
early in the history of the universe when  potential, or vacuum,
energy was  the dominant component of the energy density.
The usual way to realise an accelerated expansion is by means
of forms of energy other than ordinary matter or radiation.
Early enough energy cannot be described in term of an ideal gas
but it should be described in terms of quantum fields.
Usually a scalar field is implemented to drive this inflationary
era and in the present section we will consider its dynamics in detail.

Let us study now the relevant equations for a homogeneous scalar field
$\phi (t)$ with effective potential $V(\phi)$ in the framework of a
FRW model, in presence of radiation, $\rho_r (t)$.
The classical equations of motion [Ellis, 1991] are given by
the Friedmann equation
\be
\label{ellis42}
H^2 + K = {1\over 3} 8\pi G
[ V(\phi) + {1\over 2} \dot\phi^2 + \rho_r ] ~,
\ee
the Raychaudhuri equation
\be
\label{ellis43}
\dot H + H^2 = {1\over 3} 8\pi G
[ V(\phi) -  \dot\phi^2 - \rho_r ] ~,
\ee
the energy conservation equation for $\phi$
\be
\label{powlaw23a}
{d\over dt}[ {1\over 2}\dot\phi^2 +  V(\phi) ]
= - 3 H \dot\phi^2 - \delta ~,
\ee
which is equivalent to the equation of motion for $\phi$, and
the energy conservation for radiation
\be
\label{powlaw23b}
{d\over dt}\rho_r = - 4 H \rho_r + \delta ~,
\ee
where $\delta$ accounts for the creation of ultra--relativistic
particles (radiation) due to the time variation of the scalar
field.
$\delta$ may be expressed as [Kolb \& Turner, 1990]
\be
\label{powlaw24}
\delta = \Gamma \dot\phi ~,
\ee
where the characteristic time $\Gamma^{-1}$ for particle creation
depends upon the interactions of $\phi$ with other fields.
As we are mainly interested in the phase of adiabatic evolution of the
scalar field, where $\Gamma << 3 H$ and thus particle creation processes
are not operative, we will neglect the $\delta$ term in what follows.
When arriving at the {\sl reheating} phase (at the very end of inflation)
the damping of the coherent oscillations of the scalar field will
lead to the creation of light particles, which, after thermalisation,
will heat the universe to a temperature appropriate for continuing
its evolution in the radiation era.

It is a common practice, when dealing with the above equations,
to assume a particular kind of rolling condition for the scalar field.
The slow--rolling approximation $|\ddot\phi| << 3 H \dot\phi$
yields $3 H \dot\phi + \partial V / \partial\phi = 0$ and this
describes the evolution of $\phi$ during inflation.
On the other hand at the end of inflation the situation reverts
itself and we have
$\ddot\phi + \partial V / \partial\phi + \Gamma \dot\phi^2 \simeq 0$.
This signals the fast--rolling evolution during reheating.

Although these are the approximations that one commonly finds in the
literature, we will here follow the analysis of Ellis [1991] and
study the dynamics of the field exactly.
Combining Eqs. (\ref{ellis42}), (\ref{ellis43}) (now without
radiation density contribution) and assuming $\dot\phi \not= 0$
we have
\be
\label{ellis48}
 V(\phi) = (8\pi G)^{-1} [\dot H + 3 H^2 + 2 K] ~,
\ee
and
\be
\label{ellis49}
\dot\phi^2(t) = (4\pi G)^{-1} [- \dot H + K] ~.
\ee
To specify a model we need only proceed as follows.
First, choose a value for $k$ (cf. Eq. (\ref{elly1}))
and the initial value for $\phi$.
Secondly we specify the time--dependent scale factor $a(t)$
(such that the right--hand side of Eq. (\ref{ellis49}) is positive)
and compute from this the Hubble parameter and its derivative.
Use then (\ref{ellis49}) to get $\dot\phi(t)$, integrate it to
find $\phi(t)$ and finally invert it to get $t(\phi)$.
Finally, plug this into (\ref{ellis48}) to get $V(\phi)$.
On the other hand,
if $\dot\phi = 0$, equation (\ref{powlaw23a}) tells us that
$\partial V / \partial\phi \simeq 0$ and thus the potential is flat.
In summary, in all cases it is possible to find (in principle)
the form of the inflationary effective potential satisfying the
classical inflationary equations (without any assumption on the way
$\phi$ `rolls')
with the desired functional form for the scale factor $a(t)$,
the chosen curvature $k$ and initial condition for the scalar field.

A specific example was given by Lucchin \& Matarrese [1985].
Taking flat spatial sections $k=0$ they chose
a power--law behaviour for the scale factor
$a(t)\propto t^p$, with constant $p>1$ [Abbott \& Wise, 1984b].
They studied the evolution of the system of classical equations
from an initial time $t_i$ with $\phi(t_i) \equiv \phi_i \not= 0$
and assumed (as it is usually done) that after a brief period of
inflation the radiation contribution to the total kinetic and potential
energy is depressed, and so we may neglect the contribution
of $\rho_r$ to the relevant equations.
With this ansatz for the scale factor we readily get
\be
\label{powlaw28}
\dot\phi^2 \simeq {p\over 4 \pi G} t^{-2} ~,
\ee
which gives
\be
\label{powlaw29}
\phi (t) \simeq \phi_i \pm \sqrt{p\over 4 \pi G} \ln (t /  t_i) ~.
\ee
Inserting the solution with the plus sign into (\ref{ellis42}) with
$K = 0$, and making use of (\ref{powlaw28}) we get
\be
\label{powlaw211}
V(\phi) \simeq {3p - 1\over 2} {p\over 4 \pi G t_i^2}
\exp \left(- 2 \sqrt{4 \pi G \over p} (\phi - \phi_i)\right) ~.
\ee
It should be said that this potential is to be considered as an
approximation (for the evolution of the inflaton until before the
time of reheating) of a more complex potential. In particular
the featureless exponential potential cannot provide an oscillatory end
to inflation. Power--law inflation, however, is of particular
interest because exact analytic solutions exist both for the dynamics
of inflation and for the density perturbations generated.

\begin{flushleft}
{\bf Slow--rolling}
\end{flushleft}

As we mentioned above, in a universe dominated by a homogeneous
scalar field $\phi$,
minimally coupled to gravity,  the equation of motion
is given by (\ref{powlaw23a}), which we may write
\be
\label{LL52}
\ddot \phi + 3 H \dot\phi + \partial V / \partial\phi = 0 ~.
\ee
The energy density and pressure
(neglecting radiation or other matter--energy components)
are given by
\be
\label{LL53}
\rho \simeq    V(\phi) + {1\over 2} \dot\phi^2  ~~ ; ~~
  p  \simeq  - V(\phi) + {1\over 2} \dot\phi^2 ~.
\ee
We then see that provided the field rolls slowly, \ie,
$\dot\phi^2 < V$, we will have $3 p < - \rho$ and,
from equation (\ref{ellis43}) or equivalently
$\ddot a = - 4\pi G (\rho + 3 p) a / 3$,
we have $\ddot a > 0$, namely, an accelerated expansion.

In most of the usually considered models of inflation there are
three conditions that are satisfied. The first is a statement
about the solution of the field equation, and says that the motion
of the field in overdamped, namely that the force term
$\partial V / \partial\phi$ balances the friction term
$3 H \dot\phi$, and so
\be
\label{LL55}
\dot\phi \simeq - {1\over 3 H} {\partial V \over \partial\phi} ~.
\ee
The following two conditions are statements about  the
form of the potential. The first one constrains the steepness
of the potential (or better, its squared)
\be
\label{LL56}
\epsilon \equiv {m_P^2 \over 16 \pi}
\left({V' \over V}\right)^2 << 1 ~,
\ee
where $' \equiv \partial / \partial\phi$.
This implies that the condition $\dot\phi^2 < V$ is well satisfied, and
$H^2 \simeq 8\pi G V(\phi) / 3$ which implies that the Hubble parameter is
slowly varying.
The third condition that is usually satisfied is
\be
\label{LL59}
\chi \equiv {m_P^2 \over 8 \pi} {V'' \over V}
{}~~ ; ~~ |\chi| << 1 ~,
\ee
and is independent of the other two conditions [Liddle \& Lyth, 1993].

\subsection{Problems inflation can solve}
\label{sec-cannot}

We have seen before that the standard model, although extremely
successful in its predictions, left some open questions
(cf. \S\ref{sec-drawbacks}).
In this section we review some common features of inflationary
cosmologies and make the way to understand how these
problems may be solved.

Let us first consider the {\sl horizon problem}.
We saw in the previous section that we may define the
accelerated expansion of the universe by the condition
$\ddot a > 0$.
We may calculate now
\be
\label{li1}
{d (H^{-1} / a) \over d t}
= - {1\over a H^2} \left( H^2 + \dot H \right)
= - {1\over a H^2} {\ddot a \over a} ~,
\ee
which, together with $\ddot a > 0$ implies
$d (H^{-1} / a) / d t < 0$, namely that the comoving
Hubble length
$r_H \equiv H^{-1} / a$
is decreasing during any accelerated expansion period.
This is the basic mechanism through which inflation solves the
horizon problem.

At the beginning of the inflationary period the comoving
Hubble length
is large and therefore perturbations on very large scales
(like our present horizon scale, and beyond) are generated causally
within $r_H$.
With the accelerated expansion $r_H$ decreases to such an extent that
its subsequent increase during the radiation and matter eras
following inflation is not enough to give it back to the length it
had before the inflationary epoch.

While $r_H$ decreases during inflation all those perturbation scales
that are fixed (in comoving coordinates) are effectively seen as
`exiting' the Hubble radius\footnote{One many times finds statements in the
	literature about perturbations exiting and entering the horizon.
	Clearly, once the particle horizon encompasses a given distance scale,
	there is causal communication and homogenisation on that scale.
	Therefore one hardly sees how this scale can ever `exit' the horizon
	and loose the causal contact it has achieved before.
	In order to avoid many headaches one should speak in terms of
	the Hubble radius, $H^{-1}$, quantity that remains nearly
	constant during an inflationary phase
	(however, I am aware that this thesis might be plagued with sentences
	where the word horizon
	(in place of the more correct `Hubble radius')
	is employed; my apologies)
	[see Ellis \& Rothman, 1993 for a pedagogical review].}.
After inflation ends, during the radiation and matter eras,
$r_H$ grows again and the Hubble radius stretches beyond
some of these perturbation scales: they are `entering' the Hubble radius.
However, by that time these scales have already had time to get
causally connected before exiting $H^{-1}$
and, \eg, produce the same level of CMB anisotropies
in the sky.

With regards to the {\sl flatness problem}, we have seen in
\S\ref{sec-drawbacks} that the evolution equation satisfied by
$\Omega$ makes it evolve away from unity (cf. Eq. (\ref{evoOmega})).
However this happens for a standard perfect fluid equation of state, with
$p / \rho = w > -1/3$ (thus, satisfying the strong energy condition).
During inflation the expansion is in general driven by a slow--rolling
scalar field, and in this case the previous condition on $w$
does not apply.
As we can see from Eqs. (\ref{LL53}), during a scalar field dominated
universe we have $w \simeq -1$ and for this value of
$w$ equation (\ref{evoOmega}) tells us that $\Omega$ will rapidly
approach unity.
Provided sufficient inflation is achieved (necessary for solving the
horizon problem, for example),  $|\Omega - 1|$ at the end
of inflation is dynamically driven to a value small enough to render
$\Omega \sim 0.2$ today, as it
is indeed estimated from, \eg, rotation--curve measurements
in spiral galaxies, and other dynamical determinations of
$\Omega$.
\footnote{One comment is in order:
many times one finds statements saying that inflation solves the flatness
problem too well, thus predicting a $\Omega$ value too close to 1 today.
Recently, however, there has been a growing interest in models of
inflation with $\Omega\not= 1$
[see, \eg, Amendola, Baccigalupi, \& Occhionero, 1995
and references therein].}

Finally, inflation solves the {\sl unwanted relics problem},
the essential reason being the following:
since the patch of the universe that we observe today was once
inside a causally connected region, presumably of the order of the
correlation length $\xi$ (see \S\ref{sec-Kibbbb} below),
the Higgs field could have been aligned throughout the
patch (as this is, in fact, the lowest energy configuration).
Thus, being no domains with conflicting Higgs orientations, the
production of defects is grossly suppressed, and
we expect of order 1 or less topological defects in our
present Hubble radius.
In other words, the huge expansion produced by an early inflationary
era in the history of the universe dilutes the abundance of
(the otherwise overproduced) magnetic monopoles or any other
cosmological `pollutant' relic\footnote{This also implies, however,
	that any primordial baryon number density will be diluted away
	and therefore a sufficient temperature from reheating
	as well as baryon--number and CP--violating interactions
	will be required after inflation [Kolb \& Turner, 1990].}.

\subsection{Generation of density perturbations}
\label{sec-GenDenPer}

One of the most important problems of cosmology is the problem
of the origin of the primordial density inhomogeneities
that gave rise to the clumpy universe where we live.
This is closely connected with the issue of initial conditions.
Before the advent of the inflationary scenario there was virtually
no idea of which processes were the responsible for the
formation of the large--scale structure we see today.
It was possible that galaxies were originated by vortex perturbations of
the metric [Ozernoi \& Chernin, 1967],
from sudden events like the explosion of stars [Ostriker \& Cowie, 1980],
or even from the formation of black holes [Carr \& Rees, 1984].

In the inflationary cosmology this situation changed. First of all,
it was understood that all those perturbations present before inflation
were rendered irrelevant for galaxy formation, since inflation washes out all
initial inhomogeneities.
Further, a finite period of accelerated expansion of the universe
naturally explains that perturbations on cosmologically interesting
scales had their origin inside the Hubble radius at some point in
the inflationary phase.

The analysis of the linear evolution of density perturbations is
normally performed by means of a Fourier expansion in modes.
Depending on the physical wavelength of the modes, relative to the
Hubble radius, the evolution splits into two qualitatively different
regimes: for perturbations of size smaller than $H^{-1}$, microphysical
processes (such as pressure support, quantum mechanical effects, etc)
can act and alter their evolution. On the contrary, when the typical
size of the perturbation is beyond the Hubble radius, their amplitude
remains essentially constant (due to the large friction term
$3 H \dot\phi$ in the equation of motion) and the only effect upon
them is a conformal stretching in their wavelength due to the
expansion of the universe.

Inflation has the means to produce scalar (density) and tensor
(gravitational waves) perturbations on cosmologically interesting
scales.
The dynamics of producing density perturbations involves the
quantum mechanical fluctuations of a scalar field in a nearly
de Sitter space.
Since the couplings of the inflaton are necessarily weak,
\eg, in order not to generate too much CMB anisotropies,
its quantum fluctuations can be estimated
by the vacuum fluctuations of a free field in a de Sitter space
[see Gibbons \& Hawking, 1977; Bunch \& Davies, 1978].
In what follows we consider the Lagrangian
\be
\label{si216}
{\cal L} =
{1\over 2} \dot\phi^2 - {1\over 2} e^{-2 H t} (\nabla\phi)^2 - V(\phi)
{}~,
\ee
in the background metric $ds^2 = dt^2 - \exp(2 H t) [dx^2+dy^2+dz^2]$.
To keep things simple we consider a massive free field,
$V(\phi) = m^2 \phi^2 / 2$.
{}From (\ref{si216}) we may derive the equation of motion, and then quantise
the system using the usual equal time commutation relations.
After Fourier decomposing the field $\phi(\vec x , t)$ we get the
equation of motion for a particular mode as follows
\be
\label{si219}
[ \partial^2_t + 3 H \partial_t + k^2 \exp (-2 H (t-t_*)) - m^2 ]
\phi_k(t) = 0 ~,
\ee
where $t_*$ is the time at which inflation began.
Upon defining $z \equiv k H^{-1} \exp (- H (t-t_*))$
and $\nu \equiv \left( \frac{9}{4} - m^2 H^{-2} \right)^{1/2}$,
equation (\ref{si219}) may be written
\be
\label{si219bis}
\left(
z^2 {\partial^2 \over \partial z^2} - 2 z {\partial \over \partial z}
+ z^2 - \nu^2 + {9\over 4}
\right) \phi_k(t) = 0 ~.
\ee
This has the form of a Bessel equation and the solutions are
\be
\label{si219bisbis}
\phi_k(t) \propto z^{3/2} H_\nu^{(1,2)}(z) ~,
\ee
where $H_\nu^{(1)}$, $H_\nu^{(2)}$ stand for Hankel functions
[Bunch \& Davies, 1978].
The most general solution is a linear combination with coefficients
$c_1$, $c_2$ satisfying $|c_1|^2-|c_2|^2=1$.
Different values for these constants lead to different vacuum states
for the quantum theory.
The mode $\phi_k(t)$ describes the evolution of a perturbation of
wavelength $\sim k^{-1} \exp(Ht)$. For very early times this wavelength
is much smaller than $H^{-1}$ and at such short distances
de Sitter and Minkowski spaces are indistinguishable.
The short wavelength limit corresponds to large $z$'s and, given that
in this limit we have $H_\nu^{1 (2)} \propto \exp (- (+) i k \Delta t)$,
we see that the choice which corresponds to positive frequency modes
in the flat space limit is given by $(c_1, c_2) = (1, 0)$.
Upon normalisation one gets
\be
\label{si220}
\phi_k(t) = {1\over 2} \sqrt{\pi\over H}
\exp( -{3\over 2} H (t-t_*) ) H_\nu^{(1)}(z) ~.
\ee
The spectrum of fluctuations of the scalar field is
$|\delta\phi|^2 \simeq  \la |\phi^2(t)| \ra$. Using the solution
(\ref{si220}) and taking the limit of large time ($z << 1$) and
$m^2 << H^2$ (and thus we have
$\nu \simeq \frac{3}{2}  - \frac{1}{3} (m/H)^2$)
we get the variance of the scalar field perturbation given by
\be
\label{si223}
|\delta\phi|^2 = {H^2 \over 4 \pi^2}
\exp \left(- {2 m^2\over 3 H} (t - t_*)\right)
\int_{H}^{H\exp(H(t - t_*))} d(\ln k)
\left( {k\over H} \right)^{2 m^2 / (3 H^2)} ~.
\ee
The upper limit of the integral is fixed by the last (the smallest)
wavelength that crosses the horizon at time $t$ during the inflationary
expansion. The lower limit takes into account that inflation
starts at time $t_*$ and therefore there is a first (maximum)
wavelength that crosses the horizon when inflation began.
{}From Eq. (\ref{si223}) we easily see that the contribution to
$|\delta\phi|$ per logarithmic interval of $k$ is given by
\be
\label{si224}
{\cal P}_\phi^{1/2} (k)
= {H \over 2 \pi}
\left( {k\over H} \right)^{m^2 / (3 H^2)}
\exp \left(- {m^2\over 3 H} (t - t_*)\right) ~,
\ee
and, in the limit $m^2 << H^2$ it reduces to well--known result
${\cal P}_\phi^{1/2} (k) \simeq H / 2 \pi$.
A measurement of the $\phi_k$'s will yield random phases, and
the distribution of the moduli will have a dispersion
$\sim \la |\delta\phi_k|^2 \ra = H^2 / (2 k^3)$.
Accordingly, the spectrum of the inflaton
(cf. Eq. (\ref{si224})) a few times after the horizon exit is given by
${\cal P}_\phi^{1/2} (k) \simeq H_{ex} / 2 \pi$
[Vilenkin \& Ford, 1982; Linde, 1982b; Starobinskii, 1982],
where $H_{ex}$ stands for the (slowly varying) Hubble parameter
at horizon exit.

\subsection{Evolution of fluctuations}
\label{sec-EvFluctu}

In this section we follow the recent review
by Liddle \& Lyth [1993]. The spectrum of the density contrast
after matter domination may be written
\be
\label{LL35}
{\cal P}_{\delta} = \left({k \over a H}\right)^4
T^2(k) \delta_H^2(k) ~,
\ee
where $T(k)$ is the linear transfer function which becomes
equal to 1 on very large scales, meaning that very little
evolution is suffered by those large scales that entered
the horizon recently, during the matter dominated
era\footnote{The correct transfer function fit comes only
after numerically solving the relevant perturbation equations,
with initial conditions specifying the relative abundance of the
different matter components, and taking into account the
free streaming of neutrinos, the diffusion of radiation on
scales smaller than Silk's scale, etc.
We will not go into the details here;
these are reviewed in, \eg, [Efstathiou, 1990].}.
The quantity $\delta_H^2(k)$ specifies the initial spectrum
of density perturbations at horizon entry (hence the subindex $_H$)
and is related to the spectrum of the initial curvature
perturbations ${\cal R}$ by
\be
\label{LL36}
\delta_H^2(k) = {4\over 25} {\cal P}_{\cal R}(k) ~.
\ee
$\delta_H^2$ is exactly equal to the value of ${\cal P}_{\delta}$
on horizon entry on scales much larger than 100 $h^{-1}$ Mpc, roughly
the size of the horizon at the time of equality between matter and
radiation, and is approximately equal to it on smaller scales.

The standard assumption is that
$\delta_H^2 \propto k^{n-1}$, corresponding to a density spectrum
$P \propto k^{n}$ with spectral index $n$.
The standard choice $n=1$ yields the `flat' spectrum, proposed
by Peebles \& Yu [1970],
Harrison [1970] and Zel'dovich [1970] as being the only one
leading to small perturbations on all
scales\footnote{Compared with the standard $n=1$ spectrum,
tilted models with $n<1$ lead to relatively less
power on small scales and have been advocated many times, in particular
to alleviate the too large pairwise galaxy velocity dispersion
on scales of order 1 $h^{-1}$ Mpc. On the other hand, recent
analyses of the {\sl COBE}--DMR data on very large angular
scales tend to indicate a `blue' spectra ($n>1$). However this may
be partially due to the (large--scale)
`tail' of the so--called Doppler peaks that, \eg,
CDM plus inflation models predict to appear on the degree scale.
See Chapter \ref{chap-texture}.}.

The curvature perturbation ${\cal R}$ of comoving hypersurfaces
is given in term of the perturbations of the inflaton field by
[Lyth, 1985; Sasaki, 1986]
\be
\label{LL537}
{\cal R} = {H \over \dot\phi} \delta\phi ~.
\ee
The spectrum of ${\cal R}$ is then given by
\be
\label{LL538}
{\cal P}_{\cal R}^{1/2} = {H \over \dot\phi} {\cal P}_{\phi}^{1/2} ~.
\ee
Now, given that the curvature perturbation ${\cal R}$ is constant after
horizon exit [Liddle \& Lyth, 1993], thus ${\cal P}_{\cal R}^{1/2}$
remains constant even though
$H / \dot\phi$ and ${\cal P}_{\phi}^{1/2}$ may vary separately.
As long as the scale is far outside the horizon we have
\be
\label{LL539}
{\cal P}_{\cal R}^{1/2} = {H_{ex}^2 \over 2\pi \dot\phi_{ex}} ~,
\ee
where $_{ex}$ subindex stands for quantities evaluated at horizon
exit (cf. the last paragraph of \S\ref{sec-GenDenPer}).
Using Eqs. (\ref{LL36}) and  (\ref{LL55})
we find the amplitude of the
adiabatic density perturbation at horizon crossing [Lyth, 1985]
\be
\label{LL518}
\delta_H^2 (k) = {32\over 75} {V_{ex}\over m_P^4}
\epsilon_{ex}^{-1} ~,
\ee
where $\epsilon_{ex}$ is the small parameter given by (\ref{LL56})
at horizon exit.
One therefore has a perturbation amplitude on re--entry that
is determined by the conditions just as it left the horizon, which
is physically reasonable.
In general one does not know the values of $V_{ex}$ and
$\epsilon_{ex}$ so one cannot use them to predict the spectral form
of the perturbations; in fact, within generic inflationary models,
these values depend on the particular length--scale of the perturbation.
In an exactly exponential inflation these two parameters
are constant (independent of the particular scale exiting the horizon
during the inflationary era) and therefore the perturbation amplitude
on horizon entry is constant and independent of the scale,
leading to the `flat' spectrum we mentioned above.

In general the spectral index of density perturbations
depends on the considered scale. However, if this dependence is
weak (at least within a cosmologically interesting range of scales)
we may define $n-1\equiv d \ln \delta_H^2 / d (\ln k)$,
which is implicitly based in the power--law dependence of $\delta_H^2$.
Differentiating equation (\ref{LL518}), and using the slow--rolling
conditions on $\chi$ and $\epsilon$, Liddle \& Lyth [1992]
derive the result
\be
\label{LL541}
n=1+2\chi_{ex}-6\epsilon_{ex}
\ee
where now the $_{ex}$ subindex refers, for definiteness, to
the moment when the observable universe leaves the horizon.

Analogously, for tensor perturbations (gravitational waves)
we can define  $n_T \equiv d \ln {\cal P}_{GW} / d (\ln k)$
and, following a similar analysis, we get\footnote{More accurate
	results for $n$ and $n_T$ are given in [Stewart \& Lyth, 1993]
	and [Kolb \& Vadas, 1994]. We will use their results
	in \S\ref{skewnessVSn} below in the framework of some interesting
	inflationary models.}
\be
\label{LL536}
n_T = - 2 \epsilon_{ex} ~.
\ee
We thus see that in order to have significant deviation from
the flat $n=1$ and $n_T=0$ spectra we necessarily need to
violate the slow--rolling conditions $\epsilon << 1$ and $|\chi| << 1$.
Nevertheless, we will see below that within many of the currently
used inflationary models, values for $n$ below 1 can be achieved.
Models leading to $n>1$ have recently been studied by
Mollerach, Matarrese \& Lucchin [1994].

\subsection{An overview of models}
\label{sec-Closett}

We here briefly review a few inflationary models;
the list is by no means exhaustive.
We will say more about some of them, as well as introduce a couple of
other popular models, in Chapter \ref{chap-primor}, when we
will study non--Gaussian features (\eg, the CMB skewness)
predicted by primordial non--linearities in the evolution of the
inflaton in the framework of these models.
We want to emphasise that although these models solve the
cosmological problems outlined in \S\ref{sec-cannot}
there is as of 1995 no convincing way to realise the
scenario [see Brandenberger, 1995; Turner, 1995 for recent reviews].

\begin{flushleft}
{\bf `Old' inflation}
\end{flushleft}

The `old' inflationary model [Guth, 1981; Guth \& Tye, 1980]
is based on a scalar field theory undergoing a first--order phase
transition [Kirzhnits \& Linde, 1976].
In this model the universe has an initial expansion state with
very high temperature, leading to a symmetry restoration in the early
universe. The effective potential $V_T(\phi)$ has a local (metastable)
minimum at $\phi = 0$ even at a very low temperature.
The potential also has a global (true) minimum at some other value,
say, $\phi = \sigma$ where the potential vanishes (in order to avoid
a large cosmological constant at present time).
In old inflation the crucial feature is a barrier in the potential
separating the symmetric high--temperature minimum from the
low--temperature true vacuum. As a result of the presence of the barrier
the universe remains in a supercooled vacuum state $\phi = 0$
for a long time. The energy--momentum tensor of such a state
rapidly becomes equal to $T_{\mu\nu} = g_{\mu\nu} V(0)$
(all other form of matter--energy rapidly redshift)
and the universe expands exponentially (inflates) until the moment
when the false vacuum decays.
As the temperature decreases with the expansion the scalar field
configuration, trapped in the false vacuum, becomes metastable
and eventually decays to the global minimum by quantum tunnelling
through the barrier.
This process leads to the nucleation of bubbles of true vacuum,
and these bubbles expand at the velocity of light converting
false vacuum to true.

Reheating of the universe occurs due to bubble--wall collisions.
Sufficient inflation was never a real concern; the problem with
this classical picture is in the termination of the false--vacuum
phase, usually referred to as the `graceful exit' problem.
For successful inflation it is necessary to convert the vacuum
energy to radiation. This is accomplished through the collision
of the bubbles.

Guth himself [Guth, 1981; see also Guth \& Weinberg, 1983]
realised that the scenario had a serious problem, in that the
typical radius of a bubble today would be much smaller than our
observable horizon. Thus the model predicted extremely large
inhomogeneities inside the Hubble radius, in contradiction with
the observed CMB radiation isotropy.
The way out was thought to be in the percolation of the generated
bubbles, which would then homogenise in a region larger than our
present horizon.
In these collisions the energy density tied up in the bubble--walls
may be converted to entropy, and in order to have a graceful exit
there must be many collisions.
But, it so happens that the exponential expansion of the background
space overwhelms the bubble growth (the volume inside the bubbles
expands only with a power--law). This prevents percolation and
the `graceful exit' problem from being solved.

\begin{flushleft}
{\bf `New' inflation}
\end{flushleft}

Soon after the original (and unsuccessful) model laid down the main idea
of the convenience of an early era of accelerated expansion,
inflation was revived by the realisation that it was possible to have
an inflationary scenario without recourse to a strongly first--order
phase transition.
Linde [1982a] and independently Albrecht \& Steinhardt [1982] put
forwards a modified scenario, the `new' inflationary model.
The starting point is a scalar field theory with a double well
`mexican hat' potential which undergoes a second--order phase
transition. $V(\phi)$ is symmetric and $\phi = 0$ is a local
maximum of the zero temperature potential.
As in old inflation, here also, finite temperature effects are
the responsible for confining $\phi$ to lay near the maximum at
temperatures greater than the critical one, $T_c$.
For temperatures below $T_c$ thermal fluctuations make
$\phi = 0$ unstable and the field evolves towards one of the global
minima, $\phi = \pm \sigma$, ruled by the equation of motion
(cf. (\ref{LL52}))
\be
\label{bra518}
\ddot \phi + 3 H \dot\phi + \partial V / \partial\phi = 0 ~.
\ee
The transition proceeds now by spinoidal
decomposition\footnote{We will be more precise about this concept
	in \S\S\ref{sec-PhaseTrans}--\ref{sec-Kibbbb}
	below [see also Mermin, 1979].}
and hence
$\phi (\vec x , t)$ will be homogeneous within one correlation length.

If the potential near the maximum is flat enough the
$\ddot \phi$ term can be neglected and the scalar field will
undergo a period of slow--rolling.
The field has both `kinetic' and `potential' energy; however, under
the slow--roll hypothesis the velocity of the Higgs field will be
slow and the potential energy will dominate, driving the
accelerated expansion of the universe.
The phase transition occurs gradually and if significant inflation
takes place huge regions of homogeneous space will be produced;
we would be living today deep inside one of these regions.

There is no graceful exit problem in the new inflationary model.
Since the spinoidal decomposition domains are established before the
onset of inflation, any boundary walls will be inflated away
outside our present Hubble radius.
Thus, our observable universe should contain less than one topological
defect produced in the transition. This is good news for the
gauge monopole problem, but also bad news for global defects and local
cosmic strings.

Reheating occurs in this model as soon as the slow--roll evolution
of the $\phi$ breaks down. The scalar field will begin
its accelerated motion towards one of the absolute minima of the
potential. In so doing it overshoots and then starts oscillating
about it. The amplitude of this oscillation is damped by the expansion
of the universe and predominantly by the couplings that $\phi$ has
with other fields. These couplings lead to the decay of $\phi$
into lighter relativistic particles that after thermalisation
reheat the universe.
However, constraints from the level of density perturbations
and CMB anisotropies generated by this model force the couplings
between $\phi$ and other fields to be extremely small. This is
not only viewed as a `fine tuning' problem difficult to explain
by any underlying particle theory, but also leads to a low reheat
temperature, which may make baryogenesis
problematical\footnote{Recently there has been an interest
in reconsidering reheating scenarios.
See  \eg ~Refs.
\cite{ASTW82, AbbFaWise82, TraschenBrande90, KoLiStar94,
ShtanovTraschenBrande94, Boyaetal, DolgovFreese95}.
Needless to say much of the power of inflation
relays in being able to find a way to re--establish the conditions
necessary for a smooth matching with the standard hot big bang model,
which is so successful in its predictions
after, say, 1 sec (nucleosynthesis) in the history of the universe.}.
This is why the scalar field responsible for the spontaneous symmetry
breaking in the minimal $SU(5)$ grand unified theory with a
Coleman--Weinberg potential was discarded as a possible candidate
for the inflaton
[Bardeen, Steinhardt \& Turner 1983; Brandenberger \& Kahn, 1984].

\begin{flushleft}
{\bf Chaotic inflation with a polynomial potential}
\end{flushleft}

A model that successfully implements inflation and that can occur
for very general types of scalar field potentials was proposed
in [Linde, 1983a].
Here the scalar field is not part of any grand unified theory, and
its presence is just to drive an era of inflation.
No specific thermal equilibrium state previous to inflation
is required, unlike what happened in `new' inflation; in fact
this was considered as a weak point, since very small coupling constants
were required in order that the generated density and CMB perturbations
be in agreement with observations,
and thus the achievement of thermal equilibrium which would drive the
inflaton to the minimum of its potential was hard to justify as a
generic initial condition.

Linde's idea solves this problem by envisioning that different
initial values for the inflaton field will be attained in different
regions of the universe (these initial values being `chaotically'
distributed). In those Hubble volumes where the conditions for
an adequate period of inflation are satisfied the problems of
the standard big bang cosmology might be solved, and they yield
a universe like ours. This view leads to a very inhomogeneous
universe, albeit on scales much larger than our present horizon.

Let us now explain the basic features of the model and see how
the field $\phi(\vec x , t)$ could be distributed in the early
universe. We know that the upper limit for an accurate classical
description of spacetime is given by the Planck scale
(times near to $t_P\sim 10^{-44}$ sec).
At that epoch the effective potential
is given with an accuracy of ${\cal O}(m_P^4)$, due to the
uncertainty principle. Thus a typical distribution of the
field in a classical universe, which emerges from the spacetime
`foam' at that time, is given by [Linde, 1990]
$V(\phi) \lsim m_P^4$ and $(\partial_\nu\phi)^2 \lsim m_P^4$.

Consider, for simplicity, the $V(\phi) = \lambda\phi^4 / 4$
model, with $\lambda << 1$.
Let us focus on those domains of the universe where the field
was initially homogeneous (on a scale larger than $H^{-1}$)
in the sense that $(\partial_\nu\phi)^2 \lsim V(\phi)$, and
sufficiently large, $\phi \gsim m_P$.
The equation of motion in such a domain is
$\ddot\phi + 3 H \dot\phi = - \lambda\phi^3$.
The energy density is $\rho\simeq V(\phi)$ and the Hubble parameter
$H\simeq \left(8\pi V(\phi) / 3 \right)^{1/2} / m_P \sim
\lambda^{1/2} \phi^2 / m_P$ is slowly varying.
The equation of motion can be expressed as
\be
\label{si29}
\ddot\phi + \sqrt{6\pi\lambda} {\phi^2 \over m_P} \dot\phi
= - \lambda\phi^3 ~.
\ee
Assuming slow--rolling (condition valid for $\phi^2 >> m_P^2 / (6\pi)$)
we get $\phi (t) = \phi_{in}
\exp \left( -\sqrt{\lambda\over 6\pi} m_P ~ t \right)$, with
$\phi_{in}$ the initial value.
Furthermore, this domain expands with a scale factor
\be
\label{si212}
a(t)
=      a_{in} \exp \left[ \int_{t_{in}}^t H(t) dt                 \right]
\simeq a_{in} \exp \left[ {\pi\over m_P^2} (\phi_{in}^2 - \phi^2(t))
\right]
{}~.
\ee
This is a quasi--exponential expansion for
$\phi >> m_P$ and by the time the slow--rolling condition
breaks down the domain will have expanded by a factor
$\sim \exp \left(\pi\phi_{in}^2 / m_P^2 \right)$.
For $\phi_{in} \gsim 4.4 m_P$ the universe expands more
than $e^{60}$ times, enough for solving the horizon and flatness
problems\footnote{Note that $\phi_{in} \gsim 4.4 m_P$ does not
contradict $V(\phi) \lsim m_P^4$, provided we take
$\lambda$ small enough, in agreement with many reasonable theories.}.
The reheating picture is similar to the previous models of inflation:
for $\phi \lsim m_P / 3$ the scalar field oscillates about the minimum
of the potential, and the vacuum energy  stored is converted into radiation.
However, there is still `fine tuning' in the sense that, \eg,
an adequate level of  CMB anisotropies constrains $\lambda$ to
take a very  small value.

\begin{flushleft}
{\bf $R^2$ inflation}
\end{flushleft}

We have seen that within chaotic models of inflation the scalar field
associated to the inflaton had not a direct connection with a
grand unified theory. Its presence was justified just by the `need'
to have an accelerated era in the early universe.
Here instead we will see how an inflaton field naturally
arises in the case when we minimally extend General Relativity.
We will concentrate in an inflationary model motivated by a
modification of Einstein gravity. The Lagrangian for
gravity is\footnote{The first model of inflation, studied by
Starobinskii [1980], was of this form.}
\be
\label{LL549}
{\cal L} = - (16\pi G)^{-1}[ R + R^2 / (6 M^2) ] ~.
\ee
The second ($\propto R^2$) part is what modifies the usual
Einstein--Hilbert first term.
The justification of the ansatz (\ref{LL549}) can be found
in [Cardoso \& Ovrut, 1993] in the framework of supergravity theories.
After a conformal transformation [Whitt, 1984] the $R^2$ term
may be eliminated and we are left with the standard--gravity Lagrangian,
but interacting with an additional scalar field (identified with $R$) given
by
\be
\label{salo725}
\phi = \left( 3 / (16\pi G) \right)^{1/2}
\ln \left( 1 + R / (3 M^2) \right)  ~,
\ee
where the new scalar field $\phi$ must satisfy the potential
[see, \eg, Liddle \& Lyth, 1993 for a review,
and Salopek, Bond \& Bardeen, 1989 for an analysis of the fluctuation
spectrum]
\be
\label{LL551}
V(\phi) = {3 M^2 \over 32 \pi G}
\left[
1 - \exp \left( - (16\pi G / 3)^{1/2} \phi \right)
\right]^2 ~.
\ee
In the regime $\phi \gsim m_P$ this potential produces the
slow--rollover evolution of the scalar field, and thus inflation
is achieved.
Approximately when the field reaches the value
$\phi_{end} \sim (8\pi G)^{-1/2}$ inflation ends and the field
oscillates. For a scale of the order of our present horizon the
small parameters of Eqs. (\ref{LL56}) and (\ref{LL59})
can be calculated yielding
$\epsilon\sim 10^{-4}$ and $\chi\sim -0.02$,
and thus a spectral index of density perturbations $n\simeq 0.96$.
The tensor mode contribution to the anisotropies is negligible in
this model.

\begin{flushleft}
{\bf Power--law inflation with an exponential potential}
\end{flushleft}

Exponential inflation is by no means a necessary condition
to solve the problems of the standard big bang cosmology.
In principle any type of accelerated expansion may solve, say,
the horizon and flatness problems, and we will here consider
the case in which the scale factor is given by $a(t) \propto t^p$,
with $p>1$  [Abbott \& Wise, 1984b].
Lucchin \& Matarrese [1985] were the first to study
the cosmological--background equations with a scale factor of this form,
and found exact solutions yielding an inflaton potential of the form
(cf. \S\ref{sec-Scalarfielddynamics})
\be
\label{powlaw211-bis}
V(\phi)=V_i\exp\left(- 2 \sqrt{4 \pi \over p} {\phi\over m_P} \right) ~.
\ee
This potential may also find a motivation in the context of
extended inflation [see, \eg, La \& Steinhardt, 1989].
Furthermore, the featureless exponential potential does not
provide a way to end the inflationary stage, say, by oscillations
of the field, and so suitable modifications or even bubble nucleation
mechanisms (\eg, in extended inflation) should be invoked.
Nevertheless, the slow--roll conditions are satisfied if $p>>1$
and we get $\epsilon = \chi / 2 = 1 / p$ independent of the scale.
The spectral indices are $n_T \simeq -2/p$ and $n \simeq 1-2/p$.
\footnote{Exact analytic solutions for both the inflaton
dynamics and for the density perturbations exist in this model.
The exact spectral indices are
$n_T = -2/(p-1)$ and $n = 1-2/(p-1)$, in very good agreement with
the slow--roll results [see, \eg, Lyth \& Stewart, 1992].}
The tensor--to--scalar ratio, $R$, of the contributions to the
variance of the CMB anisotropies is  $R \simeq 12 / p$.

It is beyond the scope of the present thesis
to give a detailed account of
all the existent models that have been proposed in the literature
in the recent past. A few more of them we will discuss in
\S\S\ref{Workedexamples}--\ref{skewnessVSn} in connection with
the CMB skewness they predict.
Some of these models have been strongly constrained by observations and
only by contorting somewhat unnaturally can they still be alive.
However, `many' others resist and this is seen by many authors as
a weakness of the whole idea. It would be desirable to clean up
a bit the `closet of models' and leave there just the best candidates
we have.

\subsection{Stochastic approach to inflation}
\label{sec-Stocha}

In \S\ref{sec-Scalarfielddynamics} we have studied the dynamics
of the inflaton field assuming it homogeneous and obeying a
classical Klein--Gordon equation.
Quantum fluctuations in the field are stretched to super--horizon
scales and behave very much like classical
perturbations\footnote{Classicality of long wavelength quantum
	fluctuations in de Sitter space has been established in
	[Guth \& Pi, 1985] in a mode--by--mode treatment.
	This is the standard picture but, although it has been
	widely used, it is not widely understood; see [Miji\'c, 1994].
	The origin of classical perturbations as a consequence of
	quantum fluctuations is a subtle subject.
	The quantum to classical transition may be studied in the
	context of quantum cosmology, by means of the Wheeler--DeWitt
	equation for the wave function of the universe in a
	minisuperspace.
	For the achievement of the classical behaviour, the system
	under study has to show not only a strong correlation between
	coordinates and momenta (\ie, {\sl behave} in a classical way),
	but also all signs of quantum interference should disappear.
	This last leads to the notion of {\sl decoherence} of the
	density matrix [Zurek, 1982; see also Halliwell, 1989;
	Padmanabhan, 1989; Gangui \etal, 1991; Castagnino \etal, 1993,
	and references therein].}.
However, the quantum nature of the scalar field leads
to very interesting physics as well,
and the stochastic approach is the natural framework to study it.
The basic idea is due to Vilenkin [1983] and Starobinskii [1986].
The approach consists in splitting the scalar field in momentum space
into long and short wavelength modes.

Starting from the Heisenberg operator equation of motion
for the scalar field, the evolution of the long wavelength
part satisfies a classical, but stochastic, equation of motion.
The quantum effects, in the form of short wavelength modes,
build up a noise term, as we will explain below.
The field satisfies
$\nabla_\nu\nabla^\nu\phi+ \partial V / \partial\phi = 0$.
The scalar field may be written
$\phi = \phi_{{\rm long}} + \phi_{{\rm short}}$, or more explicitly
\be
\label{si234}
\phi = \phi_{{\rm long}}(\vec x , t) +
{1\over (2\pi)^{3/2}} \int d^3k ~ \Theta (k-\epsilon a H)
\left(
a_k\phi_k(t)e^{-i \vec k \cdot \vec x} +
a^{\dagger}_k\phi^*_k(t)e^{i \vec k \cdot \vec x}
\right) ~,
\ee
where $\phi_{{\rm long}}(\vec x , t)$ contains only modes
such that $k << a H$;
$a^{\dagger}_k$, $a_k$ are the usual creation and annihilation
operators, $\epsilon$ is a constant much smaller than 1, and
$\Theta$ stands for the step function.
Usually inflation takes place in regions where the scalar field
potential is not very steep and thus the short wavelength part satisfies
the massless equation $\nabla_\nu\nabla^\nu\phi_{{\rm short}} = 0$.
Thus the short wavelength modes can be taken as
\be
\label{si236}
\phi_k = {H\over \sqrt{2k}}
\left( {1\over a H} + {i\over k} \right)
\exp (ik/aH) ~.
\ee
{}From this Starobinskii [1986] derives the equation of motion for
the long wavelength `coarse--grained' part which, in the
slow--rolling approximation, reads
\be
\label{si237}
\dot\phi_{{\rm long}}(\vec x , t) =
- {1\over 3 H} {\partial V \over \partial\phi_{{\rm long}}}
+ f (\vec x , t) ~,
\ee
where the spatial gradient of $\phi_{{\rm long}}$ has been neglected
since it is subdominant for modes $k << a H$ (and hence the evolution
of the coarse--grained field can be followed in each domain independently),
and where
\be
\label{si238}
f (\vec x , t) = {1\over (2\pi)^{3/2}}
\int d^3k ~ \delta (k-k_s) \dot k_s
\left(
a_k\phi_k e^{-i \vec k \cdot \vec x} +
a^{\dagger}_k\phi^*_k e^{i \vec k \cdot \vec x}
\right) ~,
\ee
where $k_s = \epsilon a H$
stands for the inverse of the coarse--grained domain radius.
The correlation function for $f$, built up by all high frequency
modes, may be computed from this last equation and is given by
\be
\label{si239}
\la f(\vec x , t) f(\vec x , t') \ra =
{H^3\over 4 \pi^2} \delta (t-t') ~,
\ee
and has a white noise spectrum in time.

According to this picture the universe is described
by the value of the field in different coarse--grained
domains of comoving size $k_s^{-1} \simeq (\epsilon a H)^{-1}$.
With the inflationary expansion (whereby the comoving radius of
these domains decrease) an initial domain of radius
$(\epsilon a_1 H_1)^{-1}$ gets divided into
${\cal O}(e^3)$ subdomains of size $\sim (\epsilon a_2 H_2)^{-1}$,
after one Hubble time--step such that $t_2 \simeq t_1 + H_1^{-1}$
(since the temporal rate of change of $\phi$ is smaller than $H$).
The magnitude of the coarse--grained field $\phi$ in these
smaller domains in determined by both the classical (convective)
and the noise (stochastic) parts appearing in the right--hand side of
Eq. (\ref{si237}).\footnote{The reader may have noted that we
	have already dropped the label $_{{\rm long}}$ of the
	coarse--grained field.}
The classical force `pushes' the field down the potential in all the
subdomains, whereas the stochastic force acts
with different strengths\footnote{The typical magnitude of the change
	produced in the field in one Hubble time is given by the
	variance $\sim H / 2\pi$, cf. Eq. (\ref{si239}).}
and arbitrary direction in different subdomains.
This process repeats itself in all subsequent Hubble time--steps.

The result of this process is that in the new subdomains the
field takes on different (uncorrelated) values:
in those where the classical `drag' overwhelms the
stochastic `push' the field goes down the potential and
inflation will eventually end, yielding a domain like our
present universe.
The fact that our present causal region was, at the time of
inflation, divided into many coarse--grained subdomains with
slightly different values for $\phi$ explains the existence of
energy--density perturbations (and eventually also the structure
originated from them).
However, in those domains where the opposite happens
(namely, where $(3 H^2)^{-1} \partial V / \partial\phi \lsim H / 2\pi$)
there will be regions where the stochastic force is greater than the
classical one and points upwards the potential. These domains
will keep on inflating and, as they expand much faster than their
non--inflating counterparts, they fill the majority of the
universe (even if at each time more and more regions end their
accelerated expansion).
This picture leads to a `stationary' universe, where there could
have been no beginning and would be no end, with some domains inflating
and others ending inflation all the time, forever: Linde's eternal
inflation picture [Linde, 1990].

At this point one could attempt to study the probability
distribution $P(\phi , t)$ for the coarse--grained field $\phi$
by means of the Fokker--Planck equation.
By doing this we are concentrating on one particular domain, say
with comoving coordinate $\vec x$.\footnote{see \eg,
	[Matarrese \etal, 1989] for a detailed study of
	the probability distribution function $P$ in the case
	of many relevant chaotic models of inflation.}
Let us consider, for definiteness, chaotic power--law inflation
with an exponential potential.
By suitably redefining a new coarse--grained field
$\phi' \equiv \phi' (\phi - \phi_{cl})$,
as a non--linear function of the difference between the old
$\phi$ and its classical value\footnote{$\phi_{cl}$ is
	the solution of Eq. (\ref{si237}) when the noise term
	is `switched off'.},
the analysis allows one to derive the dispersion of
$\phi'$ around its mean value $\la\phi'\ra$.
This dispersion depends on the scale in question, namely, on the
time a particular scale left the horizon during the inflationary
expansion. It turns out that for scales smaller than our present
horizon (relevant for structure formation) the actual value of
$\phi'$ is very close to $\la\phi'\ra$, and consequently
the distribution of the field $\phi$ is a narrow Gaussian to a
very good approximation and is peaked at the classical
trajectory $\phi_{cl}$.
Instead, if we look at scales much larger than our horizon, the
dispersion of $\phi'$, non--linearly related to the coarse--grained
field $\phi$, can be of the same order as its mean value.
Therefore, the linearisation of this non--linear
$\phi' \equiv \phi' (\phi - \phi_{cl})$ relation does not apply
anymore and, as a consequence of this, a highly non--Gaussian
distribution for $\phi$ is predicted on very large scales
[see Mollerach \etal, 1991 for details].

In the previous paragraph we considered correlations of the
coarse--grained field at different times but at the same point.
Now we will concentrate also on spatial correlations which could be
derived from the Langevin--type equation (\ref{si237}).
Because of the time dependence of the Hubble parameter it has
been proposed that a more fundamental time variable
is given by $\alpha \propto \ln (a) = \int H(t) dt$
[Starobinskii, 1986; see also Chapter \ref{chap-primor}].
{}From the definition of the noise (cf. Eq. (\ref{si238}))
we can compute
\bea
\label{si252}
\lefteqn{
\la f(\vec x_1 , \alpha_1) f(\vec x_2 , \alpha_2) \ra =
{1\over 2\pi^{2}}
\int_0^{\infty} dk k^2 ~ \delta \left(k-k_s(\vec x_1 , \alpha_1)\right)
                       ~ \delta \left(k-k_s(\vec x_2 , \alpha_2)\right)
        }
\nonumber \\
&  &
\times
{d k_s \over d\alpha}(\vec x_1 , \alpha_1)
{d k_s \over d\alpha}(\vec x_2 , \alpha_2)
\phi_k(\alpha_1) \phi^*_k(\alpha_2) j_0(k |\vec x_1 - \vec x_2|) ~,
\eea
which yields the correlation in the noise for spatially separated points.
Now we may compute the spatial correlation for the
fluctuations in the field, defined as
$\delta\phi (\vec x , \alpha) = \phi (\vec x , \alpha) -
\phi_{cl} (\alpha)$. Given that $\phi_{cl}$ is solution of
equation (\ref{si237}) (now written down in terms of the
new time variable $\alpha$) without noise term, we have
\be
\label{si25boh}
\delta\phi (\vec x , \alpha) = \int_0^\alpha
f (\vec x , \alpha') d\alpha' ~,
\ee
and from this
\be
\label{si253}
\la\delta\phi (\vec x_1 , \alpha)
   \delta\phi (\vec x_2 , \alpha') \ra =
\int_0^\alpha d\alpha_1 \int_0^{\alpha'} d\alpha_2
\la f(\vec x_1 , \alpha_1 ) f(\vec x_2 , \alpha_2 ) \ra ~.
\ee
Using (\ref{si252}), (\ref{si253}) may be cast as
\be
\label{si254}
\la\delta\phi (\vec x_1 , \alpha)
   \delta\phi (\vec x_2 , \alpha') \ra =
{1\over 2\pi^{2}}
\int_{k_s(0)}^{{\rm min} (k_s(\alpha) , k_s(\alpha'))}
dk k^2
\phi_k(\alpha_k) \phi^*_k(\alpha'_k) j_0(k |\vec x_1 - \vec x_2|) ~,
\ee
where $\alpha_k$ corresponds to the time when $k = \epsilon a H$.
{}From Eq. (\ref{si236}), in the case of short wavelength modes,
we have
$\phi_k(\alpha_k) \simeq i H \left(\phi(\alpha_k)\right)
\exp( i\epsilon ) / (\sqrt{2} k^{3/2})$, and
Eq. (\ref{si254}) may be cast as
\be
\label{si255}
\la\delta\phi (\vec x_1 , \alpha)
   \delta\phi (\vec x_2 , \alpha') \ra =
{1\over (2\pi)^{2}}
\int_{k_s(0)}^{ k_s( {\rm min} (\alpha , \alpha') ) }
{dk \over k}
H(\phi(\vec x_1 , \alpha_k))
H(\phi(\vec x_2 , \alpha'_k))
j_0(k |\vec x_1 - \vec x_2|) ~.
\ee
This is a function of the value the Hubble parameter
takes on when the scale $k^{-1}$ is equal to the coarse--grained
radius at each of the points.

We saw before that if we want to concentrate on
scales of the order of, and smaller than, our present horizon
the distribution of values of the coarse--grained field
$\phi$ turns out to be highly peaked around the classical trajectory
$\phi_{cl}$. Then,
for these scales it is a good approximation to take for $H$
the value it would have were the field just described by
$\phi_{cl}$.
To be specific, let us consider again the case of an exponential potential
$V(\phi) \propto \exp (- \lambda \sqrt{8\pi G} \phi )$,
where $\lambda^2 = 2 / p$ (cf. Eq. (\ref{powlaw211})).
In this case the classical solution of Eq. (\ref{si237})
is $\phi_{cl} (\alpha) = \phi_0 - \lambda (8\pi G)^{-1/2}\alpha$,
and so we have
$H(\phi_{cl} (\alpha)) \propto  \exp (- \lambda^2 \alpha / 2)$
and also
$k = \epsilon a_0 H_0 \exp [(1 - \lambda^2 / 2) \alpha_k]$,
where $H_0 = H(\alpha = 0)$, etc.
We finally get [Mollerach \etal, 1991]
\be
\label{si256}
\la\delta\phi (\vec x_1 , \alpha)
   \delta\phi (\vec x_2 , \alpha') \ra =
{H_0^2\over (2\pi)^{2}}
\int_{k_s(0)}^{ k_s( {\rm min} (\alpha , \alpha') ) }
{dk \over k}
\left({k\over k_s(0)}\right)^{-\lambda^2 / (1 - \lambda^2 / 2)}
j_0(k |\vec x_1 - \vec x_2|) ~.
\ee
We see from this that in the case $p >> 1$, namely when
$\lambda \to 0$ (when power--law inflation approaches de Sitter
inflation) we recover the scale invariant spectrum of fluctuations
with amplitude $H_0 / 2\pi$.
However, for an arbitrary value of $\lambda$ in the range
$0<\lambda < \sqrt{2}$ we have a spectral index
of density fluctuations $n = 1 - 2\lambda^2 / (2 - \lambda^2) < 1$.
This  spectral index `tilts' the spectrum, with the net effect of
transferring more power on large scales,
as is the benchmark of power--law inflation [Lucchin \& Matarrese, 1985].

We will come to use extensively the stochastic approach to inflation
reviewed in this section again in Chapter \ref{chap-primor}.
Then we will perform a second--order perturbative expansion around
the classical solution to the Langevin--type equation for the
coarse--grained scalar field, the ultimate aim being the computation
of higher--order correlation functions for the perturbations
in the inflaton.
Upon horizon re--entry of the relevant perturbation scales,
these correlations will be passed on to the peculiar
gravitational potential and, via the Sachs--Wolfe effect, to
the CMB temperature anisotropies.

\section{Topological defects}
\label{sec-topo}
\markboth{Chapter 2. ~THEORIES OF THE EARLY UNIVERSE}
		      {\S ~2.2. ~TOPOLOGICAL DEFECTS}

A central concept of particle physics theories attempting to unify
all the fundamental interactions is the concept of symmetry breaking.
As the universe expanded and cooled down,
first the gravitational interaction, and subsequently
all other known forces: the strong, the weak and the electromagnetic
force, would have begun adopting their own identities.
In the context of the standard hot big bang theory the
spontaneous breaking of fundamental symmetries is realised as a
phase transition in the early universe.
Such phase transitions have several exciting cosmological consequences
and thus provide an important link between particle physics and cosmology.

There are several symmetries which are expected to break down in the
course of time.
In each of these transitions the space--time gets `oriented' by the presence
of a hypothetical force field called the `Higgs field', pervading all the
space. This field orientation signals the transition from a state
of higher symmetry to a final state where the system under consideration
obeys a smaller group of symmetry rules.
As an every--day analogy we may consider the transition from liquid
water to ice; the formation of the crystal structure ice (where water
molecules are arranged in a well defined lattice), breaks the symmetry
possessed when the system was in the higher temperature liquid phase,
when every direction in the system was equivalent.
In the same way, it is precisely the orientation in the Higgs field
which breaks the highly symmetric state between particles and forces.

Kibble [1976] made use of a model in which the phase transition
proceeds by the formation of uncorrelated domains that subsequently
coalesce, leaving behind relics in the form of defects.
Such relic `flaws' are unique examples of incredible amounts of energy
and this feature attracted the minds of many cosmologists.
In the expanding universe, widely separated regions in space have not had
enough time to `communicate' amongst themselves and are therefore not
correlated, due to a lack of causal contact.
It is therefore natural to suppose that different regions ended up
having arbitrary orientations of the Higgs field and that, when they
merged together, it was hard for domains with very
different preferred directions to adjust themselves and fit smoothly.

Most interestingly, unlike other proposed mechanism
for the generation of cosmological observable features,  topological
defects can be reproduced in the laboratory! This is possible thanks to
 `scaling' properties with which these defects are endowed.
In fact, when all relevant lengths are uniformly scaled down,
experimentalist have within their reach a physically equivalent situation
as that in the early universe but with all the advantages of a manageable
laboratory experiment.
In 1985, Zurek [1985]  proposed using the transition from normal to
superfluid $^4$He to test the Kibble mechanism. His idea was that after a
rapid quench, defects would form throughout the system.
By studying their
formation,
interesting hints for cosmology could be
	discovered\footnote{Note
	however that the analogy is not complete; there are some
	differences between the two classes of systems.
	In condensed matter systems the dynamics is in general
	non--relativistic and friction dominated, whereas in flat
	spacetime defect motions are highly relativistic, even in the
	case of being in presence of a surrounding plasma that may
	eventually damp somewhat its dynamics.}.
Initially the superfluid transition of helium turned out to
be hard to deal with; temperatures of order $\sim$2K were necessary and
very extreme laboratory conditions were required.
However, people managed to perform similar experimental tests by resorting to
particular organic compounds called `liquid crystals' [de Gennes, 1974].
These substances undergo transitions at temperatures ranging from
10$^\circ$ to 200$^\circ$ centigrade and generate structure easily
detectable with the naked eye or with a microscope
[Chuang \etal, 1991; Bowick \etal, 1994].
After an initial transient, the created network of defects evolves in a
self--similar manner and, although dynamically changing sizes,
always looks the same (in a statistical sense).

Recently McClintock and co--workers
[Hendry \etal, 1994; see also Yurke, 1994]
have succeeded in carrying out Zurek's original idea.
By making a fast adiabatic expansion through the critical density in
$^4$He they were able to observe copious production of quantised
vortices (the condensed--matter analogue of cosmic strings).
These different tests (both with liquid crystals and helium)
provide a kind of `experimental confirmation' of cosmological
topological defect theory, increasing the credibility in these ideas.

These `non--conventional' theoretical models
for generating large--scale structure have been very well studied and
highly detailed simulations have been performed to test and
confront their predictions against a large bulk of
observational material
[see Vilenkin \& Shellard, 1994 for an updated review].
Not only can they account for a host of astrophysical data,
but they also do it in a way so transparent and economical that they
seriously challenge the so far `standard' inflationary model.
One severe testing ground for these models is given by the cosmic
microwave background anisotropy test.
According to the big bang theory, the extremely high energies reigning
during the first instants of time in the early universe
favoured the close interaction between matter and radiation.
Perturbations in this primordial plasma remained imprinted in the
radiation that decoupled from matter around a hundred thousand years
after the bang.
These signatures in the relic radiation are of key importance in trying to
unveil the actual mechanism behind the origin of the large--scale structure
surrounding us today.
Any proposed `seed' for structure formation
should not produce too much anisotropy
in the CMB radiation in order to be consistent with current experiments.
This turns out to be one of the most stringent tests on the models.

{\sl Cosmic strings}
can account for the formation of large--scale filaments and
sheets, observed galaxy formation rates and galactic magnetic fields.
They also generate peculiar velocities on large scales,
and are consistent with the statistical properties
of the CMB anisotropies measured by {\sl COBE}--DMR on large angular scales.
{\sl Texture} simulations are able to reproduce galaxy--galaxy correlation
functions, get clusters of galaxies significantly clustered on the
correct scales and reproduce large--scale streaming motions as observed.
Moreover, this scenario predicts the existence of density peaks that are
much more numerous than expected from the standard inflationary mechanism of
generation of structure. This in turn leads to earlier galaxy formation
and could help in explaining the existence of very old objects, like quasars.

\subsection{Phase transitions and finite temperature field theory}
\label{sec-PhaseTrans}

Phase transitions are known to occur in the early universe.
Examples are the quark to hadron (confinement) transition, which QCD
predicts at an energy around 1 GeV, and the electroweak phase transition
at about 250 GeV.
Within grand unified theories, aiming to describe the physics beyond
the standard model, other phase transitions are predicted to occur
at energies of order $10^{15}$ GeV; during these, the Higgs field
tends to fall towards the minima of its potential while the overall
temperature of the universe decreases as a consequence of the expansion.

A familiar theory to make a bit more quantitative the above
considerations is the $\lambda |\phi|^4$ theory,
\be
\label{lambda4}
{\cal L} = {1\over 2} |\partial_\mu\phi|^2 +
{1\over 2} m_0^2 |\phi|^2
- {\lambda \over 4!} |\phi|^4 ~,
\ee
with $m^2_0 > 0$.
The second and third terms on the right hand side yield the usual
`mexican hat' potential for the complex scalar field.
For energies much larger than the critical temperature, $T_c$,
the fields are in the so--called `false' vacuum: a highly symmetric
state characterised by a vacuum expectation value
$\la | \phi | \ra = 0$.
But when energies decrease the
symmetry is spontaneously broken: a new `true' vacuum develops and the
scalar field rolls down the potential and sits onto
one of the degenerate new minima. In this situation the
vacuum expectation value becomes $\la | \phi | \ra^2 = 6 m_0^2 / \lambda$.

Research done in the 1970's in finite--temperature  field theory
[Weinberg, 1974; Dolan \& Jackiw, 1974; Kirzhnits \& Linde, 1974]
has led to the result that
the temperature--dependent effective potential can be written down as
\be
\label{VfiniT}
V_T( | \phi | ) =
-{1\over 2} m^2(T) |\phi|^2  + {\lambda \over 4!} |\phi|^4
\ee
with
$T_c^2 = 24 m_0^2 / \lambda $,  $m^2(T) = m_0^2 (1 - T^2 / T_c^2)$,
and $\la | \phi | \ra^2 = 6 m^2(T) / \lambda$.
We easily see that when $T$ approaches $T_c$ from below
the symmetry is restored, and again we have $\la | \phi | \ra = 0$.
In condensed--matter jargon, the transition described above
is second--order, or also first--order proceeding by
spinoidal decomposition (caused by a rapid quench in the system)
[Mermin, 1979].\footnote{In a first--order phase transition the order
	parameter (\eg, $\la | \phi | \ra$ in our case) is not continuous.
	It may proceed by bubble nucleation
	[Callan \& Coleman, 1977; Linde, 1983b]
	or by spinoidal decomposition [Langer, 1992].
	Phase transitions can also be continuous second--order
	processes. The `order' depends sensitively on the ratio
	of the coupling constants appearing in the Lagrangian.}

\subsection{The Kibble mechanism}
\label{sec-Kibbbb}

The model described in the last subsection is an example in which the
transition may be second--order.
As we saw, for temperatures much larger than the critical
one the vacuum expectation value of the scalar field vanishes
at all points of space, whereas for $T < T_c$ it evolves smoothly
in time towards a non vanishing $\la | \phi | \ra$.
Both thermal and quantum fluctuations influence the new value
taken by  $\la | \phi | \ra$
and therefore it has no reasons to
be uniform in space.
This leads to the existence of domains wherein the
$\la | \phi (\vec x) | \ra$ is coherent and regions where it is not.
The consequences of this fact are the subject of this section.

Phase transitions can also be first--order
proceeding via bubble nucleation.
At very high energies the symmetry breaking potential
has $\la | \phi | \ra = 0$ as the only vacuum state. When the temperature
goes down to $T_c$ a set of vacua, degenerate to the previous one,
develops. However this time the transition is not smooth as before, for
a potential barrier separates the old (false) and the new (true) vacua.
Provided the barrier at this small temperature is high enough,
compared to the thermal energy present in the system, the field
$\phi$ will remain trapped in the false vacuum state even for small
($< T_c$) temperatures. Classically, this is the complete picture.
However, quantum tunnelling effects can liberate the field from
the old vacuum state, at least in some regions of space:
there is a probability per unit time and volume in space
that at a point $\vec x$ a bubble of true vacuum will nucleate.
The result is thus the formation of bubbles of true vacuum
with the value of the field in each bubble being independent of
the value of the field in all other bubbles.
This leads again to the formation of domains where the fields are
correlated, whereas no correlation exits between fields belonging
to different domains.
Then, after creation the bubble will expand at the speed of light
surrounded by a `sea' of false vacuum domains.
As opposed to second--order phase transitions, here the nucleation
process is extremely inhomogeneous and
$\la | \phi (\vec x) | \ra$ is not a continuous function of time.

Let us turn now to the study of correlation lengths and their
r\^ole in the formation of topological defects.
One important feature in determining the size of the
domains where $\la | \phi (\vec x) | \ra$ is coherent
is given by the spatial correlation of the field $\phi$.
Simple field theoretic considerations [see, \eg, Ref. \cite{cope}]
for long wavelength fluctuations of $\phi$ lead to
different functional behaviours for the
correlation function $G(r) \equiv \la \phi(r_1)\phi(r_2) \ra$, where
we noted $r = |r_1 - r_2|$.
What is found depends radically on the whether the wanted correlation
is computed between points in space separated by a distance $r$
much smaller or much larger than a characteristic length
$\xi^{-1} = m(T) \simeq \sqrt{\lambda} ~ |\la\phi\ra |$, known as
the {\sl correlation length}.
We have
\be
\label{llave}
G(r) \simeq
\cases{
{T_c \over 4 \pi r} \exp (- {r\over \xi}) ~~~~~~~~  r >> \xi      \cr
{}~~~                                                               \cr
{T^2 \over 2 r^2}  ~~~~~~~~~~~~~~~~~~~                r << \xi ~.      }
\ee

This tells us that domains of size $ \xi \sim m^{-1}$ arise where
the field $\phi$ is correlated.
On the other hand, well beyond $\xi$ no correlations exist and thus
points separated apart by $r >> \xi$ will belong to domains with
in principle arbitrarily different orientations of the Higgs field.
This in turn leads, after the merging of these domains in a
cosmological setting, to the existence of defects, where field
configurations fail to match smoothly.

However, when $T \to T_c$ we have $m\to 0$ and so $\xi\to\infty$,
suggesting perhaps that for all points of space the field
$\phi$ becomes correlated. This fact clearly violates causality.
The existence of particle horizons $H^{-1}$ in cosmological models constrains
microphysical interactions over distances beyond this causal domain.
Therefore we get an upper bound to the correlation length as
$\xi < H^{-1} \sim t$.

The general feature of the existence of uncorrelated domains
has become known as the Kibble mechanism [Kibble, 1976]
and it seems to be generic to most types of phase transitions.

\subsection{A survey of topological defects}
\label{sec-ASurv}

Different models for the Higgs field lead to the formation of a whole
variety of topological defects, with very different characteristics and
dimensions.
Some of the proposed theories have symmetry breaking patterns leading
to the formation of `domain walls' (mirror reflection discrete
symmetry): incredibly thin planar surfaces
trapping enormous concentrations of mass--energy which separate
domains of conflicting field orientations,
similar to two--dimensional sheet--like structures found in ferromagnets.
Within other theories, cosmological fields get distributed in such a way
that the old (symmetric) phase gets confined into a finite region of space
surrounded completely by the new (non--symmetric) phase. This situation
leads to the generation of defects with linear geometry called
`cosmic strings'.
Theoretical reasons suggest these strings (vortex lines) do not have
any loose ends in order that the two phases not get mixed up.
This leaves infinite strings and closed loops as the only possible
alternatives for these defects to manifest themselves in the early
universe\footnote{`Monopole' is another possible topological defect;
	we  defer its discussion to the next subsection.
	Cosmic strings bounded by monopoles is yet
	another possibility in GUT phase transitions of
	the kind, \eg, ${\bf G}\to {\bf K}\times U(1)\to {\bf K}$.
	The first transition yields monopoles carrying a magnetic charge
	of the $U(1)$ gauge field, while in the second transition the
	magnetic field in squeezed into flux tubes connecting
	monopoles and antimonopoles
	[Langacker \& Pi, 1980].}.

With a bit more abstraction scientists have even conceived other
(semi) topological defects, called `textures'. These are
conceptually simple objects, yet, it is not so easy to imagine
them for they are just global field configurations living on
a three--sphere vacuum manifold (the minima of the effective
potential energy), whose non linear evolution perturbs spacetime.
Turok [1989] was the first to realise that many
unified theories predicted the existence of peculiar Higgs field
configurations known as (texture) knots, and that these could be of
potential interest for cosmology.
Several features make these defects interesting.
In contrast to domain walls and cosmic strings, textures
have no core and thus the energy is more evenly distributed over space.
Secondly, they are unstable to collapse and it is precisely this last
feature which makes these objects cosmologically relevant, for this
instability makes texture knots shrink to a microscopic size, unwind
and radiate away all their energy.
In so doing, they generate a gravitational field that perturbs the
surrounding matter in a way which can seed structure formation.

Let us now explore the conditions for the existence of
topological defects. It is widely accepted that the final goal
of particle physics is to provide a unified gauge theory
comprising strong, weak and electromagnetic interactions
(and some day may be also gravitation, if we really want to go beyond the
Planck scale).
This unified theory is to describe the physics at very high
temperatures, when the age of the universe was slightly bigger than
the Planck time.
At this stage, the universe was in a state with the highest possible
symmetry, described by a symmetry group {\bf G},
and the Lagrangian modeling the system of all possible
particles present should be invariant under the action of the
elements of {\bf G}.

As we explained before, the form of the finite temperature
effective potential of the system is subject to variations during
the cooling down evolution of the universe.
This leads to a chain of phase transitions whereby some of the
symmetries present in the beginning are not present anymore
at lower temperatures.
The first of these transitions may be described as
{\bf G}$\to${\bf H}, where now {\bf H}
stands for the new (smaller) unbroken symmetry group ruling
the system.
This chain of symmetry breakdowns eventually ends up with
$SU(3)\times SU(2)\times U(1)$, the symmetry group underlying the
`standard model' of particle physics, as it should be.

The smaller group {\bf H} contains elements {\bf g} of the
original group that, when acting on the field $\phi$, they
leave it with its expectation value, roughly,
{\bf g}$\phi = \phi$.
However, we are interested in the vacuum manifold, as it is
this the one that tells us something about the state
of minimum energy, where the fields most probably sit down.
In topology theory the manifold of degenerate vacuum states is
identified with the coset space ${\cal M} =$ {\bf G}$/${\bf H}.
Field configurations that are left invariant when acted upon by the
elements of ${\cal M}$ are precisely those that minimise
the free energy.

The importance of the study of the vacuum manifold
lies in the fact that it is precisely the topology of ${\cal M}$
what determines the type of defect that will arise.
Homotopy theory tells us how to map ${\cal M}$ into physical space
in a non--trivial way, and what ensuing defect will be produced.
For instance, the existence of non contractable loops in ${\cal M}$
is the requisite for the formation of  cosmic strings.
In formal language this comes about whenever we have
the first homotopy group
$\pi_1 ({\cal M}) \neq$ {\bf 1}, where {\bf 1} corresponds to
the trivial group.
If the vacuum manifold is disconnected we then have
$\pi_0 ({\cal M}) \neq$ {\bf 1}, and domain walls are predicted to form
in the boundary of these regions where the field $\phi$
is away from the minimum of the potential.
Analogously, if $\pi_2 ({\cal M}) \neq$ {\bf 1} it follows that the
vacuum manifold contains non contractable two--spheres, and
the ensuing defect is a monopole.
Textures arise when ${\cal M}$ contains non contractable
three--spheres
and in this case it is the third homotopy group, $\pi_3 ({\cal M})$,
the one that is non trivial.

\subsection{Local and global monopoles and domain walls}
\label{sec-monoANDdo}

Generically topological defects will be produced
if the conditions for their existence are met. Then
for example if the unbroken group {\bf H} contains a disconnected
part, like an explicit $U(1)$ factor
(something that is quite common in many phase transition
schemes discussed in the literature), monopoles will be
left as relics of the transition. This is due to
the fundamental theorem on the second homotopy group of
coset spaces [Mermin, 1979],
which states that for a simply--connected covering group
{\bf G} we have\footnote{The isomorfism between two groups is
	noted as $\cong$.
	Note that by using the theorem we therefore can reduce the
	computation of $\pi_2$ for a coset space to the computation
	of $\pi_1$ for a group.
	A word of warning: the focus in this thesis is on the physics and
	the mathematically--oriented reader should bear this in mind,
	especially when we will become a bit sloppy with the notation.
	In case this happens, consult the book [Steenrod, 1951]
	for a clear exposition of these matters.}
\be
\pi_2({\bf G} / {\bf H}) \cong \pi_1({\bf H}_0) ~,
\ee
with ${\bf H}_0$ being the component of the unbroken group
connected to the identity.
Then we see that since monopoles are associated with
unshrinkable surfaces in {\bf G}$/${\bf H}, the previous
equation implies their existence if {\bf H} is
multiply--connected.
The reader may guess what the consequences are for
GUT phase transitions: in grand unified theories a semi--simple
gauge group {\bf G} is broken in several stages down to
{\bf H} $= SU(3)\times SU(1)$. Since in this case
$\pi_1({\bf H}) \cong {\cal Z}$,
the integers, we have $\pi_2 ({\bf G} / {\bf H}) \neq$ {\bf 1}
and therefore gauge monopole solutions exist [Preskill, 1979].

Monopoles are yet another example of stable topological defects.
Their formation stems from the fact that the vacuum expectation
value of the symmetry breaking Higgs field has random orientations
($\la\phi^a\ra$ pointing in different directions in group space)
on scales greater than the horizon.
One expects therefore to have a probability of order unity that a
monopole configuration will result after the phase transition
(cf. the Kibble mechanism).
Thus, about one monopole per Hubble volume should arise and
we have for the number density $n_{monop} \sim 1 / H^{-3} \sim
T_c^6 / m_P^3$,
where $T_c$ is the critical temperature, when the transition occurs.
We also know the entropy density at this temperature,
$s \sim T_c^3$, and so the monopole to entropy ratio
is $n_{monop} / s \simeq 100 (T_c / m_P)^3$.
In the absence of non--adiabatic processes after monopole creation
this constant ratio determines their present abundance.
For the typical value  $T_c\sim 10^{14}$ GeV we
have  $n_{monop} / s \sim 10^{-13}$. This estimate leads
to a present $\Omega_{monop} h^2 \simeq 10^{11}$, for the
superheavy monopoles $m_{monop}\simeq 10^{16}$ GeV that are
created\footnote{These are the actual figures for a gauge $SU(5)$ GUT
	second--order phase transition. Preskill [1979] has shown that in
	this case monopole antimonopole annihilation is not effective
	to reduce their abundance. Guth \& Weinberg [1983] did the case
	for a first--order phase transition and drew qualitatively similar
	conclusions regarding the excess of monopoles.}.
This value contradicts standard cosmology and the presently most
attractive way out seems to be to allow for an early period of
inflation: the massive entropy production
will hence lead to an exponential decrease of the initial
$n_{monop} / s$ ratio, yielding $\Omega_{monop}$ consistent
with observations.\footnote{The
	inflationary expansion reaches an end in the so--called
	reheating process, when the enormous vacuum energy driving
	inflation is transferred to coherent oscillations of the inflaton
	field. These oscillations will in turn be damped by the creation
	of light particles whose final fate is to thermalise and reheat
	the universe (cf. \S\ref{sec-Closett}).}
In summary, the broad--brush picture
one has in mind is that of a mechanism that could solve the
monopole problem by `weeping' these unwanted relics out of our sight,
to scales much bigger than the one that will eventually become our
present horizon today.

Note that these arguments do not apply for global monopoles as
these (in the absence of gauge fields) possess long--range
forces that lead to a decrease of their number in comoving
coordinates. The large attractive force between global monopoles and
antimonopoles leads to a high annihilation probability and
hence monopole over--production does not take place.
Simulations performed by Bennett \&  Rhie [1990] showed
that global monopole evolution rapidly settles into a scale
invariant regime with only a few monopoles per horizon
volume at all times.

Given that global monopoles do not represent a danger
for cosmology one may proceed in studying their observable
consequences. The gravitational fields of global monopoles
may lead to matter clustering and CMB anisotropies. Given an
average number of monopoles per horizon of $\sim 4$,
Bennett \& Rhie [1990] estimate a scale invariant spectrum of
fluctuations $( \delta\rho / \rho )_H \sim 30 G \tilde\eta^2$ at horizon
crossing\footnote{The spectrum of density fluctuations on smaller
scales has also been computed.
They normalise the spectrum at $8 h^{-1}$ Mpc and agreement
with observations lead them to assume that galaxies are clustered
more strongly than the overall mass density, this implying a
`biasing' of a few [see Bennett, Rhie \& Weinberg, 1993 for
details].}.
In a subsequent paper they simulate the large--scale
CMB anisotropies and, upon normalisation with {\sl COBE}--DMR,
they get roughly $G \tilde\eta^2 \sim 6 \times 10^{-7}$ in
agreement with a GUT energy scale $\tilde\eta$ [Bennett \& Rhie, 1993].

Let us concentrate now on domain walls, and briefly try to
show why they are not welcome in any cosmological context
(at least in the simple version we here consider -- there is
always room for more complicated (and contrived) models).
If the symmetry breaking pattern is appropriate at least
one domain wall per horizon volume will be formed.
The mass per unit surface of these two-dimensional objects
is given by $\sim \lambda^{1/2} \tilde\eta^3$, where $\lambda$ as usual
is the coupling constant in the symmetry breaking potential
for the Higgs field.
Domain walls are generally horizon--sized and therefore their
mass is given by $\sim \lambda^{1/2} \tilde\eta^3 H^{-2}$. This
implies a mass energy density roughly given by
$\rho_{DW}\sim \tilde\eta^3 t^{-1}$ and we may readily see now
how the problem arises: the critical density goes as
$\rho_{crit} \sim  t^{-2}$  which implies
$\Omega_{DW}(t) \sim (\tilde\eta / m_P)^2 \tilde\eta t$.
Taking a typical GUT value for $\tilde\eta$ we get
$\Omega_{DW}(t\sim 10^{-35}{\rm sec}) \sim 1$ {\sl already} at the
time of the phase transition. It is not hard to imagine that
today this will be at variance with observations; in fact we get
$\Omega_{DW}(t \sim 10^{18}{\rm sec}) \sim 10^{52}$. This
indicates that models where domain walls are produced
are tightly constrained, and the general feeling is
that it is best to avoid them altogether [see Kolb \& Turner, 1990
for further details].

\subsection{Are defects inflated away?}
\label{sec-topoandinfla}

It is important to realise the relevance that the Kibble's mechanism
has for cosmology; nearly every sensible grand
unified theory (with its own symmetry breaking pattern) predicts
the existence of defects.
We have seen in \S\ref{sec-cannot} how an early era of inflation
helped in getting rid of the unwanted relics.
One could well wonder if the very same Higgs field
responsible for breaking the symmetry
would not be the same one responsible for driving  an era of inflation,
thereby diluting the density of the relic defects.
This would get rid not only of (the unwanted) monopoles and
domain walls but also of any other (cosmologically appealing) defect.
Let us sketch why this actually does not occur
(cf. \cite{BrandenbergerRev}).
Take first the symmetry breaking potential of Eq. (\ref{VfiniT})
at zero temperature and add to it
a harmless $\phi$--independent term $3 m^4 / (2\lambda)$. This will not
affect the dynamics at all. Then we are led to
\be
\label{VfiniT2}
V(  \phi  ) =
{\lambda \over 4!} \left(
\phi^2 - {\tilde\eta }^2
\right)^2 ~,
\ee
with $\tilde\eta = ( 6 m^2 / \lambda )^{1/2}$
the symmetry breaking energy scale,
and where for the present heuristic digression we just took a real
Higgs field.
Consider now the equation of motion for $\phi$,
\be
\label{Higgapprox}
\ddot \phi \simeq - {\partial V\over\partial\phi }
= {\lambda\over 3!} \phi^3 - m^2 \phi
\approx - m^2 \phi ~,
\ee
for $\phi << \tilde\eta$ very near the false vacuum of the effective mexican
hat potential and
where, for simplicity, the expansion of the universe and
possible interactions of $\phi$ with other fields were neglected.
The typical time scale of the solution is $\tau\simeq m^{-1}$.
For an inflationary epoch to be effective we need
$\tau >> H^{-1}$, \ie, a sufficiently large number of e--folds
of slow--rolling solution. Note, however, that after some
e--folds of exponential expansion the curvature term in
the Friedmann equation becomes subdominant and we have
$H^2 \simeq 8\pi G ~V(0) / 3 \simeq (2\pi m^2 / 3 )(\tilde\eta / m_P)^2$.
So, unless $\tilde\eta  > m_P$, which seems unlikely for a GUT phase
transition, we are led to $\tau << H^{-1}$ and therefore
the amount of inflation is not enough for getting rid of
the defects generated during the transition by hiding them
well beyond our present horizon.

\subsection{Cosmic strings}
\label{sec-comi}

Cosmic strings are without any doubt the topological defect
most thoroughly studied, both in cosmology and solid--state
physics (vortices).
The canonical example, describing flux tubes in superconductors,
is given by the Lagrangian
\be
\label{lagraCS}
{\cal L} =  -{1\over 4} F_{\mu\nu} F^{\mu\nu}
+ {1\over 2} |D_\mu\phi|^2
- {\lambda \over 4!} \left( |\phi |^2 - {\tilde\eta }^2 \right)^2 ~,
\ee
with $F_{\mu\nu} = \partial_{[\mu}A_{\nu ]}$, where $A_{\nu}$ is
the gauge field and the covariant derivative is
$D_\mu = \partial_\mu + i e A_{\mu}$, with $e$ the gauge coupling
constant.
This Lagrangian is invariant under the action of the
Abelian group ${\bf G}=U(1)$, and the spontaneous breakdown of
the symmetry leads to a vacuum manifold ${\cal M}$ that is a circle,
$S^1$, \ie, the potential is minimised for
$\phi = \tilde\eta\exp (i\theta)$, with arbitrary $0\leq\theta\leq 2\pi$.
Each possible value of $\theta$ corresponds to a particular
`direction' in the field space.

Now, as we have seen earlier, due to the overall cooling down of
the universe, there will be regions where the scalar field
rolls down to different vacuum states.
The choice of vacuum is totally independent for regions
separated apart by one correlation length or more, thus leading
to the formation of domains of size $\xi\sim \tilde\eta^{-1}$.
When these domains coalesce they give rise to edges in the
interface.
If we now draw a imaginary circle around one of
these edges and the angle $\theta$ varies by $2\pi$ then
by contracting this loop we reach a point where we cannot go any
further without leaving the manifold ${\cal M}$. This is
a small region where the variable $\theta$ is not defined and,
 by continuity,
the field should be $\phi = 0$.
In order to minimise the spatial gradient energy these small
regions line up and form a line--like defect called cosmic string.

The width of the string is roughly
$m_\phi^{-1} \sim (\sqrt{\lambda} \tilde\eta)^{-1}$, $m_\phi$ being the
Higgs mass. The string mass per unit length, or tension, is
$\mu \sim \tilde\eta^2$. This means that for GUT cosmic strings,
where $\tilde\eta\sim 10^{16}$ GeV, we have $G\mu \sim 10^{-6}$.
We will see  below that the dimensionless combination $G\mu$,
present in all signatures due to strings,
is of the right order of magnitude for rendering these defects
cosmologically interesting.

There is an important difference between global and gauge (or local)
cosmic strings: local strings have their energy confined mainly
in a thin core, due to the presence of gauge fields $A_\mu$ that
cancel the gradients of the field outside it. Also these gauge
fields make it possible for the string to have a quantised
magnetic flux along the core.
On the other hand, if the string was
generated from the breakdown of a {\sl global} symmetry there are no
gauge fields, just Goldstone bosons, which, being massless, give
rise to long--range forces. No gauge fields can compensate
the gradients of $\phi$ this time and therefore there is an
infinite string mass per unit length.

Just to get a rough idea of the kind of models studied in the
literature, consider the case ${\bf G} = SO(10)$ that is broken to
${\bf H} = SU(5)\times {\cal Z}_2$.  For this pattern we have
$\Pi_1({\cal M}) = {\cal Z}_2$, which is clearly non trivial and therefore
cosmic strings are formed [Kibble \etal, 1982].\footnote{In the
	analysis one uses the
	fundamental theorem stating that, for a simply--connected Lie
	group {\bf G} breaking down to {\bf H}, we have
	$\pi_1({\bf G} / {\bf H}) \cong \pi_0({\bf H})$;
	see [Hilton, 1953].}

\subsection{Observational features from cosmic strings}
\label{sec-comicsigna}

Let us finish this brief account of cosmic strings by
providing just a quick description of their remarkable
cosmological features.

Many of the proposed observational tests for the existence of
cosmic strings are based on their gravitational interactions.
In fact, the gravitational field around a straight static string
is very unusual [Vilenkin, 1981].
A test particle in its vicinity feels no Newtonian
attraction; however the existence of a deficit angle
$\Delta = 8\pi G \mu$ makes
the topology of space around the string that of a cone.
Two particles initially at rest while the string is far away,
will suddenly begin moving towards each other after the string has
passed between them. Their head--on velocities will be proportional
to $\Delta$.
Hence, the moving string will built up a {\sl wake} of particles behind
it that may eventually form the `seed' for accreting
more matter into sheet--like structures.

Also, the peculiar topology around the
string makes it act as a cylindric gravitational lens that may
produce double images of distant light sources, \eg, quasars.
The angle between the two images produced by a typical GUT string
would be $\propto G\mu$ and of order of a few seconds of arc,
independent of the impact parameter and
with no relative magnification between the images.

The situation gets even more interesting when we allow the string
to have small--scale structure, called wiggles, as in fact simulations
indicate.
The wiggles modify its effective mass per unit length, $\tilde\mu$,
and also built up a Newtonian attractive term in the velocity boost
inflicted on nearby test particles.
These wiggles also produce inhomogeneities in the wake of accreting
matter and may lead to the fragmentation of the structure.
The `top--down' scenario of structure formation thus follows
naturally in a universe with strings.

If cosmic strings really exist, the anisotropies in the CMB
they produce would have a characteristic signature.
One such effect comes about due to the Doppler shift
that background radiation suffers when a string intersects the
line of sight.
The conical topology around the string will produce a differential
redshift of photons passing on different sides of it,
$\D \approx 8 \pi G \tilde\mu v \gamma$, with
$\gamma = (1-v^2)^{-1/2}$ the Lorentz factor and $v$ the velocity
of the moving string.
This `stringy' signature was first studied by
Kaiser \& Stebbins [1984] and Gott [1985].
We will return to it in Chapter \ref{sec-kurtostrings} where we will
also give a more detailed description of this effect, and implement it
within an analytic model for the study on the CMB excess kurtosis
parameter.

\subsection{Global textures}
\label{sec-texuuu}

Whenever a global non--Abelian symmetry is spontaneously and
completely broken (hopefully at a grand unification scale), global
defects called textures are generated.
Theories where this global symmetry is only partially broken
do not lead to global textures, but instead to
global monopoles and non--topological textures.
As we already mentioned global monopoles do not suffer
the same constraints as their gauge counterparts
(cf. \S\ref{sec-monoANDdo}): essentially,
having no associated gauge fields they do not possess magnetic
charge and therefore cannot be accelerated by the magnetic
field in our galaxy, unlike the gauge monopoles.
Thus there will be no `evaporation' flux of monopoles from
the halo of the galaxy and the Parker limit [Parker, 1970]
does not apply.
On the other hand, non--topological textures are a generalisation
that allows the broken subgroup {\bf H} to contain non--Abelian
factors. It is then possible to have $\pi_3$ trivial as in,
\eg, $SO(5)\to SO(4)$ broken by a vector, for which case
we have ${\cal M} = S^4$, the four--sphere [Turok, 1989].
Having explained this, let us concentrate in global topological
textures from now on.

Textures, unlike monopoles or cosmic strings, are not
well localised in space. This is due to the fact that the field
remains
in the vacuum everywhere, in contrast to what happens for
other defects, where the field leaves the vacuum manifold precisely
where the defect core is.
Since textures do not possess a core, all the energy of the field
configuration is in the form of field gradients.
This fact is what makes them interesting objects {\sl only}
when coming from global theories: the presence of gauge
fields $A_\mu$ could (by a suitable reorientation) compensate the
gradients of $\phi$ and yield $D_\mu\phi = 0$, hence
canceling out (gauging away) the energy of the
configuration\footnote{This does not imply, however, that
the classical dynamics of a gauge texture is trivial. The
evolution of the $\phi$--$A_\mu$ system will be determined
by the competing tendencies of the global field to unwind and
of the gauge field to compensate the $\phi$ gradients. The result
depends on the characteristic size $L$ of the texture:
in the range $m_\phi^{-1} << L << m_A^{-1} \sim (e\tilde\eta)^{-1}$
the behaviour of the gauge texture resembles that of the global
texture, as it should, since in the limit $m_A$ very small
($e\to 0$) the gauge texture turns into a global one
[Turok \& Zadrozny, 1990].}.

One feature endowed by textures that really makes these
defects peculiar is their being unstable to collapse.
The initial field configuration is set at the phase transition,
when $\phi$ develops a nonzero vacuum expectation value.
$\phi$ lives in the vacuum manifold ${\cal M}$ and winds
around ${\cal M}$ in a  non--trivial way on scales greater
than the correlation length, $\xi \lsim t$.
The evolution is determined by the nonlinear dynamics
of $\phi$.
When the typical size of the defect becomes
of the order of the horizon, it collapses on itself.
The collapse continues until eventually the size of the defect
becomes of the order of $\tilde\eta^{-1}$, and at that point the
energy in gradients is large enough to raise the field
from its vacuum state.
This makes the defect unwind, leaving behind a trivial
field configuration.
As a result $\xi$ grows to about the
horizon scale, and then keeps growing with it.
As still larger scales come across the horizon knots are constantly
formed, since the field $\phi$ points in different directions
on ${\cal M}$ in different Hubble volumes.
This is the scaling regime for textures, and when it holds
simulations show that one should expect to find of order
0.04 unwinding collapses per horizon volume per Hubble time
[Turok, 1989].
However, unwinding events are not the most frequent feature
[Borrill \etal, 1994], and when one considers random field
configurations without an unwinding event the number
raises to about 1 collapse per horizon volume per Hubble time.
We will be using these results from the simulations in Chapter
\ref{chap-AnaModTex} when we will consider an analytic
texture model.

During the radiation era, and when the correlation length is
already growing with the Hubble radius, the texture field
has energy density
$\rho_{texture}\sim (\nabla\phi)^2 \sim \tilde\eta^2 / H^{-2}$, and
remains a fixed fraction of the total density
$\rho_{c} \sim t^{-2}$ yielding
$\Omega_{texture} \sim 8 G \tilde\eta^2$. Thus we do not need
to worry about textures dominating the universe.

Needless to say, the texture collapse generates perturbations
in the metric of spacetime. These in turn will affect the photon
geodesics leading to CMB anisotropies, the clearest possible
signature to probe the existence of these exotic objects being the
appearance of hot and cold {\sl spots} in the microwave maps.
Due to the scaling behaviour mentioned above, the density
fluctuations induced by textures on any scale at horizon
crossing are given by $(\delta\rho / \rho )_H  \sim 8 G \tilde\eta^2$.
CMB temperature anisotropies will be of the same amplitude.
Numerically--simulated maps, with patterns smoothed over
$10^\circ$ angular scales, by Bennett \& Rhie [1993]
yield, upon normalisation to the {\sl COBE}--DMR
data, a dimensionless value $G \tilde\eta^2 \sim 10^{-6}$, in
good agreement with a GUT phase transition energy scale.

\subsection{Evolution of global textures}
\label{sec-texevol}

We mentioned earlier that the breakdown of any non--Abelian
global symmetry led to the formation of textures. The simplest
possible example involves the breakdown of a global
$SU(2)$ by a complex doublet $\phi^a$, where the latter may
be expressed as a four--component scalar field, \ie,
$a=1\ldots 4$.
We may write the Lagrangian of the theory much in the same
way as was done in Eq. (\ref{lagraCS}), but now we drop the
gauge fields (thus the covariant derivatives become partial
derivatives).
Let us take the symmetry breaking potential as follows,
$V(  \phi  ) = {\lambda \over 4} \left( |\phi|^2 - {\tilde\eta }^2
\right)^2$. The situation in which a global $SU(2)$ in broken
by a complex doublet with this potential $V$ is
equivalent to the theory where $SO(4)$ is broken by a
four--component vector to $SO(3)$, by making $\phi^a$
take on a vacuum expectation value.
We then have the vacuum manifold ${\cal M}$ given by
$SO(4) / SO(3) = S^3$, namely, a three--sphere with
$\phi^a\phi_a = \tilde\eta^2$.
As $\pi_3 (S^3) \not= {\bf 1}$
(in fact, $\pi_3 (S^3) = {\cal Z}$)
we see we will have non--trivial solutions
of the field $\phi^a$ and global textures will arise.


As usual, variation of the action with respect to the
field  $\phi^a$ yields the equation of motion
\be
\label{phieqn}
{\phi^b}'' + 2 {a' \over a} {\phi^b}' - \nabla^2 \phi^b =
- a^2 {\partial V \over\partial\phi^b } ~,
\ee
where primes denote derivatives with respect to conformal time
and $\nabla$ is computed in comoving coordinates.
When the symmetry in broken three of the initially four
degrees of freedom go into massless Goldstone bosons
associated with the three directions tangential to the
vacuum three--sphere. The `radial' massive mode that remains
($m_\phi \sim \sqrt{\lambda}\tilde\eta$)
will not be excited, provided we concentrate on length scales
much larger than $m_\phi^{-1}$.

To solve for the dynamics of the field $\phi^b$,
two different approaches have been implemented in the literature.
The first one faces directly the full equation (\ref{phieqn}),
trying to solve it numerically.
The alternative to this exploits the fact that, at temperatures
smaller than $T_c$, the field is constrained to live in the true vacuum.
By implementing this fact via a Lagrange multiplier\footnote{In fact,
in the action the coupling constant $\lambda$ of the `mexican hat'
potential is interpreted as the Lagrange multiplier.}
we get
\be
\label{phieqnsigma}
\nabla^\mu\nabla_\mu\phi^b =
- {\nabla^\mu\phi^c\nabla_\mu\phi_c \over \tilde\eta^2} \phi^b ~~ ; ~~
\phi^2 = \tilde\eta^2  ~,
\ee
with $\nabla^\mu$ the covariant derivative operator.
Eq. (\ref{phieqnsigma}) represents a non--linear sigma model
for the  interaction of the three massless modes
[Rajaraman, 1982].
This last approach  is only valid when probing length scales
larger than the inverse of the mass $m_\phi^{-1}$.
As we mentioned before, when this condition is not met the
gradients of the field are strong enough to make it leave
the vacuum manifold and unwind.

The approach (cf. Eqs. (\ref{phieqnsigma})) is suitable for
analytic inspection. In fact, an exact flat space solution was found
assuming a spherically symmetric ansatz. This solution
represents the collapse and subsequent conversion of a texture knot
into massless Goldstone bosons,
and is known as the  spherically symmetric  self--similar (SSSS)
exact unwinding solution.
We will say no more here with regard to the solution, but
just refer the interested reader to the original articles
[see, \eg, Turok \& Spergel, 1990; Notzold ,1991].
In what follows we will be mainly interested in considering the
results of the numerical simulations from Eq. (\ref{phieqn}),
the reason being that these simulations take full account also of
the energy stored in gradients of the field, and not just
in the unwinding events. We will apply the results of the
simulations performed in [Durrer \& Zhou, 1995] in
Chapter \ref{chap-texture} below.

\chapter{CMB anisotropies}
\label{chap-CMBanis}
\markboth{Chapter 3. ~CMB ANISOTROPIES}
         {Chapter 3. ~CMB ANISOTROPIES}

The standard method of characterising the CMB fluctuation
spectrum is by means of a multipole moment expansion:
current experiments measure the distribution
of temperatures in the sky, $\Delta T / T (\gg)$, where
$\gg$ points in a certain direction.
Similar to what one does in Fourier analysis, namely, the study
of a function by means of its expansion in plane waves,
in our case we expand any given function defined on the sky
(like $\Delta T / T$) in spherical harmonics, $Y_\ell^m$.
Since we are only able to measure mean values for the
stochastic distribution of temperature fluctuations,
and since physical quantities should
not depend on the particular direction we are looking at
(isotropy assumption on very large scales),
we see that at the end of the day only the multipole index $\ell$
will be relevant.\footnote{We are necessarily
	being a bit loose at
	this stage in trying to give a brief introduction. A detailed
	account of these matters will be given in Chapter
	\ref{chap-statistics}.}

The temperature autocorrelation function
tells us about possible anisotropies in the radiation coming
from points in the sky separated by an angle $\vartheta$, and
is given by

\be
\label{poccc}
\la C_2(\vartheta) \ra
= \langle{\Delta T\over T}({\gg}){\Delta T\over T}({\gg}')\ra
= {1\over 4\pi}\sum_\ell(2\ell+1){\cal C}_\ell P_\ell(\cos\vartheta)
\ee

\noindent
where $\la\cdot\ra$ means averaging on the sky and
$\gg\cdot\gg' = \cos\vartheta$.

The quantity $\ell (\ell + 1) {\cal C}_{\ell}$ is usually referred to as
the angular power spectrum.
With this definition, an initial distribution of density fluctuations
with $n = 1$ produces a flat ($\ell$--independent) spectrum on
large angular scales ($\ell$ small), namely, a `plateau' in the
$\ell (\ell + 1) {\cal C}_{\ell}$ vs. $\ell$ plot.
Notice that both inflation and topological defect models lead to
approximately scale invariant spectra on large scales; therefore,
observations on large scales alone cannot by themselves
select one, and not the other, of these competing models.

Interestingly enough, going to smaller scales it turns out that
power spectra predicted from inflationary and defect models
are not similar: the Doppler peaks expected on angular scales
$\sim 1^\circ$ might be a helpful tool to discriminate between them;
see Chapter \ref{chap-texture}.

On even smaller scales the details of the process of
recombination become important:
the last scattering surface has a finite thickness, which is due to
the noninstantaneous recombination of hydrogen, and to the residual
ionisation level remaining.
The finite width of this shell smooths out fluctuations on scales
$\sim 8 \Omega^{-1/2}$ arcmin or smaller.

The detection and detailed analysis of temperature anisotropies
on different angular scales provide a unique measure of the
primordial density fluctuations from which large--scale structure
evolved, as well as a direct probe of the as yet uncertain values
of the cosmological parameters.
In the following sections we will enumerate the different
sources of temperature anisotropies as well as the physical
processes responsible for their suppression, together with
a brief explanation of the physics behind them.

\section{Contributions to $\D$}
\label{sec-contributions}
\markboth{Chapter 3. ~CMB ANISOTROPIES}
      {\S ~3.1. ~CONTRIBUTIONS TO $\D$}

Several are the authoritative reviews on this subject
[\eg, Efstathiou, 1990; White \etal, 1994],
and we refer the reader to them for more detailed presentations of the
items treated here.
In this section we will briefly account for the main effects that
may leave their imprint in the CMB temperature anisotropies.

\begin{flushleft}
{\bf Dipole anisotropy }
\end{flushleft}

The {\it dipole} anisotropy was for a long time,
until the {\sl COBE} detection,
the only measured temperature variation.
Both `intrinsic' and `extrinsic' contributions are present
in the maps
and that is why it is source of large uncertainties.
Essentially, the extrinsic part is

\be
\D \simeq {v\over c} \cos\theta ~,
\ee

\noindent
where $\theta$ is
the angle between the direction of motion and that of
observation,
and is mainly due to our peculiar motion with respect to
the comoving rest frame of the CMB radiation.
The intrinsic dipole on the other hand is expected to be two orders of
magnitude smaller and
of comparable amplitude to the quadrupole.\footnote{On the same footing,
the amplitude of the extrinsic quadrupole is a factor $\sim \frac{v}{c}$
that of the extrinsic dipole.}

This dipole anisotropy was detected and from its trace in the $\D$ maps
both $v$ and the direction of our peculiar motion can be determined.
The best estimate to date is from {\sl COBE} which gives a speed of
$365 \pm 18 ~{\rm km} ~ {\rm sec}^{-1}$ [Smoot \etal, 1991].
When this velocity is decomposed and
{}~i) the motion of the Earth around the Sun,
ii) the motion of the Sun in the galaxy and, finally,
iii) the motion of the Milky Way towards Andromeda in the Local Group (LG)
are subtracted, the motion of our LG may be determined to be
$\sim 600 ~{\rm km} ~ {\rm sec}^{-1}$
towards the direction (in galactic coordinates)
$\ell \simeq 270^\circ$, $b \simeq 30^\circ$.\footnote{For a
comprehensive introduction to these matters consult
Padmanabhan [1993].}

\begin{flushleft}
{\bf Sachs--Wolfe effect}
\end{flushleft}

The second effect, and the most relevant one on very large scales,
is the {\sl Sachs--Wolfe} effect [Sachs \& Wolfe, 1967].
When focusing on these perturbations we are facing rather
`simple' physics:
since the scales involved are much larger than the size of the horizon at
last scattering the analysis is to a great extent
transparent to the microphysics of recombination, and consequently
also to some of the still unknown cosmological parameters, like
$\Omega_B$,
$\Omega_{CDM}$,
$\Omega_{HDM}$,
$\Omega_\Lambda$,
$h$,
and
ionisation history.\footnote{On the other hand results for these scales
{\sl are} sensitive to other parameters such as relative (amplitude)
contributions to $\D$ from energy density (scalar metric),
gravitational wave (tensor metric), isocurvature scalar and other
primordial fluctuations, and
to the shape of the initial fluctuation spectra (as parameterised by the
spectral indexes of the different components).}

The physics behind Sachs--Wolfe fluctuations is straightforward: matter
perturbations distort space, and therefore also distort geodesics of
photons that were last scattered at the recombination era.
The net effect can be expressed as $\D \sim - \Phi $.
Since the perturbations in the gravitational potential $\Phi$
are time independent in the linear regime, we only need to worry about
$\Phi$ on the last scattering surface.
Blueshift and redshift cancellation effects ensure that distortions
to the CMB photons on their way to us are unimportant.\footnote{Note
however that we expect perturbations to go non--linear in the end,
and therefore intervening clumps will undoubtedly play a r\^ole --
see below.}

\newpage

\begin{flushleft}
{\bf Intrinsic fluctuations}
\end{flushleft}

In this paragraph we will consider fluctuations intrinsic to the radiation
field itself on the last scattering surface.
These vary according to whether the primordial fluctuations are
adiabatic or isocurvature in nature.

{\sl Adiabatic} perturbations are those for which the densities in matter and
radiation (and by the equivalence principle also all other components
of the energy density)
are perturbed so that the entropy per baryon is the same as in the
unperturbed state.\footnote{For an initial
state to be adiabatic (or isoentropic) it is necessary that,
for any two types of
components, say A and B, the ratio of the number density of the A particles
to the B particles, $n_A/n_B$ be a constant independent of position.
The origin of this terminology draws back to the universe with just radiation
and plasma content, wherein $n_{rad}/n_B$ is a measure of the entropy per
baryon.}

Since the radiation field energy density goes like
$\rho_{rad} \propto T^4$
(and remembering that the density contrast is defined as
$\delta\equiv\frac{\delta\rho}{\rho}$),
we have
$\D = \frac{1}{4} \delta_{rad} = \frac{1}{3} \delta_{mat}$, where the last
equality follows from the adiabatic condition. Now, since the CMB photons
start off their trip during matter domination, one has
$\delta\rho =  \rho_{mat} \delta_{mat} +
\rho_{rad} \delta_{rad} \simeq \rho \delta_{mat}$.
Thus we get
$\D = \frac{1}{3} \frac{\delta\rho}{\rho}$
for adiabatic perturbations.

On the other hand, {\sl isocurvature} perturbations do not affect the total
energy density and are therefore characterised by $\delta\rho = 0$
[see, \eg, Kolb \& Turner, 1990 for a review].
They correspond to perturbations in the form of the local equation of state.
Let us consider a two--fluid system, \eg, a non--relativistic CDM component
plus radiation. Then, we have  $\rho = m_{cdm} n_{cdm} + \rho_{rad}$
and thus
$\D = - \frac{1}{4} (\rho_{cdm}/\rho_{rad}) (\delta n_{cdm}/n_{cdm})$.
We call $s$ the entropy density ($s \propto T^3$ is proportional to
the number density of relativistic particles)
and $n_{cdm}$ the number density of CDM particles, thus
$s/n_{cdm}$ is a measure of the entropy per particle.

It turns out to be convenient to work with the following
perturbation in the number density
$\delta(n_{cdm}/s)/(n_{cdm}/s) = \delta(n_{cdm})/(n_{cdm}) - 3  \D$,
and with this we get

\be
\label{deltatisocur}
\D = - {1\over 4}
{\delta(n_{cdm}/s) \over   (n_{cdm}/s) }
\left(
{
(\rho_{cdm} / \rho_{rad})
\over
(1 + \frac{3}{4} \rho_{cdm} / \rho_{rad})
}\right).
\ee

Now, at late times the universe becomes matter dominated
$\rho_{cdm} >> \rho_{rad}$, and thus
$\D \simeq - \frac{1}{3} \delta(n_{cdm}/s)/(n_{cdm}/s)$.
When a given  mode becomes smaller than the horizon, microphysical
processes play a r\^ole and fluctuations in the local pressure can generate
fluctuations in the energy density. From that moment onwards the difference
between adiabatic and isocurvature is irrelevant.
Further, the perturbation in the number density becomes a density
perturbation of the same amplitude
$\delta(n_{cdm}/s)/(n_{cdm}/s) \sim \delta\rho / \rho$
yielding finally
$\D \simeq - \frac{1}{3} \delta\rho / \rho$.

Inflationary models  are normally invoked as examples
of a scenario for the generation of
adiabatic perturbations, by means of quasi--de Sitter
fluctuations
in the inflaton.
On the other hand, cosmic strings provide an example of isocurvature
perturbations (of course, not the only one, but an interesting one):
during the phase transition when these defects are
produced, some of the radiation energy density goes into the formation
of the network of strings. Thus, the total energy density remains
unaltered while a change in the effective equation of state of the
matter takes place in those regions where the defects arise
(we will say more on strings and their
influence on the CMB anisotropies in \S\ref{sec-introstrings}).

\begin{flushleft}
{\bf Acoustic oscillations and the Doppler peaks}
\end{flushleft}

On smaller scales, say $\sim 1^\circ $, one is effectively
probing those perturbations with characteristic scales of order
of the horizon distance, or smaller than it, at recombination.
Microphysical processes play definitely an important r\^ole now.

Baryons and photons are still tightly coupled through
Compton scattering and may be studied as a single fluid.
Oscillations in a given mode of the density perturbations
are expected when pressure forces dominate over gravity.
While in the case of CDM the fluid is pressureless and thus
always gravity wins, in generic fluids the behaviour of an
arbitrary mode will depend strongly on the sound speed in the
particular fluid.
This latter in turn determines the Jeans length
$\lambda_J$, which
is precisely that perturbation wavelength for which the
pressure and gravity forces exactly balance
(refer to \S\ref{sec-Gravinstab}): near recombination
we have $\lambda_J \sim 2\pi c_s H^{-1}$, with $c_s\sim 1/\sqrt{3}$.
Once the wavelength of a given mode falls below the Jeans length,
pressure forces take over and therefore acoustic oscillations
appear.

These oscillations arise both in the density and velocity
perturbations, and together lead to the existence of peaks in the
$\ell (\ell + 1) {\cal C}_\ell$ vs. $\ell$ plot as remnants of the time
at which matter and radiation decoupled.
Even though the terminology might be somewhat misleading,
these features in the angular power spectrum are known under
the name of `Doppler peaks', although {\sl no} Doppler effect
is taking place since the components are oscillating together
and little or no velocity difference exists between them.

\begin{flushleft}
{\bf Sunyaev--Zel'dovich fluctuations}
\end{flushleft}

Non primeval in origin but at the root            
of spectral distortions detected in a few reach clusters,
Sunyaev--Zel'dovich fluctuations have been very well studied
[see, \eg, Zel'dovich \& Sunyaev, 1969;
Sunyaev \& Zel'dovich, 1970].
A generic feature in structure formation scenarios is the existence of
large regions of tenuous hot gas, left behind in the process of
clumping and subsequent fragmentation of matter.
The Sunyaev--Zel'dovich effect arises as a consequence of the CMB photons
passing through these clouds of hot electrons, where they are
(inverse Compton) scattered and get  distorted.
The effect may be expressed
as an equivalent temperature perturbation as a function of the
wavelength, and on the Rayleigh--Jeans side  yields

\be
\D \sim - 2 k_B T_e / (m_e c^2) ~,
\ee

\noindent
where
$k_B$ is the Boltzmann constant, and $T_e$ and $m_e$ refer to the
electron's temperature and mass, respectively.

\begin{flushleft}
{\bf Secondary anisotropies: the Vishniac effect}
\end{flushleft}

If the binding energy of the gaseous components of newly bound structures
is thermalised, then temperature fluctuations may result on
sub--arcminute scales.
We will see below that primary fluctuations might be greatly suppressed on
these scales due to the smearing associated with the finite thickness of
the last scattering surface, and thus newly--generated secondary
anisotropies may dominate.
While in principle there are several sources in second--order
perturbation theory, there is a contribution, coming from a product
of velocities and densities, that dominates over all others
[\eg, Hu \etal, 1994; Dodelson \& Jubas, 1995]; this is known as
the Vishniac effect [Vishniac, 1987].
This contribution turns out
to be negligible for standard ionisation histories; it is however
important whenever we allow for late reionisation, as infall of
baryons into the CDM perturbation wells plus the growth of the
perturbation in the intervening period makes this a relevant
source of anisotropies.

\newpage

\section{Suppression of anisotropies}
\label{sec-suppression}
\markboth{Chapter 3. ~CMB ANISOTROPIES}
{\S ~3.2. ~SUPPRESSION OF ANISOTROPIES}

\begin{flushleft}
{\bf Width of the last scattering surface}
\end{flushleft}

We explained before (\eg, in \S\ref{sec-TherEvol}) how
the expansion of the universe makes the mean energy density
gradually to decrease, and that this produces the decoupling
of radiation from matter (due to the lower abundance of partners
to interact in order to keep the equilibrium).
Now, it may well happen that the last scattering surface was in fact
not sudden, \ie, that it took some time for the majority of photons to
freely stream away from their interactions with matter.
This would be the most plausible situation and thus
we see that this surface actually has a `width', which means that
every time we look in some direction on the sky we are in fact detecting
a bunch of photons that were emitted from different points along the line
of sight.
This implies a kind of averaging of the fluctuations,
and a destructive interference between the modes with
wavelengths smaller than the thickness of the last
scattering surface, with the result of a net
`washing out' of perturbations on very small angular scales, of
order $8 \Omega^{-1/2}$ arcmin.

\begin{flushleft}
{\bf Silk damping}
\end{flushleft}

When focusing on very small scales it becomes apparent that
baryons and radiation are in reality imperfectly coupled.
Photons possess a mean free path $\lambda_C$ in the baryons
due to Compton scattering.
Due to the frequent collisions, suffered by a large part
of the photons, they randomly walk through the
baryons with the result that hot and cold spots get mixed.
Fluctuations only remain in a relatively small unscattered
fraction of them, causing a nearly exponential decrease
in amplitude as the characteristic scale of the perturbation
becomes smaller than the diffusion length
$\lambda_D\sim\sqrt{N}\lambda_C$.
Photons
tend to stream away of the otherwise growing perturbations and
in so doing they smooth the distribution of fluctuations
(photon diffusion).
This effect ends up by suppressing the level of
anisotropies on very small scales, and is known as the Silk damping
[Silk, 1968].

Both the finite thickness of the last scattering surface and
Silk damping are the most efficient mechanisms for the
suppression of anisotropies on very small angular scales.
We may see this in Figure \ref{steinfig} below;
these effects are the responsible for the suppression
of the angular power spectrum for $\ell > 1000$.

\begin{flushleft}
{\bf Early reionisation}
\end{flushleft}

At last scattering the mean free path of the photons goes like
$\lambda_C\propto (x_e n_b)^{-1}$. $\lambda_C$ increases due to the
decrease in the ionisation fraction $x_e$ during recombination
[Hu, Sugiyama \& Silk, 1995].
{}From this we see that the diffusion scale is dependent on the
ionisation history (as well as on the number density in
baryons, $n_b$) and therefore a delay in last scattering,
as might happen if early reionisation occurs, may
lead to an extended photon diffusion with the resulting
destruction of anisotropies, specially at the degree scale.
The acoustic Doppler peaks are the dominant feature on these
small scales and the net effect of reionisation
would be a reduction (if not a complete suppression) of them.
On the other hand, for large angular scales we do not expect
the ionisation history to play any relevant r\^ole. Therefore,
the normalisation of the CMB spectrum on large scales
is safe from these uncertainties.

\section{Adiabatic fluctuations as source of $\D$}
\label{sec-adiabatic}
\markboth{Chapter 3. ~CMB ANISOTROPIES}
{\S ~3.3. ~ADIABATIC FLUCTUATIONS AS SOURCE OF $\D$}

The dominant contribution to large scale fluctuations comes from the
Sachs--Wolfe effect. Sachs \& Wolfe [1967]
in their classic paper on linear perturbations around an
Einstein--de Sitter background universe, derived a general formula for the
CMB anisotropy associated with scalar and metric perturbations.
In this perturbed universe model each photon is assumed to have
been emitted at conformal time $\eta_r$ (at recombination) and detected
by us at $\eta_0$ (present time).\footnote{In Ref.\cite{Sachs+Wolfe67},
recombination was taken as instantaneous. We have seen that the actual
width of the last scattering surface is not relevant but for very
small scales.}

The temperature fluctuation, to first order in the perturbing quantity,
may be expressed as

\be
\label{sw}
\D = \Phi            \bigg|^0_r -
     {\bf v} \cd \gg \bigg|^0_r -
\frac{1}{2} \int^{\eta_0}_{\eta_r} d \eta ~
h_{\mu\nu,0}[x^\alpha(\eta)] ~
\gamma^\mu \gamma^\nu ,
\ee

\noindent
where $x^\alpha(\eta) = \gamma^\alpha (\eta_0 - \eta)$
is the unperturbed photon geodesic, and $\gamma^\mu = (1,\gg)$, with
$\gg$ pointing along the line of sight.
In the above equation a uniform source was assumed
(thus, no intrinsic fluctuations so far -- see below),
and we also used the correspondence
$h_{00} = 2 \Phi$ relating the metric perturbation to the gravitational
potential.

The three terms in Eq. (\ref{sw}) are identified as
(i) the gravitational redshift (that may also depend on the position)
due to the difference between $\Phi$'s at recombination and now;
(ii) Doppler shifts, due to peculiar motions of both emitter and receiver;
and (iii) a term sensible to time dependences of the metric perturbation.
In practice, one is not interested in anisotropies produced by peculiar
motions of the observer which create a pure dipole anisotropy on the sky,
whereas we will see in Chapter \ref{chap-texture} that perturbations in
the peculiar velocity of the fluid components is key for the development
of the Doppler peaks.
The time dependence of $h_{\mu\nu}$ is relevant for models with
gravitational waves, and therefore the integral term plays an important
r\^ole.
Similarly, for $\Lambda$--dominated or non--flat models in general,
(iii) is relevant, as well.

In the Einstein--de Sitter model and if perturbations in the metric
arise only due to density perturbations, then the integrand is essentially
$4 \partial\Phi / \partial\eta$.
If we consider just growing mode
density perturbations in the linear regime, $\Phi$ is constant in
time\footnote{This applies
for perfect fluids in a flat space regardless of whether
the universe is dominated by matter or radiation.
It may be most easily seen from the Poisson equation, and
remembering the fact that the density contrast grows
with the scale factor as
$\delta \propto a$ ($\propto a^2$) for matter (radiation)
domination.}
and therefore the integral vanishes.
On the other hand, there have been scenarios proposed
(such as those where the occurrence of a `late--time phase transition'
greatly influences the total temperature anisotropies)
where it is in fact the integrated term the interesting one.
Also, this term will
become relevant when studying the (mildly) non--linear growth of
perturbations and their influence on CMB non--Gaussian features,
cf. Chapter \ref{chap-integrated} below.

At this point we should stress that the result of Eq. (\ref{sw})
has not (yet) the usual form that one usually meets when considering
the Sachs--Wolfe effect, even in the absence of tensor metric fluctuations.
In fact we still need to take into account the intrinsic anisotropies in the
radiation field, cf. \S\ref{sec-contributions}.
For adiabatic perturbations,
we found $\D = \frac{1}{3} \frac{\delta\rho}{\rho}$ and,
recalling that an overdensity generates a larger Newtonian potential
($\frac{\delta\rho}{\rho} \sim  2 \Phi$), we see there is a partial
cancellation, yielding the familiar result $\D \sim  \frac{1}{3} \Phi$,
where we assumed $\Phi = 0$ at the time of observation, and where the
Doppler and integrated terms in (\ref{sw}) were not explicitly written
down.

With regards to $\D$ coming from isocurvature fluctuations,
we have seen in \S\ref{sec-contributions} how a positive
density perturbation
(and the corresponding $\Phi$) is compensated by a negative fluctuation
in the temperature of the CMB radiation.
Thus, this perturbation {\sl adds} to the Sachs--Wolfe one, and this yields
relatively large $\D$, as to be compared with the corresponding adiabatic
models, for the same perturbation in the density.
We will not go into further details but just refer the interested reader
to the reviews in [White, Scott \& Silk, 1994] and
[Stebbins, 1993] for nice accounts of these matters.

\section{Non--linear effects}
\label{sec-nonlinear}
\markboth{Chapter 3. ~CMB ANISOTROPIES}
        {\S ~3.5. ~NON--LINEAR EFFECTS}

We have seen how linear perturbations in a flat, matter dominated FRW
universe led Sachs and Wolfe to derive their well--known formula.
Now, what would happen if we considered a different cosmological model
and/or we let the initially small perturbations go non--linear and
form large--scale density inhomogeneities?
Rees \& Sciama [1968] were the first to note
that these evolving mass concentrations would make the gravitational
potential $\Phi$ vary with time, even for a fixed space point.
Thus, radiation travelling through these intervening clumps will
be perturbed and the integral in Eq. (\ref{sw})
will depend on what lies along the photon's path.

Recently, using ray tracing techniques through N--body CDM
simulations, Tuluie \& Laguna [1995] showed that {\sl two} effects
related to the variations of the gravitational potential are
relevant on small scales.
The studied effects are the intrinsic evolution of $\Phi$
(\eg, changes in the depth of the potential wells)
during the time the photons take to transverse the protostructure,
and the change in the gravitational potential as the large--scale
structures formed (like clusters) move across the microwave sky.
They find that both these effects can be of similar amplitude,
and amount to a level of anisotropies of order
$10^{-7} \lsim \D \lsim 10^{-6}$ on scales of $\sim 1^\circ$.

We will revisit integrated effects due to a varying
gravitational potential during the mildly non--linear growth of
perturbations in Chapter \ref{chap-integrated} below,
when we will calculate the post--recombination contribution
to the CMB three--point temperature correlation function.

\section{Overview}
\label{sec-over}
\markboth{Chapter 3. ~CMB ANISOTROPIES}
                   {\S ~3.6. ~OVERVIEW}

\begin{figure}[tbp]
  \begin{center}
    \leavevmode
    \epsfxsize = 9.5cm
    \epsfysize = 9.5cm
    \epsffile{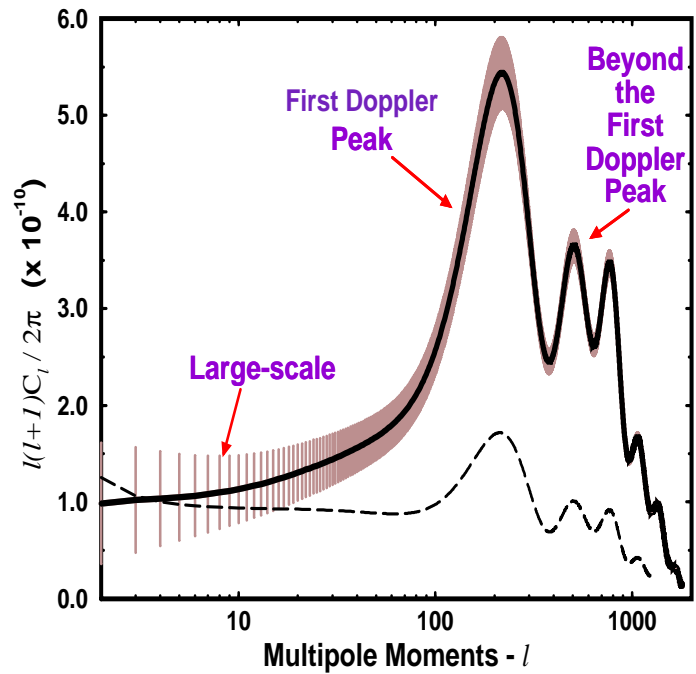}
  \end{center}
\caption{{\sl Angular power spectra
predicted within inflationary cosmologies [from Steinhardt, 1995].
The spectra are for CDM and
cosmological parameters $h=0.5~,~\Omega_B=0.05$, with
cosmological constant $\Lambda=0$.}}
\label{steinfig}
\end{figure}

\noindent
Although we have not introduced so far the necessary
notation and machinery for giving a `quantitative' account of the many
contributions to the CMB temperature anisotropies
on the various angular scales, let us at least qualitatively
depict what currently is though to be the
`standard model' for the fluctuations in the relic radiation.

In Figure \ref{steinfig} we show the generic prediction for the
angular power spectrum from standard inflation plus CDM on all
relevant angular scales, namely, from
the very large ($\ell$ small) to the smallest scales
(large $\ell$'s) currently probed by observations.
In this figure,
the upper solid curve comes from a primordial (power--law)
spectrum of density fluctuations, with spectral index
$n=1$ on large angular scales (those scales which
suffered little or no evolution at all) and no
gravitational wave contribution.
Vertical hashmarks give an idea of the theoretical
uncertainties inherent to the CMB anisotropies which
go under the name `cosmic variance' and, as we see from
the plot, are specially relevant on large angular scales.
The lower curve (dashed)
is for $n=0.85$ and shows the apparent `tilt' in the
otherwise flat Sachs--Wolfe plateau; this fact leads to
spectra with more power on large scale (as to be
compared with the `flat' Harrison--Zel'dovich spectrum
$n=1$).
This curve also
shows how the corresponding spectrum gets modified when
the amplitude on large scales is built up by both scalar and
tensor modes (50--50 mixture at the quadrupole level in the
plot).
For scales well within the horizon at recombination
(as is the case for $\ell \gsim 150$) gravitational waves
redshift away and thus the overall amplitude gets
equally reduced (as to be compared with the case with
solely scalar contribution to the anisotropies).

\begin{figure}[t]
  \begin{center}
    \leavevmode
    \epsfxsize = 8cm
    \epsfysize = 8cm
    \epsffile{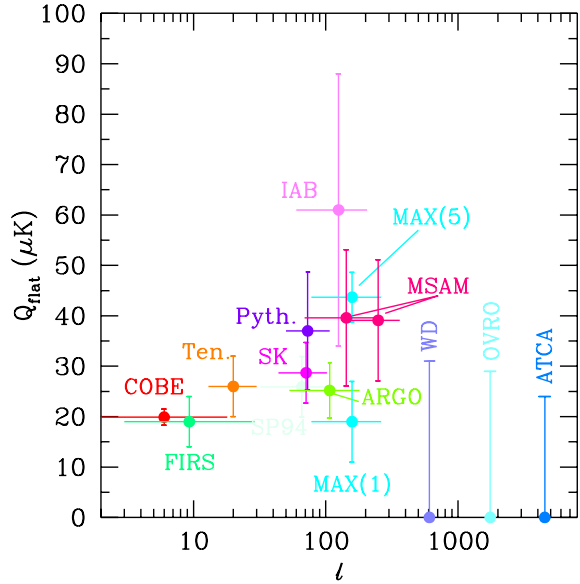}
  \end{center}
\caption{{\sl Amplitude of the fluctuations in different
experiments, as a function of the scale (multipole
$\ell\sim \theta^{-1}$) [from Scott \etal, 1995].
${\cal Q}_{flat}$ is the best--fitting amplitude of
a flat power spectrum, given by
$\ell (\ell + 1) {\cal C}_\ell =
(24\pi / 5)({\cal Q}_{flat}/T_0)^{2}$.}}
\label{Scottetalfig}
\end{figure}

\newpage

While on large angular scales the observational
results are well settled, on smaller scales
the situation is not as clear.
In Figure \ref{Scottetalfig} we show the present
status regarding observations.
As it is clear from this figure there is some
indication around scales $\ell \sim 200$ that the
level of anisotropies grows, hinting for the existence
of the Doppler peaks [Scott, Silk \& White, 1995].
However, it is also apparent from the not always compatible
data points and vertical error bars that presently
nothing can be definitely concluded. More, independent
pieces of observations are thus needed to corroborate
the present experimental results and to fill in the gaps.

\chapter{CMB statistics}
\label{chap-statistics}
\markboth{Chapter 4. ~CMB STATISTICS}
         {Chapter 4. ~CMB STATISTICS}

\section{Dealing with spherical harmonics}
\label{sec-dealing}
\markboth{Chapter 4. ~CMB STATISTICS}
{\S ~4.1. ~DEALING WITH SPHERICAL HARMONICS}

In the following sections we will compute correlations of temperature
anisotropies. The pattern of these anisotropies is distributed
on the CMB sky two--sphere and, as such, it will prove useful to expand
the temperature fluctuation field $\D$
(as we may do with any function that lives on a sphere)
in {\sl spherical harmonics}.

It is conventional to define the spherical harmonics in terms of the
associated Legendre functions as follows
\be
Y_\ell^m(\gg) \equiv Y_\ell^m(\theta,\phi) =
\sqrt{
{2 \ell + 1 \over 4 \pi}{(\ell - m)!\over (\ell + m)! }
} ~
P_\ell^m(\cos\theta) ~ e^{i m \phi}
\ee
with $\ell$ and $m$ integers fulfilling the relations $\ell \ge 0$ and
$\vert m \vert \leq \ell$.

As in the expression above we will often use the angles
$\theta, \phi$ or the unit vector
$\gg = (\sin\theta \cos\phi , \sin\theta \sin\phi , \cos\theta)$
interchangeably;
when working with temperature anisotropies $\gg$ will be taken as the
line--of--sight direction on the sky.

Being the product of $e^{i m \phi}$, which form a complete set of orthogonal
functions in the index $m$ on the interval $0\leq\phi\leq 2\pi$, and
$P_\ell^m(\cos\theta)$, which form a similar set in the index $\ell$ for
fixed $m$ on the interval $-1\leq \cos\theta \leq 1$, it is clear that
the $Y_\ell^m$ will form an analogous set on the surface of the unit sphere
in the two indices $\ell$, $m$.

Under complex conjugation they change as
$Y_\ell^{-m} = (-1)^m {Y_\ell^{m}}^*$.
The orthonormality condition is
\be
\label{ortho}
\int d\Omega_{\gg} {Y_\ell^{m}}^*(\gg) Y_{\ell'}^{m'}(\gg) =
\delta_{\ell \ell'} \delta_{m m'}
\ee
where the differential solid angle is given by
$d\Omega_{\gg} = \sin\theta ~ d\theta ~ d\phi$.
These properties, together with the completeness relation,
allow one to expand any
square--integrable function on the unit sphere, \eg,  $\Delta$, as
\be
\label{Delta}
\Delta(\gg) = \sum_{\ell=0}^\infty \sum_{m=-\ell}^\ell
a_\ell^m Y_\ell^m(\gg)
\ee
where the complex quantities $a_\ell^m$ are given by
\be
\label{alm}
a_\ell^m = \int d\Omega_{\gg} {Y_\ell^{m}}^*(\gg)  \Delta(\gg).
\ee

We will be interested in the situation in which
 $\Delta$ is actually $\D$, the CMB temperature fluctuation, and in this case
we will call $a_\ell^m$ the multipole moments.
These are relevant quantities as in general predictions
of a theory are expressed in terms of predictions for the $a_\ell^m$.

Also useful in this work is the spherical harmonics addition theorem
\be
\label{addition}
P_\ell(\cos\theta) =  {4\pi \over 2\ell + 1}
\sum_{m=-\ell}^\ell
{Y_\ell^{m}}^*(\theta_1,\phi_1) Y_{\ell}^{m}(\theta_2,\phi_2)
\ee
This theorem applies to any spherical triangle on the unit sphere,
with sides $\theta_1$, $\theta_2$ and $\theta$, and with the angle
$|\phi_1 - \phi_2|$ opposite to $\theta$, so that
(being $\go \equiv (\theta_1,\phi_1)$ and $\gt \equiv (\theta_2,\phi_2)$)
we have
$ \got  \equiv \cos\theta = \cos\theta_1 \cos\theta_2 +
\sin\theta_1 \sin\theta_2 \cos |\phi_1 - \phi_2|$.

The spherical harmonics as defined above are connected with the bases
of the even (or single-valued) representations of the rotation
group on the sphere.\footnote{The odd (double-valued) representations
arise for half--integer $\ell$ and the corresponding bases are
called {\sl spinors}.}

We will see below that, when dealing with correlations,
it will prove convenient to define the rotationally symmetric quantities
$Q_\ell^2 = \sum_{m=-\ell}^\ell \vert a_\ell^m \vert^2$, which we will call
the {\sl multipole coefficients}.\footnote{It is standard in the literature
to work in terms of ensemble averages, in order to get theoretical
predictions from different models. In that case one computes the
quantities ${\cal C}_\ell = \la Q_\ell^2 \ra / (2\ell +1)$ and it is
actually to these   ${\cal C}_\ell$'s
that one assigns the name {\sl multipole coefficients}.}

That the above defined quantities are rotationally symmetric
can be seen from the fact that a suitable average over the $m$
index was performed.
Given the expansion in spherical harmonics of any function
\be
f(\go) = \sum_{\ell} \sum_{m} a_\ell^m Y_\ell^m(\go),
\ee
one could wonder how the multipole moments change for a rotated version
of the same function, say $f(\gt)$, written in terms of the {\sl same}
$Y_\ell^m(\go)$'s.
That is, we express
$f(\gt) = \sum_{\ell} \sum_{m} ~^{(old)}\!a_\ell^m Y_\ell^m(\gt)$
and, by recalling the relation
\be
Y_\ell^m(\gt) = \sum_{m'} D^\ell_{m' m} Y_\ell^{m'}(\go)
\ee
we get
$f(\gt) = \sum_{\ell} \sum_{m'} ~^{(new)}a_\ell^{m'} Y_\ell^{m'}(\go)$
where we defined
\be
{}~^{(new)}a_\ell^{m'} = \sum_{m} D^\ell_{m' m} ~^{(old)}a_\ell^m.
\ee

This means that after rotation the new $a_\ell^{m}$ are just a
linear combination of the old ones, with $\ell$ index fixed.
We see then that a rotation on the sphere will not change the
multipole coefficients $Q_\ell^2$, which depend only
on $\ell$.\footnote{The so--called {\sl $D$--functions} are in fact
$(2\ell +1) \times (2\ell +1)$ unitary matrices $D^\ell$, and they form
a $(2\ell +1)$--dimensional irreducible representation of the groups
$SO(3)$ and $SU(2)$.
See Ref.\cite{Nyborg+Froyland} for a clear exposition
of their properties.}

Now going back to the defining expression for the $Q_\ell^2$, and
baring in mind what we discussed above, we see that
no matter how we rotate a given function the different $\ell$
contributions to it are independent and do not mix: say, if we
focus on the dipole ($\ell = 1$) contribution (the linear part), and
we rotate, it will never pick up quadratic ($\ell = 2$),
cubic ($\ell = 3$) or any other $\ell \neq 1$ terms.

\section{The CMB two--point correlation function}
\markboth{Chapter 4. ~CMB STATISTICS}
{\S ~4.2. ~THE CMB TWO--POINT CORRELATION FUNCTION}

Our main aim below will be to compute the CMB three--point temperature
correlation function. The very existence of non--vanishing higher--order
correlation functions on large angular scales
is due to possible departures from a Gaussian
behaviour of the primordial peculiar gravitational potential
(cf. Chapter \ref{chap-primor}).
Also, even  for the case of primordial Gaussian curvature
fluctuations, the mildly
non--linear gravitational evolution gives rise to a non--vanishing
three--point correlation function and
we will see in Chapter \ref{chap-integrated}
below what predictions we get for this {\sl integrated} effect.

As a first step we will define the connected two-- and three--point
correlation functions of the CMB temperature anisotropies
as measured by a given observer,
\ie, on a single microwave sky. Let us then define the temperature fluctuation
field $\D(\vec x ;\gg) \equiv (T(\vec x ;\gg) - T_0(\vec x))/T_0(\vec x)$,
where $\vec x$ specifies the position of the observer, the unit vector
$\gg$ points in a given direction from $\vec x$ and
$T_0(\vec x) \equiv \int {\dO_{\gg}\over 4\pi} T(\vec x ;\gg)$
represents the mean temperature of the CMB measured by that observer
(\ie, the {\em monopole} term).

The angular two--point correlation function $C_2(\vec x; \alpha)$ measured
by an observer placed in $\vec x$ is the average product of
temperature fluctuations in two directions $\gg_1$ and $\gg_2$ whose angular
separation is $\alpha$; this can be written as
\be
\label{2cor}
C_2 (\vec x; \alpha) =
\int {\dO_{\gg_1}\over 4\pi} \int {\dO_{\gg_2}\over 2\pi}
\delta\bigl(\gg_1 \cdot \gg_2 - \cos \alpha\bigr)
\D(\vec x ;\gg_1) \D(\vec x ;\gg_2) \ .
\ee
As well known, in the limit $\alpha \to 0$ one recovers the CMB variance
$C_2 (\vec x ) = \int {\dO_{\gg}\over 4\pi} [\D(\vec x ;\gg)]^2$.
Expanding the temperature fluctuation in spherical harmonics
\be
\label{expansion}
\D(\vec x ;\gg) = \sum_{\ell=1}^\infty \sum_{m=-\ell}^\ell a_\ell^m (\vec x)
{\cal W}_\ell Y_\ell^m(\gg),
\ee
and writing the Dirac delta function as a
completeness relation for Legendre polynomials $P_\ell$
\be
\label{completeness}
\delta\bigl(\cos\theta_1 - \cos\theta_2\bigr) =
\sum_{\ell = 0}^\infty (\ell + \frac12 )
P_\ell (\cos\theta_1) P_\ell (\cos\theta_2),
\ee
we easily arrive at the expression
\be
\label{2cor'}
C_2 (\vec x; \alpha) =
{1 \over 4\pi} \sum_{\ell} P_\ell (\cos \alpha) Q_\ell^2 (\vec x)
{\cal W}^2_\ell \ ,
\ee
where $Q_\ell^2 = \sum_{m=-\ell}^\ell \vert a_\ell^m \vert^2$.

It is standard to compare theory and experiments through the
$a_\ell^m$. However real measurement of temperature anisotropies
are limited by finite resolution and specific strategies are employed
by the many current experimental settings.
How to account in our definitions for this modeling of the real world?
It has become conventional to include this piece of {\sl dirty physics}
by the use of a {\sl filter}, limiting the range in $\ell$
(and hence, the angular scale) to which the experiment is sensitive, in
the form of a {\sl window function}.
In the previous expressions ${\cal W}_\ell$ represents in fact the
window function of the specific experiment.
Setting ${\cal W}_0 = {\cal W}_1=0$ automatically
accounts for both monopole and dipole subtraction.

Let us remind the reader that the subtraction techniques
employed in current anisotropy experiments do not allow us to get a
value for the monopole.
Moreover, the dipole contribution present in the maps
(essentially $\D \simeq \frac{v}{c} \cos\theta$, where $\theta$ is
the angle between the direction of motion and that of observation --
see \S\ref{sec-contributions})\footnote{This contributes
with a $\ell = 1$ term in the expansion in spherical harmonics.}
is considered to be mainly due to our peculiar motion with respect to
the comoving rest frame of the CMB radiation.

We will often concentrate below on large angular scales.
In the case of the {\sl COBE}--DMR experiment the window function
for $\ell \geq 2$ is
${\cal W}_{\ell}\simeq\exp\left[-\frac12 \ell({\ell}+1)\sigma^2\right]$,
where $\sigma$ is the dispersion of the antenna--beam profile, which
measures the angular response of the detector [\eg, Wright \etal, 1992].
In some cases the quadrupole term is also subtracted from the maps
[\eg, Smoot \etal, 1992];
in this case we also set ${\cal W}_2=0$.\footnote{When the
first analyses of the DMR maps were performed it was noted
that the quadrupole was also easy target of second--order Doppler effects,
and these could contaminate the former with a {\sl rms} kinematic quadrupole
of order $1.3~ \mu K$. This, together with the high intrinsic theoretical
uncertainty (cosmic variance, see below) of the lower order multipoles,
led the team to subtract, not only the mean and dipole, but also the
quadrupole from the maps. It was afterwards recognised that this in fact
makes little difference in the maximum likelihood analyses done for
estimating the spectral index $n$ and the $Q_{rms-PS}$
normalisation [\eg, G\'orski \etal, 1994],
and it is standard in current literature to state both fittings, with
and without $\ell = 2$ contribution.}

\section{The CMB three--point correlation function}
\label{sec-3pointfun}
\markboth{Chapter 4. ~CMB STATISTICS}
{\S ~4.3. ~THE CMB THREE--POINT CORRELATION FUNCTION}

The analogous expression for the angular three--point correlation function
is obtained by taking the average product of
temperature fluctuations in three directions $\gg_1$, $\gg_2$ and $\gg_3$
with fixed angular separations $\alpha$ (between $\gg_1$ and $\gg_2$), $\beta$
(between $\gg_2$ and $\gg_3$) and $\gamma$ (between $\gg_1$ and $\gg_3$);
these angles have to satisfy the obvious inequalities $\vert \alpha - \gamma
\vert \leq \beta \leq \alpha + \gamma$.
One then has
\bea
\label{3cor}
\lefteqn{
C_3 (\vec x; \alpha,\beta,\gamma) =
\int {\dO_{\gg_1}\over 4\pi} \int_0^{2\pi} {d \varphi_{12} \over 2 \pi}
\int_{-1}^1 d\cos \vartheta_{12}
\delta\bigl(\cos \vartheta_{12} - \cos \alpha\bigr)
\int_{-1}^1 d\cos \vartheta_{23}
}
\nonumber \\
&  &
\times \delta\bigl(\cos \vartheta_{23} - \cos \beta\bigr)
\!\int_{-1}^1 \!\! d\cos \vartheta_{13}
\delta\bigl(\cos \vartheta_{13} - \cos \gamma\bigr)
\D(\vec x ;\gg_1) \D(\vec x ;\gg_2) \D(\vec x ;\gg_3),
\nonumber \\
&  &
\eea
where $\cos \vartheta_{\alpha\beta} \equiv \gg_\alpha \cdot \gg_\beta$
and $\varphi_{12}$ is the azimuthal angle of $\gg_2$ on the plane
orthogonal to $\gg_1$.
The above relation can be rewritten in a form
analogous to Eq.(\ref{2cor}), namely
\bea
\label{3cor'}
\lefteqn{
C_3 (\vec x; \alpha,\beta,\gamma) = N(\alpha,\beta,\gamma)
\int {\dO_{\gg_1}\over 4\pi} \int {\dO_{\gg_2}\over 2\pi} \int
{\dO_{\gg_3}\over 2} \delta\bigl(\gg_1 \cdot \gg_2 - \cos \alpha\bigr)
}
\nonumber \\
&  &
\times
\delta\bigl(\gg_2 \cdot \gg_3 - \cos \beta\bigr)
\delta\bigl(\gg_1 \cdot \gg_3 - \cos \gamma\bigr)
\D(\vec x ;\gg_1) \D(\vec x ;\gg_2) \D(\vec x ;\gg_3) \ ,
\eea
where $N(\alpha,\beta,\gamma) \equiv \sqrt{1 - \!\cos^2\!\alpha -
\!\cos^2\!\beta - \!\cos^2\!\gamma + \!2
\cos\alpha \cos\beta \cos\gamma}$.\footnote{
To show that the latter expression is correctly normalised one can
expand the delta functions in Legendre polynomials (\ref{completeness}),
use the spherical harmonics addition theorem (\ref{addition}), and
note the relation $\sum_\ell (2 \ell +1) P_\ell(x) P_\ell(y) P_\ell(z)
= {2 \over \pi} (1-x^2-y^2-z^2+2xyz)^{-1/2}$
(see, \eg, \cite{Prudnikov}).}
Setting $\alpha=\beta=\gamma=0$ in these general expressions one obtains the
CMB {\em skewness} $C_3 (\vec x ) = \int {\dO_{\gg}\over 4\pi}
[\D(\vec x ;\gg)]^3$. Also useful are the {\em equilateral} three--point
correlation function [\eg, Falk \etal, 1993] and the
{\em collapsed} one [\eg, Hinshaw \etal, 1994 and references therein],
corresponding to the choices
$\alpha=\beta=\gamma$, and $\alpha=\gamma$, $\beta=0$, respectively.
Alternative statistical estimators, more suited to discriminate bumpy
non--Gaussian signatures in noisy data, have been recently introduced by
Graham \etal ~[1993].
In all the above formulas, full--sky coverage was assumed, for simplicity.
The effects of partial sky coverage on some of the statistical quantities
considered here are discussed in detail by
Scott, Srednicki \& White [1994]
(see also Ref.\cite{Bunn+Co94}).

Following the procedure used above for $C_2(\vec x;\alpha)$, we
can rewrite the three--point function in the form
\bea
\label{3cor''}
\lefteqn{
C_3 (\vec x; \alpha,\beta,\gamma) = N(\alpha,\beta,\gamma) {\pi \over 2}
\sum_{\ell_1,\ell_2,\ell_3}\sum_{m_1,m_2,m_3}
a_{\ell_1}^{m_1} a_{\ell_2}^{m_2}
{a_{\ell_3}^{m_3}}^* {\cal W}_{\ell_1} {\cal W}_{\ell_2} {\cal W}_{\ell_3}
}
\nonumber \\
&  &
\times
\sum_{j,k,\ell} \sum_{m_j,m_k,m_\ell} P_j(\cos\alpha) P_k(\cos\beta)
P_\ell(\cos\gamma) {\cal H}_{j\ell\ell_1}^{m_j m_\ell m_1}
{\cal H}_{kj\ell_2}^{m_k m_j m_2}
{\cal H}_{k\ell\ell_3}^{m_k m_\ell m_3} \ ,
\eea
where the coefficients ${\cal H}_{\ell_1\ell_2\ell_3}^{m_1 m_2 m_3}
\equiv \int \dO_{\gg} {Y_{\ell_1}^{m_1}}^*(\gg) Y_{\ell_2}^{m_2}(\gg)
Y_{\ell_3}^{m_3}(\gg)$, which can be easily expressed in terms of
Clebsch--Gordan coefficients [\eg, Messiah 1976], are only non--zero if
the indices $\ell_i$, $m_i$ ($i=1,2,3,$) fulfill the relations:
$\vert \ell_j - \ell_k \vert \leq  \ell_i \leq \vert \ell_j + \ell_k \vert$,
$\ell_1 + \ell_2 + \ell_3 = even$ and $m_1 = m_2 + m_3$.
We may occasionally in what follows make use of a related quantity
$\bar {\cal H}_{\ell_1\ell_2\ell_3}^{m_1 m_2 m_3} \equiv
(-1)^{m_1} {\cal H}_{~\ell_1\ell_2\ell_3}^{-m_1 m_2 m_3}$ which has the
further advantage of being invariant under circular permutation of its
columns $\left(^{m_i}_{\ell_i}\right)$.
In fact, this is apparent
from the following expression in terms of the
{\sl `3j' symbols}\footnote{This expresses, for the particular case
of the spherical harmonics, a more general theorem, called the
{\sl Wigner--Eckart} theorem.}:
\be
\label{goodh}
\bar {\cal H}_{\ell_1\ell_2\ell_3}^{m_1 m_2 m_3} =
\left( (2\ell_1 +1)(2\ell_2 +1)(2\ell_3 +1) / 4\pi \right)^{1/2}
\left(^{\ell_1~\ell_2~\ell_3}_{0~~0~~0}\right)
\left(^{\ell_1~\,\;\ell_2~\,\;\ell_3}_{m_1~m_2~m_3}\right) ~.
\ee

The collapsed three--point function measured by the
observer in $\vec x$ reads
\be
\label{skew}
C_3 (\vec x;\alpha) = {1 \over 4 \pi}
\sum_{\ell_1,\ell_2,\ell_3}\sum_{m_1,m_2,m_3}
P_{\ell_1}(\cos\alpha)
a_{\ell_1}^{m_1} a_{\ell_2}^{m_2}
{a_{\ell_3}^{m_3}}^* {\cal W}_{\ell_1} {\cal W}_{\ell_2} {\cal W}_{\ell_3}
{\cal H}_{\ell_3\ell_2\ell_1}^{m_3 m_2 m_1}
\ee
which, for $\alpha=0$, gives a useful expression for the skewness.

\section{Predictions of a theory: ensemble averages}
\label{sec-ensemble}
\markboth{Chapter 4. ~CMB STATISTICS}
{\S ~4.4. ~PREDICTIONS OF A THEORY: ENSEMBLE AVERAGES}

To obtain definite predictions for the statistics described above,
one needs to exploit the random nature of the multipole moments
$a_\ell^m$.
These coefficients tell us about the underlying theory for the
generation of temperature fluctuations and, depending on the case, will
be taken as following a Gaussian statistics (in which case predictions
are fully specified by giving the two--point function for the $a_\ell^m$'s)
or, more generally, a non--Gaussian one (and in this case also the
bispectrum will be needed,\footnote{Of course, for a complete description
	we would need all higher--order correlations
	$\la a_{\ell_1}^{m_1} \ldots a_{\ell_N}^{m_N} \ldots \ra$.
	However, in what follows we concentrate in the first of
	these combinations, namely, the bispectrum.}
as we will show below).
In either case, the $a_\ell^m$'s are stochastic random variables of
position, with zero mean and rotationally invariant variance depending
only on $\ell$: $\la a_\ell^m (\vec x) \ra = 0$ and
\be
\label{angspectrum}
\la a_{\ell_1}^{m_1}(\vec x) {a_{\ell_2}^{m_2}}^*(\vec x) \ra =
\delta_{\ell_1\ell_2} \delta_{m_1 m_2} {\cal C}_{\ell_1}
\ ,
\ee
where brackets $\la\cdot\ra$ stand for an average over the ensemble of
possible universes.

Within standard inflationary models, fluctuations are Gaussian and
therefore knowing the ${\cal C}_{\ell}$'s is enough for characterising
the angular fluctuation spectrum. The ${\cal C}_{\ell}$'s however
are not trivial to compute, specially on small scales (large $\ell$)
where they encode crucial information regarding the
cosmological parameters.
On the other hand, on large scales where the Sachs--Wolfe contribution to
the anisotropy is likely to dominate the signal, we have
$\D(\vec x ;\gg) = {1 \over 3} \Phi( \vec x + r_0 \gg)$,
with $r_0=2/H_0$ the horizon distance and $H_0$ the Hubble constant,
and the multipole coefficients are given by
\be
\label{clsandphi}
{\cal C}_\ell \propto \int_0^\infty dk k^2 P_\Phi(k) j_\ell^2(k r_0)
\ ,
\ee
where $P_\Phi(k)$ is the gravitational potential power--spectrum
and $j_\ell$ is the $\ell$--th order spherical Bessel
function.\footnote{The possible infrared
divergence of this expression for $\ell=0$ has no practical effect on
observable quantities, since the monopole is always removed.}

For large scales we can make the
approximation $P_\Phi(k) \propto k^{n-4}$, where $n$ corresponds to the
primordial index of density fluctuations (\eg, $n=1$ is the Zel'dovich,
scale invariant case), in which case [\eg, Bond \& Efstathiou, 1987;
Fabbri, Lucchin \& Matarrese, 1987]
\be
\label{cl}
{\cal C}_\ell = {{\cal Q}^2\over 5}
{\Gamma(\ell+n/2-1/2)\Gamma(9/2-n/2)\over\Gamma(\ell+5/2-n/2)\Gamma(3/2+n/2)},
\ee
with ${\cal Q} = \la Q_2^2 \ra^{1/2}$ the {\em rms} quadrupole, which
is simply related to the quantity $Q_{rms-PS}$ defined by
[Smoot \etal, 1992]:
${\cal Q} =\sqrt{4\pi} Q_{rms-PS}/T_0$.

Notice that for the quadrupole we get
${\cal C}_{\ell = 2} = {\cal Q}^2 / 5$, and this depends on the
spectral index $n$ (\ie, the normalisation of the fluctuation spectrum
depends on the form of the spectrum).
If we instead take $n=1$ we get
$\ell (\ell + 1) {\cal C}_{\ell} = 6 {\cal Q}^2 / 5$, which is independent
of the multipole index $\ell$.
This is often referred to as a `flat' spectrum, since
gravitational potential fluctuations (as well as the amplitude of the
density contrast at horizon crossing) are independent of the
scale.\footnote{Here it is useful to remember the handy relation
between the comoving wavenumber $k$ and the index $\ell$:
$\ell \simeq (6,000 h^{-1} {\rm Mpc})k$.}

Roughly speaking the amplitude for ${\cal C}_{\ell}$ is determined
by fluctuations on angular scales $\theta \sim \pi/\ell$
radians.\footnote{This correspondence applies for small $\ell$
(\ie, large scales). For smaller scales, however, it turns out
that $\theta \sim 60^\circ /\ell$ is more appropriate as a
rule--of--thumb.}
As we mentioned in Chapter \ref{chap-CMBanis},
a plot of $\ell (\ell + 1) {\cal C}_{\ell}$ is usually referred to as
the angular power spectrum.
With this definition an exact $n = 1$ initial spectrum,
assuming the relevant scales suffered virtually no evolution after
becoming smaller than the horizon (as is the case for very large scales),
produces a flat spectrum
(a `plateau' in the $\ell (\ell + 1) {\cal C}_{\ell}$ vs.
$\ell$ plot).

Inflationary theories predict tensor metric fluctuations, in addition to
(and statistically independent of) scalar modes.
As a consequence, the total multipole moment is actually the sum of
two orthogonal\footnote{To first order in the perturbations, scalar and
tensor modes are uncoupled, and thus evolve independently from each other.}
contributions, which are ordinarily noted as
${\cal C}^{(T)}_{\ell}$ and ${\cal C}^{(S)}_{\ell}$
for tensor and scalar, respectively.
Tilted ($n<1$) models, as are predicted by \eg, power--law inflation,
some versions of intermediate inflation, natural inflation
and a few other models,
imply more power (relative to a flat spectrum) on large scales
and consequently enhance the contribution from long wavelength
tensor modes (gravitational waves)\footnote{The exception being natural
	inflation where the level of gravitational wave contribution
	in negligible [Freese \etal, 1990].}.
This is of relevance for
the correct normalisation of the spectrum on large scales, as it is the
actual case after the detection of anisotropies in the {\sl COBE}--DMR
maps.

\chapter{Primordial non--Gaussian features}
\label{chap-primor}
\markboth{Chapter 5. ~PRIMORDIAL NON--GAUSSIAN FEATURES}
         {Chapter 5. ~PRIMORDIAL NON--GAUSSIAN FEATURES}

It has been realised that inflationary models also predict
small deviations from Gaussianity.
Every sensible inflation model will produce some small but
non--negligible non--Gaussian effects, both
by the self--interaction of the {\em inflaton} field,
and by the local back--reaction of its self--gravity.
It is our interest here to quantify this prediction,
with the largest possible generality, so that observational results
on the CMB angular three--point function can be used as a further test
on the nature of the primordial perturbation process.

Unfortunately, we will find that single--field inflation models
generally imply mean values for the skewness which are well below
the cosmic {\sl rms} skewness of a Gaussian field, which confirms and
generalises earlier results based on a simple toy--model
[Falk, Rangarajan \& Srednicki, 1993; Srednicki, 1993].

Falk \etal ~ [1993] first gave a quantitative estimate of the size of
non--Gaussian effects through a calculation of the three--point CMB
correlation function from perturbations generated in a simple model,
where the inflaton has cubic self--interactions.

Here the problem is considered in a totally
general and self--consistent way. We use the stochastic approach
to inflation [\eg, Starobinskii, 1986;
Goncharov, Linde \& Mukhanov, 1987; see also \S\ref{sec-Stocha}], as
a technical tool to self--consistently account for all second--order effects
in the generation of scalar field fluctuations during inflation and their
giving rise to curvature perturbations. We also properly account for the
non--linear relation between the inflaton fluctuation and the peculiar
gravitational potential. Our derivation moreover removes a non--realistic
restriction to purely de Sitter background made by Falk \etal ~ [1993],
which is especially important when non--Zel'dovich perturbation spectra
are considered.

The technical derivation in the frame of stochastic inflation of the
connected
two-- and three--point function (or its Fourier counterpart, the bispectrum),
for the inflaton field first and the local gravitational potential next, is
reported in \S\ref{sec-StochasticStuff}.
{}From this we compute (\S\ref{sec-angularbispectrum})
the CMB temperature anisotropy bispectrum and apply it to
the collapsed three--point function.
\S\ref{sec-skewness}
deals with the zero--lag limit of the three--point function:
the skewness, for which we provide a universal inflationary expression,
which we then specialise to some of the most popular inflaton potentials:
exponential [Lucchin \& Matarrese, 1985], quartic and quadratic
[Linde, 1983, 1985] as well as a simple potential for hybrid inflation
[Linde, 1994].
Interestingly enough, this last model allows for a spectral index
greater than one.\footnote{Note
that inflation is able to produce perturbation spectra with essentially all
values of $n$ around unity [\eg, Mollerach, Matarrese \& Lucchin 1994], as
it can be needed to match the {\em COBE}-DMR data [Smoot \etal, 1992] with
observations on smaller scales.}
In \S\ref{sec-skewcosmicvariance}
we discuss the effects of the cosmic variance on the possibility of
observing a non--zero mean value for the skewness.
As a by--product of this
analysis we compute the `dimensionless' {\sl rms} skewness for a large range
of the primordial spectral index of density fluctuations $n$.
Finally, we provide a brief discussion in \S\ref{sec-skewDiscu}


\section{Stochastic inflation and the statistics of the gravitational
         potential}
\label{sec-StochasticStuff}
\markboth{Chapter 5. ~PRIMORDIAL NON--GAUSSIAN FEATURES}
{\S ~5.1. ~THE STATISTICS OF $\Phi$}

In order to take into account all the effects contributing to a non--vanishing
primordial three--point correlation function of $\Phi$, we will perform
the computation in two steps. In \S\ref{sec-inflatonbispectrum}
we compute the three--point
function for the inflaton field perturbation $\delta \phi$. The most
convenient way to perform this calculation is in the frame of the stochastic
approach to inflation [Starobinskii, 1986; Goncharov \etal, 1987],
which naturally takes into account all the multiplicative effects in the
inflaton dynamics that are responsible for the non--Gaussian features.
Then, in \S\ref{sec-3pointgravpot}
we compute the extra--contribution to the three--point
function of the gravitational potential that arises due to the non--linear
relation between $\Phi$ and $\delta \phi$. This effect has been previously
noticed by Barrow \& Coles [1990] and Yi \& Vishniac [1993].

\subsection{The inflaton bispectrum}
\label{sec-inflatonbispectrum}

To study the dynamics of the inflaton, we will apply the stochastic approach.
This is based on defining a coarse--grained inflaton field
$\phi(\vec x,\alpha)$, obtained by
suitable smoothing of the original quantum field over a scale larger
than the Hubble radius size, whose dynamics is described by
a multiplicative Langevin--type equation (cf. \S\ref{sec-Stocha}).
This is obtained by adding
to the classical equation of motion a Gaussian noise term whose amplitude is
fixed by the {\sl rms} fluctuation of the scalar field at Hubble radius
crossing,
\be
\label{lange}
{\partial \phi(\vec x, \alpha) \over \partial\alpha}=
- {m_P^2 V'(\phi) \over 8\pi V(\phi)} +
{H(\phi) \over 2\pi} \eta(\vec x, \alpha) \ ,
\ee
where $V(\phi)$ is the inflaton potential, primes denote differentiation
with respect to $\phi$ and $m_P$ is the Planck mass.
The Hubble parameter here should be consistently calculated from the local
energy density of the coarse--grained inflaton.
The noise term $\eta$ has zero
mean and autocorrelation function [\eg, Mollerach \etal, 1991]
\be
\label{auto}
\la
\eta (\vec x, \alpha)
\eta (\vec x', \alpha')
\ra
=
j_0 ( q_s(\alpha) \vert \vec x - \vec x'\vert)
\delta (\alpha -\alpha') \ .
\ee
The use of the time variable $\alpha = \ln \left(a/a_*\right)$ in this
equation has been motivated by Starobinskii [1986], who noticed
that $\alpha$ accounts for the possible time dependence of the Hubble
parameter. This is particularly relevant when general,
\ie ~non--de Sitter, inflation is studied. We also defined the coarse--grained
domain size through the comoving wave--number
$q_s(\alpha) \equiv \epsilon H(\alpha) a(\alpha)$, with $\epsilon$
a number smaller than one, $H(\alpha) \equiv H(\phi_{cl}(\alpha))$, with
$\phi_{cl}(\alpha)$ the homogeneous classical solution of the Langevin equation
(\ie, that obtained with the noise term `switched' off). Finally the
scale factor $a(\alpha)$ is obtained by integration of $H(\alpha)$.

The stochastic dynamics of the coarse--grained field
within a single coarse--graining domain (\ie, for $\vec x = \vec x'$)
can be studied in terms of the Fokker--Planck equation for the probability
distribution function of $\phi$.
In our case, instead, since we are interested also in spatial correlations
of the field, we will solve directly the Langevin equation above.
To the aim of computing the three--point function of $\phi$, a second--order
perturbative expansion around the classical solution
is enough. We will require that the potential $V(\phi)$ is a
smooth function of its argument, which translates into
requiring well defined values for the {\sl steepness} of the potential
$X(\alpha) \equiv X(\phi_{cl}(\alpha)) \equiv m_P V'(\phi_{cl}) /
V(\phi_{cl})$
and its derivatives [Turner, 1993] throughout the range of relevant scales.
Apart from this requirement we keep the analysis general; only at the end
we will apply our results to some specific inflationary potentials
(\S\S\ref{Workedexamples}--\ref{skewnessVSn}).
We first expand $V(\phi)$ around $\phi_{cl}$, up to second order in
$\delta\phi(\vec x,\alpha) \equiv \phi(\vec x,\alpha) - \phi_{cl}(\alpha)$,
$V(\phi) = V(\phi_{cl}) + V'(\phi_{cl})
\delta\phi + \half V''(\phi_{cl}) \delta\phi^2 + \cdots$. Replacing this into
the Langevin equation we obtain
\be
\label{eq1}
{\partial \delta\phi(\vec x,\alpha) \over\partial\alpha} = A(\alpha)
\delta\phi(\vec x,\alpha) + B(\alpha) \delta\phi^2(\vec x,\alpha) +
\left[ D_1(\alpha) + D_2(\alpha) \delta\phi(\vec x,\alpha) \right]
\eta (\vec x,\alpha) \ ,
\ee
where we have used
$\partial \phi_{cl}/ \partial\alpha = - m_P^2 V'(\phi_{cl}) /
(8\pi V(\phi_{cl}))$. In Eq.(\ref{eq1}) we defined
\be
A = -{m_P\over 8\pi} X'
{}~ ; ~
B = -{m_P\over 16\pi} X''
{}~ ; ~
D_1 = {H(\alpha) \over 2\pi}
{}~ ; ~
D_2 = {H(\alpha)X \over 4 \pi m_P} \ .
\ee
Let us now split the field perturbation as $\delta\phi =
\delta\phi_1 + \delta\phi_2$ of
first and second order, respectively. We can also define a rescaled
variable $\tilde \eta \equiv (H(\alpha)/2\pi) \eta$.
We then find
\be
\label{eq94}
\delta\phi_1(\vec x ,\alpha) = X(\alpha)\int_0^\alpha
d\alpha' X^{-1}(\alpha') \tilde \eta (\vec x ,\alpha')
\ee
\be
\delta\phi_2(\vec x ,\alpha) = X(\alpha)
\int_0^{\alpha} d\alpha'
\left[
{B(\alpha')\over X(\alpha')} \delta\phi_1^2 (\vec x ,\alpha') +
{1\over 2 m_P}
\delta\phi_1 (\vec x ,\alpha') \tilde \eta (\vec x ,\alpha')
\right] \ .
\ee
Let us now calculate the connected three--point correlation function of
$\delta\phi$. The lowest order non--vanishing contribution
reads
\bea
\label{eqale24}
\lefteqn{\la  \delta\phi (\vec x_1, \alpha_1)
\delta\phi (\vec x_2, \alpha_2)
\delta\phi (\vec x_3, \alpha_3) \ra =
}
\nonumber \\
&  &
\!X(\alpha_3) \!\int_0^{\alpha_3}\! d\alpha'
\left[{B(\alpha')\over X(\alpha')}
\la \delta\phi_1 (\vec x_1, \alpha_1) \delta\phi_1 (\vec x_3, \alpha')
\ra
\la \delta\phi_1 (\vec x_2, \alpha_2) \delta\phi_1 (\vec x_3, \alpha')
\ra\! +\! [\vec x_1\!\leftrightarrow\!\vec x_2 ] \right]
\nonumber \\
&  &
+ {X(\alpha_3)\over 2 m_P}\! \int_0^{\alpha_3}\! d\alpha'
\left[ \la \delta\phi_1 (\vec x_1, \alpha_1) \delta\phi_1 (\vec x_3, \alpha')
\ra \la \delta\phi_1 (\vec x_2, \alpha_2) \tilde \eta  (\vec x_3, \alpha')
\ra\! +\! [\vec x_1\!\leftrightarrow\!\vec x_2 ]
\right]
\nonumber \\
&  & + ~2 \times 2 ~terms \ .
\eea
The term proportional to $X(\alpha_3)/ 2 m_P$ in the r.h.s. of this
equation can be recast in the form
\be
\label{eq92}
{X(\alpha_2)X(\alpha_3)\over 2 m_P}\!
\int_0^{\alpha_{min}}\!\! d\alpha'
X^{-1}(\alpha')\!
{H^2(\alpha')\over (2\pi)^2}\!
\la
\delta\phi_1 (\vec x_1, \alpha_1)
\delta\phi_1 (\vec x_3, \alpha')
\ra
{}~j_0 (q_s(\alpha')\vert\vec x_2 -\vec x_3\vert)
\ee
where we defined $\alpha_{min}\equiv {\rm min}[\alpha_3,\alpha_2]$.
We need now to compute the $\delta\phi$ autocorrelation function.
To this aim, recalling that $\alpha$ is the time when the perturbation
wavelength
$\sim a/q$ equals the size of the coarse--graining domain,
we change the integration variable in Eq. (\ref{eq94})) from $\alpha$ to
$q = \sqrt{8\pi V(\alpha)/ 3}~ q_* e^{\alpha} / H_* m_P$.
The subscript $*$ denotes quantities evaluated at the time
when we start to solve the Langevin equation; this is chosen in such a way
that the patch of the universe where we live is homogeneous on a scale
slightly above our present horizon
[see, \eg, the discussion by Mollerach \etal, 1991].
We then find
\be
\la
\delta\phi_1 (\vec x_1, \alpha_1)
\delta\phi_1 (\vec x_3, \alpha')
\ra
=
{1 \over 2\pi^2}
\int_{q_*}^{q_{min}(q_1,q' )} dq q^2 P(q)
{X(q_1)X(q')\over X(q)X(q)}
j_0 (q \vert \vec x_1 - \vec x_3\vert)
\ee
where we defined
$P(q) \equiv {1\over 2} q^{-3} H^2(q)$ where $\alpha = \alpha (q)$.
Using this we can rewrite Eq. (\ref{eq92}) as
\bea
\lefteqn{
{X(q_1)X(q_2)X(q_3)\over 2 m_P}
\int {d^3q'\over (2\pi)^3}
\int {d^3q\over (2\pi)^3}
{P(q') P(q)\over X^2(q)}
	}
\nonumber \\
&  &
\times
W(q'; q_{min}(q_3,q_2))
W(q ; q_{min}(q_1,q' ))
e^{i \vec q \cdot (\vec x_1  - \vec x_3 )}
e^{i {\vec q}' \cdot (\vec x_2  - \vec x_3 )}
\eea
where we defined the filter function $W(q;q_i) \equiv\Theta
(q-q_*)-\Theta (q-q_i)$ ($\Theta$ is the Heaviside function).

A similar analysis can be performed for the term proportional to
$B(\alpha)$ in Eq. (\ref{eqale24}). We get
\bea
\lefteqn{
X(q_1)X(q_2)X(q_3)
\int_{q_*}^{q_3} {dq'\over q'} B(q') X(q')
\int {d^3q''\over (2\pi)^3}
\int {d^3q'''\over (2\pi)^3}
{P(q'')\over X^2(q'')} {P(q''')\over X^2(q''')}
	}
\nonumber \\
&  &
\times
W(q'''; q_{min}(q_2,q'))
W(q'' ; q_{min}(q_1,q'))
e^{i {\vec q}'' \cdot (\vec x_1  - \vec x_3 )}
e^{i {\vec q}''' \cdot (\vec x_2  - \vec x_3 )} \ .
\eea
So far we have been working in configuration space. In order to obtain
the gravitational potential at Hubble radius crossing during the
Friedmann era, it is convenient to Fourier transform the
coarse--grained inflaton fluctuation. One has
$\delta\phi (\vec x, \alpha(q)) = (2\pi)^{-3}
\int d^3k~ \delta\phi (\vec k) \Theta (q - k)
e^{i \vec k \cdot \vec x }$,
where $\delta\phi (\vec k)$ denotes the Fourier transform of the
full scalar field at the time $\alpha(q)$.
This follows from the fact that at the time $\alpha(q)$
the only modes $k$ that contribute to the coarse--grained variable
are those which have already left the inflationary horizon, namely
$k < q$. We can then obtain the Fourier transform
$\delta\phi(\vec k, \alpha(q)) = \delta\phi (\vec k) \Theta (q - k)$,
and, in the limit $k \to q^-$ (or equivalently $q \to k^+$, that is, when
we consider the horizon--crossing time of the given scale),
we simply have $\delta\phi(\vec k, \alpha(k)) = \delta\phi(\vec k)$.
In other words, at the time $\alpha(k)$ the Fourier transform of the
coarse--grained variable coincides with that of the full field.
Using these results in Eq. (\ref{eqale24}) we finally obtain the
inflaton bispectrum through
\bea
\label{field3}
\lefteqn{\la
\delta\phi (\vec k_1, \alpha(k_1))
\delta\phi (\vec k_2, \alpha(k_2))
\delta\phi (\vec k_3, \alpha(k_3)) \ra = (2\pi)^3 \delta^3
(\vec k_1 + \vec k_2 + \vec k_3) P(k_2) P(k_3)
}
\nonumber \\
&  &
\times {X(k_1)\over X(k_2)X(k_3)}
\left[
{ X^2(k_2)\Theta(k_2-k_3) + X^2(k_3)\Theta(k_3-k_2) \over 2 m_P}
+ 2 \int_{k_*}^{k_1} {dq'\over q'} B(q') X(q')
\right]
\nonumber \\
&  &
+ \{\vec k_1\!\leftrightarrow\!\vec k_2\}
+ \{\vec k_1\!\leftrightarrow\!\vec k_3\} \ .
\eea

\subsection{The three--point function of the gravitational potential}
\label{sec-3pointgravpot}

We want now to compute the three--point function of the peculiar
gravitational potential, $\la\Phi(r_0\go)\Phi(r_0\gt)\Phi(r_0\gth) \ra$.
The approximate constancy of the gauge--invariant quantity
$\zeta$ outside the horizon
[\eg, Bardeen, Steinhardt \& Turner, 1983] allows to
obtain the gravitational potential during the matter--dominated
era, given the value of $\delta\phi$ during inflation.
During inflation one has $\zeta(\vec x,\alpha) \simeq -
\delta\phi(\vec x,\alpha)/(\partial\phi/\partial\alpha)$, which is usually
interpreted as a
linear relation between $\zeta$ and $\delta\phi$. However, when calculating the
three--point function of $\Phi$ one cannot disregard the second--order effects
coming from the fluctuations of $\partial\phi/\partial\alpha$.

Recalling that
\be
{\partial\phi\over\partial\alpha} =
-{m_P^2\over 8\pi}{V'(\phi)\over V(\phi)}\simeq
-{m_P\over 8\pi} \left[ X(\alpha) + X'(\alpha) \delta\phi \right]
\ee
one gets
\be
\label{grapot}
\zeta(\vec x)= {8\pi\over m_P X(\alpha)}
\left(\delta\phi(\vec x ,\alpha)
- {X'(\alpha)\over X(\alpha)} \delta\phi^2 (\vec x,\alpha)
\right) \ .
\ee
This equation is expressed in configuration space;
the Fourier transform of the first term inside the brackets is just
$\delta\phi(\vec k,\alpha)$. For the second term we get a
convolution of the type $\int d^3p \delta\phi(\vec p)
\delta\phi(\vec k - \vec p)$.
For the scales of interest one has $\Phi(\vec k) \simeq -
3\zeta(\vec k)/5$.
Then, adding the two above contributions and evaluating the expression
at the horizon crossing time we get
\begin{eqnarray}
\label{grapot2}
\Phi(\vec k)\!=\! {24\pi \over 5 m_P X(\alpha(k))}
\!\!\left[\!
\delta\phi(\vec k , \alpha(k)\!)\!
- {X'(\!\alpha(k)\!)\over X(\!\alpha(k)\!)}
\!\!\int\!\!\! {d^3p\over (2\pi)^3}
\delta\phi(\vec p , \alpha(k))
\delta\phi(\vec k\! -\! \vec p , \alpha(k))
\!\right].
\end{eqnarray}
{}From this equation we calculate

\newpage

\bea
\label{bispe}
\lefteqn{ \la \Phi(\vec k_1)\Phi(\vec k_2)\Phi(\vec k_3) \ra =
{ \left(24\pi / 5 m_P \right)^3 \over X(k_1)X(k_2)X(k_3) }
\la \delta\phi(\vec k_1)\delta\phi(\vec k_2)\delta\phi(\vec k_3) \ra +
	}
\nonumber \\
&  &\!
+
{ \left(-24\pi / 5 m_P \right)^3 X'(k_1)
\over
X^2(k_1)X(k_2)X(k_3)
}
\int {d^3p\over (2\pi)^3}
\la
\delta\phi(\vec p )
\delta\phi(\vec k_1 - \vec p)
\delta\phi(\vec k_2 )
\delta\phi(\vec k_3 )
\ra
\nonumber \\
&  &\!
+ \{\vec k_1\!\leftrightarrow\!\vec k_2\}
+ \{\vec k_1\!\leftrightarrow\!\vec k_3\}
\eea
Using the fact that, at horizon crossing, $\la\delta\phi(\vec k_1)
\delta\phi(\vec k_2 ) \ra = (2\pi)^3 P(k_1) \delta^3(\vec k_1 + \vec k_2)$,
we can write
\be
\label{new}
\la \Phi(\vec k_1)\Phi(\vec k_2) \ra = (2 \pi)^3 \delta^3(\vec k_1 +
\vec k_2) f^2(\alpha(k_1)) P(k_1) \ ,
\ee
with $f(\alpha(k)) \equiv 24 \pi/5 m_P X(\alpha(k))$.
Finally, using Eqs. (\ref{new}) and (\ref{field3}) we find
\bea
\label{bispe3}
\lefteqn{
\la \Phi(\vec k_1)\Phi(\vec k_2)\Phi(\vec k_3) \ra =
\left({24\pi\over 5 m_P}\right)^3
(2\pi)^3 \delta^3(\vec k_1 + \vec k_2 + \vec k_3)
{P(k_2)\over X^2(k_2)}
{P(k_3)\over X^2(k_3)}
}
\nonumber \\
&  &\!
\times
\left\{
{X^2(k_2)\Theta(k_2-k_3) + X^2(k_3)\Theta(k_3-k_2) \over 2 m_P}
-
{2X'(k_1)X(k_2)X(k_3)\over X^2(k_1)}
+
\right.
\nonumber \\
&  &
\left.
+
2\int_{k_*}^{k_1} {dq'\over q'} B(q') X(q')
\right\}
 + \{\vec k_1\!\leftrightarrow\!\vec k_2\}
 + \{\vec k_1\!\leftrightarrow\!\vec k_3\}
\eea
We are interested in considering perturbation modes that left the horizon
about 60 e--foldings before the end of the inflationary epoch. In the explicit
examples below $X(k)$ turns out to be a slowly varying function of $k$. We
therefore approximate the steepness
$X(k) \sim X_{60}$ in what follows. Then, to the lowest
non--vanishing order, the three--point correlation function is
\bea
\label{eqqq}
\lefteqn{\la \Phi (\vec k_1) \Phi (\vec k_2) \Phi (\vec k_3) \ra =
{1 \over f_{60}}\left[
{X_{60} \over 2 m_P}
-
{2 X'_{60} \over X_{60}}
+
{2\over X_{60}}\int_{k_*}^{k_{60}} {dq\over q}B(\alpha(q)) X(\alpha(q))
\right]
	}
\nonumber \\
&  &
\times ~
(2\pi)^3
\delta^3 (\vec k_1 + \vec k_2 + \vec k_3)
\left[
P_{\Phi}(k_1) P_{\Phi}(k_2) + P_{\Phi}(k_2) P_{\Phi}(k_3)
+ P_{\Phi}(k_3) P_{\Phi}(k_1)
\right]
\eea
with $f_{60}\equiv 24\pi / 5 m_P X_{60}$.
Using the expression for $P(q)$ we obtain the power--spectrum for the
peculiar gravitational potential
\be
\label{powerspec}
P_{\Phi}(k) = 2\pi^2 f_{60}^2 k^{-3} H^2(\alpha (k))/ 4 \pi^2 \simeq
{1 \over 2} f_{60}^2 H_{60}^2 k^{-3} (k/k_*)^{n-1},
\ee
up to possible presence of small logarithmic
corrections.
In this equation we used the fact that the variance per $\ln k$
of the perturbation in the $\zeta$ variable,
${\cal P}_\zeta =
k^3 P_\zeta / (2\pi^2) \propto k^{n-1}$ is given at horizon crossing by
$\sim (2\pi)^{-2}H^4/\dot\phi^2$.
By using the equation of motion for the
inflaton field, $2 \dot H \simeq - 8\pi \dot\phi^2 / m_P^2$
we then have
$H^4/\dot\phi^2 \propto H^4/\dot H \propto t^{-2} \propto H^2$
(with $a\propto t^p$ and $p>1$ during inflation), and so
$H^2 \propto k^{n-1}$, as was in fact used in the second equality in
Eq. (\ref{powerspec}) above.

\section{The CMB angular bispectrum}
\label{sec-angularbispectrum}
\markboth{Chapter 5. ~PRIMORDIAL NON--GAUSSIAN FEATURES}
{\S ~5.2. ~THE CMB ANGULAR BISPECTRUM}

Now we can obtain the two-- and three--point correlation functions for $\Phi$
in configuration space by inverse Fourier transforming Eqs. (\ref{new}) and
(\ref{eqqq}).
We then get the following general expressions for the
mean two-- and three--point functions of the primordial gravitational
potential
\be
\label{newer}
\la \Phi(r_0 \go) \Phi(r_0 \gt) \ra = {9  \over 4 \pi}
\sum_{\ell\ge 0} (2 \ell+1) P_\ell(\gg_1\cdot\gg_2) {\cal C}_\ell
\ee
and
\bea
\label{eq717}
\lefteqn{
\la \Phi(r_0 \go) \Phi(r_0 \gt) \Phi(r_0 \gth) \ra
= {81  \over  (4\pi)^2} \Phi_3
\sum_{j, \ell \ge 0} (2j+1)(2\ell+1) {\cal C}_j {\cal C}_\ell
        }
\nonumber \\
&  &
\times
\bigl[ P_j(\go \cdot \gth) P_\ell(\go \cdot \gt) +
       P_j(\gt \cdot \go) P_\ell(\gt \cdot \gth) +
       P_j(\gth \cdot \go) P_\ell(\gth \cdot \gt) \bigr] \ ,
\eea
where now the {\sl rms} quadrupole (cf. Eq. (\ref{cl}))
for the multipole coefficients is given by
\be
\label{quad}
{\cal Q}^2 = {8 \pi^2 H_{60}^2 \over 5 m_P^2 X_{60}^2}
{\Gamma (3-n)\Gamma (3/2+n/2)\over\Gamma^2(2-n/2)\Gamma (9/2-n/2)}
\ee
and
\be
\Phi_3 =
{1 \over f_{60}}
\left[
{X_{60} \over 2 m_P}
-
{2 X'_{60} \over X_{60}}
+
{2\over X_{60}} \int_{k_*}^{k_{60}} {dq\over q}
B(\alpha(q)) X(\alpha(q))
\right]
\ee
In order to obtain Eqs. (\ref{newer}) and (\ref{eq717}), we
had to use the definition of ${\cal C}_\ell$. The one given in
\S\ref{sec-ensemble}
differs from the results obtained here because of the window function
$W(k)$, appearing in the inflationary expressions; this can be however
neglected if we account for the oscillating behaviour of $j_\ell$ for large
arguments and for the fact that $j_\ell\to 0$ for small arguments (that helps
in cancelling the lower divergence).

\begin{figure}[tbp]
\begin{center}
\leavevmode
{\hbox %
{\epsfxsize = 7cm \epsffile{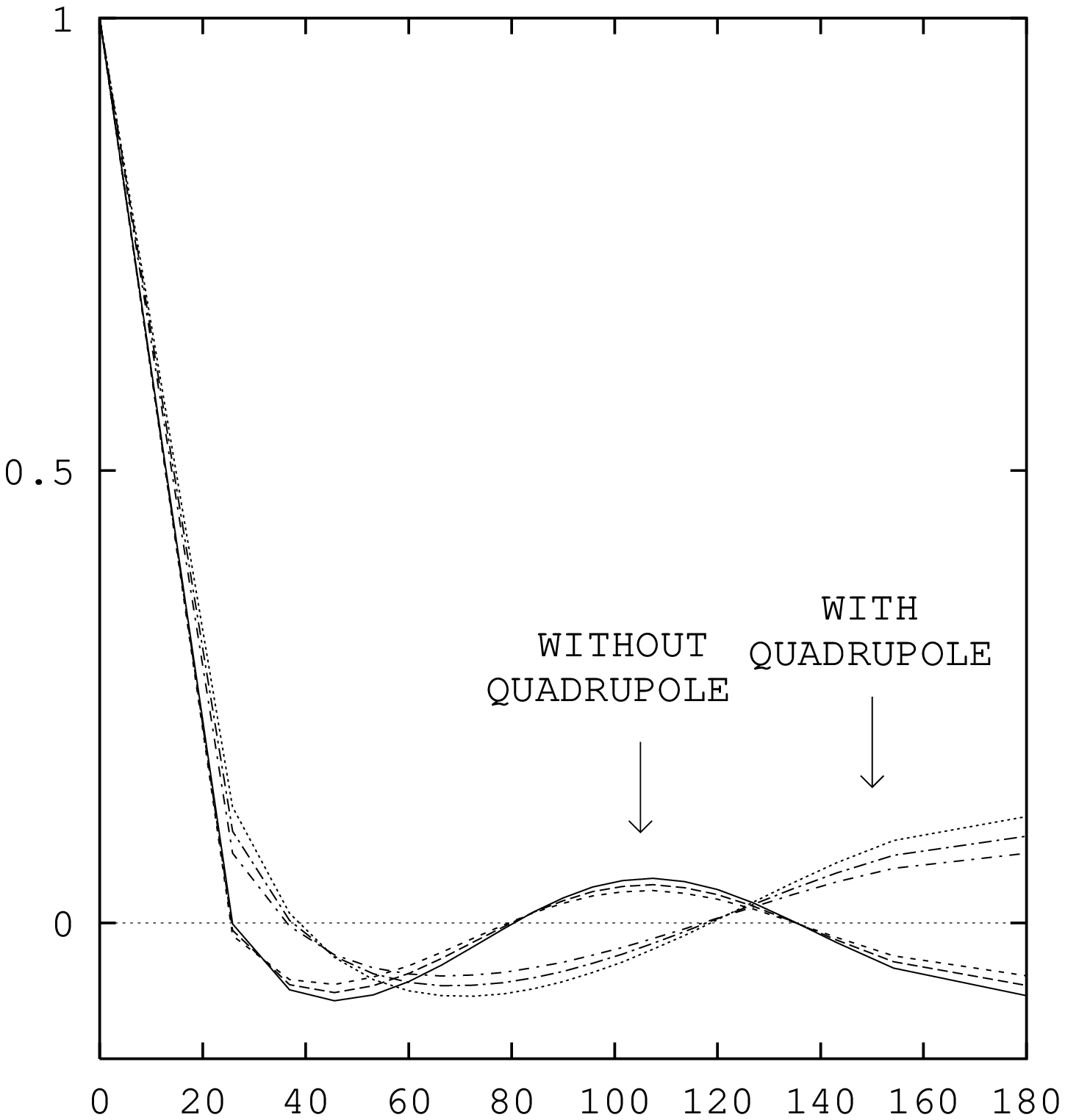} }
{\epsfxsize = 7cm \epsffile{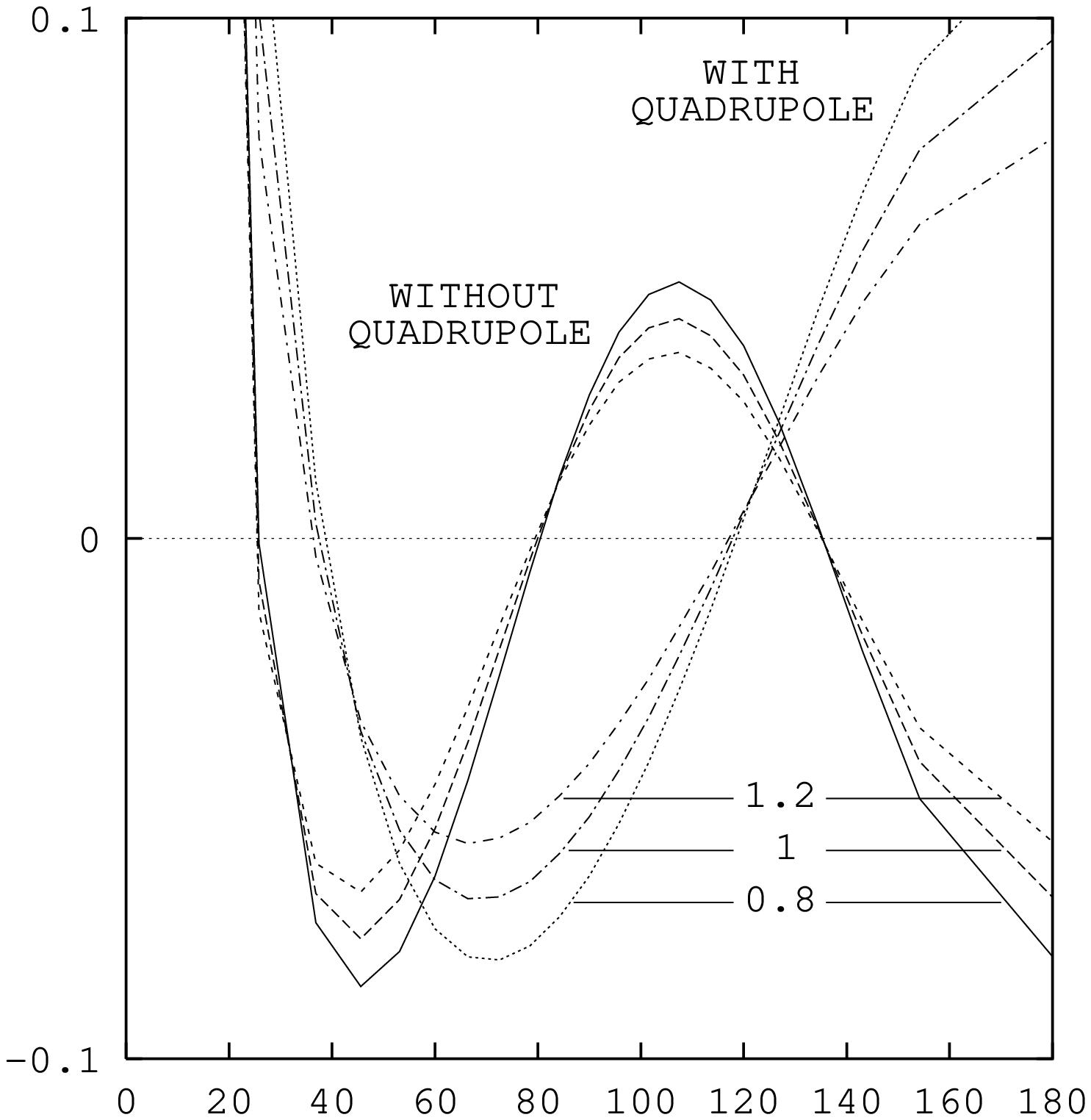} } }
\end{center}
\caption{{\sl Collapsed three--point function (with and without the
quadrupole contribution), normalised to the skewness, as a function of the
angular separation $\alpha$ (left panel). The different curves refer to three
typical values of the primordial spectral index: $n=0.8,~1,~1.2$.
The symbols in the left panel are the same as in the right one, which shows a
vertical expansion of the same plot.}}
\label{Figu1-Coll}
\end{figure}

The equations above allow us to recover Eq. (\ref{angspectrum})
for the angular spectrum, and the following expression for
the angular bispectrum,
\be
\la a_{\ell_1}^{m_1} a_{\ell_2}^{m_2} {a_{\ell_3}^{m_3}}^* \ra
= 3 \Phi_3 \bigl[ {\cal C}_{\ell_1} {\cal C}_{\ell_2} +
{\cal C}_{\ell_2} {\cal C}_{\ell_3} +
{\cal C}_{\ell_3} {\cal C}_{\ell_1}
\bigr] {\cal H}_{\ell_3 \ell_1 \ell_2}^{m_3 m_1 m_2} \ .
\ee
Replacing the latter expression into Eq. (\ref{3cor''}) we obtain the general
form of the mean three--point correlation function. Some simplifications occur
for the collapsed three--point function, for which we obtain
\bea
\label{eq57}
\lefteqn{
\la C_3(\alpha) \ra
= {3 \over (4\pi)^2} \Phi_3
\sum_{\ell_1,\ell_2,\ell_3}
P_{\ell_1}(\cos\alpha) {\cal F}_{\ell_1 \ell_2 \ell_3}
        }
\nonumber \\
&  &
\times
(2 \ell_1 +1) (2 \ell_2 +1) (2 \ell_3 +1)
\bigl[ {\cal C}_{\ell_1} {\cal C}_{\ell_2} +
{\cal C}_{\ell_2} {\cal C}_{\ell_3} +
{\cal C}_{\ell_3} {\cal C}_{\ell_1} \bigr]
{\cal W}_{\ell_1} {\cal W}_{\ell_2} {\cal W}_{\ell_3}
\ ,
\eea
where the coefficients ${\cal F}_{\ell_1 \ell_2 \ell_3} \equiv
(4\pi)^{-2}\int\dO_{\gg} \int\dO_{\gg'} P_{\ell_1}(\gg\cdot\gg') P_{\ell_2}
(\gg\cdot\gg') P_{\ell_3}(\gg\cdot\gg')$ may be suitably expressed in terms of
products of factorials of $\ell_1$, $\ell_2$ and $\ell_3$, using standard
relations for Clebsch--Gordan coefficients, by noting that
${\cal F}_{k \ell m} = \left(^{k~\ell~m}_{0~0~0}\right)^2$
[cf. Messiah, 1976].
The CMB mean skewness $\la C_3(0) \ra$ immediately follows from the
above equation for $\alpha=0$. A plot of the angular dependence of the
collapsed three--point function above, normalised to the skewness, is reported
in Figures \ref{Figu1-Coll},
for typical values of the spectral index $n$.

\section{The CMB skewness}
\label{sec-skewness}
\markboth{Chapter 5. ~PRIMORDIAL NON--GAUSSIAN FEATURES}
{\S ~5.3. ~THE CMB SKEWNESS}

In what follows we will restrict our analysis to the mean
skewness as given by Eq. (\ref{eq57}), for $\alpha=0$.
In the numerical calculation below we will assume that the higher
multipoles are weighted by a $7^\circ\llap.5$ FWHM beam,
resulting in $\sigma=3^\circ\llap.2$ [\eg, Wright \etal, 1992].
To normalise our predictions we will consider the {\sl rms} quadrupole
obtained by [Seljak \& Bertschinger, 1993] through a Maximum--Likelihood
analysis, namely $Q_{rms-PS} = (15.7\pm 2.6) \exp[0.46 (1-n)]\mu K$
[see also Scaramella \& Vittorio, 1993 and
Smoot \etal, 1994].\footnote{Note that this value
of $Q_{rms-PS}$ assumes a multivariate Gaussian distribution function,
accounting for both the signal and the noise; while in principle
one should repeat the analysis consistently with the assumed statistics of
the temperature perturbations, the quasi--Gaussian nature of our
fluctuation field allows to extrapolate this Maximum--Likelihood
estimate without sensible corrections.}
For the models with $n\lsim 1$, considered below, this expression should be
multiplied by an extra factor $[(3-n)/(14-12n)]^{1/2}$ to
account for the effective decrease in the estimated value of $Q_{rms-PS}$
due to the contribution of gravitational waves [Lucchin, Matarrese \&
Mollerach, 1992]. For the mean temperature we take the FIRAS determination
$T_0=2.726\pm 0.01 K$ [Mather \etal, 1994].

\begin{figure}[tbp]
  \begin{center}
    \leavevmode
    \epsfxsize = 6cm      
    \epsfysize = 5.33 cm  
    \epsffile{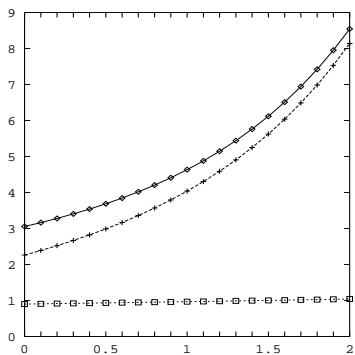}
  \end{center}
\caption{{\sl Geometrical factors ${\cal I}_p(n)$ as a function of $n$.
Crosses are for ${\cal I}_{3/2}(n)$, where the quadrupole term is removed.
Diamonds correspond to ${\cal I}_{3/2}(n)$ including the quadrupole
contribution. Open boxes correspond to ${\cal I}_{2}(n)$. For the latter
case the difference in including or removing the quadrupole contribution
is ${\cal O}(10^{-2})$, which is not perceivable in this graph.}}
\label{geomet-factors}
\end{figure}

To estimate the amplitude of the non--Gaussian character of the fluctuations
we will consider the `dimensionless' skewness
${\cal S}_1 \equiv \la C_3(0) \ra / \la C_2 (0)\ra ^{3/2}$.
Alternatively, if we want our results to be independent of the normalisation,
we may also define the ratio ${\cal S}_2 \equiv \la C_3(0)\ra/\la C_2
(0)\ra^2$, as suggested by the hierarchical aspect of our expression
for the skewness.
By writing the multipole coefficients as
${\cal C}_\ell = { {\cal Q}^2 \over 5 } \tilde {\cal C}_\ell$
(\ie, we factor out the amplitude as given by the {\sl rms} quadrupole),
we obtain
\be
\label{eq63bis}
{\cal S}_1 =
{ \sqrt{45\pi}\over 32 \pi^2} {\cal Q} X_{60}^2
\left[1-4 m_P {X'_{60}\over X^2_{60}} + {\cal G}\right] {\cal I}_{3/2}(n)
\nonumber
\ee
and
\be
{\cal S}_2 =
\label{eq63}
{15 \over 16 \pi } X_{60}^2
\left[ 1-4 m_P {X'_{60}\over X^2_{60}} + {\cal G} \right] {\cal I}_2(n)
\ee
with
${\cal G} = 4 m_P X_{60}^{-2}
\int_{k_*}^{k_{60}} (dq/q) B(\alpha(q)) X(\alpha(q))$
and where we also defined the spectral index--dependent geometrical factor
\be
\label{eq64}
{\cal I}_p(n)\! =\! {{1\over 3}
\sum_{\ell_1,\ell_2,\ell_3}(2\ell_1\!+\!1)(2\ell_2\!+\!1)(2\ell_3\!+\!1)
\bigl[\tilde {\cal C}_{\ell_1} \tilde {\cal C}_{\ell_2}\! +\!
\tilde {\cal C}_{\ell_2} \tilde {\cal C}_{\ell_3}\! +\!
\tilde {\cal C}_{\ell_3} \tilde {\cal C}_{\ell_1}\bigr]
{\cal W}_{\ell_1}\! {\cal W}_{\ell_2}\! {\cal W}_{\ell_3}
{\cal F}_{\ell_1,\ell_2,\ell_3}
\over \left[\sum_\ell (2l+1) \tilde {\cal C}_\ell {\cal W}_\ell^2 \right]^p}
\ee
where the exponent $p$ in the denominator takes values
$3/2$ and $2$ for ${\cal S}_1$ and ${\cal S}_2$, respectively.
We see clearly from Eqs. (\ref{eq63bis}) and (\ref{eq63}) that
${\cal S}_2$ is {\sl independent} of the normalisation, whereas ${\cal S}_1$
it is not.
The numerical factors ${\cal I}_p(n)$ in Eq. (\ref{eq64}) are
plotted in Figure \ref{geomet-factors}
for different values of the primordial index.

\subsection{Worked examples: inflationary models}
\label{Workedexamples}

Let us now specialise our general expressions to some simple inflationary
models.

\begin{flushleft}
{\bf Exponential potential}
\end{flushleft}

Let us first consider power--law inflation driven by the exponential
potential $V(\phi) = V_0 \exp\left(-\lambda\sqrt{8\pi G}\phi\right)$,
with $\lambda < \sqrt{2}$ [Lucchin \& Matarrese, 1985].
In this case  the power--spectrum is an exact power--law
with $n = 1 - 2\lambda^2 / (2 - \lambda^2)$.
We note in passing that the right spectral dependence of the
perturbations can be recovered using the above stochastic approach
[Mollerach \etal, 1991].
For this model we find $X = - \sqrt{8\pi}\lambda $,
whose constant value implies $A = B = 0$.
We then have
\be
{\cal S}_1
=
{3 \lambda\over 4} ~ {H_{60}\over m_P}~{\cal I}_{3/2}(n)~
\left[
{\Gamma (3-n)\Gamma (3/2+n/2)\over\Gamma^2(2-n/2)\Gamma (9/2-n/2)}
\right]^{1/2}
{}~;~ \ \ \ \
{\cal S}_2 = {15 \over 2 } ~\lambda^2
{}~{\cal I}_{2}(n) \ .
\ee
The {\sl COBE}--DMR
results constrain the amplitude of $H_{60}$. For the case $n =
0.8$ we have $H_{60}/m_P = 1.8 \times 10^{-5}$. This gives
${\cal S}_1 =  9.7 \times 10^{-6}$ and
${\cal S}_1 =  1.1\times 10^{-5}$, without and with the
quadrupole contribution respectively, while ${\cal S}_2 =  1.3$ in both
cases.

\begin{flushleft}
{\bf Quartic potential}
\end{flushleft}

Consider now the potential ${1\over 4}\lambda \phi^4$ for chaotic
inflation [Linde, 1983, 1985].
For this model $X = 4 m_P/\phi$ and therefore
$m_P B \sim m_P^2 X'' \ll X$. Thus we can take ${\cal G}\sim 0$
in the coefficients of Eqs. (\ref{eq63}) and (\ref{eq63bis})
for the dimensionless skewness parameters.
By using the slow--roll solution for the inflaton we may write
$X \simeq
4 m_P \phi_{60}^{-1}\left[1+\left(
\sqrt{\lambda\over 6\pi} m_P / H_{60}
\right) \ln(k/k_{60})\right]$ where the logarithmic correction to
the scale invariant power--spectrum is small.
Inflation ends when the inflaton takes the value
$\phi_{end} \simeq \sqrt{2/3\pi} ~m_P$ implying
$\phi_{60} \simeq \sqrt{60/\pi} ~m_P$.
In this case the spectral index is $n \simeq 1$.
We find
\be
{\cal S}_1
=
\sqrt{2 \lambda \over \pi}~{\phi_{60}\over m_P}
{}~{\cal I}_{3/2}(1)
{}~;~ \ \ \ \ \
{\cal S}_2
=
{30 \over \pi }\left({m_P\over \phi_{60}}\right)^2
{}~{\cal I}_{2}(1) \ .
\ee
{\em COBE}--DMR
constrains the value of $\lambda$ to be $\lambda\simeq 1.4 \times
10^{-13}$. We get ${\cal S}_1 = 5.2\times 10^{-6}$ and
${\cal S}_1 =  6.0\times 10^{-6}$ without and with the quadrupole contribution
respectively, while ${\cal S}_2 = 0.5$.

\begin{flushleft}
{\bf Quadratic potential}
\end{flushleft}

Another simple potential for chaotic inflation is
$\frac12 m_{\phi}^2 \phi^2$ [Linde, 1983, 1985].
In this case $X \simeq
2 m_P \phi_{60}^{-1}\left[1+\left(
m_P^2 / 4\pi \phi_{60}^2
\right)
\ln(k/k_{60})\right]$,
$\phi_{60} \simeq \sqrt{30/\pi} ~m_P$
and $n \simeq 1$. We find
\be
{\cal S}_1
=
{3\over 2\sqrt{\pi}}
{}~{m_{\phi}\over m_P}
{}~{\cal I}_{3/2}(1)
{}~;~ \ \ \ \ \
{\cal S}_2
=
{45 \over 4\pi } \left({m_P\over \phi_{60}}\right)^2
{}~{\cal I}_{2}(1)  \ ,
\ee
with $m_{\phi}/m_P \simeq 1.1 \times 10^{-6}$ from the
{\em COBE}--DMR
normalisation. We get ${\cal S}_1 = 3.9 \times 10^{-6} $ and
${\cal S}_1 = 4.5\times  10^{-6} $ without and with the quadrupole
contribution respectively, while ${\cal S}_2 = 0.4$.

\begin{flushleft}
{\bf Hybrid inflation model}
\end{flushleft}

Finally, let us consider a model of hybrid inflation recently
proposed by Linde [1994]. Inflation happens in this model during the
slow--roll evolution of the inflaton in the effective
potential $V(\phi)= V_0 + {1\over 2}m^2\phi^2$, where $V_0 = M^4 / 4$
is a cosmological--constant--like term.
At a given time, when the inflaton field takes the value $\phi=M$,
its coupling
with a second scalar field triggers a second--order phase
transition of the latter (whose vacuum energy density is responsible for
the cosmological constant term), which makes inflation end.
An interesting prediction of this model is that the spectral index of
density fluctuations for wavelengths which left the
horizon while $V_0>m^2 \phi^2/2$ is larger than unity, namely
$n \simeq 1 + 2 m^2 m_P^2 /(8\pi V_0)$ [Mollerach \etal, 1994].
For this potential we find $X = m_P m^2 \phi / V_0$, implying ${\cal G} = 0$.
We then obtain
\be
{\cal S}_1
=
{\sqrt{3}\over 2}
{\phi_{60}\over m_P}
{m^2\over M^2}
\!\left( 1 - {M^4\over m^2 \phi^2_{60} } \right)
{\cal I}_{3/2}(n)
\left[
{\Gamma (3-n)\Gamma (3/2+n/2)\over\Gamma^2(2-n/2)\Gamma (9/2-n/2)}
\right]^{1/2}
\ee
\be
{\cal S}_2
=
{15\over\pi }
{m_P^2 m^4 \phi_{60}^2 \over M^8}
\!\left( 1 - {M^4\over m^2 \phi^2_{60} } \right)
{\cal I}_{2}(n)
\nonumber
\ee
The slow--roll solution for the inflaton gives
$\phi_{60} = M \exp [60~ m^2 m_P^2/(8\pi V_0)]$. Choosing a value
$n=1.1$, the {\em COBE}--DMR results constrain the free parameters to be
$M \simeq 1.3 \times 10^{-4} m_P$ and $m \simeq 10^{-8} m_P$.
In this case we find ${\cal S}_1 = - 1.1\times 10^{-5}$
and ${\cal S}_1 = - 1.3\times 10^{-5}$
without and with the quadrupole contribution, while
${\cal S}_2 = - 1.6$. Note that these results suggest some correlation
between the sign of the skewness and the value of the spectral index
$n$.

\subsection{On the correlation between $n$ and the sign of the skewness}
\label{skewnessVSn}

Let us now explore somewhat more this correlation
between the sign of the skewness and the value of the spectral index
$n$.
We are here mainly concerned with single--field
inflationary potentials. Some of the most popular models
are characterised by their simplicity and universality (such as the
quadratic and quartic chaotic potentials we already treated),
by their being exact solutions of the
equations of motion for the inflaton (like power--law  and intermediate
inflation), or by their particle physics motivation (as natural
inflation with an axion--like potential).

In contrast with these simpler models where one has just one relevant
parameter more general potentials, with more freedom, were also
considered in the literature. One example of this is the polynomial potential
[Hodges \etal, 1990]
which for an adequate choice of the parameters was found
to lead to broken scale invariant spectra on a wide range of scales
with interesting consequences for large scale structure.

One should also worry about initial conditions
[Goldwirth \& Piran, 1992].
While for single--field models the only effect of kinetic terms
consists in slightly changing the initial value of $\phi$ (leaving
invariant the phase space of initial field values leading to sufficient
inflation), for models with more than
one scalar field initial conditions can be important,
\eg, for double inflation
[Starobinskii, 1985; Kofman \etal, 1985; Silk \& Turner, 1987]
(with two stages of inflation, each one dominated by a different
inflaton field) leading to primordial non--Gaussian
perturbations on cosmological interesting scales.
Within the latter models, however, the question of how
probable it is that a certain initial configuration will be realised in
our neighbouring universe should be addressed.
More recently other examples of interesting multiple--field models with
broken scale invariance have also been considered
[see, \eg, Salopek, 1992].
Here all scalar fields contribute to
the energy density and
non--Gaussian features are produced when the scalar fields pass over the
interfaces of continuity of the potential.
Extension of the single--field stochastic approach developed in
\S\ref{sec-StochasticStuff}
for calculating the CMB angular bispectrum generated
through Sachs--Wolfe effect from primordial curvature perturbations
in the inflaton in order to include many scalar fields is therefore
needed.

It is also of interest here to calculate the accurate form of the
spectrum of primordial perturbations (\eg, by finding the value of the
spectral index) at the moment the scales relevant for our study left
the Hubble radius.
Our expression ${\cal S}_1$ for the dimensionless skewness is accurate
up to second order in perturbation theory (the first non vanishing
order) and therefore we
should calculate $n$ at least to the same order.
We will borrow the notation for
the slow--roll expansion parameters from
[Kolb \& Vadas, 1994].
These are defined\footnote{It is easy to relate these parameters to
	the steepness and its derivatives. In fact
	$\epsilon = X^2 / (16\pi)$,
	$\eta  = \epsilon + m_P X' / (8\pi)$ and
	$\xi  =  \epsilon + 3 m_P X' / (8\pi) + m_P^2 X'' / (4\pi X)$.
	Of course, the value for $\epsilon$ coincides with that defined
	in Eq. (\ref{LL56}), whereas $\eta = -\epsilon + \chi$,
	cf. Eq. (\ref{LL59}).}
as
$\epsilon(\phi) = m_P^2 \left( H'(\phi)/ H(\phi) \right)^2 / 4 \pi $,
$\eta(\phi)  =  m_P^2  H''(\phi) / (4\pi H(\phi) ) $ and
$\xi(\phi)  = m_P^2 H'''(\phi) / ( 4\pi H'(\phi))$.
In the slow-roll approximation $\epsilon$ and $\eta$ are less than one.
The same is not true in general for $\xi$ and this may cause consistency
problems when this term is incorrectly neglected
(see the discussion in Ref.\cite{KolbVadas94}).
We will see below examples of this.

Let ${\cal Q}_S^2$ (${\cal Q}_T^2$) be the contribution of the scalar
(tensor) perturbation to the variance of the quadrupole temperature
anisotropy.  The complete second--order expressions for the tensor
to scalar ratio\footnote{If the relations were derived for density
	fluctuation amplitudes rather than for the temperature
	anisotropies there would be a 12.5 instead of the 14 in
	the expression for $R$ below.}
and for the spectral index are given respectively by
$R \equiv {\cal Q}_T^2 / {\cal Q}_S^2
\simeq 14\epsilon\left[1-2C(\eta-\epsilon )\right]$
and
\be
\label{12}
n  = 1 -  2\epsilon\left[2-\frac{\eta}{\epsilon}+4(C+1)\epsilon -
(5C+3)\eta + C\xi (1+2(C+1)\epsilon - C\eta) \right],
\ee
evaluated at $\phi\simeq\phi_{60}$.
In this equation
$C\equiv -2+\ln 2 +\gamma\simeq -0.7296$ and $\gamma=0.577$ is
the Euler--Mascheroni constant
[Kolb \& Vadas, 1994; see also Stewart \& Lyth, 1993].

In the rest of this subsection we will treat three different
realisations of the inflationary idea. After a brief introduction
to each of these models\footnote{The models presented here were not
	described before in this thesis. In this sense, the present
	subsection can be regarded also as a complement to the survey
	we performed in \S\ref{sec-Closett}}
we compute the CMB skewness (cf. Eq. (\ref{eq63bis}))
and the relation to the spectral index.
Let us now turn to the examples.

\begin{flushleft}
{\bf Natural inflation}
\end{flushleft}

To begin with let us consider the {\sl natural} inflationary scenario. First
introduced in [Freese, Frieman \& Olinto, 1990]
this model borrows speculative ideas from
axion particle physics [Kolb \& Turner, 1990].
Here the existence of disparate mass
scales leads to the explanation of why it is physically attainable to
have potentials with a height many orders of magnitude below its width
[Adams, Freese \& Guth, 1991],
as required for successful inflation where usually self coupling
constants are fine--tuned to very small values.

This model considers a Nambu--Goldstone (NG)
boson, as arising from a spontaneous symmetry breakdown of a global
symmetry at energy scale $f\sim m_P$, playing the r\^ole of the inflaton.
Assuming there is an additional explicit symmetry breaking phase at mass scale
$\Lambda\sim m_{GUT}$ these particles become pseudo NG bosons and a periodic
potential due to instanton effects arises.
The simple potential (for temperatures $T\leq\Lambda$) is of the form
$V(\phi) = \Lambda^4 [1+\cos(\phi / f)]$.

For us this $V(\phi)$ constitutes an axion--like
model with the scales $f$ and $\Lambda$ as free parameters.
It is convenient to split the parameter space into two regions.
In the $f >> m_P$ zone the whole inflationary period happens in the
neighbourhood of the minimum of the potential, as may be clearly seen
from the slow--rolling equation [Steinhardt \& Turner, 1984]
$\vert\cos (\phi / f)\vert \leq 24 \pi / (24 \pi + (m_P / f)^2 ) $
which is only violated near $\phi_{end} \simeq \pi f$, and from the small
value of the steepness
$\vert X \vert = {m_P \over f } \tan ({\phi\over 2 f})$
for $\phi$ smaller than $\phi_{end}$ \footnote{A method for
	reconstructing the inflationary potential
        (applied in particular to our present model) was developed in
        [Turner, 1993];
        see also [Liddle \& Turner, 1994],
        for a second--order generalisation of this reconstruction
        process.}.
Thus by expanding $V(\phi)$ around the minimum it is easy to see the
equivalence between this potential and the quadratic one
$V \sim m^2 (\phi - \pi f)^2 / 2$ with $m^2 = \Lambda^4 / f^2$.
We have already studied the latter in \S\ref{Workedexamples}
and we will add nothing else here.
On the other hand, let us consider the other regime,
where $f \lsim m_P$. Reheating--temperature considerations place a lower
limit on the width $f$ of the potential. For typical values of the model
parameters involved, a temperature $T_{reh} \lsim 10^8$GeV is attained.
GUT baryogenesis via the usual out--of--equilibrium decay of X--bosons
necessitates instead roughly $T_{reh} \sim 10^{14}$GeV
(the mass of the gauge bosons) for successful reheating
[Steinhardt \& Turner, 1984].
Thus the final temperature is not high enough to create them from the
thermal bath.
Baryon--violating decays of the field and its products
could be an alternative to generate the observed asymmetry if taking
place at $T_{reh} > 100$GeV, the electroweak scale.
This yields the constraint $f \gsim 0.3 m_P$
[Freese, 1994], implying $n\gsim 0.6$ (see below).
Attractive features of this model include the possibility of having a
density fluctuation spectrum with extra power on large scales. Actually
for $f\lsim 0.75~m_P$ the spectral index may be accurately expressed as
$n\simeq 1 - m_P^2 / 8\pi f^2$.\footnote{Recently Kolb \& Vadas [1994]
	have given
        second--order accurate expressions for $1-n$, $n_T$ and $R$
        (ratio of the tensor to scalar contributions to density
        perturbations). Applied to our present model this importantly
        changes the relation between $R$ and $1-n$. However the
        second--order correction to $n$
        ($\sim 0.2 (m_P^2 / 8\pi f^2)^2\sim 6\times 10^{-3}$ for our
        choice of $f$) is not relevant.}
This tilt in the spectrum as well as the negligible
gravitational wave mode contribution to the CMB anisotropy might lead to
important implications for large scale structure
[Adams \etal, 1993].

Let us take now $f\simeq 0.446~m_P$ corresponding to $n\simeq 0.8$.
Slow--rolling requirements are satisfied provided the accelerated
expansion ends by $\phi_{end}\simeq 2.78 f$, very near the minimum of the
potential. Furthermore, the slow--rolling solution of the field equations
yields
the value of the scalar field 60 e-foldings before the end
of inflation
$\sin(\phi_{60}/2 f) \simeq \exp(- 15 m_P^2 / 4\pi f^2 )$
where we approximated $\phi_{end} \simeq \pi f$.
We find\footnote{In Eq. (\ref{eq63bis})
	an additional contribution to the
        squared brackets of the form ${\cal G} = 4 m_P
        \int_{k_*}^{k_{60}} (dq/q) B(\alpha(q)) X(\alpha(q))$
        was found (the scale $k_*$ signals the time when we start to solve
        the Langevin equation, corresponding to a patch of the universe
        homogeneous on a scale slightly above our present horizon).
        As we saw, ${\cal G}$ turns out to be always negligible or
        zero for the specific models so far considered. As we will see below
        also here
        this term yields a value much smaller than the others explicitly
        written in (\ref{eq63bis}); \eg, within the
        frame of natural inflation ${\cal G}$ can be worked out easily
        yielding  $\vert {\cal G}\vert  \sim 1.1\times 10^{-6} (m_P / f)^4 $,
        much smaller than
        $ X_{60}^2 - 4 m_P X_{60}' \simeq 2 (m_P / f)^2$
        ($f\sim m_P$, see below).} (cf. Eq.(\ref{eq63bis}))
\be
X_{60}^2 - 4 m_P X_{60}' = (m_P / f)^2 [2+\sin^2(\phi_{60} / 2 f)]
                            (1-\sin^2(\phi_{60} / 2 f) )^{-1}
                            \simeq 2 (m_P / f)^2.
\ee
Two years of data by {\sl COBE}--DMR [Bennett \etal, 1994]
are not yet enough to separately
constrain the amplitude of the quadrupole and the spectral index. A
maximum likelihood analysis yields
$Q_{rms-PS} = 17.6 \exp[0.58 (1-n)]\mu K$.
By making use of Eq. (\ref{quad}) for the {\sl rms} quadrupole,
{\sl COBE}--DMR
results constrain the value of the free parameter
$\Lambda \simeq 1.41\times 10^{-4} m_P$.
We find ${\cal S}_1 \simeq 3.9\times 10^{-5}$,
a rather small signal for the non--Gaussian amplitude of the fluctuations.

\begin{flushleft}
{\bf Intermediate inflation}
\end{flushleft}

We will now study a class of universe models where the scale factor
increases at a rate intermediate between power--law inflation --as
produced by a scalar field with exponential potential
 -- and the standard de Sitter inflation.
Barrow [1990] shows that it is possible to parameterise these
solutions by an equation of state with pressure $p$ and energy density
$\rho$ related
by $\rho + p = \gamma \rho^{\lambda}$, with $\gamma$ and $\lambda$
constants.
The standard perfect fluid relation is recovered for $\lambda = 1$
leading to the $a(t)\sim e^{H_{inf} t}$ ($H_{inf}$ constant during
inflation) solution of the
dynamical equations when the spatial curvature $k = 0$ and $\gamma = 0$,
while  $a(t)\sim t^{2/3\gamma}$ for $0<\gamma< 2/3$.
This non--linear equation of state (and consequently the two limiting
accelerated expansion behaviours) can be derived from a scalar field with
potential $V = V_0 \exp (- \sqrt{3\gamma}\phi)$.
On the other hand for $\lambda > 1$ we have
$a(t)\propto \exp (A t^f)$ ({\sl intermediate} inflation) with $A > 0$ and
$0 < f \equiv 2 (1 - \lambda) / (1 - 2\lambda) < 1$ and again in this
last case it is possible to mimic the matter source with that produced
by a scalar field $\phi$, this time with potential
\be
\label{IntPot}
V(\phi) =
{8 A^2\over (\beta + 4)^2} \left({m_P\over \sqrt{8\pi}}\right)^{2+\beta}
\left({\phi\over \sqrt{2 A \beta}}\right)^{-\beta}
\left(
6 - {\beta^2 m_P^2 \over 8\pi}\phi^{-2}
\right)
\ee
with $\beta = 4 (f^{-1} - 1)$.

The equations of motion for the field in a $k = 0$ FRW universe
may be expressed by
$3 H^2 = 8\pi ( V(\phi) + \dot\phi^2 /2 ) / m_P^2 $
and
$\ddot\phi + 3 H \dot\phi = - V'$.
Exact solutions for these equations with the potential of Eq. (\ref{IntPot})
are of the form
[Barrow, 1990; Barrow \& Saich, 1990; Barrow \& Liddle, 1993]
\be
H(\phi) = A f (A \beta / 4\pi )^{\beta /4} (\phi /m_P)^{-\beta /2}
{}~~ ; ~~
\phi(t) = (A \beta / 4\pi )^{1/2} t^{f/2} m_P ~.
\ee
Solutions are found for all $\phi > 0$ but only for
$\phi^2 > (\beta^2 / 16 \pi) m_P^2$
we get $\ddot a > 0$ (\ie, inflation).
In addition $\beta > 1$ is required to ensure that the accelerated
expansion occurs while the scalar field rolls (not necessarily
slowly) down the potential, in the region to the right of the maximum
(as it is generally the case).

\begin{figure}[tbp]
\vspace{-2cm}
\begin{center}
\leavevmode
{\hbox %
{\epsfxsize = 7cm \epsffile{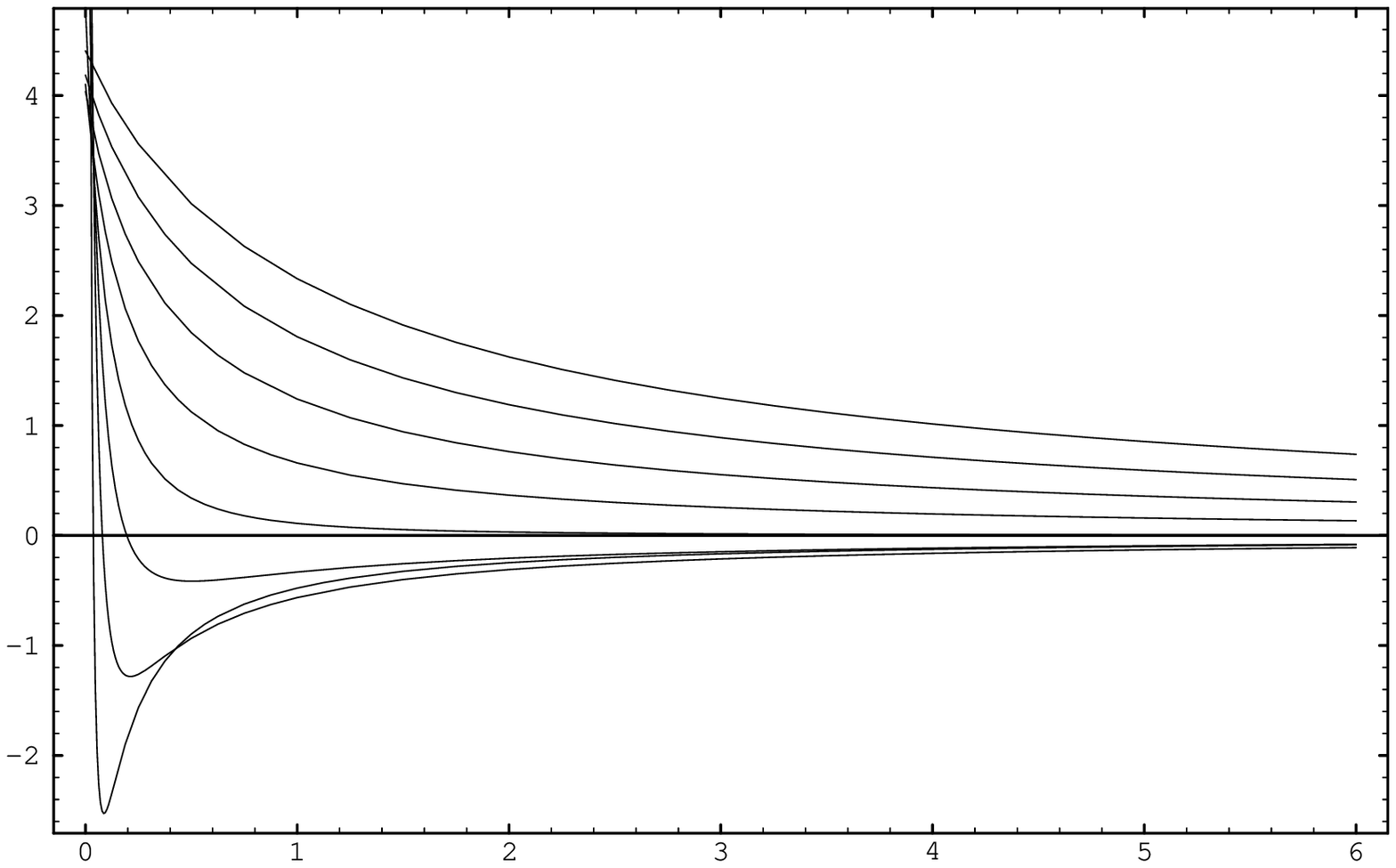} }
{\epsfxsize = 7cm \epsffile{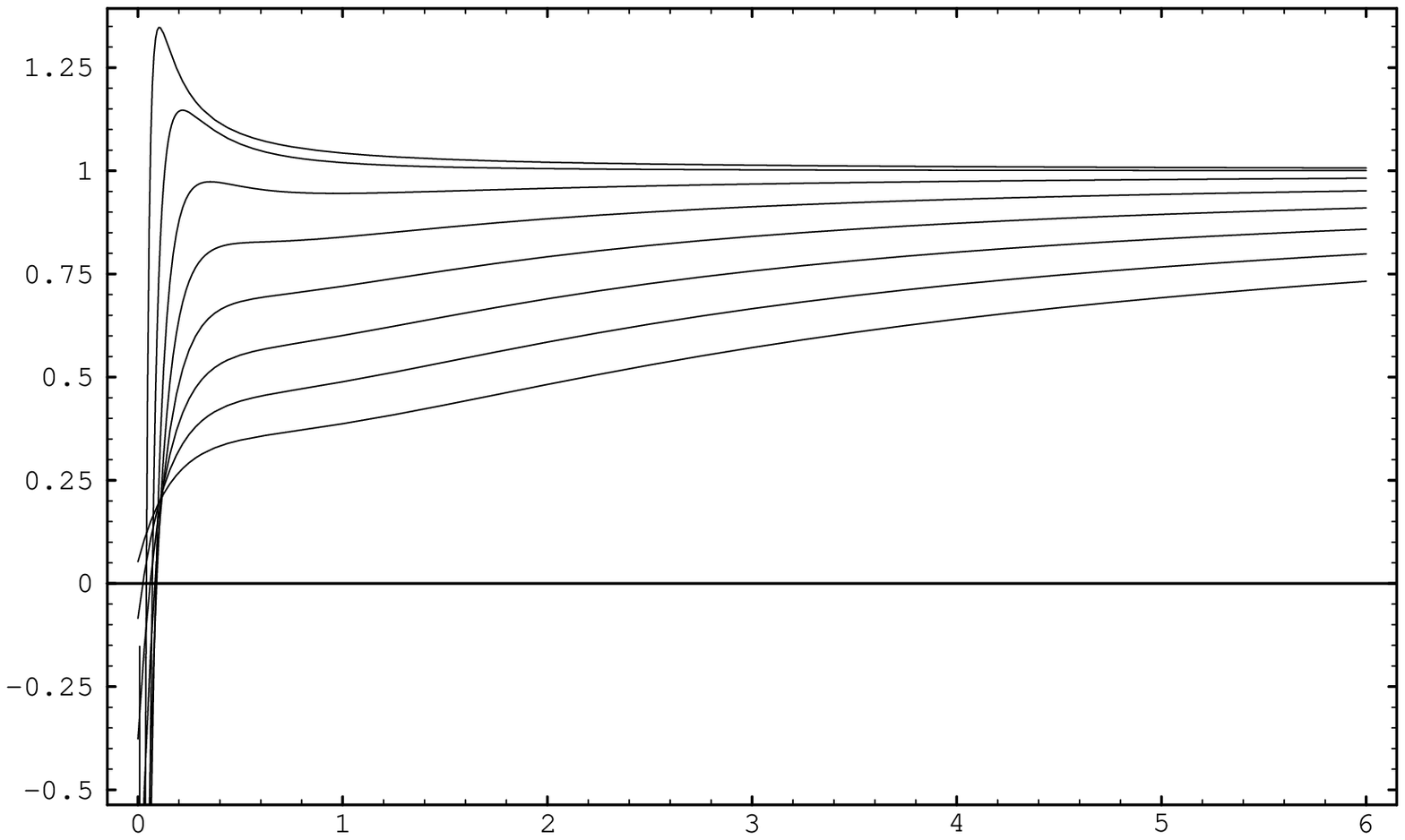} } }
\end{center}
\vspace{-2.5cm}
\caption{(a) {\sl Dimensionless skewness ${\cal S}_1$
(in units of the quadrupole ${\cal Q}$)
as a function of $d\equiv \phi^2 - \beta^2 / 16\pi$, the
value of the squared of the field beyond the minimum allowed,
for different choices of $\beta$. Curves from bottom to top
correspond to $\beta$ from 1 to 8 (left panel).}
(b) {\sl Spectral index
as a function of $d$. Now curves from
top to bottom are those corresponding to $\beta$ from 1 to 8 (right panel).}}
\label{cas1}
\end{figure}

{}From the full potential (\ref{IntPot}) we may compute the value of
the dimensionless skewness ${\cal S}_1$ (for convenience we will be taking
the field $\phi$ normalised in Planck mass units from here on)
\be
\label{11}
{\cal S}_1 =
0.17 {\cal Q}   \left[
{\beta (\beta - 4)\over \phi^2} + {4\beta^4 (\beta - 1) -
192\pi \beta^2 (\beta - 6)\phi^2 \over \phi^2 (48\pi \phi^2 - \beta^2)^2 }
\right] ,
\ee
evaluated at $\phi\simeq\phi_{60}$.
In Eq.(\ref{11}) we took ${\cal I}_{3/2}(n)\simeq 4.5$ (from
Eq. (\ref{eq63bis})), that is the case for the specific examples we discuss
below. A plot of this quantity as a function of
$d\equiv \phi^2 - \beta^2 / 16\pi > 0$
(the value of the squared of the field beyond the minimum allowed)
for different values of $\beta$ is
given in Figure \ref{cas1} (left panel).
Note that both positive and negative values of
${\cal S}_1$ are therefore allowed just by modifying the choice of
$\beta$. Another generic feature is the rapid decrease of the
non--Gaussian amplitude for increasing values of the field beyond
$\beta / \sqrt{16\pi}$ (\ie, $d>0$). Clearly this is because
for large $\phi$ we approach the slow--roll region where the steepness
becomes increasingly small.

Similar calculations may be done for the spectral index.
The explicit expression of $n$
calculated from (\ref{12}) for
our potential (\ref{IntPot})
is complicated and not very illuminating.
Figure \ref{cas1} (right panel)
illustrates the variation of $n$ as a function of $d$
(\ie, the scale dependence of the spectral index)
for different values of the parameter $\beta$.

\begin{figure}[tbp]
\vspace{-2cm}
  \begin{center}
    \leavevmode
    \epsfxsize =  7cm  %
    \epsfysize =  8cm  %
    \epsffile{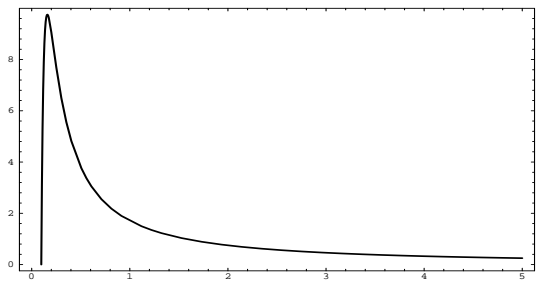}
  \end{center}
\vspace{-3cm}
\caption{{\sl Inflaton potential $V(\phi)$ for the parameter choice
$\beta = 1.2$ ($f = 0.77$).
The field is taken in units of $m_P$, while the potential is normalised in
units of $(10^{-1} m_P A^{1/f})^2$.}}
\label{cas2}
\end{figure}

These two figures show that for acceptable values of the spectral index
very small amplitudes for ${\cal S}_1$ are generally predicted.
As an example we consider $\beta = 1.2$. This ansatz yields a negative
${\cal S}_1$ being $\phi_{60} \simeq 0.37$ the value of the field that
maximises $\vert {\cal S}_1\vert \simeq 2.25 {\cal Q}$.
We show in Figure \ref{cas2}
the form of the inflaton potential (\ref{IntPot}) for this particular
$\beta$. We see that for the scales that exit the Hubble radius 60
e--foldings before the end of inflation\footnote{This model provides
	no natural end to the inflationary
        expansion. Therefore we could invoke arguments similar to those
        found in the literature [\eg, Barrow \& Maeda, 1990]
        within which suitable modifications of the potential or bubble
        nucleation (\eg, within an extended inflation model) would be
        responsible for the end of inflation.}
the value of the inflaton, $\phi_{60}$, is located in the steep region
beyond the maximum of the potential.
The value of the spectral index associated with this choice of $\beta$ is
$n\simeq 1.29$ (a 3\% below the first--order result $n\simeq 1.34$).
This value in excess of unity for the scales under
consideration yields a spectrum with less power on large scales
(compared with a Harrison--Zel'dovich one) making the long wave length
gravitational wave contribution to the estimated quadrupole subdominant.
The slow-roll parameters
for this scale are $\epsilon = 0.16$, $\eta = 0.37$ and
$\xi = 1.01$\footnote{Although the slow--rolling assumption is not
        necessary for finding solutions of the  field equations, it is
        instructive to focus also on this limit as it represents the attractor
        solution for large enough time.}.
While the first two are smaller than one, $\xi$ is not and so cannot be
considered an expansion variable on the same footing as $\epsilon$ and
$\eta$. Terms proportional to $\xi$ (and therefore non negligible)
in Eq. (\ref{12})
are those not included in the second--order analysis done for
the first time in Ref.\cite{StewartLyth93}.
Taking the {\em COBE}--DMR normalisation we finally get
${\cal S}_1 \simeq - 4.4\times 10^{-5}$.

Let us now consider $\beta = 7$. Now the potential falls to zero as a
power--law much more rapidly than in the previous case.
If we take $n\simeq 0.8$ we see from Figure \ref{cas1} (right panel)
that $d\simeq 6.05$
($\phi_{60}\simeq 2.65$) corresponding to the
slow--rolling region of the potential\footnote{In this region we can
        effectively neglect the term $\propto\phi^{-2}$
        in the parenthesis of Eq. (\ref{IntPot}) leading to a simplified
        form for $V$.}.
For the parameters we find the following  values: $\epsilon = 0.13$,
$\eta = 0.17$ and $\xi = 0.27$.
{}From these we see that the first--order result
($n\simeq 0.82$)
is
2.5 \%
above the full second--order one.
In this case we get
${\cal S}_1 \simeq 0.50 {\cal Q}$.
Now gravitational waves contribute substantially to the detected
quadrupole. Actually we have
${\cal Q}_T^2 / {\cal Q}_S^2 \simeq 1.99$
[Liddle \& Lyth, 1993; Barrow \& Liddle, 1993; Kolb \& Vadas, 1994]
(while we would have had $\sim 1.89$ up to first order).
Thus the estimated quadrupole should be multiplied by a factor
$(1 + 1.99)^{-1/2}$ to correctly account for the tensor mode
contribution. Finally, we get
${\cal S}_1 \simeq  7.5\times 10^{-6}$.

\begin{flushleft}
{\bf Polynomial potential}
\end{flushleft}

We are interested in considering a potential of the form
\be
\label{PolyPot}
V(\phi) = A (\frac{1}{4}\phi^4 + \frac{\alpha}{3}\phi^3 +
\frac{\beta}{8}\phi^2 ) + V_0
\ee
where for convenience $\phi$ is written in Plack mass units and $A$,
$\alpha$ and $\beta$ are dimensionless parameters. Translation
invariance allows us to omit the linear $\phi$ contribution to $V$.
A detailed analysis of a potential of the form (\ref{PolyPot}) was done
by Hodges \etal ~ [1990].
Parameter space diagrams were constructed and
regions where non--scale invariance was expected were isolated. Here we
will just summarise what is necessary for our study.

Taking $\beta > 8\alpha^2 / 9$ ensures that $\phi = 0$ is the global
minimum and therefore $V_0 = 0$.
If we further require $\beta > \alpha^2 $ then no false vacua are
present.
Scalar curvature perturbations are conveniently expressed in terms of the
gauge--invariant variable $\zeta$
[Bardeen, Steinhardt \& Turner, 1983; Liddle \& Lyth, 1993].
The power spectrum associated with it, assuming slow--roll evolution of
the scalar field in the relevant region of the potential, is given by
${\cal P}_{\zeta}^{1/2}(k)\propto H^2 / \dot\phi$
evaluated at horizon crossing time.
Equivalently ${\cal P}_{\zeta}^{1/2}(k)\propto V^{3/2} / (m_P^3 V') $ which
suggests that regions of the parameter space  where the slope of the
potential goes through  a minimum or a maximum will be of interest as far
as broken scale invariance is concerned.
The presence of these extrema in $V'$ is guaranteed by taking
$\alpha^2 < \beta < 4\alpha^2 / 3$, with $\alpha < 0$ in order for the
scalar field to roll down the potential from the right.
We will thus concentrate our study in the vicinity of  an inflection
point $\phi_f$, \ie, near the curve $\beta = \alpha^2$ (but still
$\beta > \alpha^2$). In this limit we have $\phi_f \simeq -\alpha /2$
and the number of e--foldings taking place in the region of approximately
constant slope about $\phi_f$ is given by
$N = -\pi\alpha^3 / \sqrt{3(\beta - \alpha^2)}$ [Hodges \etal, 1990].
Fixing $N=60$ as the number required for achieving sufficient inflation,
the interesting parameters ought to lie close to
the curve
$\beta \simeq \alpha^2 + \pi^2 \alpha^6 / (3 N^2)
       \simeq \alpha^2 + y \alpha^6 $
with $y \simeq 9.1\times 10^{-4}$.

We find
$ X_{60}^2 - 4 X_{60}' = 192~ y ~(9 y - \alpha^{-4}) /
[\alpha^2(6 y + \alpha^{-4})^2]$
where, again, we are normalising the field in Planck mass
units, $y \simeq 9.1\times 10^{-4}$ and we have
evaluated the field for $\phi_{60} = \phi_f \simeq -\alpha /2$.

\begin{figure}[tbp]
\vspace{-2cm}
\begin{center}
\leavevmode
{\hbox %
{\epsfxsize = 7cm \epsffile{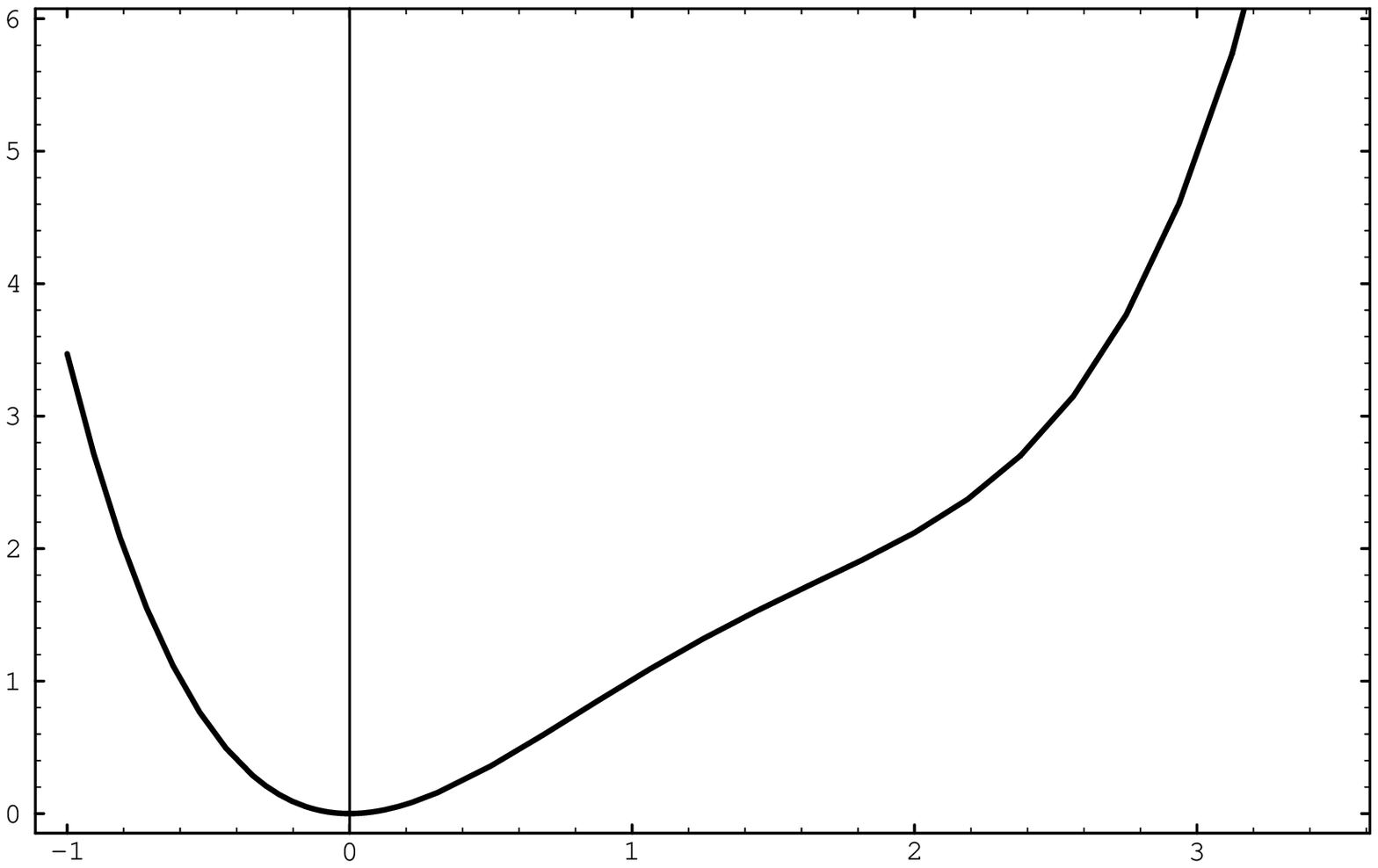}}{\epsfxsize=7cm \epsffile{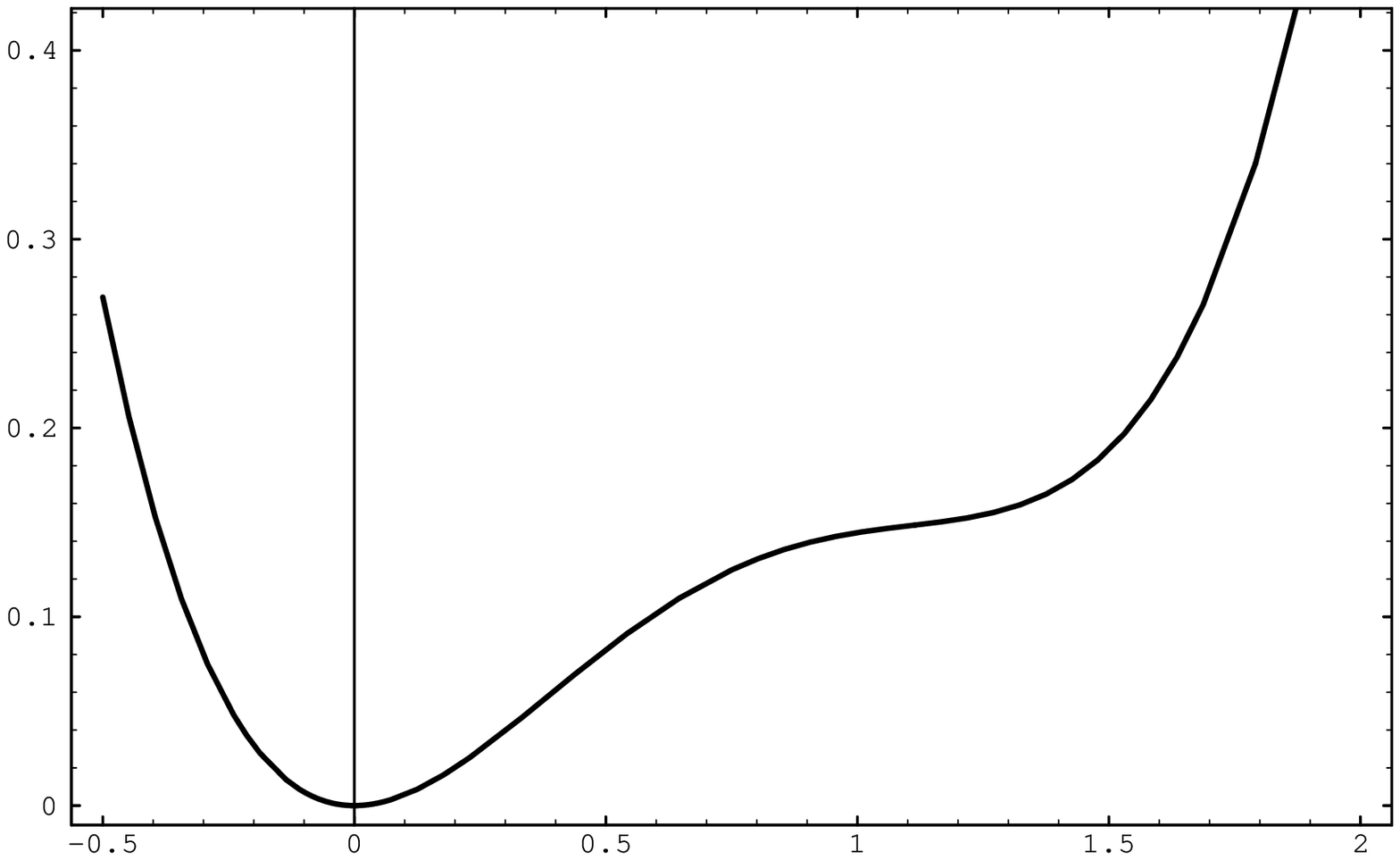} }}
\end{center}
\vspace{-3cm}
\caption{{\sl Polynomial potential $V(\phi)$ as a function of the
scalar field ($\phi$ is taken in Planck mass units). The overall
normalisation parameter $A$ is taken to be one;
(a) for $\alpha = -3.69$ and $\beta = 15.92$ (left panel);
(b) for $\alpha = -2.24$ and $\beta =  5.13$ (right panel).}}
\label{cas3}
\end{figure}

Variation of $\alpha$ in the allowed range results in both
positive and negative values for ${\cal S}_1$. Clearly
$\beta > 10\alpha^2 / 9$ yields ${\cal S}_1 > 0$.
Although in this region of the parameter space the value of $\alpha$ that
makes ${\cal S}_1 $ maximal corresponds to $\alpha = -5.44$, this value
conflicts with  the requirement  $\beta < 4\alpha^2 / 3$ for the
existence of an inflection point. We take instead $\alpha = -3.69$ which
makes ${\cal S}_1 $
the largest possible one and at the same time agrees within
a few percent with approximating $\phi_f\simeq -\alpha /2$.
Then $\beta \simeq \alpha^2 + y \alpha^6 \simeq 15.92 $. This guarantees
we are effectively exploring the neighbourhood of the curve
$\beta = \alpha^2$ in parameter space.
A plot of the potential for these particular values is given in
Figure \ref{cas3} (left panel).
The slow--roll
parameters in this case have values: $\epsilon = 5.94\times 10^{-3}$,
$\eta = 5.84\times 10^{-3}$ and $\xi = 1.32\times 10^{-1}$.
By using Eq. (\ref{12}) we get $n\simeq 0.99$.

Let us consider now the case where $\alpha^2 < \beta < 10\alpha^2 / 9$.
This choice of potential
parameters leads to ${\cal S}_1 < 0$. The non--Gaussian
signal gets maximised for $\alpha = -2.24$ and $\beta = 5.13$.
Figure \ref{cas3} (right panel) shows the potential in this case.
Note the resemblance between this form of the potential and that of the
hybrid [Linde, 1994]
model $V_0 + m^2\phi^2 /2$ for $V_0$ dominating and in the
vicinity of the flat region.
In that case the same sign of the dimensionless skewness
(cf. \S\ref{Workedexamples})
and `blue' spectra [Mollerach \etal, 1994], $n>1$, were predicted.
For this value of $\alpha$ we get: $\epsilon = 9.33\times 10^{-4}$,
$\eta = 6.76\times 10^{-3}$ and $\xi = 2.76$.
We get $n\simeq 1.01$.

We can now use Eq. (\ref{quad}) for the quadrupole to find the overall
normalisation constant $A$ of the potential.
Bennett \etal ~ [1994] get a best fit
$Q_{{\it rms}-PS} = (17.6\pm 1.5)\mu K$ for $n$ fixed to one, as it is our
present case to a very good approximation.
Thus, for $\alpha = -3.69$ we get $A=2.87\times 10^{-12}$
and ${\cal S}_1 \simeq 1.1\times 10^{-6}$
while for $\alpha = -2.24$ we get $A=5.88\times 10^{-12}$
and ${\cal S}_1 \simeq -1.9\times 10^{-6}$.

The above two values for $n$ show that the departure from scale invariance is
actually very small (in fact, this is because we are
exploring a very narrow range of scales).
Note also the relatively large value of $\xi$ in the
last case compared with $\epsilon$ and $\eta$. This tells us
that the terms proportional to $\xi$ are non negligible in general.
Also the rather small amplitudes for ${\cal S}_1$ agree with
previous numerical analyses [Hodges \etal, 1990] in which adherence to the
correct level of anisotropies in the CMB radiation under the simplest
assumptions of inflation, like slow--rolling down with potential
(\ref{PolyPot}), practically precludes any observable non--Gaussian
signal.

\section{The r.m.s. skewness of a Gaussian field}
\label{sec-skewcosmicvariance}
\markboth{Chapter 5. ~PRIMORDIAL NON--GAUSSIAN FEATURES}
{\S ~5.4. ~THE RMS SKEWNESS OF A GAUSSIAN FIELD}

It was first realised by Scaramella \& Vittorio [1991] that detecting a
non--zero three--point function or skewness for temperature
fluctuations in the sky cannot be directly interpreted as a signal for
intrinsically non--Gaussian perturbations. In fact, even a Gaussian
perturbation field has non--zero chance to produce a non--Gaussian sky
pattern. This problem is related to what is presently known as
{\sl cosmic variance},
and is particularly relevant for fluctuations on large angular
scales, \ie, for low--order multipoles of the temperature fluctuation field.
One way to quantify this effect is through the {\sl rms} skewness
of a Gaussian field  $\la C_3^2(0) \ra _{Gauss}^{1/2}$.
It is easy to find
\be
\la C_3^2(0) \ra _{Gauss} =
3\int^1_{-1}d \cos\alpha \la C_2(\alpha) \ra ^3 ~.
\ee
Scaramella \& Vittorio [1991] and, more
recently, Srednicki [1993], focused on the most popular case
of a scale--invariant spectrum, $n=1$ and
the {\em COBE}--DMR window function, corresponding to a Gaussian with
dispersion $\sigma=3^\circ\llap.2$ [\eg, Wright \etal, 1992]. We will be
interested here in the same quantity, but for various values of $n$.
\begin{figure}[tbp]
  \begin{center}
    \leavevmode
    \epsfxsize = 6cm      
    \epsfysize = 5.33 cm  
    \epsffile{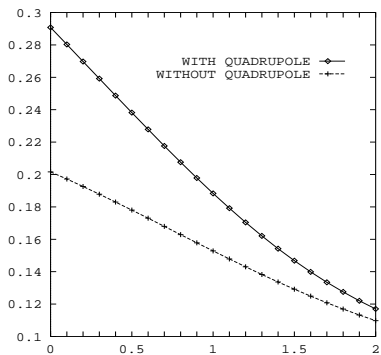}
  \end{center}
\caption{{\sl Normalised {\sl rms} skewness of a Gaussian temperature
fluctuation field as a function of the spectral index $n$, both including
and removing the quadrupole contribution.}}
\label{Figu2-CV}
\end{figure}
In Figure \ref{Figu2-CV}
we have plotted the normalised {\sl rms} skewness
$\la C_3^2(0) \ra_{Gauss}^{1/2} / \la C_2(0)\ra ^{3/2}$
as a function of the spectral index, both including and removing the
quadrupole: in both cases this ratio is in the range $0.1 - 0.3$ for
interesting values of $n$. The values obtained for $n=1$, both with and
without the $\ell=2$ contribution, are identical to those given by
Srednicki [1993], who adopted the same smoothing angle (the slightly
different definition of window function cannot affect our dimensionless
skewness ratios).
As a rough criterion, we can
conclude that, in order to detect primordial non--Gaussian signatures,
$\la C_3(0) \ra$ must be at least of the same order as $\la C_3^2(0)
\ra_{Gauss}^{1/2}$.

\section{Discussion}
\label{sec-skewDiscu}
\markboth{Chapter 5. ~PRIMORDIAL NON--GAUSSIAN FEATURES}
{\S ~5.5. ~DISCUSSION}

The results reported in the previous sections seem to preclude any chance of
actually obtaining observable non--Gaussian signals at least in the frame
of inflation, unless one resorts to more complicated multiple--field
models
[\eg, Allen, Grinstein \& Wise, 1987; Kofman \etal, 1990; Salopek, 1992].
However, it should be stressed that all the results reported in
\S\ref{sec-angularbispectrum}
would apply to any large--scale anisotropy where the temperature fluctuation
can be obtained by a local perturbative calculation (this is not the
case, for instance, for secondary anisotropies, which are to be ascribed
to an integrated effect -- see Chapter \ref{chap-integrated}):
under these conditions the {\sl hierarchical} form of Eq. (\ref{eq717}),
where the bispectrum is a sum of products of two power--spectra, holds.

As we have seen, single--scalar--field inflationary models generally
lead to very small skewness ratios
${\cal S}_1 = \la C_3(0) \ra/ \la C_2(0) \ra^{3/2}$, so that their
non--Gaussian features cannot be distinguished from the cosmic {\sl rms}
skewness.
Both for the exponential potential and for the quadratic and quartic ones
the skewness can be estimated by taking just the first term in
the brackets of Eqs. (\ref{eq63bis}) and (\ref{eq63}).
An upper limit on its magnitude is then obtained by requiring that they
give rise to an accelerated universe expansion (\ie, inflation), which
provides a constraint on the steepness of the potential, $X^2 < 24\pi$.
This corresponds to ${\cal S}_1 \lsim  10^{-4}$ and ${\cal S}_2 \lsim 20$.

Notwithstanding the non--zero value of the skewness, any actual
detection of this signal is hardly distinguishable from
the cosmic variance noise in which it is embedded: the
intrinsic limitation induced by our impossibility of making
measurements in more than one universe.
In fact, the overall coefficient of ${\cal S}_1$ is generically
much smaller than the dimensionless {\sl rms} skewness calculated from an
underlying Gaussian density field. Removing the quadrupole contribution
to reduce the cosmic {\sl rms} skewness
does not change this main conclusion, because the predicted mean skewness
would also be reduced by a comparable factor.
Things get even worse if we take into account the `sample variance' due
to the galactic cut in the maps [Scott \etal, 1994].
Therefore, any possible skewness detection on the
{\sl COBE} scale is most probably due to these statistical effects than to
any primordial non--Gaussian feature.
In this sense the quasi--Gaussian inflationary predictions for the CMB
anisotropies are in full agreement with the recent analysis of the
three--point function and the skewness from {\sl COBE}--DMR data
[Hinshaw \etal, 1994, 1995; Smoot \etal, 1994].

\chapter{Integrated effects}
\label{chap-integrated}
\markboth{Chapter 6. ~INTEGRATED EFFECTS}
         {Chapter 6. ~INTEGRATED EFFECTS}

\section{Introduction}
\label{sec-integrated0}
\markboth{Chapter 6. ~INTEGRATED EFFECTS}
{\S ~6.1. ~INTRODUCTION}

After the detection by {\it COBE}--DMR [\eg, Smoot \etal, 1992; Bennett
\etal, 1994] of anisotropies in the cosmic microwave background,
several other measurements of anisotropies at different angular scales
have been announced. This new set of observations has provided an
extra tool to improve our understanding of the structure formation
process. As we have already mentioned in previous sections, the usual
analysis in terms of the two--point correlation
function of the {\it COBE} data fixes the {\sl rms} amplitude of the
primordial perturbations. A comparison with measurements at smaller
angular scales encodes information about the matter content of the
universe and other cosmological parameters to which these are
sensitive. Furthermore, an analysis of the three--point correlation
function of the data, as it has recently been performed for the {\it
COBE}--DMR first-- and two--year data [Hinshaw \etal, 1994, 1995], can
provide useful clues about the statistical properties of cosmological
perturbations. Further analyses probing the statistical nature of
temperature fluctuations have been performed on the {\sl COBE}--DMR
maps by
Smoot \etal  [1994], Kogut \etal  [1994] and Torres \etal  [1995].
As we already pointed out,
these studies are of particular relevance as they could help to
distinguish among inflation and topological defects as the source
of the primordial fluctuations.

We saw in previous Chapters that inflationary models
predict in general a quasi--Gaussian distribution of density
perturbations.
This was due to the fact that they are originated by the
quantum fluctuations of a very weakly coupled scalar field.
The effect of the small non--linearities in the inflaton
dynamics (cf. \S\ref{sec-StochasticStuff})
and the mean value of the resulting imprints on the
three--point function of the gravitational potential
[Falk \etal, 1993; Gangui \etal, 1994; Gangui, 1994] were
several orders of magnitude smaller than the typical values expected
for a particular realisation of an ensemble of Gaussian universes
[Scaramella \& Vittorio, 1991; Srednicki, 1993].
As the anisotropies observed at large angular scales are
essentially determined by the fluctuations in the gravitational
potential, the CMB three--point correlation function at the scales
probed by {\it COBE} is also expected to be Gaussian.
An analysis of
the {\sl rms} skewness of inflationary models giving rise to
isocurvature baryon perturbations, by Yamamoto and Sasaki [1994],
yields values smaller than those obtained for curvature perturbation
models.
On the other hand, topological defects are the typical example
of non--Gaussian distributed perturbations. However, for many
observations, the relevant object is not the individual effect of each
particular defect, but the superposition of many of them, which
results in a nearly Gaussian pattern [\eg, Gott \etal, 1990; Coulson
\etal, 1994; Gangui \& Perivolaropoulos, 1995]. We will discuss this in
Chapter \ref{sec-kurtostrings} when we will consider an analytic model
for the generation of CMB kurtosis from cosmic strings.

Moreover, it has been
shown [Scherrer \& Schaefer, 1994] that, for a wide class of models
leading to non--Gaussian density perturbations, the corresponding
temperature fluctuation field, as induced by the Sachs--Wolfe effect,
is nearly Gaussian thanks to the Central Limit Theorem.
Thus, a
detailed analysis of the predictions for various models, taking into
account all the relevant effects, is necessary to find out the best
tools to distinguish among them.

The analysis of the three--point correlations in the {\it COBE}--DMR
two--year anisotropy maps shows evidence for a non--vanishing signal
in the data but at a level consistent with a superposition of
instrumental noise and Gaussian CMB fluctuations [Hinshaw \etal, 1995].
As the noise level will diminish rapidly with additional data,
in the four--year map the sensitivity is expected to be limited by the
cosmic variance.

Even for the case of primordial Gaussian curvature fluctuations, the
non--linear gravitational evolution gives rise to a non--vanishing
three--point correlation function of the CMB. It has been argued by
Luo \& Schramm [1993] that the amplitude of this effect is several
orders of magnitude larger than that predicted by Falk \etal ~[1993]
for a cubic self--interacting inflaton model, and
with a similar angular dependence. A more recent estimate of this
effect on the skewness of the CMB by Munshi, Souradeep \& Starobinskii
[1995] finds that the amplitude of the gravitational and inflationary
non--linearity contributions is comparable for typical inflation
models. We will provide in this Chapter a detailed and more general
analysis of the possible observational consequences of the
non--linear gravitational growth of initially Gaussian perturbations
on the CMB three--point function.

\section{Contribution to the CMB three--point correlation function
         from the Rees--Sciama effect}
\label{sec-integrated1}
\markboth{Chapter 6. ~INTEGRATED EFFECTS}
{\S ~6.2. ~CONTRIBUTION TO THE THREE--POINT FUNCTION...}

At large angular scales the anisotropies in the CMB are given by
[\eg, Mart{\'\i}nez--Gonz\'alez, Sanz \& Silk, 1990;
see also Chapter \ref{chap-CMBanis}]
\begin{equation}
\frac{\Delta T}{T}(\hat \gamma)
= \frac{1}{3}\Phi(\hat\gamma \eta_0,\eta_r)
 + 2 \int_{\eta_r}^{\eta_0} d \eta
\frac{\partial}{\partial \eta} \Phi ({\bf x},\eta) \bigg|_{ {\bf x}=
\hat \gamma (\eta_0 - \eta)},
\label{rs}
\end{equation}
where the first term represents the well--known Sachs--Wolfe effect
[Sachs \& Wolfe, 1967], while the second one corresponds to the
Rees--Sciama, or integrated Sachs--Wolfe effect [Rees \& Sciama, 1968].
In the previous formula $\hat \gamma$ denotes a direction in the sky
and $\eta$ is the conformal time, with $\eta_0$ and $\eta_r$ the
present and recombination times, respectively. The contribution of the
Rees--Sciama effect to the total anisotropy is small in a flat
matter--dominated universe because the gravitational potential keeps
constant in time within the linear regime. It gives however a
non--vanishing contribution when the non--linear evolution of
perturbations is taken into account. This non--linear contribution
will generate a non--vanishing three--point function, even for
primordial Gaussian perturbations in the energy density.

Let us start off from the Poisson equation, which in comoving
coordinates reads
$\nabla^2\Phi({\bf x},\eta) = 4\pi G \rho a^2 \delta({\bf x},\eta)$.
We will perform a perturbative analysis and therefore write, \eg,
$\delta({\bf x},\eta) = \delta_1({\bf x},\eta) + \delta_2({\bf x},\eta)
+ \dots$. Let us now outline how to get the first two
contributions to the gravitational potential
(\ie, $\Phi({\bf x},\eta)= \Phi_1({\bf x})+\Phi_2({\bf x},\eta)$).
We will normalise the scale factor such as $a_0 = 1$ (today) and the
distance to the horizon is given by $\eta_0 = 3 t_0 = 2 / H_0$, in a flat
matter domination era ($a(\eta) = (\eta/\eta_0)^2$).
Thus, the background energy density is
$\rho = (3 H_0^2 / 8\pi G) a^{-3}$, and the Poisson equation may be cast
as $\nabla^2\Phi({\bf x},\eta) = (3 H_0^2 / 2 a) \delta({\bf x},\eta)$.
Fourier transforming this we readily get
\be
\label{firstorder}
\Phi_1({\bf k}) = - {3 H_0^2 \over 2 k^2}
{ \delta_1({\bf k},\eta) \over a(\eta) },
\ee
to lowest order.
In linear perturbation theory we have
$\delta_1({\bf k},\eta) \propto a(\eta)$ and so we get the well--known
result that $\Phi_1$ is time independent\footnote{As we have already
mentioned, this conclusion also applies in a radiation dominated regime}.

The second--order quantities may be expressed in terms of first--order ones
and so we have
\be
\delta_2({\bf x},\eta) = \frac{5}{7}\delta_1^2 +
\frac{2 a}{3 H_0^2} \vec\nabla\delta_1\cdot\vec\nabla\Phi_1 +
\frac{2}{7} (\frac{2 a}{3 H_0^2})^2
{\Phi_1}_{,ij} {\Phi_1}^{,ij}.
\ee
After Fourier transforming and straightforward algebra we get
(upon use is made once again of the Poisson equation)
the second--order gravitational potential [Peebles, 1980; Fry, 1984]
\begin{equation}
\Phi_2({\bf k},\eta)=
-{a(\eta)\over 21 H_0^2 k^2}
\int \frac{d^3{\bf k}'}{(2\pi)^3} \Phi_1({\bf k}-{\bf k}') \Phi_1({\bf
k}')\big(3 k^2 k'^2 +7 k^2 {\bf k}\cdot{\bf k}' -10({\bf k}\cdot{\bf
k'})^2\big) ~.
\label{phi}
\end{equation}

The three--point correlation function for points at three arbitrary
angular separations $\alpha$, $\beta$ and $\gamma$ is given by the
average product of temperature fluctuations in all possible three
directions with those angular separations among them. The general
expression is given in \S\ref{sec-3pointfun}.
In this section, for
simplicity, we will restrict ourselves to the {\it collapsed} case,
corresponding to the choice $\alpha=\beta$ and $\gamma=0$, that is one
of the cases analysed for the {\it COBE}--DMR data by Hinshaw \etal ~
[1994, 1995] (the other is the {\it equilateral} one,
$\alpha=\beta=\gamma$). The collapsed three--point correlation
function of the CMB is given by
\begin{equation}
C_3(\vec x;\alpha) \equiv \int {d\Omega_{\hat \gamma_1}\over 4\pi}
                          \int {d\Omega_{\hat \gamma_2}\over 2\pi}
\D (\hat\gamma_1) \D^2(\hat\gamma_2)
\delta(\hat\gamma_1\cdot\hat\gamma_2 -\cos \alpha).
\label{skew1}
\end{equation}
For $\alpha=0$, we recover the well--known expression for the
skewness, $C_3(0)$. By expanding the temperature fluctuations in
spherical harmonics, cf. Eq. (\ref{expansion}),
we can write the collapsed three--point function as in equation
(\ref{skew}).
Then, by taking the mean value  we get
\be
\label{missing}
\la C_3 (\alpha)\ra = {1 \over 4 \pi}
\sum_{\ell_1,\ell_2,\ell_3}\sum_{m_1,m_2,m_3}
P_{\ell_1}(\cos\alpha)
\la a_{\ell_1}^{m_1} a_{\ell_2}^{m_2} {a_{\ell_3}^{m_3}}^* \ra
{\cal W}_{\ell_1} {\cal W}_{\ell_2} {\cal W}_{\ell_3}
{\cal H}_{\ell_3\ell_2\ell_1}^{m_3 m_2 m_1} ~.
\ee
The mean angular bispectrum predicted by a given model can be obtained
from\footnote{For a general discussion of the properties of the angular
bispectrum see, \eg, Luo [1994].}
\begin{eqnarray}
\langle a_{\ell_1}^{m_1} a_{\ell_2}^{ m_2} {a_{\ell_3}^{m_3}}^*
\rangle&=&
\int d\Omega_{\hat \gamma_1} d\Omega_{\hat \gamma_2}
d\Omega_{\hat \gamma_3}
{Y_{\ell_1}^{m_1}}^*(\hat \gamma_1)
{Y_{\ell_2}^{m_2}}^*(\hat \gamma_2)
{Y_{\ell_3}^{m_3}}(\hat \gamma_3)\nonumber\\
&&\langle \D(\hat \gamma_1) \D(\hat \gamma_2)
\D(\hat \gamma_3)\rangle ~.
\label{bis}
\end{eqnarray}
The contribution from the Sachs--Wolfe term has been
computed by Falk \etal ~[1993], Gangui \etal ~[1994] and Gangui
{}~[1994], accounting for primordial non--linearities due to inflaton
self--interactions. The leading contribution coming from the
Rees--Sciama effect is obtained from
\begin{eqnarray}
\label{leading}
\lefteqn{
\langle \D(\hat \gamma_1) \D(\hat \gamma_2)
\D(\hat \gamma_3)\rangle =}\nonumber\\&&
 \frac{2}{9}
\big\langle
\Phi_1(\hat \gamma_1 \eta_0)
\Phi_1(\hat \gamma_2 \eta_0) \int_{\eta_r}^{\eta_0} \! d\eta
\frac{\partial}{\partial
\eta} \Phi_2 ({\bf x},\eta)\bigg|_{{\bf x}=\hat \gamma_3 (\eta_0
-\eta)} \big\rangle \!
+ \! (\hat \gamma_1 \leftrightarrow \hat \gamma_3) \!
+ \! (\hat \gamma_2 \leftrightarrow \hat \gamma_3).
\end{eqnarray}
At the same order in perturbation theory there are also
non--vanishing contributions of the type $\frac {1}{27} \langle
\Phi_1(\hat \gamma_1 \eta_0) \Phi_1(\hat \gamma_2 \eta_0) \Phi_2 (\hat
\gamma_3\eta_0,\eta_r) \rangle$; compared to the non--local
Rees--Sciama terms, these terms are however suppressed by about two
orders of magnitude, because of both the $\eta$--dependence of
$\Phi_2$ and the different numerical factors.\footnote{One might
think that contributions of the kind $\sim
\langle \Phi_1 (\hat \gamma_1 \eta_0) \Phi_1(\hat \gamma_2 \eta_0)
        \Phi_2 (\hat \gamma_3\eta_0,\eta_r) \rangle$
should, on the contrary, dominate the signal since, as no integration
over the photon's path is involved, the three spots on the last scattering
surface --among which we compute correlations-- would be closer and therefore
`more correlated'
than in the situation in which two spots
(whose positions are, say, $\hat\gamma_1 \eta_0$ and $\hat\gamma_2 \eta_0$)
lay on the surface and
the third one ($\hat\gamma_3 (\eta_0 - \eta)$) sweeps the trajectory of the
photon towards us (and therefore would be further away from the other two),
resulting in a `loss of correlation' that would eventually
suppress the signal.
However, this is not the case for we should also remember the time
dependence in $\Phi_2$ (cf. Eq. (\ref{phi})).
It is precisely the increase in the scale factor
(of order $\sim 1000$ for standard scenarios) what counter--balances the
above mentioned effect and
(together with the factor 6 difference in numerical
coefficients:  $\frac {1}{27} \rightarrow \frac {2}{9}$)
makes the term in Eq. (\ref{leading})  the dominant one.}
Using the analytic
expression for $\Phi_2$ from Eq. (\ref{phi}), a straightforward
computation leads to
\begin{eqnarray}
\lefteqn{
\langle \D(\hat \gamma_1) \D(\hat \gamma_2) \D(\hat \gamma_3)
\rangle =}\nonumber\\&&
-\frac{1}{189} \int \frac{d^3 {\bf k}_1}{(2\pi)^3}\int \frac{d^3
{\bf k}_2}{(2\pi)^3} \int_{\eta_r}^{\eta_0} d\eta \eta
e^{i{\bf k}_1\cdot \hat \gamma_1 \eta_0} e^{i{\bf k}_2\cdot \hat
\gamma_2 \eta_0} e^{-i({\bf k}_1 +{\bf k}_2)\cdot\hat \gamma_3
(\eta_0 -\eta)}
\nonumber\\&&
\left(3 (k_1^2\! +\! k_2^2)\! +\! 7 ({\bf k}_1\! +\! {\bf
k}_2)^2 - \frac{10}{({\bf k}_1\! +\! {\bf k}_2)^2}\left((({\bf k}_1\!
+\! {\bf k}_2)\! \cdot \!{\bf k}_1)^2 \!+\!(({\bf k}_1\! + \! {\bf
k}_2)\! \cdot \!{\bf k}_2)^2\right)\right)\nonumber\\ &&
 P_\Phi(k_1) P_\Phi(k_2) + (\hat \gamma_1 \leftrightarrow
\hat \gamma_3) +(\hat \gamma_2 \leftrightarrow \hat \gamma_3),
\end{eqnarray}
with $P_\Phi(k)$ the gravitational potential power spectrum.
Using this expression and well--known integral relations for spherical
harmonics, the three angular integrations in Eq. (\ref{bis}) can be
performed and we obtain for the bispectrum
\begin{eqnarray}
\lefteqn{
\langle a_{\ell_1}^{ m_1} a_{\ell_2}^{ m_2} {a_{\ell_3}^{m_3}}^*
\rangle=
-\frac{4}{63 \pi^2} \int d^3 {\bf k}_1 \int d^3 {\bf k}_2
\int_{\eta_r}^{\eta_0} d\eta \eta \sum_{j_1,j_2,n_1,n_2}
i^{\ell_1+\ell_2-j_1-j_2}{\cal H}^{n_1 n_2 m_3}_{j_1 j_2 \ell_3}
}\nonumber \\
&&
j_{\ell_1}(k_1 \eta_0) j_{\ell_2} (k_2 \eta_0) {Y_{\ell_1}^{m_1}}^*
(\hat {\bf k}_1) {Y_{\ell_2}^{m_2}}^*(\hat {\bf k}_2)
j_{j_1}(k_1(\eta_0 -\eta)) j_{j_2}(k_2(\eta_0 -\eta)){Y_{j_1}^{n_1}}
(\hat {\bf k}_1) \nonumber \\
&&{Y_{j_2}^{n_2}}^*(\hat {\bf k}_2) P_{\Phi}(k_1)
P_\Phi (k_2)\left(\frac{20 k_1^2 k_2^2 +8 ({\bf k}_1 \cdot{\bf
k}_2)^2 +14 (k_1^2 + k_2^2) {\bf k}_1 \cdot{\bf k}_2}{k_1^2 + k_2^2
+2 {\bf k}_1 \cdot {\bf k}_2}\right),
\end{eqnarray}
where $j_\ell$ are spherical Bessel functions of order $\ell$. Now it
is possible to make the integrations in $d\Omega_{\hat {\bf k}}$ and
perform the summation over $m_i$ in Eq. (\ref{missing}), using relations
among Clebsh--Gordan coefficients\footnote{In particular, we use the
identity $\sum_{m=-\ell}^{\ell}\langle \ell ~2k ~m ~0 | \ell ~m
\rangle = (2 \ell + 1 ) \delta_{k0}$.} to obtain for the collapsed
three--point function
\begin{eqnarray}
\label{skewf}
\lefteqn{
\la C_3 (\alpha)\ra =
- {A^2 \over 63 \pi^2 32 \eta_0^6}
\sum_{\ell_1 \ell_2 \ell_3} (2\ell_1\!+\!1)(2\ell_2\!+\!1)
{\cal W}_{\ell_1}{\cal W}_{\ell_2} {\cal W}_{\ell_3}
 \langle \ell_1 \ell_2 0 ~0| \ell_3 0\rangle^2
P_{\ell_1}(\cos\alpha)
}\nonumber\\
&&\int_0^\infty d w_1 w_1^{n-3}
\int_0^\infty d w_2 w_2^{n-3} \int_{\eta_r/\eta_0}^1 d z ~\frac{1-z}{z}~
J_{\ell_1+\frac{1}{2}} (w_1)
J_{\ell_1+\frac{1}{2}} (w_1 z)\nonumber \\
&&J_{\ell_2+\frac{1}{2}} (w_2) J_{\ell_2+\frac{1}{2}} (w_2 z)
 \left( 10 (w_1^2 +w_2^2) +
5 {(w_1^2 - w_2^2)^2 \over w_1 w_2}
\ln\bigg|{w_1 - w_2 \over w_1 + w_2}
\bigg|\right) .
\end{eqnarray}
One can readily see from the last expression the advantages of our
approach: any primordial spectrum may be studied, as we have $P_\Phi
(k) \equiv A (k \eta_0)^{n-4}$.
To find the amplitude $A$ we just use Eq. (\ref{powerspec})
and substitute Eq. (\ref{quad}) in it, to get
\be
\label{powerspecbis}
P_{\Phi}(k) \simeq
{36 k^{-3} \over 5} \left({k\over k_*}\right)^{n-1}
{4\pi\over T_0^2} Q_{rms-PS}^2
{\Gamma^2(2-n/2)\Gamma(9/2-n/2) \over \Gamma(3-n)\Gamma(3/2+n/2)} ~,
\ee
where we also used ${\cal Q} =\sqrt{4\pi} Q_{rms-PS}/T_0$.
Now recall that the wave number $k_*$ refers to the largest scales
crossing the horizon today, and so
$k_*\simeq a_0 H_0$. As we are taking $a_0 = 1$ today and
$\eta_0 = 3 t_0 = 2/H_0$ for matter dominated regime, we have
$k_* = 2/\eta_0$.
Thus, we are normalising the amplitude of the spectrum to the
$Q_{rms-PS}$ value determined by {\it COBE}--DMR through the relation
\be
A/\eta_0^3=
{ 36~ 2^{1-n} 4 \pi Q_{rms-PS}^2 \over 5 T_0^{2} }
{\Gamma^2(2-n/2)\Gamma(9/2-n/2) \over \Gamma(3-n) \Gamma(3/2+n/2)}.
\ee
Moreover, the
expansion in multipoles allows us to specify any chosen window
function ${\cal W}_\ell$ as well as to subtract the desired multipole
contributions from our expression, and thus match the different
settings of various experiments (provided that we stay far enough from
the Doppler peak). The integrals in Eq.(\ref{skewf}) can be
numerically evaluated to obtain the prediction for the collapsed
three--point correlation function produced through the Rees--Sciama
effect, to be measured by the particular experiment.

\section{The r.m.s. collapsed three--point function}
\label{sec-integrated2}
\markboth{Chapter 6. ~INTEGRATED EFFECTS}
{\S ~6.3. ~THE RMS COLLAPSED THREE--POINT FUNCTION}

We turn now to the computation of the cosmic variance associated to
the collapsed three--point correlation function. If the fluctuations
in the CMB temperature were Gaussian, the mean value of the
three--point correlation function over the ensemble of observers would
be zero. However, as we are able to perform observations in just one
particular sky, this prediction comes with a theoretical error bar, or
`cosmic variance', that indicates the typical values expected for
particular realisations of the Gaussian process. Only non--vanishing
values larger than this can be interpreted as a signal of intrinsic
non--random phases in the distribution of CMB temperature
fluctuations. The expected amplitude of the cosmic variance $\langle
C_3^2(\alpha)\rangle$ at a particular angular scale can be computed
from Eq. (\ref{skew1})
\begin{eqnarray}
\langle C_3^2(\alpha) \rangle  &\!\equiv &\!
\int {d\Omega_{\hat \gamma_1}\over 4\pi}
\int {d\Omega_{\hat \gamma_2}\over 2\pi}
\int {d\Omega_{\hat \gamma_3}\over 4\pi}
\int {d\Omega_{\hat \gamma_4}\over 2\pi}
\delta(\hat\gamma_1\cdot\hat\gamma_2 -\cos \alpha)
\delta(\hat\gamma_3\cdot\hat\gamma_4 -\cos \alpha) \nonumber\\&&
\langle \D(\hat\gamma_1)\D^2(\hat\gamma_2)
\D(\hat\gamma_3)\D^2(\hat\gamma_4)\rangle ~.
\label{cv}
\end{eqnarray}
We assume that the multipole coefficients are Gaussian
distributed random variables with angular spectrum
given by Eqs. (\ref{angspectrum}) and (\ref{cl})
as it follows from $P_\Phi(k) \propto k^{n-4}$.
Using standard combinatorial properties, we can compute
$\langle C_3^2(\alpha)\rangle_{Gauss}$ from Eq. (\ref{cv})
\begin{eqnarray}
\label{rmscoll}
\lefteqn{\langle C_3^2(\alpha)\rangle_{Gauss}=
{2\over (4\pi)^3}
\sum_{\ell_1 \ell_2 \ell_3}  {\cal C}_{\ell_1}{\cal C}_{\ell_2}
{\cal C}_{\ell_3} {\cal W}_{\ell_1}^2 {\cal W}_{\ell_2}^2
{\cal W}_{\ell_3}^2 (2\ell_1 +1)
(2\ell_2+1)} \nonumber\\&&
P_{\ell_1}(\cos \alpha)(P_{\ell_1}(\cos \alpha)
+P_{\ell_2}(\cos \alpha)+ P_{\ell_3}(\cos \alpha))
\langle \ell_1 \ell_2 0 ~0 |\ell_3 0\rangle^2.
\end{eqnarray}
This last expression takes into account only the effect of the cosmic
variance on the collapsed three--point function; one can easily modify
it to allow for the experimental noise, by adding its contribution to
the angular spectrum.

\section{Results for the {\sl COBE}--DMR experiment}
\label{sec-integrated3}
\markboth{Chapter 6. ~INTEGRATED EFFECTS}
{\S ~6.4. ~RESULTS FOR THE {\sl COBE}--DMR EXPERIMENT}

\begin{figure}[tbp]
  \begin{center}
    \leavevmode
    \epsfxsize = 8cm 
    \epsfysize = 7cm 
    \epsffile{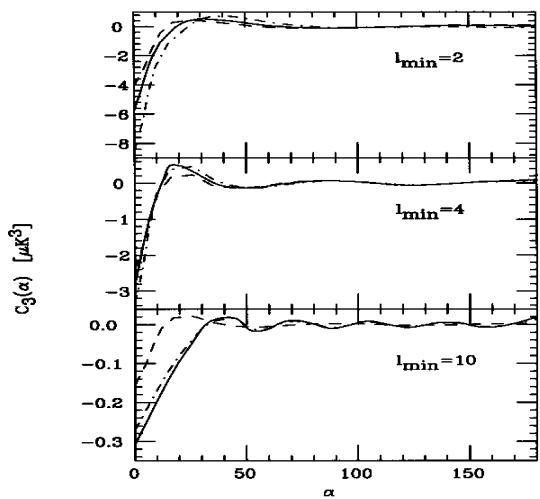}
  \end{center}
\caption{{\sl The collapsed three--point function
${\rm C}_3(\alpha) \equiv T_0^3 \la C_3(\alpha)\ra$
(where $\la C_3(\alpha)\ra$ is given
in the text), in $\mu K^3$ units, as
predicted by the Rees--Sciama effect, vs. the angular scale $\alpha$.
The dot--dashed line curves refer to the $n=0.7$ spectral index,
the solid line ones to $n=1$ and the dashed line ones to $n=1.3$.
The top, middle and bottom panels represent the cases
$l_{min}=2,~4,~10$ respectively.}}
\label{rs1}
\end{figure}

We can now apply our formalism to the {\it COBE}--DMR measurements. We
consider a Gaussian window function, $W_\ell \simeq
\exp(-\frac{1}{2}\ell(\ell+1)\sigma^2)$, with dispersion of the
antenna--beam profile $\sigma=3^\circ .2$ [Wright \etal, 1992].
Figure \ref{rs1}
shows the expected collapsed three--point function arising from the
Rees--Sciama effect, for three different values of the primordial
spectral index $n=0.7, ~1, ~1.3$. Note that all three types of
perturbation spectra can be easily generated in the frame of inflation
models [\eg, Mollerach, Matarrese \& Lucchin, 1994, and references
therein]. In all cases the perturbation amplitude has been normalised
to two--year {\it COBE}--DMR data, using the $\ell=9$ multipole
amplitude, $a_\ell =8~\mu {\rm K}$, according to the procedure
proposed by G\'orski \etal ~[1994], which leads to $Q_{rms-PS}= 24.2,
{}~19.5, ~15.9 ~\mu {\rm K}$, for $n=0.7, ~1, ~1.3$, respectively. The
top panel is obtained by subtracting from the map only the dipole
contribution. In the central panel also the quadrupole and the
octopole have been removed. In the bottom panel all the multipoles up
to $\ell=9$ have been subtracted. This procedure allows easy
comparison with the recent analysis of the two--year data by the
{\it COBE} team [Hinshaw \etal, 1995].

%

\begin{figure}[tbp]
  \begin{center}
    \leavevmode
    \epsfxsize = 8cm 
    \epsfysize = 7cm 
    \epsffile{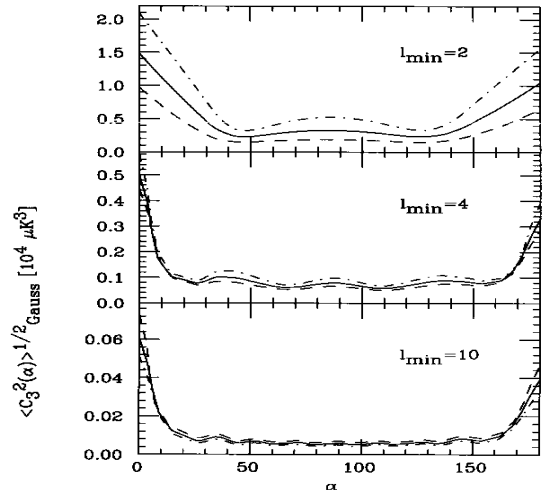}
  \end{center}
\caption{{\sl The {\sl rms} collapsed three--point function
$\langle {\rm C}_3^2(\alpha)\rangle^{1/2}_{Gauss} \equiv
T_0^3 \langle C_3^2(\alpha)\rangle^{1/2}_{Gauss}$
(where $\langle C_3^2(\alpha)\rangle^{1/2}_{Gauss}$  is given
in the text), in $10^4~\mu K^3$ units, as arising from the cosmic
variance of Gaussian fluctuations. Symbols are as in the previous figure.
Note that the top, middle and bottom panels, representing the cases
$l_{min}=2,~4,~10$ respectively, have different vertical scales.}}
\label{rs2}
\end{figure}

Figure \ref{rs2} contains the corresponding
plots for the {\sl rms} collapsed three--point function obtained from
our initially Gaussian fluctuation field,
$\langle C_3^2(\alpha)\rangle_{Gauss}^{1/2}$.

%

So far we have assumed full--sky
coverage; the effect of partial coverage, due to the cut at Galactic
latitude $\vert b \vert < 20^\circ$, increases the cosmic variance and
can be approximately taken into account by multiplying it by a factor
$1.56$ [Hinshaw \etal, 1994]. An analytic estimate [Srednicki, 1993;
Scott, Srednicki \& White, 1994] gives a somewhat smaller value, 1.23.
Note that the subtraction of low order multipoles leads to a strong
decrease both in the signal and in the cosmic variance. For all
considered multipole subtractions and angular separations, the
expected signal stays typically three orders of magnitude below the
cosmic variance, which makes the effect undetectable. The results are
found to be weakly dependent on the considered spectral index. For
comparison, the amplitude of the collapsed three--point function
produced by non--linearities in the inflaton dynamics [see Gangui et
al., 1994] is about a factor of ten smaller than that arising from the
Rees--Sciama effect. We can also evaluate the skewness parameter
${\cal S}_2 \equiv \langle (\Delta T/T)^3 \rangle / \langle (\Delta
T/T)^2\rangle^2$, which has the advantage of being normalisation
(\ie, $Q_{rms-PS}$) independent for the considered effect
(cf. \S\ref{sec-skewness});
in the case,
\eg, of only dipole subtraction and $n=1$, ${\cal S}_2 \approx - 5$.
Note that, in the same case, the heuristic estimate by Luo \& Schramm
[1993] leads to ${\cal S}_2 \approx 18$, while the method by
Munshi \etal ~[1995] yielded ${\cal S}_2 \approx -1$
(slight underestimate -- now revised to
${\cal S}_2 \approx -5$ ($-2$) with (without)
$\theta_{FWHP} = 10^\circ$ smoothing).
The corresponding {\sl rms} skewness parameter, arising from the cosmic
variance of a scale--invariant Gaussian field, is ${\cal S}_2^{rms}
\approx 1.3 \times 10^4 ~(Q_{rms-PS}/19.5~\mu {\rm K})^{-1}$.

One may try to increase the amplitude of the signal--to--cosmic
variance ratio, by referring to experimental settings, such as FIRS
[\eg, Ganga \etal, 1993] and Tenerife [\eg, Hancock \etal, 1994],
which are more sensitive to intermediate angular scales. However, the
little increase of this ratio is not enough to make it detectable;
furthermore, one has to bear in mind that the Tenerife measurement is
further affected by a large sample variance.

Finally, let us stress that our calculations, reported in
[Mollerach, Gangui, Lucchin \& Matarrese, 1995],
apply to measurements of
CMB anisotropies on scales large enough to be unaffected by the
Doppler peak. The possible extension of this effect to smaller scales
is discussed by Munshi \etal ~[1995], who however argue that also on
those scales the cosmic variance largely overcomes the intrinsic
non--Gaussian signal.

\chapter{On the CMB kurtosis from cosmic strings}
\label{sec-kurtostrings}
\markboth{Chapter 7. ~ON THE CMB KURTOSIS FROM COSMIC STRINGS}
         {Chapter 7. ~ON THE CMB KURTOSIS FROM COSMIC STRINGS}

\section{Introduction}
\label{sec-introstrings}
\markboth{Chapter 7. ~ON THE CMB KURTOSIS FROM COSMIC STRINGS}
                                      {\S ~7.1. ~INTRODUCTION}

We have seen in Chapter \ref{chap-theories}
how the strong link between cosmology and particle physics led to
the generation of mainly two classes of theories
which attempt to provide physically
motivated solutions to the problem of the origin of structure in the
universe.
According to the one class of theories, based on inflation
[\eg, Linde, 1990],
primordial inhomogeneities in the energy density arose from zero--point
quantum fluctuations of a scalar field during an
epoch of superluminal expansion of the universe.
These fluctuations may be shown to obey Gaussian statistics to a very
high degree and to lead to an approximately scale invariant power spectrum
of density perturbations.
According to the second class of theories, those based on topological
defects [\eg, Vilenkin \& Shellard, 1994; Hindmarsh \& Kibble, 1994],
primordial fluctuations were produced by a superposition of seeds made of
localised energy density trapped during a symmetry breaking phase transition in
the early universe. Topological defects with linear geometry are known as
cosmic strings and may be shown to be consistent with standard cosmology,
unlike their pointlike (monopoles) and planar (domain walls)
counterparts which require dilution by inflation to avoid overclosing the
universe. Cosmic strings are predicted to form during a phase transition in
the early universe by many but not all grand unified theories.

Cosmic strings can explain  the formation of large scale filaments and
sheets [Vachaspati, 1986; Stebbins \etal, 1987;
Perivolaropoulos \etal, 1990;
Vachaspati \& Vilenkin, 1991; Vollick, 1992; Hara \& Miyoshi, 1993],
galaxy formation at epochs $z\sim 2-3$
[Brandenberger \etal, 1987] and galactic
magnetic fields [Vachaspati, 1992b].
They can also account for peculiar velocities on large scales
[Vachaspati, 1992a; Perivolaropoulos \& Vachaspati, 1994],
and are consistent with the amplitude, spectral index
[Bouchet \etal, 1988; Bennett, Stebbins \& Bouchet, 1992;
Perivolaropoulos, 1993a; Hindmarsh, 1994]
and the statistics [Gott \etal, 1990;
Perivolaropoulos, 1993b; Moessner \etal, 1994;
Coulson \etal, 1994; Luo, 1994; Gangui \& Perivolaropoulos, 1995]
of the CMB anisotropies measured by {\sl COBE}
on angular scales of order $\theta\sim 10^\circ$.
Strings may also leave their imprint on the CMB mainly in three different ways.
The best studied mechanism for producing temperature fluctuations on the
CMB by cosmic strings is the Kaiser--Stebbins effect
[Kaiser \& Stebbins, 1984; Gott, 1985].
According to this effect, moving long strings present between the time
of last scattering $t_{ls}$ and today produce
(due to their deficit angle [Vilenkin, 1981])
discontinuities in the CMB temperature between photons reaching the
observer through opposite sides of the string.
Other mechanisms through which cosmic strings may produce CMB
fluctuations are based, \eg,
on potential fluctuations on the last scattering surface, by inducing
density and  velocity fluctuations to the matter surrounding  moving
strings; and on the Doppler effect, by producing peculiar velocities in
the ambient matter formed by the photon's last scatterers.

It was recently shown [Perivolaropoulos, 1994]
how, by superposing the effects
of these three mechanisms at all times from $t_{ls}$ to today, the power
spectrum of the total temperature perturbation
may be obtained. It turns out that (assuming standard recombination) both
Doppler and potential fluctuations at the last scattering surface
 dominate over
post--recombination effects on angular scales below $2^\circ$.
However this is {\sl not} the case for very large scales (where we will be
focusing in the present Chapter) and this justifies our neglecting the
former two sources of CMB anisotropies. The main effect of these neglected
perturbations is an increase of the Gaussian character of the fluctuations
on small angular scales. Thus, in what follows we will be considering
exclusively anisotropies generated through the Kaiser--Stebbins effect.

The main assumptions of the model were explained in
[Perivolaropoulos, 1993a].
The magnitude of the discontinuity is proportional not only to the deficit
angle but also to the string velocity $v_s$ and depends on the relative
orientation between the unit vector along the string ${\hat s}$ and
the unit photon wave--vector ${\hat\gamma}$.
It is given by [Stebbins, 1988]
\begin{equation}
\label{steb1}
{{\Delta T}\over T}=\pm 4\pi G\mu \gamma_s v_s \, {\hat\gamma}
   \cdot ({\hat v_s}\times {\hat s}) ,
\end{equation}
where $\gamma_s$ is the relativistic Lorentz factor and the sign changes when
the string is crossed.
Also, long strings within each horizon have random velocities, positions
and orientations.
In Eq. (\ref{steb1}), $G$ is Newton's constant and units are chosen such
that $\hbar = c = 1$, implying $G = m_P^{-2}$; $\mu \simeq m_{GUT}^2$
is the effective mass per unit length of the wiggly string.
$G\mu\simeq 10^{-6}$ is needed for consistency of the string scenario with
both a physically realisable GUT phase transition and a correct level
of CMB anisotropies.

In what follows we will be mainly concerned with the computation of
higher order correlation functions for CMB temperature anisotropies.
In this Chapter, in addition to calculating the excess
kurtosis parameter for the string--induced CMB temperature perturbations,
we also show (with explicit formulae) how to estimate
the intrinsic uncertainties with which the non--Gaussian signal ought
to be confronted, when worked out at very large angular scales.
Although the final result is somewhat disappointing (in the fact that
the signal cannot be resolved) we think it is interesting to show
explicit computations (even under simplifying assumptions within an
analytic approach) of these string--induced higher order correlation
functions.
In the following section we show how to construct the general q--point
function of CMB anisotropies at large angular scales produced through the
Kaiser--Stebbins effect. Explicit calculations are performed for the
four--point function and its zero--lag limit, the kurtosis.
Next, we calculate the (cosmic) variance for the kurtosis assuming Gaussian
statistics for arbitrary value of the spectral index and compare it with
the string predicted value (\S\ref{sec-rmsKurto}).
Finally, in \S\ref{sec-KurtoDisscu} we briefly discuss our results.

\vspace{18pt}
\section{The four--point temperature correlation function}
\label{sec-FourPointStrings}
\markboth{Chapter 7. ~ON THE CMB KURTOSIS FROM COSMIC STRINGS}
  {\S ~7.2. ~THE FOUR--POINT TEMPERATURE CORRELATION FUNCTION}

We discretise the time between $t_{ls}$ and today by a set of $N$ Hubble
time--steps $t_i$ such that $t_{i+1} = 2 \, t_i$, \ie, the horizon gets
doubled in each time--step.
For a redshift $z_{ls}\sim 1400$ we have
$N \simeq \log_{2}[(1400)^{3/2}] \simeq 16$.
In the frame of the multiple impulse approximation [Vachaspati, 1992a]
the effect of the
string network on a photon--beam is just the linear superposition of
the individual effects, taking into account compensation
[Traschen \etal, 1986; Veeraraghavan \& Stebbins, 1990; Magueijo, 1992],
that is, only those strings within a horizon distance from the beam
inflict perturbations to the photons.
According to the previous description of the Kaiser--Stebbins effect,
the {\sl total} CMB temperature shift
in the $\gg$ direction may be approximately written as the sum of
all single impulses on the microwave photons caused by
each of the strings, over all Hubble time--steps $N$ between
the time of last scattering and today [Perivolaropoulos, 1993a]
\be
\label{1}
\D(\gg) = 4\pi G \mu \gamma_s v_s
\sum_{n=1}^{N}\sum_{m=1}^{M} \beta^{mn}(\gg) ,
\ee
where $\beta^{mn}(\gg)$ gives us information about the velocity
$v^{mn}$ and orientation $s^{mn}$
of the $m$th string at the $n$th Hubble time--step and
may be cast as
$\beta^{mn}(\gg) =  \gg\cdot\hat R^{mn} $, with
$\hat R^{mn} = v^{mn} \times s^{mn} =
( \sin\theta^{mn}\cos\phi^{mn} ,
  \sin\theta^{mn}\sin\phi^{mn} ,
  \cos\theta^{mn}                ) $
a unit vector whose direction varies randomly according to the also
random  orientations and velocities in the string network.
In Eq. (\ref{1}), $M$ denotes the mean number of strings per horizon
scale, obtained from simulations [Allen \& Shellard, 1990; Bennett \&
Bouchet, 1988] to be of order $M \sim 10$.

\begin{figure}[tbp]
  \begin{center}
    \leavevmode
    \epsfxsize = 7cm  %
    \epsfysize = 6cm  %
    \epsffile{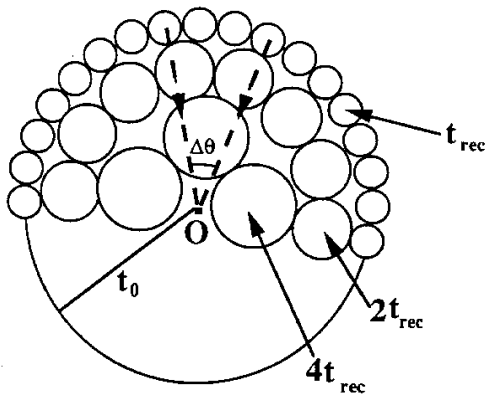}
  \end{center}
\caption{{\sl Two photon--beam paths separated by an angle $\Delta\theta$,
starting off their trip at the time of last scattering
(noted $t_{rec}$ in the plot) and being detected by an observer {\sl O}
at $t_{0}$. The horizon in three Hubble time--steps
is shown. The effects of strings during the first two time--steps
are uncorrelated for the particular angular beam separation shown.}}
\label{FigLeandros}
\end{figure}

Given an arbitrary angle on the sky $\Delta \theta$,
we will see now how to compute
the number of perturbations (impulses) common to a pair of photon beams
(see Figure \ref{FigLeandros}).
We are focusing here in the stringy--perturbations inflicted on photons
after the time of last scattering, on their way to us.
Recall we have discretised the time such that the horizon increases
by a factor 2 with each time--step. Thus, at a particular step
$p$ (with $1\leq p \leq N \simeq 16$) we have $t_p = 2^p t_{ls}$.
For a spatially flat, matter dominated universe
($a_t \equiv a(t) \propto t^{2/3}$) and in the small angle limit,
we may write
$\theta_{t_p} \simeq 1.5^\circ (z(t_p) / 1400)^{-1/2}$,
for the apparent angular size of the horizon at $t_p$.
Then, the ratio of the apparent horizon at time $t_p$ to that at
last scattering is given by
\be
\label{rev1}
{ \theta_{t_p} \over \theta_{ls} } \simeq
\left({ a_{t_p} \over a_{ls} }\right)^{1/2} =
\left({ t_p     \over t_{ls} }\right)^{1/3} = 2^{p/3}.
\ee

Let us now take an arbitrary angle $\alpha_{12}$ on the sky.
Consider a pair of photons with angular separation
$\alpha_{12} = \arccos(\got)$ {\sl greater} than $\theta_{t_p}$, for
a particular value of $p$ [here $\go$ ($\gt$) is the line--of--sight
direction of the first (second) photon].
Both of these photons will suffer temperature perturbations due to the
presence of $\sim M$ strings per Hubble time--step.
These perturbations, however, cannot be causally correlated at
$t_p$, since the size of the causal regions surrounding each of these
 photons is smaller than the angle $\alpha_{12}$ between them.
Thus, existing no overlapping of their horizons, each photon will be
perturbed by different strings, and therefore the perturbations will be
uncorrelated.

Given the angle $\alpha_{12}$ it is easy to compute the number of steps
$N_{uncorr}$ during which the perturbations on these photons are
{\sl uncorrelated}:
just substitute $\theta_{t_p}$ by $\alpha_{12}$ in  Eq. (\ref{rev1})
to get
$N_{uncorr}(\alpha_{12}) = 3 \log_{2}(\alpha_{12}/\theta_{ls})$,
which is valid for $\alpha_{12} \geq \theta_{ls} \simeq 1.5^{\circ}$.
Obviously, the {\sl larger} the angular scale $\alpha_{12}$,
the {\sl more} steps $N_{uncorr}$ will be needed for the angular size
of the horizon to overtake this scale.
If we now define $\alpha_{12}' \geq 0$ such that
$\alpha_{12} = \alpha_{12}' + \theta_{ls}$, subtract $N_{uncorr}$ from
the maximum number of steps $N$ of the discretisation of
the time along the photon path, and finally multiply the result by the
scaling number of strings per Hubble time--step $M$, we get
\be
\label{rev2}
{\cal N}_{corr}(\alpha_{12}) \equiv M [ N - N_{uncorr}] =
    M [ N - 3 \log_{2}( 1 + {\alpha_{12}\over\theta_{ls}})],
\ee
where ${\cal N}_{corr}(\alpha_{12})$ is the number of
{\sl correlated} `kicks' inflicted on a pair of photons on a scale
$\alpha_{12}$, after this scale has entered the horizon and up to the
present epoch.
In this equation, $\alpha_{12} \geq 0$ and primes were dropped.
In other words: ${\cal N}_{corr}(\alpha_{12})$ counts the total
number of correlated kicks onto the  pair of photons  after the time
at which the angle $\alpha_{12}$ between their lines--of--sight
{\sl fits} in the horizon scale.

Much in the same way, kicks inflicted on three photon--beams will be
uncorrelated at time $t_{p'}$ if any one of the three angles between any two
directions (say, $\alpha_{12}$, $\alpha_{23}$, $\alpha_{13}$)
is greater than $\theta_{t_{p'}}$, the size of the apparent horizon at time
$t_{p'}$.
So, in this case, we will be summing (cf. Eq.(\ref{1}))
over those Hubble time--steps $n$
greater than $p$, where $p$ ($>p'$) is the time--step when the condition
$\theta_{t_p} = {\rm max}[\alpha_{12}, \alpha_{23}, \alpha_{13}]$ is satisfied.
The same argument could be extended to any number of photon--beams (and
therefore for the computation of the q--point function of temperature
anisotropies)\footnote{A general expression (suitable to any angular
scale and to any source of temperature fluctuations) for the
three--point correlation function was given in [Gangui, Lucchin, Matarrese
\& Mollerach, 1994]. Although of
much more complexity, similar analysis may be done for the four--point
function and a completely general expression in terms of the angular
trispectrum may be found.}.

Let us now study the correlations in temperature anisotropies and
put the above considerations on more quantitative grounds, by
writing the mean q--point correlation function.
Using Eq. (\ref{1}) we may express it as
\be
\label{qpoint}
\la
{{\Delta T}\over T}(\hat\gamma_1) ... {{\Delta T}\over T}(\hat\gamma_q)
\ra
=
\xi^q \sum_{n_1,...,n_q=1}^N
\sum_{m_1,...,m_q=1}^M
\la \beta_1^{m_1 n_1} ... \beta_q^{m_q n_q} \ra
\ee
where $\hat\gamma_1 ...\hat\gamma_q$
are unit vectors denoting directions in the sky
and where $\beta_j^{m_j n_j}= {\hat\gamma_j}\cdot {\hat R^{m_j n_j}}$, with
${\hat\gamma_j}=
(\sin \theta_j \cos \phi_j,\sin\theta_j \sin \phi_j, cos\theta_j)$.
In the previous equation we called
$\xi \equiv 4 \pi G \mu \gamma_s v_s$.
We are here defining the ensemble average in the usual way; namely,
$\la \D(\gg_1) \D(\gg_2) \ra$ represents the average of
$\D(\gg_1) \D(\gg_2)$, for fixed directions ($\gg_1$, $\gg_2$) over a
large number of realisations of the CMB sky.
By assuming ergodicity, this is equivalent to performing a spatial average
over the whole sky sphere, with fixed angle
$\alpha_{12} = \arccos(\got)$, for a single CMB realisation.
Furthermore, we may supplement our definition
with the property that also $\la\cdot\ra = 0$ during those time--steps when
the impulses on different photon--beams are uncorrelated.
Thus, $\la\cdot\ra$ accounts also for compensation effects,
according to which at
sufficiently early times no string could have perturbed both photon--beams
whose causal regions were much smaller than the angle between them.
For simplicity we may always choose a coordinate system such that
$\theta_1=0$, $\phi_1 = 0$ (\ie, $\hat\gamma_1$ lies on the $\hat z$ axis)
and $\phi_2 = \pi/2$.
The seemingly complicated sum (\ref{qpoint}) is in practice
fairly simple to calculate because of the large number of terms that vanish
due to lack of correlation, after the average is taken. The calculation
proceeds by first splitting the product
$\beta_1^{m_1 n_1}...\beta_q^{m_q n_q}$ into all
possible sub--products that correspond to correlated kicks
at each expansion time--step
(\ie, $\beta^{m_j n_j}$'s with the same pair ($m_j, n_j$) corresponding
to the same string perturbing both beams),
and then evaluating the ensemble
average of each sub--product by integration over all directions of
${\hat R}^{m_j n_j}$.
To illustrate this technique we evaluate the two--, three-- and four--point
functions below.

Let us start off with the {\it two--point function}.
Having a correlated pair of beams in  $\go$ and $\gt$ directions from a
particular time--step $p$ onwards means simply that
$\hat R^{m_1 n_1} = \hat R^{m_2 n_2} \Leftrightarrow
n \equiv n_1 = n_2 > p \, , \, m \equiv m_1 = m_2$; otherwise the
${\hat R}^{m_j n_j}$'s
remain uncorrelated and there is no contribution to the mean
two--point function.
Therefore we will have
\be
\label{3}
\D(\gg_1) \D(\gg_2) = \xi^2
\sum_{n=p}^{N}\sum_{m=1}^{M} (\go\cdot\hat R^{mn})(\gt\cdot\hat R^{mn}) ,
\ee
where we wrote just the
correlated part on an angular scale $\alpha_{12} = \arccos(\got)$.
The uncorrelated part on this scale will vanish when performing the
ensemble average $\la \cdot \ra$, as we explained before.
Thus the mean two--point function may be written as
\be
\la
{{\Delta T}\over T}({\hat \gamma_1})
{{\Delta T}\over T}({\hat \gamma_2}) \ra =
\xi^2 \la \beta_1 \beta_2 \ra {\cal N}_{corr} (\alpha_{12}) ,
\ee
with ${\cal N}_{corr}(\alpha_{12})$ given in (\ref{rev2}).
Since we may always take
\be
\beta_1 = (0,0,1)\cdot {\hat R} ~~ ; ~~
\beta_2 = (\sin \alpha_{12}, 0, \cos \alpha_{12}) \cdot {\hat R} ~,
\ee
with
${\hat R}\equiv {\hat R}^{m_1 n_1}=
(\sin \theta \cos \phi,\sin\theta \sin \phi, cos\theta)$,
$\la {{\Delta T}\over T}({\hat\gamma_1})
     {{\Delta T}\over T}({\hat\gamma_2}) \ra$
may be calculated  by integrating over
$(\theta, \phi)$ and dividing by $4\pi$. The result is
\be
\la
{{\Delta T}\over T}({\hat\gamma_1}){{\Delta T}\over T}({\hat\gamma_2})
\ra
= \xi^2 {{\cos \alpha_{12}}\over 3} {\cal N}_{corr}(\alpha_{12}) ,
\ee
which may be shown [Perivolaropoulos, 1993a] to lead to a slightly tilted scale
invariant spectrum on large angular scales.

The {\it three--point function} may be obtained in a similar way.
In our analytic approach however,
the superimposed kernels of the distribution turn out to be
symmetric with respect to positive and negative perturbations and
therefore no mean value for the three--point correlation function arises
(in particular also the skewness is zero).
Similar conclusions were derived
from exact lattice solutions to the Nambu equations, for string evolution
and reconnection according to the Smith \& Vilenkin [1987] scheme,
adapted for evolving the network in an expanding universe
[\eg, Coulson \etal, 1994].
Notice also, that the effects of cosmic string loops that are ignored in the
present study are  expected to introduce a small skewness on scales of a few
arcseconds or smaller
(the typical angular scale of a string loop on the last
scattering surface is a fraction of an arcsecond).
On the other hand, the four--point function is easily found and a mean value
for the excess kurtosis parameter [Gangui, 1995] may be predicted,
as we show below.

In the case of the {\it four--point function} we could find terms
where (for a particular time--step) the photon--beams in
directions $\go$, $\gt$ and $\gth$ are all correlated amongst themselves,
but {\sl not} with
the beam $\gf$, which may be taken sufficiently far apart from the
other three directions.
In such a case we have
\be
\label{2}
\la \D(\gg_1) \D(\gg_2) \D(\gg_3) \D(\gg_4) \ra \longrightarrow
\la \D(\gg_1) \D(\gg_2) \D(\gg_3) \ra \,\,  \la \D(\gg_4) \ra
\ee
and (cf. the discussion in the previous paragraph) this contribution vanishes.
Another possible configuration we could encounter is the one in which the
beams are correlated two by two, but {\sl no} correlation exists between
the pairs (for one particular time--step).
This yields three distinct possible outcomes, \eg,
$\la \D(\gg_1) \D(\gg_2) \ra \, \la \D(\gg_3) \D(\gg_4) \ra $,
and the other two obvious combinations.
The last possibility is having all four photon--beams fully correlated
amongst themselves and this yields
$\la \D(\gg_1) \D(\gg_2) \D(\gg_3) \D(\gg_4) \ra_c $,
where the subscript $c$ stands for the {\it connected} part.

In a way completely analogous to that for the two--point function,
the correlated part for the combination of four beams gives
\be
\label{4}
\D(\gg_1) \D(\gg_2) \D(\gg_3) \D(\gg_4) = \xi^4
\sum_{n=p}^{N}\sum_{m=1}^{M}
(\go \cd\hat R^{mn})(\gt\cd\hat R^{mn})
(\gth\cd\hat R^{mn})(\gf\cd\hat R^{mn})
\ee

Now we are ready to write the full mean four--point function as
\bea
\label{5}
\lefteqn{
\la \D(\gg_1) \D(\gg_2) \D(\gg_3) \D(\gg_4) \ra
         }
\nonumber \\
&  &
=\!\! \left[\!
\la \D(\gg_1) \D(\gg_2) \ra \la \D(\gg_3) \D(\gg_4)\ra\! +\!2 {\rm terms}
\right]\!\! + \!\! \la \D(\gg_1) \D(\gg_2) \D(\gg_3) \D(\gg_4) \ra_c
\nonumber \\
&  &
=  3 \left[ {1\over 3} \xi^2 {\cal N}_{corr}(\theta) \cos(\theta) \right]^2  +
\la \D(\gg_1) \D(\gg_2) \D(\gg_3) \D(\gg_4) \ra_c
\eea
where for simplicity we wrote this equation using the same scale $\theta$
for all directions on the sky (first term after the equality sign);
after all, we will be
interested in the zero--lag limit, in which case we will have
$\theta \to 0$.
By using {\sl Mathematica} [Wolfram, 1991] it is simple to find that
the second term includes just the sum of twenty--one combinations of
trigonometric functions depending on
$\theta_i,\phi_i$, with $i=1,2,3,4$, the spherical angles for
the directions $\gg_i$ on the sky (cf. Eq. (\ref{4})).
These terms are the only ones which
contribute non--vanishingly after the integration over the angles
$\theta^{mn},\phi^{mn}$ is performed (assuming ergodicity).

When we take the zero--lag limit (\ie, aiming for the kurtosis)
the above expression gets largely simplified. After normalizing
by the squared of the variance
$\sigma^4 = [{1\over 3} \xi^2 {\cal N}_{corr}(0)]^2$
and subtracting the disconnected part we
get the excess kurtosis parameter
\be
\label{6}
{\cal K} = {1\over \sigma^4}
\la \D^4(\gg_1) \ra   - 3
\simeq 1.125 \left[ {10\over M} \right] \times 10^{-2} .
\ee
Note  that for values of the scaling solution $M$ increasing (large
number of seeds) the actually small non--Gaussian signal
${\cal K}$ gets further depressed, as it could be expected from the
Central Limit Theorem.

\vspace{18pt}
\section{The r.m.s. excess kurtosis of a Gaussian field}
\label{sec-rmsKurto}
\markboth{Chapter 7. ~ON THE CMB KURTOSIS FROM COSMIC STRINGS}
       {\S ~7.3. ~THE RMS EXCESS KURTOSIS OF A GAUSSIAN FIELD}

The previous section was devoted to the computation of ${\cal K}$, the
excess kurtosis parameter, as predicted by a simple analytic model in the
framework of the cosmic string scenario.
We might ask whether this particular non--Gaussian signal has any chance
of actually being unveiled by current anisotropy experiments.
Needless to say, this could provide a significant probe of the structure
of the relic radiation, and furthermore give us a hint on the possible
sources of primordial perturbations.

However, as it was realised some time ago
[Scaramella \& Vittorio, 1991;
see also \eg, Abbott \& Wise, 1984, Srednicki, 1993],
the mere detection of a
non--zero higher order correlation function (\eg, the four--point function)
or its zero--lag limit (\eg,  the kurtosis)
cannot be directly interpreted as a signal for intrinsically non--Gaussian
perturbations.
In order to tell whether or not a particular measured value for the
kurtosis constitutes a significant evidence of a departure from Gaussian
statistics, we need to know the amplitude of the non--Gaussian pattern
produced by a {\it Gaussian} perturbation field. Namely, we need to know
the {\sl rms} excess kurtosis of a Gaussian field.

Let us begin with some basics. Let us denote the kurtosis
$ K \equiv \int {\dO_{\gg}\over 4\pi} \D^4(\gg)$
and assume that the underlying statistics is Gaussian,
namely, that the
multipole coefficients $a_\ell^m$ are Gaussian distributed.
Thus, the ensemble average for the kurtosis is given by the well--known
formula: $\la K \ra = 3 \sigma^4$, where $\sigma^2$ is the mean two--point
function at zero--lag, \ie, the CMB variance as given by
\be
\label{10}
\sigma^2 \equiv \la C_2 (0) \ra =
{1 \over 4\pi} \sum_{\ell} (2\ell + 1) {\cal C}_{\ell}
{\cal W}^2_\ell ~,
\ee
where the ${\cal C}_\ell$ coefficients
are also dependent on the value
for the primordial spectral index of density fluctuations $n$ and are
given by the usual expression in terms of Gamma functions,
cf. Eq. (\ref{cl}).
In the previous expression ${\cal W}_\ell$ represents the window function
of the specific experiment.
In the particular case of the {\sl COBE}--DMR experimental setup we have,
for $\ell \geq 2$,
${\cal W}_{\ell}\simeq\exp\left[-\frac12 \ell({\ell}+1)(3.2^\circ)^2\right]$,
where $3.2^\circ$ is the dispersion of the antenna--beam profile.

However, $\la K \ra$ is just the mean value of the distribution and
therefore we cannot know its real value but within some error bars.
In order to find out how probable it is to get this value after a set of
experiments (observations) is performed, we need to know the variance of
the distribution for $K$.
In other words, we ought to know how peaked the
distribution is around its mean value $\la K \ra$.
The width of the distribution is commonly parameterised by
what is called the cosmic variance of the kurtosis:
$\sigma_{CV}^2 = \la K^2 \ra - \la K \ra^2$.
It is precisely this quantity what attaches
theoretical error bars to the actual value for the kurtosis.
Therefore, we may heuristically express
the effect of $\sigma_{CV}^2$ on  $K$  as follows:
$K \simeq \la K \ra \pm \sigma_{CV}$,  at one sigma level
(a good approximation in the case of a narrow peak).
Rearranging factors we may write this expression in a way
convenient for comparing it with ${\cal K}$:
${\cal K}_{CV} = K  / \sigma^4 - 3 \simeq
\pm \sigma_{CV} / \sigma^4$,
where ${\cal K}_{CV}$ is the excess kurtosis parameter (assuming Gaussian
statistics) purely due to the cosmic variance.
Not only is ${\cal K}_{CV}$ in general non--zero, but its magnitude
increases with the theoretical uncertainty $\sigma_{CV}$.
This gives a fundamental threshold that must be overcome by any
measurable kurtosis parameter.
Namely, unless our predicted value for ${\cal K}$
{\it exceeds} ${\cal K}_{CV}$ we will not be able to tell confidently
that any measured value of  ${\cal K}$ is due to
string--induced non--Gaussianities.

Let us now compute the cosmic variance of the kurtosis explicitly.
We begin by calculating $\la K^2 \ra$ as follows
\be
\label{12nnn}
\la K^2 \ra =
\int {\dO_{\gg_1}\over 4\pi} \int {\dO_{\gg_2}\over 4\pi}
\la \D^4(\gg_1) \D^4(\gg_2) \ra .
\ee
By assuming Gaussian statistics for the temperature perturbations
$\D(\gg)$ we may make use of standard combinatoric relations, and get
\bea
\label{13}
\lefteqn{
\la \D^4(\gg_1) \D^4(\gg_2) \ra  = \,
9 \, \la\D^2(\gg_1)\ra^2 \la\D^2(\gg_2)\ra^2
         }
\nonumber \\
&  &
+ \, 72 \, \la\D^2(\gg_1)\ra \la\D^2(\gg_2)\ra \la\D(\gg_1)\D(\gg_2)\ra^2
+ \, 24 \, \la\D(\gg_1)\D(\gg_2)\ra^4 .
\eea
As we know, the ensemble averages are rotationally invariant and
therefore
$\la\D^2(\gg_1)\ra =  \la C_2 (0) \ra \equiv \sigma^2 $
is independent of the direction $\gg_1$.
Plug this last equation into Eq. (\ref{12nnn}) and we get
\be
\label{14}
\la K^2 \ra = 9\sigma^8
     +  36\sigma^4  \int^1_{-1}d\cos\alpha \, \la C_2(\alpha) \ra^2
     +  12 \int^1_{-1}d\cos\alpha \, \la C_2(\alpha) \ra^4 .
\ee
The above integrals may be solved numerically. Then, using this result
in the expression for $\sigma_{CV}^2$ we get ${\cal K}_{CV}$, the value
for the excess kurtosis parameter of a Gaussian field.

It is also instructive to look at Eq. (\ref{14}) in some more detail, so
that the actual dependence on the spectral index becomes clear.
By expanding the mean two--point correlation functions
within this expression in terms of spherical harmonics and after some
long but otherwise straightforward algebra we find
\bea
\label{37}
\lefteqn{
{\cal K}_{CV} =
\left[ \,
72 \, {
\sum_{\ell} (2\ell + 1){\cal C}_{\ell}^2 {\cal W}_\ell^4
\over
[ \sum_{\ell} (2\ell + 1){\cal C}_{\ell} {\cal W}_\ell^2 ]^2}
\right.
}
\nonumber \\
&  &
+ \,
\left.
24 \,
{
\{
\prod_{i=1}^{4}\sum_{\ell_i}\sum_{m_i=-\ell_i}^{\ell_i}
{\cal C}_{\ell_i} {\cal W}_{\ell_i}^2
\}
\left(
\sum_L 4\pi\,
\bar {\cal H}_{\ell_1,\,\,\ell_2\,,\,\,\,\, L}^{m_1, m_2, m_3+m_4}
\bar {\cal H}_{\ell_3,\,\,\ell_4\,,\,\,\,\,\,\, L}^{m_3, m_4, -m_3-m_4}
\right)^2
\over
[ \sum_{\ell} (2\ell + 1){\cal C}_{\ell} {\cal W}_\ell^2  ]^4
}
\right]^{1/2}
\eea
where the coefficients
$\bar {\cal H}_{\ell_1,\,\ell_2,\,\ell_3}^{m_1, m_2, m_3}
\equiv \int \dO_{\gg} {Y_{\ell_1}^{m_1}}(\gg) Y_{\ell_2}^{m_2}(\gg)
Y_{\ell_3}^{m_3}(\gg)$ are those defined previously in
Eq. (\ref{goodh}).
In the above equation the $n$--dependence is hidden inside the multipole
coefficients ${\cal C}_{\ell}$ (cf. Eq.(\ref{cl})).

Eq. (\ref{37}) shows an analytic expression for computing ${\cal K}_{CV}$
which, in turn, represents a fundamental threshold for any given
non--Gaussian signal.
For interesting values of the spectral index (say, between
$0.8\lsim n \lsim 1.3$) we find no important variation in
${\cal K}_{CV}$, being its value consistent with the Monte--Carlo
simulations performed by Scaramella \& Vittorio [1991].
These authors concentrated on a Harrison--Zel'dovich spectrum, considered
also the quadrupole contribution and took a slightly different dispersion
width for the window function ($3.0^\circ$ in their simulations).

\begin{figure}[tbp]
  \begin{center}
    \leavevmode
    \epsfxsize = 6cm  %
    \epsfysize = 6cm  %
    \epsffile{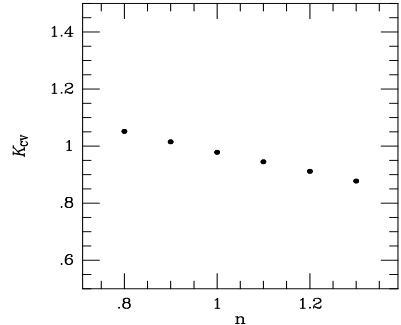}
  \end{center}
\caption{{\sl Excess Kurtosis parameter of a Gaussian temperature fluctuation
 field as function of the spectral index; $\ell_{max} = 15$ and the
 quadrupole contribution is subtracted.}}
\label{kurtofig}
\end{figure}

We solved Eq. (\ref{37}) numerically.
We subtracted the quadrupole contribution and
performed the sums up to a $\ell_{max} = 15$;
the exponential suppression of the ${\cal W}_{\ell}$ window functions
makes higher $\ell$ contribution to the sums negligible.
Our analytic analysis expresses ${\cal K}_{CV}$ as a ratio of averages,
rather than the average of a ratio as in [Scaramella \& Vittorio, 1991],
and therefore exact agreement between their and our result
should not be expected.
Nevertheless, we get the same order of magnitude for spectral index $n = 1$.
We also checked that a different $n$ does not change very much
the value of ${\cal K}_{CV}$.
We explored the cosmologically interesting range $0.8\leq n \leq 1.3$.
In Figure \ref{kurtofig}
we plot ${\cal K}_{CV}$ versus the spectral index $n$, for
$\ell_{max} = 15$ and quadrupole contribution subtracted;
the inclusion of the quadrupole in our results would do no more
than slightly rise the whole set of points. This is not a new feature and
simply reflects the intrinsic theoretical uncertainty of the lowest
order multipoles.
Note the small rate of variation of ${\cal K}_{CV}$ with $n$ and
that, as expected, ${\cal K}_{CV}$ takes larger values for smaller
spectral indexes.
This is clearly due to the fact that a small $n$ generates more power on
large scales (\ie, small $\ell$), and precisely these scales are the ones
that contribute the most to the cosmic variance of the kurtosis field.

\vspace{18pt}
\section{Discussion}
\label{sec-KurtoDisscu}
\markboth{Chapter 7. ~ON THE CMB KURTOSIS FROM COSMIC STRINGS}
                                        {\S ~7.4. ~DISCUSSION}

In the foregoing sections we showed how to implement the multiple impulse
approximation for perturbations on a photon beam
(stemming from the effect of the cosmic string network)
in the actual construction of higher order correlations for the CMB
anisotropies.
We then focused on the four--point function and on the excess kurtosis
parameter, finding for the latter a value ${\cal K}\sim 10^{-2}$.

We also found explicit  formulae for calculating the {\sl rms} excess
kurtosis ${\cal K}_{CV}$
predicted to exist even for a Gaussian underlying field, and showed its
dependence on the primordial spectral index of density fluctuations.
This constitutes the {\sl main} source of theoretical uncertainty at large
angular scales, but {\sl not} the only one: also the `sample variance'
[\eg, Scott \etal, 1994] makes the signal hard to detect, since we should
note that {\sl no} complete sampling of the CMB sky is available, not
even the {\sl COBE} one, where galactic latitudes  $\vert b\vert < 20^\circ$
are cut out from the maps.
These uncertainties completely hide the cosmic string signature:
this is clearly seen from the results of \S\ref{sec-rmsKurto},
where it was shown
that ${\cal K}_{CV}$ is two orders of magnitude larger than ${\cal K}$.
The results of our analytic approach presented in
\S\ref{sec-FourPointStrings}
fully agree with those of [Coulson \etal, 1994], where they found
little evidence for non--Gaussianity in their simulated CMB maps,
for a variety of smoothing angles.
We conclude therefore that statistical effects of the kind just
mentioned would be the most probable responsible of any kurtosis
signal detected on {\sl COBE} scales.
The results presented in this Chapter have been published in
[Gangui \& Perivolaropoulos, 1995].

\chapter{Analytic modeling of textures}
\label{chap-AnaModTex}
\markboth{Chapter 8. ~ANALYTIC MODELING OF TEXTURES}
         {Chapter 8. ~ANALYTIC MODELING OF TEXTURES}

The analytic modeling of topological defects, like cosmic strings
and textures, captures the essential features of their cosmological
predictions.
Analytic models are complementary to full numerical simulations of
defect evolution and
CMB temperature anisotropy generation,
and they rely on these simulations for fixing some free parameters
present in the models.
However, large--scale anisotropies alone can only very barely
discriminate between competing models of structure formation.
In the present Chapter we include work {\sl in progress} towards
studying to what extent CMB non--Gaussian features
further constrain topological defects models.
We derive an analytic expression
for the collapsed three--point correlation function in the
CMB temperature anisotropies for
{\sl textures}, as well
as for the {\sl rms} collapsed  three--point function (cosmic variance)
associated with it.
Applying our results to the {\sl COBE}--DMR two--year data,
preliminary estimates seem to indicate that
the CMB non--Gaussianities predicted by this model
are consistent with the theoretical uncertainties present in the
maps, and therefore large scale anisotropies
{\sl alone} can not place further constraints on these models.

\section{Introduction}
\label{sec-TextuIntrud}
\markboth{Chapter 8. ~ANALYTIC MODELING OF TEXTURES}
                            {\S ~8.1. ~INTRODUCTION}

One of the most compelling issues calling for thorough study
regards the possibility of discriminating the real
source of density perturbations leading to large--scale structure
formation. Competing models are  inflationary scenarios and
topological defects (cf. Chapter \ref{chap-theories}),
these latter formed as the consequence
of a symmetry breaking phase transition in the early evolution of our
universe.

The CMB radiation carries a
valuable information of that early epoch when matter and photons
were tightly coupled.
A useful way of trying to discriminate between models is
given by the structure in the relic radiation predicted by them.
However, production of anisotropy maps from numerical
simulations of the `defect' field equations
turns out to be a fabulous problem, and therefore some analytic
insight is desirable.
This is the reason why people turned to consider
analytic (although simplified) models for anisotropy generation
by defects, wherein the physics is more transparent and computations
can be pursued straightforwardly [Perivolaropoulos, 1993;
Magueijo, 1995; Gangui \& Perivolaropoulos, 1995].
Although it is clear that full--range numerical simulations will have
the last word regarding these and other observable signatures
of defects, in the meantime progress with these simplified models
will hint many of their important features.

Defect induced large--scale structure formation has been well studied
recently [see, \eg, Vilenkin \& Shellard, 1994; Pen \etal, 1994]
and their predictions confront successfully against a large bulk of
observations.
The amplitude of the anisotropies is directly related to
the symmetry breaking energy scale, and normalisation to
the {\sl COBE}--DMR detection makes these seeds compatible with a
physically realisable GUT phase transition.

However, more and valuable information can be readily extracted from
the CMB maps. Recently Hinshaw \etal ~[1995] analysed the
two--year {\sl COBE}--DMR data and found evidence for
a departure from Gaussian statistics, although consistent with the
level
of cosmic variance.
Many are the models wherein  non--Gaussian features can arise:
in non--standard inflationary models, for example, primordial
non--Gaussian features are easily generated
[Falk \etal, 1994; Gangui \etal, 1994; Gangui, 1994;
see also Chapter \ref{chap-primor}].
Furthermore, the post recombination mildly non--linear evolution
of perturbations also predicts a non--Gaussian signature through
the integrated Sachs--Wolfe effect
[Luo \& Schramm, 1993; Munshi \etal, 1995; Mollerach \etal, 1995;
see also Chapter \ref{chap-integrated}].
Here we will focus on the three--point correlation
function of the CMB anisotropies. In particular, it turns out that a
useful {\sl tool} to quantify the departure from Gaussian
statistics is given by the {\sl collapsed} three--point function
(cf. \S\ref{sec-3pointfun}), and we
will work exclusively with this latter in the present Chapter.

One might think that, due to the high level of theoretical noise
inherent to the lowest order multipoles probed by {\sl COBE},
very little can be said regarding non--Gaussian features in the
CMB radiation on large scales.
However, based on the analysis of the equilateral and collapsed
three--point functions, Hinshaw \etal ~[1995] were able to
rule out the primordial isocurvature baryon (PIB)
power--law inflationary model considered by Yamamoto \& Sasaki [1994]:
in this model the predicted mean skewness vanishes and
the {\sl rms} skewness (cosmic variance)
is several orders of magnitude below the corresponding level
for standard adiabatic models.
These authors suggest that the cosmic variance of the full
three--point function would be tiny as well.
This would make the theoretical error bars for the three--point
function predicted by this PIB model even smaller that the actual
data from the maps, hence running into conflict with the
result of the {\sl COBE}--DMR analysis.

Likewise, it is interesting to check whether the
non--Gaussian signatures predicted by some recently proposed
analytic defect models are in agreement with the analyses of the
maps, and if not what the constraints are.
This is the main aim of the present study, which, we reiterate,
is in progress \cite{SilyYo}.
Let us now give a couple of definitions to set up the basic framework.

The three--point correlation function for points at three arbitrary
angular separations $\alpha$, $\beta$ and $\gamma$ is given by the
average product of temperature fluctuations in all possible three
directions with those angular separations among them. The general
expression is given in Eq. (\ref{3cor'}).
Here, for simplicity, we will restrict ourselves to the
{\it collapsed} case,
corresponding to the choice $\alpha=\beta$ and $\gamma=0$  (one
of the cases analysed for the {\it COBE}--DMR data by
Hinshaw \etal ~[1994, 1995]).
The collapsed three--point correlation function of the CMB is given by
\begin{equation}
C_3(\alpha) \equiv \int
{d\Omega_{\hat \gamma_1}\over 4\pi}
\int
{d\Omega_{\hat \gamma_2}\over 2\pi}
\D (\hat\gamma_1) \D^2 (\hat\gamma_2)
\delta(\hat\gamma_1\cdot\hat\gamma_2 -\cos \alpha).
\label{skew1-spot}
\end{equation}
For $\alpha=0$, we recover the well--known expression for the
skewness, $C_3(0)$. By expanding the temperature fluctuations in
spherical harmonics
$\D (\gg ) = \sum_{\ell,m} a_{\ell}^{m} Y_{\ell}^{ m} (\gg )$,
we can write the collapsed three--point function as
\begin{equation}
\label{coll}
C_3(\alpha) =\frac{1}{4\pi} \sum_{\ell_1,\ell_2,\ell_3,m_1,m_2,m_3}
P_{\ell_1}(\cos\alpha)
a_{\ell_1 }^{m_1} a_{\ell_2}^{ m_2} a_{\ell_3}^{ m_3}
{\cal W}_{\ell_1} {\cal W}_{\ell_2} {\cal W}_{\ell_3}
\bar {\cal H}_{\ell_1 \ell_2 \ell_3}^{m_1 m_2 m_3} ~,
\end{equation}
where as usual
${\cal W_\ell}$ represents the window function of the particular
experiment and we use the notation of Eq. (\ref{goodh}).

Predictions for different models usually come as
expressions for the angular bispectrum
$\la a_{\ell_1 }^{m_1} a_{\ell_2}^{ m_2} a_{\ell_3}^{ m_3} \ra$.
As we saw in Chapter \ref{sec-kurtostrings},
within the analytic model for cosmic strings
considered there, the bispectrum simply vanishes, and so does
the three--point correlation function. In that case we are left
just with the amplitude of the {\sl rms} collapsed three--point
function.
For textures the situation is different, and it is one of the aims
of the present study to show
the variation of the collapsed function with the angle separation,
as well as that of the {\sl rms} collapsed three--point
function.

\section{Texture--spot statistics}
\label{sec-TexSpotStat}
\markboth{Chapter 8. ~ANALYTIC MODELING OF TEXTURES}
                {\S ~8.2. ~TEXTURE--SPOT STATISTICS}

One of the most serious complications in predicting
CMB fluctuations in defect theories is the necessity of
evolving the simulations over a large number of expansion
times.
Scaling helps in simplifying this issue in that if the defect
is assumed to scale (and textures {\sl do} scale)
then the defect network looks statistically the same at any time,
once its characteristic length is normalised to the horizon.
Then, by dividing the sky cone into cells corresponding to
expansion times and horizon volumes, one can restrict the study to the
CMB pattern induced in horizon--size boxes during one Hubble time.
Such studies have been performed by, \eg, Borrill \etal ~[1994] and
Durrer \etal ~[1994]; in the latter paper, in particular, a
`scaling--spot--throwing' process was implemented numerically,
with CMB spots derived from a self--similar and spherically symmetric
model of texture collapses
[Turok \& Spergel, 1990; Notzold, 1991; see also \S\ref{sec-topo}].

The implementation (within the texture scenario) of an analytic
approach for the computation of the multipole coefficients
${\cal C}_\ell$ and the full two--point function,
as well as for their cosmic variances,
was performed in [Magueijo, 1995].
The model allows to study texture--induced spots of arbitrary
shapes and relies on a hand--full of `free' parameters to be
specified by numerical simulations, like the number density
of spots, ${\bf n}$, the scaling size, $d_s$, and
the brightness factor of the particular spot, $a_n$, which tells
us about its temperature relative to the mean sky temperature
(in particular, whether it is a hot or a cold spot).
In the following we will briefly review the necessary basic
features of the model and then move on directly to the specific
study of interest to us, namely
the collapsed three--point function predicted by the model, and
its cosmic variance. We make a short discussion at the end
where we summarise what was found so far.

Let us start off discussing the spots statistics.
By dividing the whole volume in boxes of
surface $dS$ and thickness $d\ell$ we may write the
volume probability density of spot generation in the sky as follows
\be
\label{ana1}
dP = { {\bf n} \over  H^{-4}} dS d\ell dt ~,
\ee
where ${\bf n}$ is the mean number of CMB spots expected in a
horizon--size box volume and in a Hubble time.
Now, we are interested in a two dimensional distribution of
spots (since observations map the two dimensional sky sphere),
and thus we may integrate the thickness $d\ell$ in an interval
of order $H^{-1}$ to get the surface probability density
\be
\label{ana1ymedio}
dP \simeq {{\bf n} \over H^{-3} } dS dt ~.
\ee
Taking into account the expansion of the universe we may express
$dS = r^2(t) d\Omega$, with
$r(t) = 3 t ((t_0 / t)^{1/3} - 1)$,
where $t$ denotes proper time and $t_0$ stands for its value today.

We discretise the time between last scattering, $t_{ls}$,
and today by a set of Hubble
time--steps $t_n$ such that $t_{n+1} = 2 \, t_n$, \ie, the horizon gets
doubled in each time--step.
Thus, it turns out to be convenient to change time variable to
$y_n(t_n) \equiv \log_{2} (t_0 / t_n)$.
For a redshift $z_{ls}\sim 1400$ we have
$y_{ls} \simeq \log_{2}[(1400)^{3/2}] \simeq 16$.
In terms of $y_n$ the surface probability density may be cast as
\be
\label{ana2pre}
dP = N(y_n) dy d\Omega,
\ee
with
\be
\label{ana2}
N(y_n) = - {8 {\bf n} \ln (2) \over 3} \left(2^{y_n/3} - 1 \right)^2 ~.
\ee

Within the model, CMB anisotropies arise from
texture--produced spots in the sky.
Thus, we may also express the anisotropies
in terms of the profiles $W_n$
of these spots, generated at time $y_n$, as follows:
$\D  = \sum_{n} a_{n} W_n (\theta_n, y_n)$.
Here, $a_{n}$ characterises the brightness of the hot/cold
$n$--th spot
[Magueijo, 1995] and is interpreted as a random variable: it is
non--Gaussian distributed and its possible mean values
are to be computed from numerical simulations
[\eg, Borrill \etal, 1994].
$\theta_n$ is the angle subtended by a scale characterising the size
of the spots\footnote{We are here allowing for arbitrary profile shapes.
For circular spots the characteristic length is of course given by
the radius.}
in the particular Hubble time step $y_n$,
as measured by a given observer.

The non--linear evolution of the global scalar field responsible for
texture formation is the source of the CMB anisotropies.
As long as the characteristic length scale of these defects is larger
than the size of the horizon the gradients of the field
are frozen in and cannot affect the surrounding matter
components\footnote{We are here simplifying the discussion by
considering the correlation length $\xi \lsim t$ to be of the order
of the particle horizon. See Ref. \cite{Vilenkin+Shellard94} for
a complete account of these matters.}.
However, when the size of the texture becomes comparable to the
Hubble radius (this latter growing linearly with cosmic time),
microphysical processes can `push' the global fields
so that they finally acquire trivial field configurations,
reduce their energy (which is radiated away as Goldstone bosons) and,
in those cases where the field was wounded in knots, they unwind.
In so doing they generate perturbations in the metric of spacetime
and hence affect also the photon geodesics.
These unwinding events are mainly characterised by one length scale,
that of the horizon at the moment of collapse, implying that the
outcome perturbations arise at a fixed rate per horizon
volume and Hubble time.

It is this `scaling' behaviour what makes texture seeds a
viable model for structure formation;
being a generic outcome of any symmetry breaking phase transition
involving a non--Abelian GUT group, if they did not collapse
not only would they be the storage of huge amounts of energy
(\eg, in the form of gradients of the global fields) but
they would also produce too much anisotropy in the CMB and
would therefore be cosmologically intolerable.
It is therefore natural to expect a similar scaling
for the pattern of the spots they produce.
A spot appearing at time $y_n$ has typically a size
$\theta^s(y_n) \simeq d_s  ~ \theta^{hor}(y_n)$, with
$\theta^{hor}(y_n)$ the angular size of the horizon at
$y_n$, and where
\be
\label{ana3}
\theta^s(y_n)
= \arcsin \left( {d_s H^{-1}\over r(t)}          \right)
= \arcsin \left( {0.5 d_s   \over 2^{y_n/3} - 1} \right) ~.
\ee
Spot anisotropies are generated by causal seeds, and therefore
their angular size cannot exceed the size of the horizon
at that time; thus we have $d_s \lsim 1$.
The scaling hypothesis implies that the profiles satisfy
$W_n(\theta_n , y_n) = W_n(\theta_n / \theta^s(y_n))$.

Combining both previous expansions for $\D$ we easily
find the expression for the multipole coefficients
$a_{\ell}^{m}$ in terms of the brightness and the profile of
the spots
\be
\label{ana4}
a_{\ell}^{m} =
{4\pi\over 2\ell + 1} \sum_n
a_n W_n^\ell {Y_{\ell}^{ m}}^* (\gg _n ) ~,
\ee
where
\be
\label{ana5}
W_n^\ell = {2\ell + 1 \over 2} \int_{-1}^{1} W_n(x) P_\ell (x) dx ~.
\ee

\section{Spot correlations}
\label{sec-SpotCorr}
\markboth{Chapter 8. ~ANALYTIC MODELING OF TEXTURES}
                       {\S ~8.3. ~SPOT CORRELATIONS}

Having derived the $a_{\ell}^{m}$ coefficients of the expansion
of $\D$ in spherical harmonics, we are able
now to compute correlations, namely the angular spectrum and
angular bispectrum predicted within this analytic model.
{}From Eq. (\ref{ana4}) we may easily calculate the quantity
\be
\label{ana41}
{1\over 2\ell + 1}\sum_{m=-\ell}^{\ell} |a_{\ell}^{m}|^2 =
{4\pi \over (2\ell + 1)^2} \sum_{n n'} a_n a_{n'}
W_n^\ell W_{n'}^\ell P_\ell (\gg_n\cdot\gg_{n'}) ~.
\ee

Now let us compute ensemble averages $\la\cdot\ra$.
Any two texture--spots located at different $\gg_n$'s
are uncorrelated, hence we have
\be
\label{ana42}
\la a_n a_{n'} W_n^\ell W_{n'}^\ell P_\ell (\gg_n\cdot\gg_{n'}) \ra
= \delta_{n n'} a_n^2 \left( W_n^\ell \right)^2 ~.
\ee
In the present model,
due to this lack of correlation, the angular part
of the surface probability density (cf. Eq. (\ref{ana2pre}))
yields just a kronecker delta and
disappears from the expression of the
angular power spectrum $\la | a_\ell^m  |^2 \ra = {\cal C}_\ell$.
In fact, from Eqs. (\ref{ana41}) and (\ref{ana42}) we get
\be
\label{ana43}
{\cal C}_\ell =  {4\pi\over (2\ell + 1)^2}
\sum_n a_n^2 \left( W_n^\ell \right)^2 ~.
\ee
The cumulative effect of the different spots on the
${\cal C}_\ell$'s (as reflected from the profiles $W_n^\ell$)
is computed from the surface probability density
(cf. Eq. (\ref{ana2pre})).
Going to the continuous limit in the time variable $y$ the above equation
may be cast as
\be
\label{ana6}
{\cal C}_\ell = \la a^2 \ra  \left({4\pi\over 2\ell + 1}\right)^2
{\cal I}_2^{\ell \ell }
\ee
where we defined
\be
\label{ana7}
{\cal I}_D^{\ell_1 \ldots \ell_D} \equiv
\int dy N(y)  W^{\ell_1}(y) \ldots  W^{\ell_D}(y) ~.
\ee
In these equations and hereafter we drop the $_n$ subindex
and consider $y$ in the continuous limit.
$\la a^2 \ra$ in Eq. (\ref{ana6}) is the mean squared value
of the spot brightness, to be computed from the distribution
\{$a_n$\} resulting from simulations.

\section{Textures and the three--point function}
\label{sec-TexAndThree}
\markboth{Chapter 8. ~ANALYTIC MODELING OF TEXTURES}
  {\S ~8.4. ~TEXTURES AND THE THREE--POINT FUNCTION}

Let us now compute the angular bispectrum. Following the same
steps as in the previous section we may express it as
\be
\label{ana8}
\la a_{\ell_1 }^{m_1} a_{\ell_2}^{ m_2} a_{\ell_3}^{ m_3} \ra =
{(4\pi)^3 \la a^3 \ra \over (2\ell_1 + 1) (2\ell_2 + 1) (2\ell_3 + 1) }
{\cal I}_3^{\ell_1  \ell_2  \ell_3}
\bar {\cal H}_{\ell_1 \ell_2 \ell_3}^{m_1 m_2 m_3} ~.
\ee
This time the angular part of Eq. (\ref{ana2pre})
is the responsible for the appearance of the factor
$\bar {\cal H}_{\ell_1 \ell_2 \ell_3}^{m_1 m_2 m_3} $.
Now we are ready to study the mean collapsed three--point function
predicted by this model. From Eq. (\ref{coll}) we get
\begin{equation}
\label{collxx}
\la C_3(\alpha) \ra =
4\pi \la a^3 \ra \sum_{\ell_1,\ell_2,\ell_3}
P_{\ell_1}(\cos\alpha)
{\cal W}_{\ell_1} {\cal W}_{\ell_2} {\cal W}_{\ell_3}
{\cal I}_3^{\ell_1 \ell_2 \ell_3}
{\cal F}_{\ell_1 \ell_2 \ell_3} ~,
\end{equation}
where ${\cal F}_{\ell_1 \ell_2 \ell_3} \equiv
\left(^{\ell_1~ \ell_2~ \ell_3}_{0~ ~0~ ~0}\right)^2$
are just the squared of the the $3j$--symbols (cf. Messiah 1976).
The factor $\la a^3 \ra$ in Eqs. (\ref{collxx}), (\ref{ana8})
deserves some comments:
a brightness distribution \{$a_n$\}
symmetric in hot and cold spots leads to a non--vanishing angular spectrum
(cf. Eq. (\ref{ana6})).
On the other hand,
and as we might have expected, such symmetric distribution
implies  $\la a^3 \ra = 0$ and therefore yields a vanishing
three--point function.
This is precisely what happens for spots
generated by spherically symmetric self--similar (SSSS) texture
unwinding events [Turok \& Spergel, 1990].

However, the properties of the
SSSS solution are not characteristic of those of more realistic
random configurations.
Borrill \etal ~[1994] have shown that randomly generated texture
field configurations produce anisotropy patterns with
very different properties to the exact SSSS solution.
Furthermore, spots generated from concentrations of gradient
energy which do not lead to unwinding events can still produce
anisotropies very similar to those generated by unwindings,
and the peak anisotropy of the random configurations turns
out to be smaller by 20 to 40 \% with respect to the SSSS
solution. Importantly enough, they also find an asymmetry
between maxima and minima of the peaks for all studied
configurations other than the SSSS.
In our context this implies that $\la a^3 \ra \not= 0$
and we will use their results for the amplitude of the spots
(the brightness)
to estimate the amplitude of the mean collapsed three--point
function below.

There is no particular form inherent to the spot profiles
obtained in the simulations.
The outcome of the random field configurations (especially
when no symmetries are assumed) may yield spots of arbitrary
distribution of brightness and shapes.
A Gaussian distribution for the spot brightness, as given by
the profile  $W(\theta / \theta_s(y)) \propto
\exp (- \theta^2 / \theta^2_s(y))$ (cf. Eq. (\ref{ana3}))
looks as a good approximation to the bell--shaped spots obtained
by Borrill \etal ~[1994].
One of the interesting conclusions of that work
regards the number of spots one
is expected to find: whereas in the SSSS model
approximately only 4 out of 100 simulations showed unwinding events,
the number density of nonunwinding events (which as we mentioned
above also lead to spots) is much greater, typically of the
order of one such event per simulation.
This implies a mean number of CMB spots per horizon--size volume
per Hubble time ${\bf n} \simeq 1$.

Estimations of the size of the spot indicate that these are
approximately 10\% of the horizon when generated, hence
the scaling size $d_s \simeq 0.1$.
As we do not have at our disposal
the maximum and minimum brightness of each of the spots
found by Borrill \etal ~[1994] in their simulations, we
will estimate $\la a^2 \ra$ and $\la a^3 \ra$
from their quoted
$\la a_{max} \ra \simeq  0.75 \times 2\pi\epsilon$ and
$\la a_{min} \ra \simeq -0.65 \times 2\pi\epsilon$,
for the random collapse (and possible unwinding) of gradient
energy on sub--horizon scales,
in order to have a realistic working example
($\epsilon = 4\pi G \tilde\eta^2$, cf. \S\ref{sec-topo}).
We have
$\la a^2 \ra \simeq 19.44 \epsilon^2$
and
$\la a^3 \ra \simeq 18.26 \epsilon^3$.

We may now apply our formalism to the {\sl COBE}--DMR measurements and
normalise the perturbations
to the two--year data, using the $\ell=9$ multipole
amplitude, ${\cal C}^{1/2}_{\ell = 9} = 8~\mu {\rm K}$,
according to the procedure proposed by G\'orski \etal ~[1994].
The mean value for the CMB collapsed three--point function
can hence be computed from Eq. (\ref{collxx}), as well as
its dependence on the angular separation
[Gangui \& Mollerach, 1995].

Let us turn now to the computation of the cosmic variance associated
to the collapsed three--point correlation function.
We proceed by first calculating the ensemble average
of the combination of six $a_\ell^m$'s, namely,
$\la a_{\ell_1 }^{m_1} \ldots  a_{\ell_6}^{ m_6}\ra$.
The actual expression is awkward to read and not very illuminating,
and we are therefore not writing it explicitly.
{}From that it is then an easy matter to compute the cosmic variance for
the collapsed three--point function, which we note as
$\sigma_{CV}^2(\alpha ) \equiv
\la C_3^2 (\alpha ) \ra - \la C_3 (\alpha )\ra^2$.
We get
\bea
\label{colltexturms}
\lefteqn{
\sigma_{CV}^2(\alpha )
=}
\nonumber\\&&
2 ~ (4\pi)^3  \la a^2 \ra^3
\!\!\sum_{\ell_1,\ell_2,\ell_3}
\!\!P_{\ell_1}(\alpha)
( P_{\ell_1}(\alpha)\! +\! P_{\ell_2}(\alpha)\! +\! P_{\ell_3}(\alpha) )
{}~{\cal F}_{\ell_1 \ell_2 \ell_3}
\prod_{\ell = 1}^{3}
{\cal I}_2^{\ell \ell}
{{\cal W}_{\ell}^2 \over 2\ell + 1 }
\nonumber\\&&
+
(4\pi)^2 ~ \la a^2 \ra \la a^4 \ra
\sum_{\ell_1,\ell_2,\ell_3,\ell_5,\ell_6}
\left[
  P_{\ell_1}^2(\alpha) +
4 P_{\ell_1}  (\alpha) P_{\ell_2}(\alpha)  +
4 P_{\ell_2}  (\alpha) P_{\ell_5}(\alpha)
\right]
\nonumber\\ &&
\times {\cal I}_2^{\ell_1 \ell_1}
{{\cal W}^2_{\ell_1}\over 2\ell_1 + 1 }
{\cal I}_4^{\ell_2\ell_3\ell_5\ell_6} ~
{\cal W}_{\ell_2}{\cal W}_{\ell_3}{\cal W}_{\ell_5}{\cal W}_{\ell_6} ~
{\cal F}_{\ell_1 \ell_2 \ell_3}{\cal F}_{\ell_1 \ell_5 \ell_6}
\nonumber\\ &&
+
(4\pi)^2 ~  \la a^3 \ra^2
\sum_{\ell_1,\ell_2,\ell_3,\ell_5,\ell_6}
\left[
  P_{\ell_1}^2(\alpha) +
4 P_{\ell_1}  (\alpha) P_{\ell_2}(\alpha)  +
4 P_{\ell_2}  (\alpha) P_{\ell_5}(\alpha)
\right]
\nonumber\\ &&
\times {{\cal W}_{\ell_1}^2\over 2 \ell_1 + 1 }
{\cal W}_{\ell_2}{\cal W}_{\ell_3}
{\cal W}_{\ell_5}{\cal W}_{\ell_6}
{\cal I}_3^{\ell_1\ell_2\ell_3}
{\cal I}_3^{\ell_1\ell_5\ell_6}
{\cal F}_{\ell_1 \ell_2 \ell_3}
{\cal F}_{\ell_1 \ell_5 \ell_6}
\nonumber\\ &&
+
4\pi ~ \la a^6 \ra
\sum_{\ell_1,\ell_2,\ell_3,\ell_4,\ell_5,\ell_6}
P_{\ell_1}(\alpha) P_{\ell_4}(\alpha) ~
{\cal W}_{\ell_1}{\cal W}_{\ell_2}{\cal W}_{\ell_3}
{\cal W}_{\ell_4}{\cal W}_{\ell_5}{\cal W}_{\ell_6}
\nonumber\\ &&
\times
{\cal I}_6^{\ell_1\ell_2\ell_3\ell_4\ell_5\ell_6}
{\cal F}_{\ell_1 \ell_2 \ell_3}{\cal F}_{\ell_4 \ell_5 \ell_6}
{}~.
\eea
The computation of $\la C_3 (\alpha ) \ra \pm \sigma_{CV}(\alpha )$
will be an estimate of the maximum band of signals predicted from our
fluctuation field, and thus allowed by the model.


\section{Discussion}
\label{sec-TexDiscuFi}
\markboth{Chapter 8. ~ANALYTIC MODELING OF TEXTURES}
                              {\S ~8.5. ~DISCUSSION}

So far we have assumed full--sky coverage in our analytic formulae.
We have already discussed previously
(\eg, in Chapter \ref{chap-integrated} for the contribution to the
three--point function from the Rees--Sciama effect),
the effect of partial coverage, due to the cut in the maps at Galactic
latitude $\vert b \vert < 20^\circ$.
As estimated in [Hinshaw \etal, 1994] this increases the cosmic variance
by a factor $\sim 1.56$, and should also be taken into account to
quantify the uncertainties.

\begin{figure}[t]
\vspace{-2cm}
  \begin{center}
    \leavevmode
    \epsfxsize = 4in 
    \epsfysize = 6in 
    \epsffile{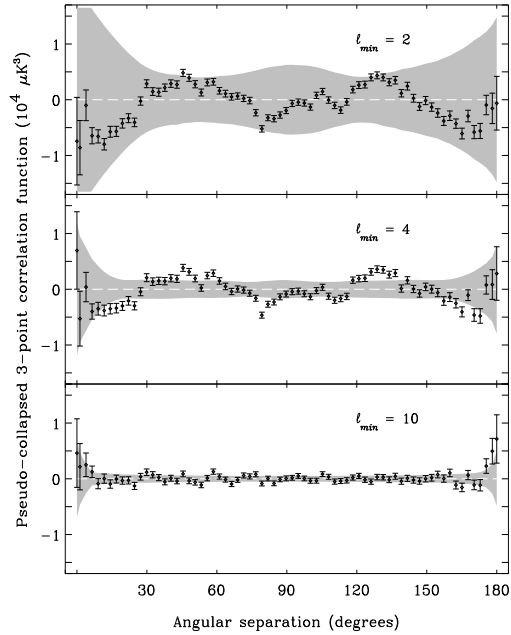}
  \end{center}
\vspace{-4cm}
\caption{{\sl The `pseudo--collapsed' three--point function
as computed from the analysis of the two--year {\sl COBE}--DMR data
[from Hinshaw \etal, 1995]. The quantity plotted is the analogue of our
$\la C_3(\alpha)\ra T_0^3$ (see the text).
Error bars represent instrument noise while the grey band
represents the {\sl rms} range of fluctuations due to
a superposition of instrument noise and cosmic variance.
Three different subtraction schemes are shown:
the top, middle and bottom panels represent the cases
$l_{min}=2,~4,~10$ respectively.}}
\label{Hin95}
\end{figure}

In Figure \ref{Hin95} we show the pseudo--collapsed three--point
correlation function, in thermodynamic temperature units,
evaluated from the two--year 53$+$90 GHz average {\sl COBE}--DMR
map containing power from the quadrupole moment and up
[from Hinshaw \etal, 1995].
The name `pseudo--collapsed' used by Hinshaw \etal ~[1994, 1995]
comes from the following fact:
whereas in the `collapsed' case discussed here
two out of the three directions
$\gg_1 , \gg_2 , \gg_3$ on the sky are exactly parallel, in the
case when analysing the actual data it turns out that
taking, say, $\gg_1$ {\sl exactly} parallel to $\gg_2$
is source of large errors.
This is due to the fact that
the temperature associated with each pixel of the map
comes with a noise contribution. The signal is thus divided into
a part with the `true' signal {\sl plus} a bias term that is a
cross--correlation of the true sky temperature with the average noise
pattern.
As one can prefigure, given the present noise level of the
DMR detector and the smallness of the true signal one is after,
the bias term dominates the behaviour of the statistics.
Thus, the strategy of taking $\gg_1$ `nearly' parallel to
$\gg_2$, while leaving $\gg_3$ arbitrary, is adopted.
This technique bypasses the above mentioned difficulties
and at the same time provides an enhancement in the statistics:
instead of taking $\gg_1$ and $\gg_2$ the same pixel, one takes
$\gg_1$  and the eight nearest neighbours of $\gg_2$, and thus one
is summing over roughly four times as many independent pixel
triples as in the true collapsed case. Needless to say, this
enhances the sensitivity [see Hinshaw \etal, 1994 for details].

The idea is to confront our results for the mean
$\la C_3(\alpha)\ra$ plus its cosmic variance `error bars'
with the experimental data as given in Figure \ref{Hin95}.
Roughly speaking, if the amplitude of
our signal plus cosmic variance is {\sl larger} than the actual data
points plus the experimental noise given in the maps, then
the analytic texture model considered in the present Chapter cannot
be constrained by the experiment. If, on the other hand,
our level of cosmic variance (on top of the collapsed three--point
signal) lies {\sl below} these points, then (based on the published
analysis of the data) something might go wrong with the analytic
model. Further study may then lead to modifications or even to the
realisation that some of the assumptions is not fully met,
thereby helping in the working out of a more realistic version
of the model.

As we mentioned in the introduction,
at the time of the writing up of this thesis,
the work described here is still {\sl in progress}.
Preliminary (and conservative) estimates seem to indicate that
the CMB mean three--point correlation function
plus the cosmic variance predicted within the analytic texture
model by Magueijo [1995] is consistent with the
{\sl COBE}--DMR two--year data reported in Figure \ref{Hin95}.
However, no definitive statement can be made, since careful checks
in our computations are still to be performed.

A similar analysis based on a recently proposed analytic model for
CMB anisotropy generation from cosmic strings
[Perivolaropoulos, 1993; Gangui \& Perivolaropoulos, 1995]
is also under way \cite{SilyYo},
and we hope to report on this work in the near future.

The prospect is interesting in itself: it would give us a hint
that, {\sl even} on large angular scales, the analysis of
non--Gaussian features in the structure of the cosmic microwave
background radiation maps provides us with a tool to constrain models,
or at least to make a consistency check of their predictions against
experimental data.

\chapter{Global textures and the Doppler peaks}
\label{chap-texture}
\markboth{Chapter 9. ~GLOBAL TEXTURES AND THE DOPPLER PEAKS}
         {Chapter 9. ~GLOBAL TEXTURES AND THE DOPPLER PEAKS}

In the present Chapter we review recent work aimed at showing how
global topological defects influence the shape of the
angular power spectrum of the CMB radiation on small scales.
As we explained before, the CMB fluctuations on large
angular scales (greater than a few degrees) are caused by perturbations in
the gravitational field via the Sachs--Wolfe effect.
However, on intermediate scales,
$0.1^\circ\lsim\theta\lsim 2^\circ$,
the main r\^ole is played by
coherent oscillations in
the baryon radiation plasma before the era of recombination.
Unless the universe is reionised at some redshift $z>50$,
these oscillations lead to the  `Doppler peaks' in the angular
power spectrum.
On even smaller scales  the anisotropies
are damped due to the finite thickness of the recombination
shell [Bond \& Efstathiou, 1984, 1987; Efstathiou, 1990],
as well as by photon diffusion during recombination, due to the
imperfect coupling between baryons and photons
(Silk damping [Silk, 1968]).

Inflation--based cold dark matter models predict the location of
the first peak to be
at $\ell \sim 220 / \sqrt{\Omega_0}$, with a height which is a few
times the amplitude of the Sachs--Wolfe `plateau' on large
scales [Steinhardt, 1995].
Here we present a corresponding study for perturbations
induced by global textures.
As we will see below, the first Doppler peak is reduced to an amplitude
comparable to that of the Sachs--Wolfe contribution, and it
is shifted to $\ell\sim 350$.
We also briefly comment on the relation between these results and
what open models predict.

\section{Acoustic oscillations and $\D$}
\label{sec-AcouOs}
\markboth{Chapter 9. ~GLOBAL TEXTURES AND THE DOPPLER PEAKS}
                  {\S ~9.1. ~ACOUSTIC OSCILLATIONS AND $\D$}

The importance of the observations of anisotropies on angular
scales ranging $0.1^\circ\lsim\theta\lsim 2^\circ$, and
smaller, cannot be over-emphasised.
Presently a host of experiments scrutinise different regions of the
sky trying to uncover the real amplitude and structure
of the relic radiation.
While the {\sl COBE}--DMR data played
undoubtedly an important r\^ole in determining the normalisation
for the angular spectra predicted by the different models, it is clear
now that the next important step towards discriminating between
competing models lies on the degree scale.
Whether it will be inflation [Steinhardt, 1993]
or defect models [Kibble, 1980]
the `final' theory, able to explain faithfully the experimental data,
is something that hopefully will not make us wait so long.
The CMB anisotropies are a powerful tool to discriminate among
these models  by  purely linear analysis.
As we saw before, they are parameterised in terms of
${\cal C}_\ell$'s, defined (cf. Chapter \ref{chap-statistics})
as the coefficients in the expansion of the angular
correlation function
\be
\label{twotwo}
\langle
C_2(\vartheta)
\rangle
= {1\over 4\pi}\sum_\ell(2\ell+1){\cal C}_\ell P_\ell(\cos\vartheta).
\ee
For a scale invariant spectrum of perturbations
$\ell (\ell + 1) {\cal C}_\ell$
is constant on large angular scales, say $\ell \stackrel{<}{\sim} 50$.
Both inflation and topological defect models lead to approximately
scale invariant spectra and therefore their shape on these scales
does not help in our aim of finding a signature to distinguish
between models.

On smaller scales the situation is more promising and we
will see below how the predictions of global textures on the
CMB angular spectrum profile depart from those expected
in the standard  CDM inflationary model.
Disregarding Silk damping, gauge invariant linear perturbation
analysis leads to [Durrer, 1994]
\begin{equation}
{\delta T\over T} = \left[
 - {1\over 4}D_g^{(r)} + V_j \gamma^j -\Psi+\Phi\right]_i^f +
\int_i^f (\Psi' - \Phi' ) d\tau  ~,
\label{dT} \end{equation}
where $\Phi$ and $\Psi$ are quantities describing the perturbations
in the geometry (Bardeen potentials)\footnote{It is well known
	that $-\Phi$ is the natural variable for the generalisation
	of the gravitational potential. Since in our analysis
	we neglect anisotropic stresses in the matter components we
	have $\Psi = -\Phi$. Hence, unlike the notation we
	used in previous Chapters, we will here {\sl stick} to the notation
	in [Kodama \& Sasaki, 1984] and interpret $\Psi$ as the
	generalised gravitational potential.}
and $\bf V$ denotes the peculiar velocity of
the radiation fluid with respect to the overall Friedmann expansion.
Primes stand for derivatives with respect to conformal time.
$D_g^{(r)}$ specifies the intrinsic density fluctuation in the
radiation fluid.
There are several gauge invariant variables which describe
density fluctuations; they all differ on super--horizon
scales but coincide inside the horizon.
Below we use another variable,
$D_r$, for the radiation density fluctuation
[Kodama \& Sasaki, 1984].
Since the coherent oscillations giving rise to the Doppler peaks
act only on sub--horizon scales, the choice of this variable
is irrelevant for our calculation.

$\Phi$ , $\Psi$ and $D_g^{(r)}$ in Eq.~(\ref{dT}) determine the
anisotropies on large angular scales\footnote{One might think that
$D_g^{(r)}$ leads {\sl just} to coherent oscillations of the baryon
radiation fluid, but this is not the case. Note that, \eg, for
adiabatic CDM models without source term one can derive
$(1/4) D_g^{(r)}= - (5/3)\Psi$ on super--horizon scales.
Since for CDM perturbations $\Phi=-\Psi$ and  $\Psi' \simeq 0$,
we see how the usual Sachs--Wolfe result  $\delta T / T =
(1/3)\Psi ({\bf x}_{rec}, t_{rec})$ is recovered.},
and have been calculated for both inflation and defect
models [Scott \etal, 1995; Bennett \& Rhie, 1993;
Pen \etal, 1994; Durrer \& Zhou, 1995].
Generically, a scale invariant spectrum is predicted and thus the
Sachs--Wolfe calculations yield mainly a normalisation for the
different models.
In this Chapter we present a computation for the Doppler
contribution from global topological defects;
in particular we perform our analysis for $\pi_3$--defects, textures
[Turok, 1989], in a universe dominated by cold dark matter.
Many of the features endowed by textures are also present
in other global defects and we believe that the analysis we present
below remains valid for these defects as well.

The Doppler contribution to the CMB anisotropies is given by
\begin{equation}
\left[
{\delta T \over T}({\vec x},\gg )
\right]^{Doppler} =
{1\over 4}D_r({\vec x}_{rec},\eta_{rec}) -
{\vec V}({\vec x}_{rec},\eta_{rec})\cdot {\hat\gamma} ~,
\label{Doppler}
\end{equation}
where ${\vec x}_{rec} = {\vec x} + {\hat\gamma} \eta_0$.
In the previous formula ${\hat\gamma}$ denotes a direction in the sky
and $\eta $ is the conformal time, with $\eta_0 $ and $\eta_{rec} $
the present and recombination times, respectively.

\section{Linear theory power spectra}
\label{sec-linearspectra}
\markboth{Chapter 9. ~GLOBAL TEXTURES AND THE DOPPLER PEAKS}
                     {\S ~9.2. ~LINEAR THEORY POWER SPECTRA}

To evaluate Eq.~(\ref{Doppler}) we  calculate $D_r $ and
${\vec V}$ at $\eta_{rec}$.
We study a two--fluid system:
baryons plus radiation, which prior to
recombination are tightly coupled, and CDM.

We start off by considering the evolution equations
%
in each fluid component
as are given in the classic paper by Kodama \& Sasaki [1984].
The evolution of the perturbation variables
in a flat background, $\Omega = 1$,  can be expressed as
\begin{equation} \begin{array}{lll}
V'_r +{a'\over a}V_r &=& k\Psi+ k{c_s^2\over 1+w}D_r \\
V'_c +{a'\over a}V_c &=& k\Psi  \\
D'_r - 3 w {a'\over a}D_r &=& (1+w) [ 3{a'\over a}\Psi-3\Phi'
   -kV_r - {9\over2} \left({a'\over a}\right)^2 k^{-1}
   (1+{w\rho_r\over \rho})V_r  ] \\
D'_c                 &=& 3{a'\over a}\Psi-3\Phi' - k V_c
   -{9\over2} \left({a'\over a}\right)^2 k^{-1}
   (1+{w\rho_r\over \rho})V_c
{}~, \end{array}
 \label{KSeq} \end{equation}
where subscripts $_r$ and $_c$ denote the baryon--radiation plasma
and CDM, respectively;
$D,~V$ are density and velocity perturbations;
$w=p_r/\rho_r$, $c_s^2=p'_r/\rho'_r$ and $\rho = \rho_r + \rho_c$.

Throughout this section we will normalise the scale factor
as $a(\eta_{eq}) = 1$, where $\eta_{eq}$ stands for the
conformal time at the time of equality between matter and
radiation (we will come back to the usual convention
$a(\eta_{0}) = 1$ at the end, when expressing our results).
For a flat universe we have
$(a' / a)^2 = 8\pi G a^2 (\rho_{rad}+\rho_{mat})/3$,
where both CDM and baryons are included in $\rho_{mat}$.
Taking into account now the different scaling behaviour
of both forms of energy with $a$, and defining
$\tau \equiv (8\pi G / 3)^{-1/2} (\rho_{eq} / 2)^{-1/2}$,
where $\rho_{eq}$ is the total energy density at equality,
we get $a(\eta)+1 = (\sqrt{2} + (\eta - \eta_{eq}) / 2 \tau )^2$,
which automatically satisfies $a_{eq}=1$, as it should.
By further requiring that $a$ vanishes at the origin
we get $\tau = \eta_{eq} / (2 (\sqrt{2} - 1))$ while
the scale factor may be cast as
\be
a(\eta) = {\eta \over \tau}
\left(1 + {1\over 4}{\eta \over \tau}\right) ~.
\ee
It is clear from this expression that for early (late) conformal
times we recover the radiation (matter) dominated phase of expansion
of the universe.

Furthermore, as both the energy density of baryons and CDM
go as $\propto a^{-3}$ we have the ratio
$\rho_{bar} / \rho_{cdm} \simeq$ const, and today yields
$\Omega_{B}$.
{}From this we may calculate
$\rho_{bar} / \rho_{rad} = (\rho_{bar} / \rho_{cdm})
(\rho_{cdm} / \rho_{rad}) \simeq \Omega_B a$.
Applying this to $w$ and $c_s^2$ of Eq. (\ref{KSeq}) we get
their time dependences as follows
\be
w     \simeq {1\over 3}(1+\Omega_B a)^{-1} ~~~ ; ~~~
c_s^2 \simeq {1\over 3}(1+\frac{3}{4}\Omega_B a)^{-1} ~.
\ee

The only place where the `seeds' (texture--seeds or whatever)
enter the system (\ref{KSeq})
is through the potentials $\Psi$ and $\Phi$.
These potentials may be split into a part coming from standard
matter and radiation, and a part due to the seeds, \eg,
$\Psi = \Psi_{(c,r)} + \Psi_{seed}$ where $\Psi_{seed},\Phi_{seed}$
are determined by the energy momentum tensor of the seed; the global
texture in our case.
Having said this, one  may easily see how the seed source terms
arise [Durrer, 1994].

It is our aim now to get rid of the velocity perturbations and
express the system (\ref{KSeq}) as two second order
equations for $D_r$ and $D_c$. For the sake of simplicity in these
intermediate steps let us neglect, for the time being,
the source terms (we will incorporate them at the end).
After straightforward algebra we find
\be
\label{VandV}
V_c = { 12 \pi G a^2 \rho_r (1 + w) V_r - k D'_c
\over
12 \pi G a^2 \rho_r (1 + w) + k^2}
\ee
and
\be
\label{oneV}
V_r =  - {
\left( \rho_c D'_c + \rho_r D'_r - 3 {a'\over a} w \rho_r D_r +
k^2 ( 12 \pi G a^2)^{-1} (1 + w)^{-1} (D'_r -
3 {a'\over a} w D_r) \right)
\over
k \left( k^2(12\pi Ga^2)^{-1}+\rho_r(1+w)+\rho_c \right)
} ~,
\ee
which express the velocity perturbations in both fluids in terms
of the intrinsic density perturbations and the background variables.
After long (and tedious) computations in which we
differentiate $V_r$ and $V_c$
from Eqs. (\ref{VandV}), (\ref{oneV}) and  we substitute back
into Eqs. (\ref{KSeq}), including this time the seed source
term, we finally derive two second order
equations for $D_r$ and $D_c$, namely
\begin{eqnarray}
D_r''+{a'\over a}[1+3c_s^2-6w+F^{-1}\rho_c]D_r'
	 -{a'\over a}\rho_cF^{-1}(1+w)D_c'
&&\nonumber \\
+4\pi Ga^2[\rho_r(3w^2-8w+6c_s^2-1)-2F^{-1}w\rho_c(\rho_r+\rho_c)
&&\nonumber\\
+\rho_c(9c_s^2-7w)+
{k^2\over 4\pi Ga^2}c_s^2]D_r -4\pi G a^2\rho_c(1+w)D_c
&=& (1+w)S~;
\hspace{1cm}
\label{dr}\\
D_c''+{a'\over a}[1+(1+w)F^{-1}\rho_r(1+3c_s^2)]D_c'
	 -{a'\over a}(1+3c_s^2)F^{-1}\rho_rD_r'
&&\nonumber \\
	-4 \pi G a^2\rho_cD_c -
	4 \pi G a^2\rho_r(1+3c_s^2)[1-2(\rho_r+\rho_c)F^{-1}w]D_r
&=& S~,
\label{dc}
\end{eqnarray}
where $F\equiv k^2(12\pi Ga^2)^{-1}+\rho_r(1+w)+\rho_c$ and
$S$ denotes a source term,
which in general is given by $S=4\pi G a^2(\rho +3p)^{seed}$.
In our case, where the seed is described by a global scalar field $\phi$,
we have $S=8\pi G a^2(\dot\phi )^2$.

{}From numerical simulations one finds that the integral of
$a^2|\dot\phi |^2$
over a shell of radius $k$ can be modeled by [Durrer \& Zhou, 1995]
\begin{equation}
 a^2\langle|\dot\phi|^2\rangle (k, \eta )\ =\
{{1\over 2} A\tilde\eta^2\over \sqrt {\eta }[1+\alpha (k\eta )
+\beta (k\eta )^2]} ~,
\label{fit}
\end{equation}
with $\tilde\eta$ denoting the symmetry breaking scale of the phase transition
leading to texture formation (cf. \S\ref{sec-topo}).
The parameters in (\ref{fit}) are
$A\sim 3.3$, $\alpha\sim -0.7/(2\pi)$ and $\beta \sim 0.7/(2\pi)^2$.
On super--horizon scales, where the source term is important,
this fit is accurate to about $10\%$.
Analytical estimates support this finding \cite{DZ}.
On small scales the accuracy reduces to a factor of 2.
By using this fit in the calculation of $D_r$ and $D_c$ from
Eqs. (\ref{dr}), (\ref{dc}) we effectively neglect the
time evolution of phases of $(\phi ')^2$;
the incoherent evolution of these phases may smear out subsequent
Doppler peaks [Albrecht \etal, 1995],
but will not affect substantially the height of the first peak.

\section{Choosing initial conditions}
\label{sec-inicon}
\markboth{Chapter 9. ~GLOBAL TEXTURES AND THE DOPPLER PEAKS}
                     {\S ~9.3. ~CHOOSING INITIAL CONDITIONS}

In defect models of large--scale structure formation
one usually assumes that the universe begins in a hot,
homogeneous state.
All perturbation variables are zero, and the primordial
fields (\eg, the global field $\phi$ in our case)
are in thermal equilibrium in the unbroken symmetry phase.
When the phase transition occurs the field $\phi$ leaves
thermal equilibrium and enters the process of phase ordering.
The large scales of interest to us get perturbed when crossing
the horizon, long after the phase transition took place.

Before the phase transition the defect fields have no
long--range correlations. The dynamics of the transition as well
as the field ordering are completely causal processes.
Isotropy and the vanishing correlation function of the
defect's stress--energy tensor on super--horizon scales
lead to the power spectrum of the source term to take
the form of white noise for $k \eta << 1$.

Following the phase transition, the distribution of the
defect field enters a `scaling' regime rapidly
[Pen \etal, 1994],
and so numerical simulations need to be
performed only within a reduced dynamical range: from `natural'
initial conditions (namely, no correlations on super--horizon
scales) and for a few expansion times. This is enough to
produce  a realistic field distribution on large scales.

Going back now to our set of
Eqs. (\ref{dr}), (\ref{dc}) we will specify our initial conditions
as follows: for a given scale $k$ we choose the initial time
$\eta_{in}$ such that the perturbation is super--horizon--sized
and the universe is radiation dominated.
In this limit the evolution equations reduce to

\begin{figure}[tbp]
  \begin{center}
    \leavevmode
    \epsfxsize = 7.5cm
    \epsfysize = 7.5cm
    \epsffile{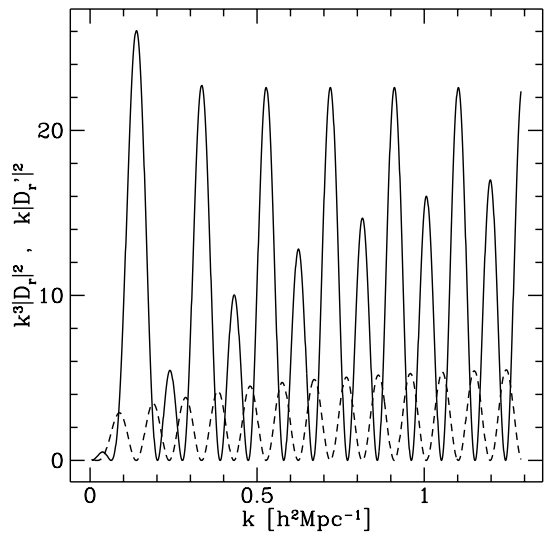}
  \end{center}
\caption{{\sl The dimensionless power spectra, $k^3|D_r|^2$ (solid line)
and $k|D'_r|^2$ (dashed line) in units of $(A\epsilon)^2$,
are shown as functions of $k$.
These are exactly the quantities which enter in the expression for the
${\cal C}_\ell$'s. We set $h=0.5~,~\Omega_B=0.05$ and $z_{rec}=1100$.}}
\label{dopp1}
\end{figure}

\newpage

\begin{eqnarray}
D_r''-{2\over \eta^2} D_r &=&{4\over 3}{A\epsilon \over \sqrt {\eta}}
{}~; \\
D_c''+{3\over \eta} D_c'-{3\over 2\eta}D_r'
-{3\over 2\eta^2}D_r  &=& {A\epsilon \over \sqrt {\eta}} ~,
\end{eqnarray}
with particular solutions
\begin{equation}
D_r\ = \ -{16\over 15}\epsilon A \eta^{3/2}\ ;\
D_c\ = \ -{4 \over  7}\epsilon A \eta^{3/2}\ .
\label{partiti}
\end{equation}
In the above equations we have introduced
$\epsilon\equiv 4\pi G\tilde\eta^2$, the only free parameter in the model.
We consider perturbations  seeded by the texture field, and
therefore it is incorrect to add a homogeneous growing mode to the
above solutions.
With these initial conditions, Eqs. (\ref{dr}), (\ref{dc}) are easily
integrated numerically, leading to the spectra for $D_r(k, \eta_{rec})$
and $D_r '(k, \eta_{rec})$ (see Figure \ref{dopp1}).

\section{Angular power spectrum from global textures}
\label{sec-textucls}
\markboth{Chapter 9. ~GLOBAL TEXTURES AND THE DOPPLER PEAKS}
     {\S ~9.4. ~ANGULAR POWER SPECTRUM FROM GLOBAL TEXTURES}

In this section we calculate the expression for ${\cal C}_{\ell}$
as function of the baryon--radiation power spectrum and its
derivative, evaluated at $\eta_{rec}$.

\begin{figure}[tbp]
  \begin{center}
    \leavevmode
    \epsfxsize = 7.5cm
    \epsfysize = 7.5cm
    \epsffile{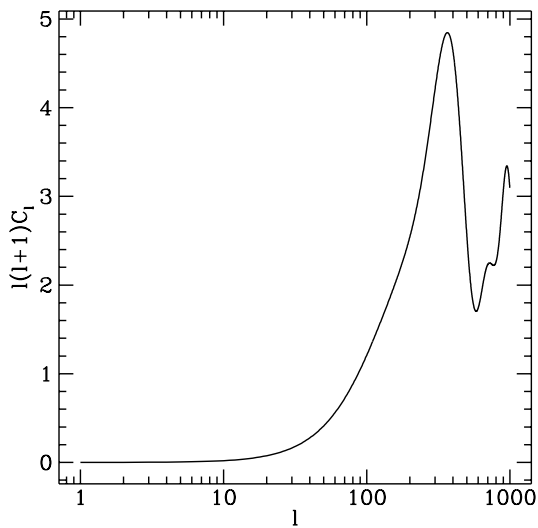}
  \end{center}
\caption{{\sl The angular power spectrum for the Doppler contribution to
the  CMB anisotropies is shown in units of $\epsilon^2$. We set
cosmological parameters $h=0.5~,~\Omega_B=0.05$ and $z_{rec}=1100$.}}
\label{dopp2}
\end{figure}

Following a standard procedure [Durrer, 1994]
we express the Fourier transform of the velocity perturbation
in the
radiation fluid as\footnote{In the previous section we have
calculated the velocity perturbations in both (coupled) fluids.
{}From them we found the intrinsic density perturbations.
In Eq. (\ref{veloo}) the velocity perturbation refers to
that intrinsic to radiation at the last scattering surface.
Given the uncertainties in the fit of the source term
(cf. (\ref{fit})) this is a good approximation.
See Ref. \cite{RuthReview} for details.}
\be
\label{veloo}
{\vec V}({\vec k}) \simeq
- {i {\vec k} D_r'({\vec k})\over k^2 (1 + w)} ~.
\ee
Now, let us Fourier transform the Doppler contribution to the
anisotropies as given by Eq. (\ref{Doppler}).
Using (\ref{veloo}) we get
\begin{equation}
\left[
{\delta T \over T}({\vec k},\eta_0,\mu )
\right]^{Doppler} =
e^{i k \eta_0 \mu}
\left(
{1\over 4}D_r({\vec k},\eta_{rec}) +
{i \mu D_r'({\vec k}, \eta_{rec}) \over k (1 + w)}
\right)
\label{Dopplerkk}
\end{equation}
where $\mu = {\hat k}\cdot\gg $.
Now, it will prove useful to perform an expansion in Legendre
polynomials as follows
\begin{equation}
\left[
{\delta T \over T}({\vec k},\eta_0,\mu )
\right]^{Doppler} =
\sum_{\ell =0}^{\infty} i^\ell \Delta_\ell ({\vec k},\eta_0)
P_\ell (\mu )~.
\label{Dopplerexpa}
\end{equation}
After some algebra we get the coefficients $\Delta_\ell$ in the
previous expansion given by
\begin{equation}
{ \Delta_\ell ({\vec k},\eta_0) \over (2\ell + 1)} =
{1\over 4}D_r({\vec k},\eta_{rec}) j_\ell ( k \eta_0) +
{ D_r'({\vec k}, \eta_{rec}) \over k (1 + w)}
j'_\ell ( k \eta_0) ~,
\label{deltalsobre}
\end{equation}
where $j_\ell$ denotes the spherical Bessel function of order $\ell$
and
$j'_\ell$ stands for its derivative with respect to the argument.
Now we claim that we may cast the Doppler contribution to
the multipole coefficients ${\cal C}_\ell$ as follows
\begin{equation}
{\cal C}_\ell^{(Doppler)} = {2\over \pi} \int dk~ k^2
{|\Delta_\ell ({\vec k},\eta_0)|^2 \over (2\ell + 1)^2} ~.
\label{claim}
\end{equation}
To prove this we make use of the ergodic hypothesis and
substitute the ensemble average by an integral over all
possible observers. Thus we may write the left--hand side
of Eq. (\ref{twotwo}) as follows
\be
\label{newway}
\la \D({\vec x}, \gg) \D^*({\vec x}, \gg') \ra =
\int d^3x \int { d\Omega_{\gg} d\Omega_{\gg'}  \over 8 \pi^2}
\D({\vec x}, \gg) \D^*({\vec x}, \gg') ~,
\ee
with the constraint $\gg\cdot\gg' = \cos\vartheta$,
and where $^*$ denotes complex conjugation.
Upon using Parseval's relation to transform the $\int d^3x$ integral
into an integral on ${\vec k}$ we see that Eq. (\ref{newway})
can be cast as
\be
\label{newway2}
{1\over (2\pi)^{3}} \int d^3k \sum_{\ell ,\ell'}
i^{\ell - \ell'}
\Delta_\ell      ({\vec k},\eta_0)
\Delta_{\ell'}^* ({\vec k},\eta_0)
\int { d\Omega_{\gg} d\Omega_{\gg'}  \over 8 \pi^2}
P_{\ell }(\mu )
P_{\ell' }(\mu ' ) ~.
\ee
Now, the last integral in Eq. (\ref{newway2}) may be written as
\be
\int { d\Omega_{\gg} d\Omega_{\gg'}  \over 8 \pi^2}
\delta (\gg\cdot\gg' - \cos\vartheta )
P_{\ell  }({\hat k}\cdot\gg )
P_{\ell' }({\hat k}\cdot\gg ' ),
\ee
where we introduced a Dirac delta to enforce the constraint on the
$\gg$'s.
Standard algebra like the one used in
Chapter \ref{chap-statistics} yields Eq. (\ref{newway})
in the form
\begin{equation}
\la C_2(\vartheta) \ra =
{1\over (2\pi)^3} \int d^3k \sum_{\ell = 0}^{\infty}
{|\Delta_\ell ({\vec k},\eta_0)|^2 \over (2\ell + 1)}
P_\ell (\cos\vartheta) ~.
\label{claimminus}
\end{equation}
If we assume now that $|\Delta_\ell ({\vec k},\eta_0)|^2$
depends only on the absolute value $k$,
we can integrate out the directions ${\hat k}$ and, upon comparison
with the expansion (\ref{twotwo}) it is easy to see that the
claimed result is achieved.
Finally, knowing the spectra for $D_r$ and $D_r'$ we calculate the
Doppler contribution to the ${\cal C}_{\ell}$'s according to
\begin{equation}
{\cal C}_{\ell} =
{2\over \pi} \int dk \left[{k^2\over 16}|D_r(k,\eta_{rec})|^2
j_{\ell}^2(k\eta_0) + (1+w)^{-2} |D_r'(k,\eta_{rec})|^2
(j_{\ell}'(k\eta_0))^2\right] ~,
\label{cldopp}
\end{equation}
where we drop the label $^{(Doppler)}$ from now on.
As we mentioned above, the source term of equations
(\ref{dr}) and (\ref{dc}) was estimated from numerical simulations
of the evolution of the global scalar field $\phi$. These simulations
give a fit for the absolute value of $\phi$ and therefore
we can say nothing with regard to phases. This also implies that
we cannot evaluate the crossed term missing in Eq. (\ref{cldopp}),
which we neglect as a first approximation to the problem.
Integrating Eq. (\ref{cldopp}), we obtain the Doppler contribution
to the angular spectrum of the CMB anisotropies
[Durrer, Gangui \& Sakellariadou, 1995]. We plot the result in
Figure \ref{dopp2}.

\section{Discussion}
\label{sec-disctextu}
\markboth{Chapter 9. ~GLOBAL TEXTURES AND THE DOPPLER PEAKS}
                                      {\S ~9.5. ~DISCUSSION}

As we see from the plot, the angular power spectrum
$\ell (\ell + 1) {\cal C}_{\ell}$  yields the Doppler peaks.
For $\ell <1000$, we find three peaks located
at $\ell=365$, $\ell=720$ and $\ell = 950$. Silk damping, which we
have not taken into account here, will decrease the
relative amplitude of the third peak with respect to the second one;
however it will not affect substantially the height of the first peak.
Our most important results regard the amplitude and position
of the first Doppler peak, for which we find
\begin{equation}
	\ell(\ell+1){\cal C}_\ell
\left|_{{~}_{\!\!  \ell\sim 360 }}
= 5\epsilon^2~.
	\right. \label{clfirstpeak}\end{equation}
It is interesting to notice that the position of the first peak is
displaced by  $\Delta \ell\sim 150$  towards smaller
angular scales than in standard inflationary models [Steinhardt, 1995].
This might be due to the difference in the growth of super--horizon
perturbations, which is $D_r \propto \eta^{3/2}$ in our case
(cf. Eqs. (\ref{partiti})),
and $D_r \propto \eta^2$ for inflationary models.

One may understand the height of the first peak from the following
analytic estimate: matching the sub--horizon with the super--horizon
solutions of Eq.~(\ref{dc}), in the matter dominated era, one finds
$D_c \sim -{2\over 5}A\epsilon(k/2\pi)^{1/2}\eta^2$.
{}From Eq.~(\ref{dr}) we then obtain in this limit
	$D_r\approx A\epsilon k^{-3/2}$.
Plugging this latter value into Eq.~(\ref{cldopp}) we get roughly
$\ell(\ell+1){\cal C}_\ell \sim (A\epsilon)^2$,
for the height of the first peak.

Let us now compare our value for the Doppler peak with the level of the
Sachs--Wolfe plateau
[Bennett \& Rhie, 1993; Pen \etal, 1994; Durrer \& Zhou, 1995]
\begin{equation}
\ell(\ell+1){\cal C}_\ell\left|_{{~}_{\!\! SW}} \sim ( 6 - 14 ) \epsilon^2.
\right. \label{SWdopeq}
\end{equation}
It is apparent from Eqs.~(\ref{clfirstpeak}) and (\ref{SWdopeq})
that the Doppler contribution from textures is
substantially smaller than for inflationary models.

Normalising with the {\sl COBE}--DMR experiment
[Smoot \etal, 1992; G\'orski \etal, 1994]
\[ \ell (\ell + 1) {\cal C}_\ell \left|_{{~}_{\!\!COBE}}
\sim 8 \times 10^{-10}\right. ~, \]
one finds the symmetry breaking energy scale when textures are
produced in the early universe,
and from this $\epsilon \sim  10^{-5}$.
This value for $\epsilon$ depends of course upon the
numerical simulations, Refs. \cite{BR,PST,DZ}.

We believe that this result, stating that  the first Doppler peak has a
height comparable to the Sachs--Wolfe plateau, is valid for all global
defects. This depends crucially on the $ 1/\sqrt{\eta}$ behavior of
$(a\dot\phi)^2={\phi'}^2$ on large scales (cf. Eq.~(\ref{fit})),
which is a generic feature of global defects:
on super--horizon scales $\dot{\phi}^2(k)$ represents white noise
superimposed on the average given by
$\la {\phi'}^2 \ra \sim
{\phi'}^2(k=0) \propto 1/\eta^2$ (the usual scaling behaviour).
The Fourier transform of ${\phi'}^2$ determines the fluctuations
on this background on a given comoving scale $\lambda$.
A patch of size $\lambda$ consists of $N=(\lambda/\eta)^3$ independent
horizon--sized volumes.
The fluctuation ${\phi'}^2(k)$ on this scale is thus proportional to
$\la {\phi'}^2 \ra / \sqrt{N} \propto 1/\sqrt{\eta}$.
Notice that this argument does not apply for local cosmic strings.

The displacement of the peaks towards smaller scales
is reminiscent of open models
[\eg, Kamionkowski, Spergel \& Sugiyama, 1994],
where the first Doppler
peak is located approximately at
$\ell_{peak} \simeq 220 / \sqrt{\Omega_0}$.
A value $\Omega_0 \simeq 0.3$ would produce
$\ell_{peak} \simeq 365$ for the first peak, as we actually find.
{\sl However}, the amplitude of the peaks within open inflationary
models is still much larger than what we find from global defects.

Similar results to those worked out in the present Chapter
were independently derived in [Crittenden \& Turok, 1995].
Following a different approach, these authors calculate the Doppler
peaks from cosmic textures in the synchronous gauge.
A main assumption of that analysis is that spatial gradients in the
scalar field are frozen outside the horizon, and therefore time
derivatives are negligible. This is not the case in the
simulations for the texture source term from Durrer \& Zhou [1995].
Apart from this we basically agree with the shape and position of
their Doppler peaks.
Regarding the height of the peaks, however, these authors seem to
obtain an amplitude similar to that predicted in standard
inflation--based models, unlike what we find.

Based on our analysis we conclude that if the existence of Doppler
peaks is indeed confirmed (recall that not all the teams agree in
their findings on the $\theta\sim 1^\circ$ scale)
and
the height of the first peak is larger than
about twice the level of the Sachs--Wolfe plateau, namely
\[ \ell(\ell+1){\cal C}_\ell \left|_{{~}_{\!\!peak}} >
        2 \times 10^{-9}\right. ~, \]
and if the first peak is positioned at $\ell<300$, {\sl then}
global topological defects are ruled out.
On the other hand, if the first Doppler peak is positioned at
$\ell \sim 350$ and its height is below the above value,
global defects are strongly favored if compared to
standard cold dark matter inflationary models.
This is a clear {\sl fingerprint}, within present
observational capabilities, to distinguish among these
two competing models of structure formation.




\begin{thebibliography}{999}
\markboth{REFERENCES}{REFERENCES}
\addcontentsline{toc}{chapter}{References}
\footnotesize






%
%
%
%
%
%
%
%
%
%
%
%
\bibitem{AbbFaWise82}
Abbott, L.F., Fahri, E. \& Wise, M.B. [1982],
{\sl Phys. Lett.} {\bf B117}, 29.
\bibitem{AbbWise84}
Abbott, L.F. \& Wise, M.B. [1984], {\sl Ap. J. Lett.} {\bf 282}, L47.
\bibitem{AbbWise84b}
Abbott, L.F. \& Wise, M.B. [1984b], {\sl Nucl. Phys.} {\bf B244}, 541.
\bibitem{Adams+Co93}
Adams, F.C., Bond, J.R., Freese, K., Frieman, J. \&
                     Olinto, A. [1993], {\sl Phys. Rev.} {\bf D47}, 426.
\bibitem{Adams+Co91}
Adams, F.C., Freese, K. \&  Guth, A.H. [1991],
                     {\sl Phys. Rev.} {\bf D43}, 965.
\bibitem{alb}
Albrecht, A., Coulson, D., Ferreira, P. \&
Magueijo, J. [1995],  Imperial Preprint /TP/94--95/30.
\bibitem{Albrecht-Steinhardt82}
Albrecht, A. \& Steinhardt, P. [1982],
{\sl Phys. Rev. Lett.} {\bf 48}, 1220.
\bibitem{ASTW82}
Albrecht, A., Steinhardt, P.J., Turner, M.S. \& Wilczek, F. [1982],
{\sl Phys. Rev. Lett.} {\bf 48}, 1437.
\bibitem{AllenGriWi87}
Allen, T.J., Grinstein, B. \& Wise, M.B. [1987],
{\sl Phys. Lett.} {\bf B197}, 66.
\bibitem{Allen+Shellard90}
Allen, B. \& Shellard, E.P.S. [1990], {\sl Phys. Rev. Lett.} {\bf 64}, 119.
\bibitem{AmBaOcchi95}
Amendola, L., Baccigalupi, C. \& Occhionero, F. [1995],
astro--ph/9504097 preprint.
\bibitem{BST83}
Bardeen, J.M., Steinhardt, P.J. \& Turner, M.S. [1983],
                {\sl Phys. Rev.} {\bf D28}, 679,1983..
\bibitem{Barrow90}
Barrow, J.D. [1990], {\sl Phys. Lett.} {\bf B235}, 40.
\bibitem{Barrow+Coles90}
Barrow, J.D. \& Coles, P. [1990], {\sl M. N. R. A. S.} {\bf 244}, 188.
\bibitem{Barrow+Liddle93}
Barrow, J.D. \& Liddle, A.R. [1993], {\sl Phys. Rev.} {\bf D47}, R5213.
\bibitem{Barrow+Maeda90}
Barrow, J.D. \& Maeda, K. [1990], {\sl Nucl. Phys.} {\bf B341}, 294.
\bibitem{Barrow+Saich90}
Barrow, J.D. \& Saich, P. [1990], {\sl Phys. Lett.} {\bf B249}, 406.
\bibitem{Bennett+Co92}
Bennett, C.L., \etal ~[1992], {\sl Ap. J. Lett.} {\bf 396}, L7.
\bibitem{Bennett+Co94}
Bennett, C.L., \etal ~[1994], {\sl Ap. J.} {\bf 430}, 423.
\bibitem{Bennett+Bouchet88}
Bennett, D.P. \& Bouchet, F.R. [1988],
{\sl Phys. Rev. Lett.} {\bf 60}, 257.
\bibitem{BR90}
Bennett, D.P. \&  Rhie, S.H. [1990], {\sl Phys. Rev. Lett.} {\bf 65}, 1709.
\bibitem{BR}
Bennett, D.P. \&  Rhie, S.H. [1993], {\sl Ap. J. Lett.} {\bf 406}, L7.
\bibitem{BRWein93}
Bennett, D.P., Rhie, S.H. \& Weinberg, D.H. [1993], preprint.
\bibitem{BennettStebbinsBouchet92}
Bennett, D.P., Stebbins, A. \& Bouchet, F.R. [1992],
{\sl Ap. J. Lett.} {\bf 399}, L5.
\bibitem{Bond88}
Bond, J.R. [1988], in {\sl The Early Universe},
Unruh, W. \& Semenoff, G., eds. (Reidel, Dordrecht).
\bibitem{BE84}
Bond, J.R. \& Efstathiou, G. [1984], {\sl Ap. J. Lett.} {\bf 285}, L45.
\bibitem{BE87}
Bond, J.R. \& Efstathiou, G. [1987], {\sl M. N. R. A. S.} {\bf 226}, 655.
\bibitem{BorrillEtAl94}
Borrill, J., \etal ~[1994], {\sl Phys. Rev.} {\bf D50}, 2469.
\bibitem{Bouchet+Co88}
Bouchet, F. R., Bennett, D. P. \& Stebbins, A. [1988],
{\sl Nature} {\bf 335}, 410.
\bibitem{BowickEtAl}
Bowick, M., Chandar, L., Schiff, E. \& Srivastava, A. [1994],
{\sl Science} {\bf 236}, 943.
\bibitem{Boyaetal}
Boyanovski, D., \etal ~[1995], Preprint PITT--09--95;
{\sl Phys. Rev.} {\bf D51}, 4419.
\bibitem{BrandenbergerRev}
Brandenberger, R. [1993], {\sl Topological Defects and Structure Formation},
EPFL lectures, Lausanne, Switzerland.
\bibitem{BrandenbergerRev95}
Brandenberger, R. [1995],
{\sl Modern Cosmology and Structure Formation},
TASI--94 Lectures at Boulder,
Donoghue, J., ed. (World Scientific, Singapore).
\bibitem{BranKahn84}
Brandenberger, R. \& Kahn, R. [1984], {\sl Phys. Rev.} {\bf D29}, 2172.
\bibitem{Brandenberger+Co87}
Brandenberger, R., Kaiser, N., Shellard, E.P.S. \& Turok, N. [1987],
{\sl Phys. Rev.} {\bf D36}, 335.
\bibitem{Bunch-Davies}
Bunch, T. \& Davies, P. [1978], {\sl Proc. R. Soc. Lon.} {\bf A360}, 117.
\bibitem{Bunn+Co94} 
Bunn, E.,Hoffman, Y. \& Silk, J. [1994], {\sl Ap. J.} {\bf 425}, 359.
\bibitem{CallanColeman}
Callan, C. \& Coleman, S. [1977], {\sl Phys. Rev.} {\bf D16}, 1762.
\bibitem{Casdoso-Ovrut}
Cardoso, G.L. \& Ovrut, B.A. [1993], CERN Preprint.
\bibitem{Carr-Rees}
Carr, B.J. \& Rees, M.J. [1984], {\sl M. N. R. A. S.} {\bf 206}, 801.
\bibitem{Carrol-Press-Turner92}
Carrol, S.M., Press, W.H. \& Turner, E.L. [1992],
{\sl Ann. Rev. Astron. and  Astrophys.} {\bf 30}, 499.
\bibitem{Castadecoh}
Castagnino, M.A., Gangui, A., Mazzitelli, F.D. \& Tkachev, I.I. [1993],
{\sl Class. Quant. Grav.} {\bf 10}, 2495.
\bibitem{ChuangEtAl}
Chuang, I., Durrer, R., Turok, N. \& Yurke, B. [1991],
{\sl Science} {\bf 251}, 1336.
\bibitem{ColesLucchin}
Coles, P. \& Lucchin, F. [1995], {\sl Cosmology: The origin and evolution
of cosmic structure} (Wiley, New York).
\bibitem{cope}
Copeland, E. [1993], in {\sl The physical universe: The interface
between cosmology, astrophysics and particle physics}, eds. Barrow, J.D.,
\etal ~(Sringer--Verlag).
\bibitem{Coulson+Co94}
Coulson, D., Ferreira, P., Graham, P. \& Turok, N. [1994],
{\sl Nature} {\bf 368}, 27.
\bibitem{ct}
Crittenden, R.G. \& Turok, N. [1995], Princeton University Preprint,
		PUPT--94--1545.
\bibitem{deGennes}
de Gennes, P. [1974],  {\sl The Physics of Liquid Crystals}
(Clarendon Press, Oxford).
\bibitem{DodeJubas1995}
Dodelson, S. \& Jubas, J.M. [1995], {\sl Ap. J.} {\bf 439}, 503.
\bibitem{DolanJackiw74}
Dolan, L. \& Jackiw, R. [1974], {\sl Phys. Rev.} {\bf D9}, 3320.
\bibitem{DolgovFreese95}
Dolgov, A. \& Freese, K. [1995],
submitted to {\sl Phys. Rev.} {\bf D}, hep--ph/9408214.
\bibitem{RuthReview}
Durrer, R. [1994], {\sl Fund. of Cosmic Physics} {\bf 15}, 209.
\bibitem{DGS}
Durrer, R., Gangui, A. \& Sakellariadou, M. [1995],
SISSA--83/95/A, astro--ph/9507035.
\bibitem{DZ}
Durrer, R. \& Zhou, Z.H. [1995], Z\"urich University Preprint,
ZH--TH19/95, astro--ph/9508016.
\bibitem{DHowZ}
Durrer, R., Howard, A. \& Zhou, Z.H. [1994],
{\sl Phys. Rev.} {\bf D49}, 681.
\bibitem{Edmonds}
Edmonds, A. R. [1960], {\sl Angular Momentum in Quantum Mechanics} ~
(Princeton University Press).
\bibitem{Efstathiou90}
Efstathiou, G. [1990], in {\sl Physics of the early Universe}
Peacock, J.A., \etal, eds. (New York: Adam Hilger).
\bibitem{Ellis91}
Ellis, G.F.R. [1991], in {\sl Proc. Banff Summer Research Institute
on Gravitation},
Mann, R. \& Wesson, P., eds. (Singapore: World Scientific).
\bibitem{EllisRothman93}
Ellis, G.F.R. \& Rothman, T. [1993], {\sl Lost Horizons}, preprint.
\bibitem{FLM87}
Fabbri, R., Lucchin, F. \& Matarrese, S. [1987], {\sl Ap. J.} {\bf 315}, 1.
\bibitem{Falk+Co93}
Falk, T., Rangarajan, R. \& Srednicki, M. [1993],
{\sl Ap. J. Lett.} {\bf 403}, L1.
\bibitem{Natural}
Freese, K., Frieman, J.A. \& Olinto, A.V. [1990],
{\sl Phys. Rev. Lett.} {\bf 65}, 3233.
\bibitem{FreeseConf93} Freese, K. [1994],
in {\sl Yamada Conference XXXVII: Evolution of the Universe and its
Observational Quest}, Tokyo, Japan 1993, preprint astro--ph/9310012.
\bibitem{Fry84}
Fry, J.N. [1984], {\sl Ap. J.} {\bf 279}, 499.
\bibitem{Ganga+Co93}
Ganga, K., Cheng, E., Meyer, S., Page, L. [1993],
{\sl Ap. J. Lett.} {\bf 410}, L57.
\bibitem{nonGauss}
Gangui, A. [1994], {\sl Phys. Rev.} {\bf D50}, 3684.
\bibitem{Rome94}
Gangui, A. [1995], in {\sl Birth of The Universe \& Fundamental Physics},
Occhionero, F., ed. (Heidelberg: Springer--Verlag). in press.
\bibitem{3point}
Gangui, A., Lucchin, F., Matarrese, S. \& Mollerach, S. [1994],
{\sl Ap. J.} {\bf 430}, 447.
\bibitem{decoh}
Gangui, A., Mazzitelli, F.D. \& Castagnino, M.A. [1991],
{\sl Phys. Rev.} {\bf D43}, 1853.
\bibitem{SilyYo}
Gangui, A. \& Mollerach, S. [1995], work in progress.
\bibitem{CSandCV}
Gangui, A. \& Perivolaropoulos, L. [1995], {\sl Ap. J.} {\bf 447}, 1.
\bibitem{Gibbons-Hawking}
Gibbons, G. \& Hawking, S. [1977], {\sl Phys. Rev.} {\bf D15}, 2738.
\bibitem{Gliner65}
Gliner, E.B. [1965], {\sl Zh. Eksp. Teor. Fiz.} {\bf 49}, 542.
                    [{\sl Sov. Phys. JETP}      {\bf 22}, 378 (1965)].
\bibitem{Gliner70}
Gliner, E.B. [1970], {\sl Dokl. Akad. Nauk SSSR} {\bf 192}, 771.
                    [{\sl Sov. Phys. Dokl.}      {\bf 15}, 559 (1970)].
\bibitem{GoldwirthPiran92}
Goldwirth, D.S. \& Piran, T. [1992], {\sl Phys. Rep.} {\bf 214}, 223.
\bibitem{Goncharov+Co87}
Goncharov, A.S., Linde, A.D.  \& Mukhanov, V.F. [1987],
                         {\sl Int. J. Mod. Phys.} {\bf A2}, 561.
\bibitem{Gorski+Co94}
G\'orski, K.M., Hinshaw, G., Banday, A.J., Bennett, C.L., Wright, E.L.,
Kogut, A., Smoot, G.F. \& Lubin, P. [1994],
{\sl Ap. J. Lett.} {\bf 430}, L89.
\bibitem{Gott+Co90}
Gott, J.R., Park, C., Juszkiewicz, R., Bies, W.E., Bennett, D.,
Bouchet, F.R. \& Stebbins, A. [1990],
{\sl Ap. J.} {\bf 352}, 1.
\bibitem{Gott85}
Gott, R. [1985], {\sl Ap. J.}, {\bf 288}, 422.
\bibitem{Graham+Co93}
Graham, P., Turok, N., Lubin, P.M. \& Schuster, J.A. [1993],
PUP--TH--1408 preprint.
\bibitem{Guth81}
Guth, A.H. [1981], {\sl Phys. Rev.} {\bf D23}, 347.
\bibitem{Guth85}
Guth, A.H. \& Pi, S.--Y. [1985], {\sl Phys. Rev.} {\bf D32}, 1899.
\bibitem{Guth-Tye80}
Guth, A.H. \& Tye, S.--H. [1980], {\sl Phys. Rev. Lett.} {\bf 44}, 631.
\bibitem{Guth-Weinberg83}
Guth, A.H. \& Weinberg, E. [1983], {\sl Nucl. Phys.} {\bf B212}, 321.
\bibitem{Halliwell1989}
Halliwell, J.J. [1989], {\sl Phys. Rev.} {\bf D39}, 2912.
\bibitem{Hancock+Co94}
Hancock, S., Davies, R.D., Lasenby, A.N., Gutierrez de la Cruz,
C.M., Watson, R.A., Rebolo, R. \& Beckman, J.E. [1994],
{\sl Nature} {\bf 367}, 333.
\bibitem{Hara+Miyoshi93}
Hara T. \& Miyoshi S. [1993], {\sl Ap. J.} {\bf 405}, 419.
\bibitem{Harrison}
Harrison, R. [1970], {\sl Phys. Rev.} {\bf D1}, 2726.
\bibitem{McClintock}
Hendry, P.C., Lawson, N.S., Lee, R.A.M., McClintock, P.V.E. \&
Williams, C.D.H. [1994], {\sl Nature} {\bf 368}, 315.
\bibitem{Hilton1953}
Hilton, P.J. [1953], {\sl Introduction to homotopy theory}
(Cambridge: Cambridge University Press).
\bibitem{Hindmarsh94}
Hindmarsh M. [1994], {\sl Ap. J.} {\bf 431}, 534.
\bibitem{Hindmarsh+Kibble95}
Hindmarsh, M. \& Kibble, T. [1994], {\sl Rep. Prog. Phys.}, in press
(preprint hep--ph/9411342).
\bibitem{Hinshaw+Co94}
Hinshaw, G., Kogut, A., G\'orski, K.M., Banday, A.J., Bennett, C.L.,
Lineweaver, C., Lubin, P., Smoot, G.F. \& Wright, E.L. [1994],
{\sl Ap. J.} {\bf 431}, 1.
\bibitem{Hinshaw+Co95}
Hinshaw, G., Banday, A.J., Bennett, C.L., G\'orski, K.M.
\& Kogut, A. [1995], {\sl Ap. J. Lett.} {\bf 446}, L67.
\bibitem{Hodges+Co90}
Hodges, H.M., Blumenthal, G.R., Kofman, L.A. \&  Primack, J.R. [1990],
{\sl Nucl. Phys.} {\bf B335}, 197.
\bibitem{HuScoSil}
Hu, W., Scott, D. \& Silk, J. [1994], {\sl Phys. Rev.} {\bf D49}, 648.
\bibitem{HSS}
Hu, W., Sugiyama, N. \& Silk, J. [1995], {\sl Nature}, in press.
\bibitem{JonesWyse85}
Jones, B. \& Wyse, R. [1985], {\sl Astron. Astrophys.} {\bf 149}, 144.
\bibitem{Kaiser+Stebbins84}
Kaiser, N. \&  Stebbins, A. [1984], {\sl Nature} {\bf 310}, 391.
\bibitem{KSS}
Kamionkowski, M., Spergel, D.N. \& Sugiyama, N. [1994],
{\sl Ap. J. Lett.} {\bf 426}, L57.
\bibitem{Kibble76}
Kibble, T.W.B. [1976], {\sl J. Phys.} {\bf A9}, 1387.
\bibitem{Kibble}
Kibble, T.W.B. [1980], {\sl Phys. Rep.} {\bf 67}, 183.
\bibitem{KibbleEtAl82}
Kibble, T.W.B., Lazarides, G. \& Shafi, Q. [1982],
{\sl Phys. Lett.} {\bf B113}, 237.
\bibitem{Kirzhnits-Linde74}
Kirzhnits, D.A. \& Linde, A.D. [1974], {\sl Sov. Phys. JETP} {\bf 40}, 628.
\bibitem{Kirzhnits-Linde76}
Kirzhnits, D.A. \& Linde, A.D. [1976], {\sl Ann. Phys.} {\bf 101}, 195.
\bibitem{KS}
Kodama, H. \&  Sasaki, M. [1984], {\sl Prog. Theor. Phys. Suppl.}
	{\bf 78}, 1.
\bibitem{Kofman+Co90}
Kofman, L.K., Blumenthal, G.R., Hodges, H. \& Primack, J.R.
[1990], in {\sl Large--Scale Structures and Peculiar
Motions in the Universe}, Latham, D.W. \& da Costa, L.N. eds.,
ASP Conference Series, Vol. 15.
\bibitem{DoubleInflation2}
Kofman, L.A., Linde, A.D. \& Starobinskii, A.A. [1985],
{\sl Phys. Lett.} {\bf B157}, 361.
\bibitem{KoLiStar94}
Kofman, L.A., Linde, A.D. \& Starobinskii, A.A. [1994],
{\sl Phys. Rev. Lett.} {\bf 24}, 3195.
\bibitem{Kogut+Co94}
Kogut, A., Banday, A.J., Bennett, C.L., Hinshaw, G., Lubin, P.
\& Smoot, G.F. 1994, {\sl Ap. J. Lett.} {\bf 439}, L29.
\bibitem{KolbSUSSP}
Kolb, E.W. [1994], in {\sl 42nd Scottish Universities Summer School
in Physics}, St. Andrews, Scotland, August 1993. in press.
\bibitem{KT90}
Kolb, E.W. \&  Turner, M.S. [1990]
               {\sl The Early Universe} (New York: Addison--Wesley).
\bibitem{KolbVadas94}
Kolb, E.W. \& Vadas, S.L. [1994], {\sl Phys. Rev.} {\bf D50}, 2479.
\bibitem{La-Steinhardt}
La, D. \& Steinhardt, P.J. [1989], {\sl Phys. Rev. Lett.} {\bf 62}, 376.
\bibitem{Langacker-Pi80}
Langacker, P. \& Pi, S.--Y. [1980], {\sl Phys. Rev. Lett.} {\bf 45}, 1.
\bibitem{Langer92}
Langer, S. [1992], in {\sl Solids far from equilibrium}, Godr\`eche, C.,
ed. (Cambridge: Cambridge University Press).
\bibitem{Liddle+Lyth92}
Liddle, A.R. \& Lyth, D.H. [1992], {\sl Phys. Lett.} {\bf B291}, 391.
\bibitem{Liddle+Lyth93}
Liddle, A.R. \& Lyth, D.H. [1993], {\sl Phys. Rep.} {\bf 231}, 1.
\bibitem{Liddle+Turner}
Liddle, A.R. \& Turner, M.S.  [1994], {\sl Phys. Rev.} {\bf 50}, 758.
\bibitem{Linde74}
Linde, A.D. [1974], {\sl Pis'ma Zh. Eksp. Teor. Fiz.} {\bf 19}, 320.
                    [{\sl JETP Lett.} {\bf 19}, 183 (1974)].
\bibitem{Linde79}
Linde, A.D. [1979], {\sl Rep. Progr. Phys.} {\bf 42}, 389.
\bibitem{Linde82}
Linde, A.D. [1982a], {\sl Phys. Lett.} {\bf B108}, 389.
\bibitem{Linde82b}
Linde, A.D. [1982b], {\sl Phys. Lett.} {\bf B116}, 335.
\bibitem{Linde83}
Linde, A.D. [1983a], {\sl Phys. Lett.} {\bf B129}, 177.
\bibitem{Linde83b}
Linde, A.D. [1983b], {\sl Nucl. Phys.} {\bf B216}, 421.
\bibitem{Linde85}
Linde, A.D. [1985], {\sl Phys. Lett.} {\bf B162}, 281.
\bibitem{Linde90}
Linde, A.D. [1990], {\sl Particle Physics and Inflationary
Cosmology} (Harwood, Chur, Switzerland).
\bibitem{hybrid}
Linde, A.D. [1994],  {\sl Phys. Rev.}  {\bf D49}, 748.
\bibitem{Lindley85}
Lindley, D. [1985], unpublished.
\bibitem{Luo94}
Luo, X. [1994], {\sl Ap. J. Lett.} {\bf 427}, L71.
\bibitem{Luo+Schramm93}
Luo, X. \& Schramm, D.N. [1993], {\sl Phys. Rev. Lett.} {\bf 71}, 1124.
\bibitem{PLI85}
Lucchin, F. \& Matarrese, S. [1985], {\sl Phys. Rev.} {\bf D32}, 1316.
\bibitem{Lucchin+Co92}
Lucchin, F., Matarrese, S. \& Mollerach, S. [1992],
{\sl Ap. J. Lett.} {\bf 401}, L49.
\bibitem{Lyth85}
Lyth, D.H. [1985], {\sl Phys. Rev.} {\bf D31}, 1792.
\bibitem{LythStewart92}
Lyth, D.H. \& Stewart, E.D. [1992], {\sl Phys. Lett.} {\bf B274}, 168.
\bibitem{Madsen-Ellis88}
Madsen, M.S. \& Ellis, G.F.R. [1988], {\sl M.N.R.A.S.} {\bf 234}, 67.
\bibitem{Magueijo92}
Magueijo, J. [1992], {\sl Phys. Rev.} {\bf D46}, 1368.
\bibitem{Magueijo95}
Magueijo, J. [1995], {\sl Phys. Rev.} {\bf D52}, 689.
\bibitem{MaMa93}
Malaney, R.A. \& Mathews, G.J. [1993], {\sl Phys. Rep.} {\bf 229}, 145.
\bibitem{Martinez+Co92}
Mart{\'\i}nez--Gonz\'alez, E., Sanz, J.L. \& Silk, J. [1992],
{\sl Phys. Rev.} {\bf D46}, 4193.
\bibitem{MatarreseEtAl89}
Matarrese, S., Ortolan, A. \& Lucchin, F. [1989],
{\sl Phys. Rev.} {\bf D40}, 290.
\bibitem{Mather+Co94}
Mather, J. C., \etal ~[1994], {\sl Ap. J.} {\bf 420}, 439.
\bibitem{Mermin79}
Mermin, M. [1979] , {\sl Rev. Mod. Phys.} {\bf 51}, 591.
\bibitem{Messiah}
Messiah, A. [1976], {\sl Quantum Mechanics}, Vol.2
(Amsterdam: North--Holland).
\bibitem{Mijic94}
Miji\'c, M. [1994], {\sl Phys. Rev.} {\bf D49}, 6434.
\bibitem{Moessner+Co94}
Moessner, R., Perivolaropoulos, L. \& Brandenberger, R. [1994],
{\sl Ap. J.} {\bf 425}, 365.
\bibitem{CMBReesSciama}
Mollerach, S., Gangui, A., Lucchin, F. \& Matarrese, S. [1995],
{\sl Ap. J.} {\bf 453}, 1.
\bibitem{Molle91}
Mollerach, S., Matarrese, S., Ortolan, A. \& Lucchin, F. [1991],
{\sl Phys. Rev.} {\bf D44}, 1670.
\bibitem{Blue}
Mollerach, S., Matarrese, S. \& Lucchin, F. [1994],
{\sl Phys. Rev.} {\bf D50}, 4835.
\bibitem{Munshi+Co95}
Munshi, D., Souradeep, T. \& Starobinsky, A.A. [1995],
preprint astro--ph/9501100.
\bibitem{Ng92}
Ng, Y.J. [1992], {\sl Int. J. Mod. Phys.} {\bf D1}, 145.
\bibitem{Nyborg+Froyland}
Nyborg, P. \& Fr\"oyland, J, (unpublished) Lecture Notes, Chapter IV:
{\sl Rotations and Angular Momentum}.
\bibitem{Notzold91}
Notzold, D. [1991], {\sl Phys. Rev.} {\bf D43}, R961.
\bibitem{Ozernoi-Chernin}
Ozernoi, L.M. \& Chernin, A.D. [1967], {\sl Astron. Zh.} {\bf 44}, 1131.
\bibitem{Ostriker-Cowie}
Ostriker, J.P. \& Cowie, L.L. [1980], {\sl Ap. J. Lett.} {\bf 243}, L127.
\bibitem{Padmanabhan1989}
Padmanabhan, T. [1989], {\sl Phys. Rev.} {\bf D39}, 2924.
\bibitem{Padmy93}
Padmanabhan, T. [1993], {\sl Structure Formation in the Universe} ~
(Cambridge: Cambridge University Press).
\bibitem{Parker70}
Parker, E.N. [1970], {\sl Ap. J.} {\bf 160}, 383.
\bibitem{Peebles80}
 Peebles, P.J.E. [1980], {\sl The Large Scale Structure of the Universe} ~
(Princeton: Princeton University Press).
\bibitem{PeeblesYu}
Peebles, P.J.E. \& Yu J.T. [1970], {\sl Ap. J.} {\bf 162}, 815.
\bibitem{PST}
Pen, U.--L., Spergel, D.N. \& Turok, N. [1994], {\sl Phys. Rev.}
{\bf D49}, 692.
\bibitem{PenziasWilson}
Penzias, A.A. \& Wilson, R.W. [1965],  {\sl Ap. J.} {\bf 142}, 419.
\bibitem{Peri93a}
Perivolaropoulos, L. [1993a], {\sl Phys. Lett.} {\bf B298}, 305.
\bibitem{Peri93b}
Perivolaropoulos, L. [1993b], {\sl Phys. Rev.} {\bf D48}, 1530.
\bibitem{Peri95}
Perivolaropoulos, L. [1994],
preprint CfA--3591, astro--ph/9402024.
\bibitem{Peri+Co90}
Perivolaropoulos, L., Brandenberger, R. \& Stebbins, A. [1990],
{\sl Phys. Rev.} {\bf D41}, 1764.
\bibitem{Peri+Vacha94}
Perivolaropoulos, L. \& Vachaspati, T. [1994],
{\sl Ap. J. Lett.}, {\bf 423}, L77.
\bibitem{Preskill79}
Preskill, J. [1979], {\sl Phys. Rev. Lett.} {\bf 43}, 1365.
\bibitem{Primack87}
Primack, J.R. [1987], in {\sl Proc. of the Intl. `Fermi' School
of Physics}, Course XCII, Varenna, 1984, Cabibbo, N. ed.
(Amsterdam: North--Holland).
\bibitem{Prudnikov}
Prudnikov, A.P., Brychkov, Yu.A. \& Marichev, O.I. [1986],
{\sl Integrals and Series}, Vol 2, Gordon and Breach Science Publishers.
\bibitem{Raffelt90}
Raffelt, G.G. [1990], {\sl Phys. Rep.} {\bf 198}, 1.
\bibitem{Rajaraman}
Rajaraman, R. [1982], {\sl Solitons and instantons}
(Amsterdam: North--Holland).
\bibitem{CMBexpmts}
Readhead, A.C.S. \&  Lawrence, C.R. [1992],
in {\sl Observations of the Isotropy of the Cosmic Microwave
Background Radiation}, Burbidge, G.,  Layzer, D. \& Sandage, A., eds.,
{\sl Ann. Rev. Astron. and Astrophys.} {\bf 30}, 653.
\bibitem{Rees+Sciama68}
Rees, M. \& Sciama, D.W. [1968], {\sl Nature} {\bf 217}, 511.
\bibitem{Rowan--Robinson}
Rowan--Robinson, M. [1985], {\sl The Cosmological Distance Ladder}
(Freeman: San Francisco).
\bibitem{Sachs+Wolfe67}
Sachs, R. \& Wolfe, A. [1967], {\sl Ap. J.} {\bf 147}, 73.
\bibitem{Salopek92}
Salopek, D.S. [1992],  {\sl Phys. Rev.} {\bf D45}, 1139.
\bibitem{SaloBondBardeen89}
Salopek, D.S., Bond, J.R. \& Bardeen, J.M. [1989],
{\sl Phys. Rev.} {\bf D40}, 1753.
\bibitem{Sasaki86}
Sasaki, M. [1986], {\sl Prog. Theor. Phys.} {\bf 76}, 1036.
\bibitem{Scaramella+Vittorio91}
Scaramella, R. \& Vittorio, N. [1991], {\sl Ap. J.} {\bf 375}, 439.
\bibitem{Scaramella+Vittorio93}
Scaramella, R. \& Vittorio, N. [1993], {\sl M.N.R.A.S.} {\bf 263}, L17.
\bibitem{Scherrer+Schaefer95}
Scherrer, R.J. \& Schaefer, R.K. [1994], OSU--TA--10/94 preprint.
\bibitem{SSW}
Scott, D., Silk, J. \&  White, M. [1995], {\sl Science} {\bf 268}, 829.
\bibitem{SampleVariance94}
Scott, D., Srednicki, M. \& White, M. [1994],
{\sl Ap. J. Lett.} {\bf 421}, L5.
\bibitem{Seljak+Bertschinger93}
Seljak, U. \& Bertschinger, E. [1993],
{\sl Ap. J. Lett.} {\bf 417}, L9.
\bibitem{ShtanovTraschenBrande94}
Shtanov, Y., Traschen, J. \& Brandenberger, R. [1995],
{\sl Phys. Rev.} {\bf D51}, 5438.
\bibitem{silk68}
Silk, J. [1968], {\sl Ap. J.} {\bf 151}, 459.
\bibitem{DoubleInflation3}
Silk, J. \& Turner, M.S. [1987],
{\sl Phys. Rev.} {\bf D35}, 419.
\bibitem{Smith+Vilenkin87}
Smith, G. \& Vilenkin, A. [1987], {\sl Phys. Rev.} {\bf D36}, 990.
\bibitem{SmithKaMa93}
Smith, M.S., Kawano, L.H. \& Malaney, R.A. [1993],
{\sl Ap. J. Suppl.} {\bf 85}, 219.
\bibitem{Smoot+Co91}
Smoot, G.F., \etal ~[1991], {\sl Ap. J. Lett.} {\bf 371}, L1.
\bibitem{Smoot+Co92}
Smoot, G.F., \etal ~[1992], {\sl Ap. J. Lett.} {\bf 396}, L1.
\bibitem{Smoot+Co94}
Smoot, G.F., Tenorio, L., Banday, A.J., Kogut, A., Wright, E.L.,
Hinshaw, G. \& Bennett, C.L. [1994], {\sl Ap. J.} {\bf 437}, 1.
\bibitem{Srednicki93}
Srednicki, M. [1993], {\sl Ap. J. Lett.} {\bf 416}, L1.
\bibitem{Starobinskii79}
Starobinskii, A.A. [1979], {\sl Pis'ma Zh. Eksp. Teor. Fiz.} {\bf 30}, 719.
                          [{\sl JETP Lett.} {\bf 30}, 682 (1970)].
\bibitem{Starobinskii80}
Starobinskii, A.A. [1980], {\sl Phys. Lett.} {\bf B91}, 99.
\bibitem{Starobinskii82}
Starobinskii, A.A. [1982], {\sl Phys. Lett.} {\bf B117}, 175.
\bibitem{DoubleInflation1}
Starobinskii, A.A. [1985], {\sl Pis'ma Zh. Eksp. Teor. Fiz.}
                {\bf 42}, 124. [{\sl JETP Lett.} {\bf 42}, 152 (1985)].
\bibitem{Starobinskii86}
Starobinskii, A.A. [1986], in {\sl Field Theory, Quantum Gravity and
Strings}, de Vega, H.J. \& Sanchez, N. eds., Lecture Notes in Physics
Vol. 246 (Berlin: Springer--Verlag).
\bibitem{Stebbins88}
Stebbins, A. [1988],  {\sl Ap. J.} {\bf 327}, 584.
\bibitem{Stebbins93}
Stebbins, A. [1993], in {\sl 16th Texas Symp. on Relativistic
Astrophysics}, Akerlof, C. \& Srednicki, M., eds.,
{\sl Ann. NY Acad. Sci.} {\bf 688}, 824.
\bibitem{Stebbins+Co87}
Stebbins, A., \etal ~[1987], {\sl Ap. J.} {\bf 322}, 1.
\bibitem{Steenrod}
Steenrod, N. [1951], {\sl Topology of Fibre Bundles}
(Princeton: Princeton University Press).
\bibitem{Stein}
Steinhardt, P.J. [1993], {\sl Class. Quantum Grav.} {\bf 10}, S33.
\bibitem{St}
Steinhardt, P.J. [1995], {\sl Cosmology at the Crossroads}, preprint.
\bibitem{ST84}
Steinhardt, P.J. \& Turner, M.S. [1984], {\sl Phys. Rev.} {\bf D29}, 2162.
\bibitem{StewartLyth93}
Stewart, E.D. \& Lyth, D.H. [1993], {\sl Phys. Lett.} {\bf B302}, 171.
\bibitem{SugiyamaEtAl93}
Sugiyama, N., Silk, J. \& Vittorio, N. [1993],
{\sl Ap. J. Lett.} {\bf 419}, L1.
\bibitem{SZ70}
Sunyaev, R.A. \& Zel'dovich, Ya.B. [1970],
{\sl Astrophys. Space Sci.} {\bf 7}, 1.
\bibitem{Torres+Co95}
Torres, S., Cay\'on, L., Mart{\'\i}nez--Gonz\'alez, E. \& Sanz, J.L.
[1995], preprint.
\bibitem{Traschen+Co86}
Traschen, J., Turok, N. \& Brandenberger, R. [1986],
{\sl Phys. Rev.} {\bf D34}, 919.
\bibitem{TraschenBrande90}
Traschen, J. \& Brandenberger, R. [1990],
{\sl Phys. Rev.} {\bf D42}, 2491.
\bibitem{Turner90}
Turner, M.S. [1990], {\sl Phys. Rep.} {\bf 197}, 67.
\bibitem{Turner93}
Turner, M.S. [1993], {\sl Phys. Rev.} {\bf D48}, 3502.;
                     {\sl ibid.} 5539.
\bibitem{Turner95}
Turner, M.S. [1995], Plenary Lecture in
{\sl 14$^{th}$ Int. Conf. on General Relativity and Gravitation},
6--12 August, Florence, Italy, to appear in the proceedings.
\bibitem{Tu}
Turok, N. [1989], {\sl Phys. Rev. Lett.} {\bf 63}, 2625.
\bibitem{TuSpe}
Turok, N. \& Spergel, D.N. [1990], {\sl Phys. Rev. Lett.} {\bf 64},
2736.
\bibitem{TuZa}
Turok, N. \& Zadrozny, J. [1990], {\sl Phys. Rev. Lett.} {\bf 65},
2331.
\bibitem{TuluieLaguna}
Tuluie, R. \& Laguna, P. [1995], {\sl Ap. J. Lett.} {\bf 445}, L73.
\bibitem{Vachaspati86}
Vachaspati, T. [1986], {\sl Phys. Rev. Lett.} {\bf 57}, 1655.
\bibitem{Vachaspati92a}
Vachaspati, T. [1992a], {\sl Phys. Lett.} {\bf B282}, 305.
\bibitem{Vachaspati92b}
Vachaspati, T. [1992b], {\sl Phys. Rev.} {\bf D45}, 3487.
\bibitem{Vachaspati+Vilenkin91}
Vachaspati, T. \&  Vilenkin, A. [1991],
{\sl Phys. Rev. Lett.} {\bf 67}, 1057.
\bibitem{Veeraraghavan+Stebbins90}
Veeraraghavan, S. \& Stebbins, A. [1990], {\sl Ap. J.} {\bf 365}, 37.
\bibitem{Vilenkin81}
Vilenkin, A. [1981], {\sl Phys. Rev.} {\bf D23}, 852.
\bibitem{Vilenkin83}
Vilenkin, A. [1983], {\sl Phys. Rev.} {\bf D27}, 2848.
\bibitem{Vilenkin-Ford82}
Vilenkin, A. \& Ford, L.H. [1982], {\sl Phys. Rev.} {\bf D25}, 1231.
\bibitem{Vilenkin+Shellard94}
Vilenkin, A. \& Shellard, E.P.S. [1994], {\sl Cosmic Strings and
other Topological Defects} ~(Cambridge: Cambridge University Press).
\bibitem{Vishniac87}
Vishniac, E.T. [1987], {\sl Ap. J.} {\bf 322}, 597.
\bibitem{Vollick92}
Vollick, D.N. [1992], {\sl Phys. Rev.} {\bf D45}, 1884.
\bibitem{WalkerEtAl91}
Walker, P.N., \etal ~[1991], {\sl Ap. J.} {\bf 376}, 51.
\bibitem{Weinberg72}
Weinberg, S. [1972], {\sl Gravitation and Cosmology} (Wiley, New York).
\bibitem{Weinberg74}
Weinberg, S. [1974], {\sl Phys. Rev.} {\bf D9}, 3357.
\bibitem{Weinberg89}
Weinberg, S. [1989], {\sl Rev. Mod. Phys.} {\bf 61}, 1.
\bibitem{White+Co94Review}
White, M., Scott, D. \& Silk, J. [1994],
{\sl Ann. Rev. Astron. and Astrophys.} {\bf 32}, 319.
\bibitem{Whitt}
Whitt, B. [1984], {\sl Phys. Lett.} {\bf B145}, 176.
\bibitem{Wolfram91}
Wolfram, S. [1991], {\sl Mathematica version 2.0},  Addison--Wesley.
\bibitem{Wright+Co92}
Wright, E.L., \etal ~[1992], {\sl Ap. J. Lett.} {\bf 396}, L13.
\bibitem{Yamamoto+Sasaki94}
Yamamoto, K. \& Sasaki, M. [1994], {\sl Ap. J. Lett.} {\bf 435}, L83.
\bibitem{Yi+Vishniac93}
Yi, I. \&  and Vishniac, E.T. [1993], {\sl Phys. Rev.} {\bf D48}, 950.
\bibitem{Yurke94}
Yurke, B. [1994], {\sl Physics World}, July issue, page 28.
\bibitem{Zeldovich70}
Zel'dovich, Ya.B. [1970], {\sl Astron. Astrophys.} {\bf 5}, 84.
\bibitem{ZS69}
Zel'dovich, Ya.B. \& Sunyaev, R.A. [1969],
{\sl Astrophys. Space Sci.} {\bf 4}, 301.
\bibitem{zurek82}
Zurek, W.H. [1982], {\sl Phys. Rev.} {\bf D26}, 1862.
\bibitem{zurek85}
Zurek, W.H. [1985], {\sl Nature} {\bf 317}, 505.

\end{thebibliography}
\end{document}